%% file: Large_Dev.TEX
\documentclass[12pt]{article}
\begin{document}
\vspace*{2cm} \noindent {\Large\bf Large Deviations in the
Spherical Model: The Rate Functions.} \vspace{\baselineskip}
\newline {\bf A.E.\ Patrick\footnote[1]{Laboratory of Theoretical
Physics, Joint Institute for Nuclear Research, Dubna 141980,
Russia e-mail: patrick@theor.jinr.ru}}
\begin{list}{}{\setlength{\rightmargin }{0 mm}
\setlength{\leftmargin }{2.5cm}} \item \rule{124mm}{0.3mm}
\linebreak {\footnotesize {\bf Abstract.} We study the spherical
model of a ferromagnet in $d$-dimensional cubes $\Omega_n$ of
volume $|\Omega_n|=n^d$ and investigate large deviations of the
magnetization of various domains $D_k\subset \Omega_n$. We focus
our attention on the low-temperature regime, $T<T_c$, and consider
domains $D_k$ of three types: $(d-1)$-dimensional layers of width
$k$, $(d-2)$-dimensional rods, and Kadanoff blocks. In the case of
layers the large-deviation probabilities decay exponentially with
$n^{d-2}$, and we obtain an explicit expression for the
corresponding rate function. When the layer width $k\ll n$, the
large-deviation probabilities are virtually independent of $k$. In
the case of rods the probabilities of large deviations exhibit
similar exponential decay, but this time it is distorted by $\log
n$ corrections. In the case of Kadanoff blocks of size $k$ the
large-deviation probabilities decay exponentially with $k^{d-2}$.}
\newline \rule[1ex]{12.4cm}{0.3mm}
\linebreak {\bf \sc key words:} {\footnotesize Critical phenomena;
equivalence classes; correlation length; Kadanoff blocks.}
\vspace{\baselineskip}
\end{list}

\section{Introduction.}
Most models studied within the theory of critical phenomena
describe the behaviour of various order parameters reasonably
well. Qualitatively predictions of mean-field and of short-range
finite-dimensional models are very similar in this respect. The
only significant discrepancy is in the values of critical
exponents. The latter predicted by 2D and/or 3D models differ
quite substantially from the set of mean-field critical exponents.

To find qualitative differences in the behaviour of various
classes of models in the entire low-temperature region (below the
critical point) one can look at the {\em probabilities of large
deviations\/}:
\[
\Pr\left[m_N\in[a,b]\right],
\]
where $m_N$ is the corresponding order parameter, $N$ is the
number of microscopic degrees of freedom in the system, and (if we
are indeed talking about large deviations) the interval $[a,b]$
does not contain the equilibrium value, $m^*$, of the order
parameter. Typically, as $N\to\infty$, large-deviation
probabilities exhibit the following asymptotic behaviour:
\[
-\frac{1}{N}\ln\Pr\left[m_N\in[a,b]\right]\sim
\min_{x\in[a,b]}R(x),
\]
where $R(x)\geq 0$ is the corresponding {\em rate function}.

Although, being the probabilities of (extremely) rare events, the
above quantities may seem unimportant they have significant
implications on, for instance, the existence/absence of such a
mysterious phenomenon as the hysteresis loop. Large-deviation
probabilities also describe the behaviour of thermodynamic systems
when it is necessary to take into account some conservation laws.
A frequently encountered example is the conservation of the number
of particles. Properties of typical configurations realising
appropriate large deviations in a system where the number of
particles is not conserved are closely related to equilibrium
properties of the same system with the conservation law.

For short-range finite-dimensional models the rate functions
$R(x)$ are always convex functions with a continuous first
derivative. On the contrary, in the low-temperature regions the
rate functions of mean-field models are not convex. Therefore, in
this respect the properties of mean-field models are unphysical.

For instance, the Hamiltonian of the Curie-Weiss model of a
ferromagnet --- the canonical mean-field model --- is given by
\[
H_N^{\rm cw}=-\frac{J}{2N}\left(\sum_{j=1}^N s_j\right)^2-h
\sum_{j=1}^N s_j,
\]
where $s_j=\pm 1$, for $j=1,2,\ldots,N$. A straightforward
calculation, see, e.g., \cite{e85}, yields the following
asymptotics
\[
-\frac{1}{N}\ln\Pr\left[\frac{1}{N}\sum_{j=1}^N s_j=\frac{k}{N}\right]\sim
F\left(\frac{k}{N}\right)-F\left(\frac{k^*}{N}\right),
\]
for $k\in\{-N,-N+2,\ldots,N-2,N\}$, where $\Pr[\cdot]$ is the
Gibbs distribution corresponding to the Hamiltonian $H_N^{\rm cw}$
and the inverse temperature $\beta$,
\[
F(x)=-\frac{\beta J x^2}{2}-\beta h
x+\frac{1-x}{2}\ln(1-x)+\frac{1+x}{2}\ln(1+x),
\]
and $k^*$ minimizes $F(\frac{k}{N})$. If $h=0$ and $\beta J>1$ the
corresponding rate function
\begin{equation}
R_{\rm cw}(x)=F(x)-\min_{y\in[-1,1]} F(y) \label{cwrf}
\end{equation}
is not convex, see Fig.\ 1. When $h\neq 0$ the function $R_{\rm
cw}(x)$ has exactly one global minimum. If $h\ (\neq 0)$ is
sufficiently small $R_{\rm cw}(x)$ also has a local minimum, which
is often interpreted as a quasi-stationary state giving rise to
the hysteresis phenomenon.

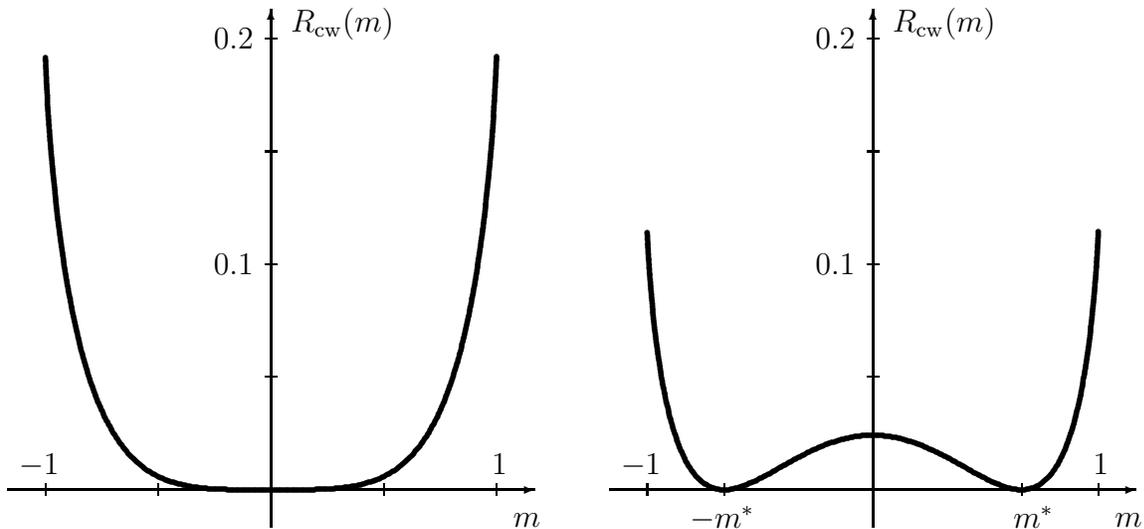
\begin{figure}
\setlength{\unitlength}{1mm}
\begin{picture}(150,70)(0,-1)
\put(0,5){\vector(1,0){70}} \put(35,0){\vector(0,1){69}}
\put(80,5){\vector(1,0){70}} \put(115,0){\vector(0,1){69}}
\put(69,1){\makebox(0,0){$m$}} \put(149,1){\makebox(0,0){$m$}}
\put(44.5,67){\makebox(0,0){$R_{\rm cw}(m)$}}
\put(124.5,67){\makebox(0,0){$R_{\rm cw}(m)$}}
\multiput(5,5)(15,0){5}{\makebox(0,0){$\rule{0.15mm}{1.5mm}$}}
\multiput(85,5)(30,0){3}{\makebox(0,0){$\rule{0.15mm}{1.5mm}$}}
\put(95.3,5){\makebox(0,0){$\rule{0.15mm}{1.5mm}$}}
\put(95.3,5){\makebox(0,0){$\rule{0.15mm}{1.5mm}$}}
\put(134.84,5){\makebox(0,0){$\rule{0.15mm}{1.5mm}$}}
\put(90.7,0){$-m^*$} \put(133.7,0){$m^*$}
\put(1.5,7){$-1$}\put(64.3,7){$1$}
\put(81.5,7){$-1$}\put(144.3,7){$1$}
\multiput(35,5)(0,15){5}{\makebox(0,0){$\rule{1.5mm}{0.15mm}$}}
\put(30,65){\makebox(0,0){$0.2$}}
\put(30,35){\makebox(0,0){$0.1$}}
\multiput(115,5)(0,15){5}{\makebox(0,0){$\rule{1.5mm}{0.15mm}$}}
\put(110,65){\makebox(0,0){$0.2$}}
\put(110,35){\makebox(0,0){$0.1$}}
\input cw_rate_f
\input cw_rate_f1
\end{picture}
\caption{The rate functions for the magnetization of the
Curie-Weiss model at $\beta=\beta_{\rm c}=1/J$ (left) and at
$\beta =1.2/J$ (right).}
\end{figure}

It is possible to consider a version of the Curie-Weiss model with
continuous random variables $s_j$. In this case the corresponding
rate function will differ from $R_{\rm cw}(x)$ given by Eq.\
(\ref{cwrf}), but certain qualitative features of $R(x)$ (for
instance, the non-convex shape in the low-temperature region) will
remain unchanged. Thus, although the rate functions $R(x)$ are not
universal quantities, some of their properties are identical
within large classes of models.

Short-range finite-dimensional lattice models can be divided into
two broad classes: continuous and discrete. A well-known exactly
solvable representative of the former class is the spherical model
of a ferromagnet defined in cubical domains $\Omega_n$ of a square
lattice $Z^d$, see \cite{bk} and the definition in Section 2. A
cube $\Omega_n$ contains $n^d\equiv N$ lattice sites. The order
parameter here is the magnetization $m_N$ and the rate function
describing the large-deviation probabilities for $m_N$ is given by
$R_{\rm sph}(x)=G(x)-G(0)$, where $G(x)=\max_{z\geq d}g(x,z)$ and
\[
g(x,z)=-\beta
J(1-x^2)(z-d)+\frac{1}{2(2\pi)^d}\int_{0}^{2\pi}\!\!\!
\ldots\int_{0}^{2\pi}\!\!\!d\omega_1\ldots
d\omega_d\,\ln\left(z-\sum_{\nu=1}^d\cos\omega_\nu\right),
\]
see \cite{p94a}. The function $R_{\rm sph}(x)$ is always convex,
see Fig.\ 2. In the low-temperature region $\beta>\beta_c$ this
function vanishes on the entire interval $x\in[-m^*,m^*]$, where
$m^*\equiv\sqrt{1-\beta_c/\beta}$ is the spontaneous magnetization
obtained by switching off a homogeneous external magnetic field
$h\downarrow0$.
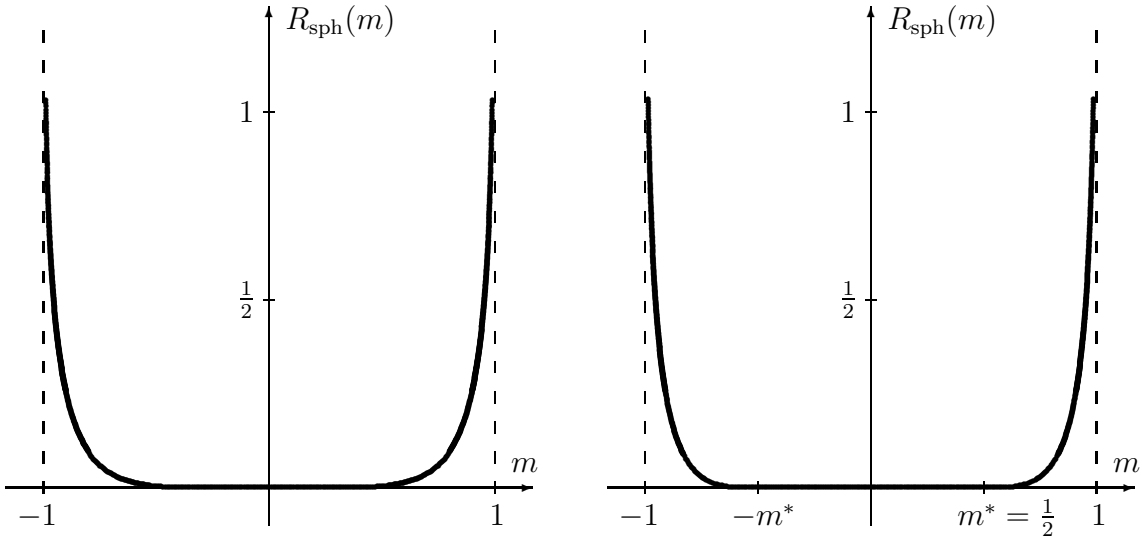
\begin{figure}
\setlength{\unitlength}{1mm}
\begin{picture}(150,70)(0,-1)
\put(0,5){\vector(1,0){70}} \put(35,0){\vector(0,1){69}}
\put(80,5){\vector(1,0){70}} \put(115,0){\vector(0,1){69}}
\put(69,8){\makebox(0,0){$m$}} \put(149,8){\makebox(0,0){$m$}}
\put(44.5,67){\makebox(0,0){$R_{\rm sph}(m)$}}
\put(124.5,67){\makebox(0,0){$R_{\rm sph}(m)$}}
\multiput(5,5)(0,4){16}{\makebox(0,0){$\rule{0.15mm}{1.5mm}$}}
\multiput(65,5)(0,4){16}{\makebox(0,0){$\rule{0.15mm}{1.5mm}$}}
\multiput(85,5)(0,4){16}{\makebox(0,0){$\rule{0.15mm}{1.5mm}$}}
\multiput(145,5)(0,4){16}{\makebox(0,0){$\rule{0.15mm}{1.5mm}$}}
\multiput(85,5)(15,0){5}{\makebox(0,0){$\rule{0.15mm}{1.5mm}$}}
\put(1.5,0){$-1$}\put(64.3,0){$1$}
\put(81.5,0){$-1$}\put(144.3,0){$1$} \put(96.5,0){$-m^*$}
\put(126.5,0){$m^*=\frac{1}{2}$}
\multiput(35,5)(0,25){3}{\makebox(0,0){$\rule{1.5mm}{0.15mm}$}}
\put(32,55){\makebox(0,0){$1$}}
\put(32,30){\makebox(0,0){$\frac{1}{2}$}}
\multiput(115,5)(0,25){3}{\makebox(0,0){$\rule{1.5mm}{0.15mm}$}}
\put(112,55){\makebox(0,0){$1$}}
\put(112,30){\makebox(0,0){$\frac{1}{2}$}}
\input sph_rate_f
\input sph_rate_f1
\end{picture}
\caption{The rate functions for the magnetization of the spherical
model at $\beta=\beta_c$ (left) and at $\beta=\frac{4}{3}\beta_c$
(right).}
\end{figure}
Switching on an arbitrarily weak magnetic field immediately leads
to a strictly convex rate function with a positive second
derivative and a unique minimum. Thus, when $h\neq0$ the spherical
model on a finite-dimensional lattice does not have
quasi-stationary states, and hence it does not exhibit any
hysteresis phenomena associated with such states.

Since the rate function $R_{\rm sph}(x)$ is equal to zero on the
entire interval $[-m^*,m^*]$ it is still necessary to find the
first non-vanishing term in the large-$n$ asymptotic expansion of
$\ln\Pr\left[m_N\in[a,b]\right]$ when
$[a,b]\cap[-m^*,m^*]\neq\emptyset$ . It turns out that in the case
of the spherical model with periodic boundary conditions this term
is given by
\[
\ln\Pr\left[m_N\in[a,b]\right]\sim -n^{d-2}\min_{x\in[a,b]}R_{\rm
sph}^{(2)}(x),
\]
where
\[
R_{\rm sph}^{(2)}(x)=2\pi^2 \beta
J\left(1-\frac{\beta_c}{\beta}-x^2\right),\quad\mbox{for }
x\in[-m^*,m^*],
\]
see Fig.\ 3. Note that, unlike $R(x)$ --- the rate function  of
the order $n^d$, lower-order rate functions (the order $n^{d-2}$
here and the order $n^{d-1}$ below) do not have to be convex
functions with a continuous first derivative.

Among discrete models the 2D Ising model on a square $n\times n$
lattice is the most frequently studied example. The order parameter here is the
magnetization $m_N$, $N\equiv n^2$. An explicit expression for the corresponding
rate function, $R_{\rm Ising}(m)$, is not known. However, it is
known that $R_{\rm Ising}(m)$ is always convex and, thus, it has the same shape
as the analogous rate function within
the spherical model shown in Fig.\ 2. In particular, for
$\beta>\beta_c$ the function $R_{\rm Ising}(m)$ vanishes on the
entire interval $[-m^*,m^*]$, where $m^*\equiv[1-\sinh^{-4}(2\beta
J)]^{1/8}$ is the spontaneous magnetization. If
$[a,b]\cap[-m^*,m^*]\neq\emptyset$, then the leading asymptotics
of the large-deviation probabilities is given by
\[
\ln\Pr\left[m_N\in[a,b]\right]\sim -n\min_{x\in[a,b]}R_{\rm
Ising}^{(1)}(x),
\]
see \cite{dks92,pf91}. It is quite remarkable that despite $R_{\rm
Ising}(x)$ is still unknown, in many cases one can find an
explicit expression for the rate function $R_{\rm
Ising}^{(1)}(x)$. For instance, it was shown in the paper
\cite{shl89} that in the case of the 2D Ising model with periodic
boundary conditions this rate function (in the notations adopted
to ours) is given by
\[
R_{\rm Ising}^{(1)}(x)=\beta w\times\left\{
\begin{array}{ll}
\sqrt{m^*-|x|},&\mbox{ for } m_0<|x|\leq m^*,\\
\sqrt{m^*-m_0},&\mbox{ for } |x|\leq m_0,
\end{array}
\right.
\]
see Fig.\ 3, where $w$ is the surface tension associated with the
droplet boundary, and $m_0$ is the magnetization value at which
the droplet shape changes from a rounded square to a ring taking
advantage of the periodic boundary conditions.

\begin{figure}
\setlength{\unitlength}{1mm}
\begin{picture}(150,70)(0,-1)
\put(0,5){\vector(1,0){70}} \put(35,0){\vector(0,1){69}}
\put(80,5){\vector(1,0){70}} \put(115,0){\vector(0,1){69}}
\put(69,8){\makebox(0,0){$m$}} \put(149,8){\makebox(0,0){$m$}}
\put(125.5,67){\makebox(0,0){$R_{\rm Ising}^{(1)}(m)$}}
\put(44.5,67){\makebox(0,0){$R_{\rm sph}^{(2)}(m)$}}
\put(5,5){\makebox(0,0){$\rule{0.15mm}{2mm}$}}
\put(65,5){\makebox(0,0){$\rule{0.15mm}{2mm}$}}
\multiput(100,5)(0,4){13}{\makebox(0,0){$\rule{0.15mm}{1.5mm}$}}
\multiput(130,5)(0,4){13}{\makebox(0,0){$\rule{0.15mm}{1.5mm}$}}
\multiput(85,5)(15,0){5}{\makebox(0,0){$\rule{0.15mm}{2mm}$}}
\put(1.5,0){$-m^*$}\put(64.3,0){$m^*$}
\put(81.5,0){$-m^*$}\put(144.3,0){$m^*$} \put(96.5,0){$-m_0$}
\put(128,0){$m_0$}
\put(15,63){\makebox(0,0){$2\pi^2\beta J (m^*)^2$}}
\put(24,61){\vector(2,-1){10}}
\put(91,56){\makebox(0,0){$\beta w \sqrt{m^*-m_0}$}}
\put(104,53.5){\vector(2,-1){10}}
\input sph_rate_f2
\input is_rate_f1
\end{picture}
\caption{The rate functions $R_{\rm sph}^{(2)}(m)$ and $R_{\rm
Ising}^{(1)}(m)$ for the magnetization of the 3D spherical and 2D
Ising models in the low-temperature region $\beta>\beta_c$.}
\end{figure}
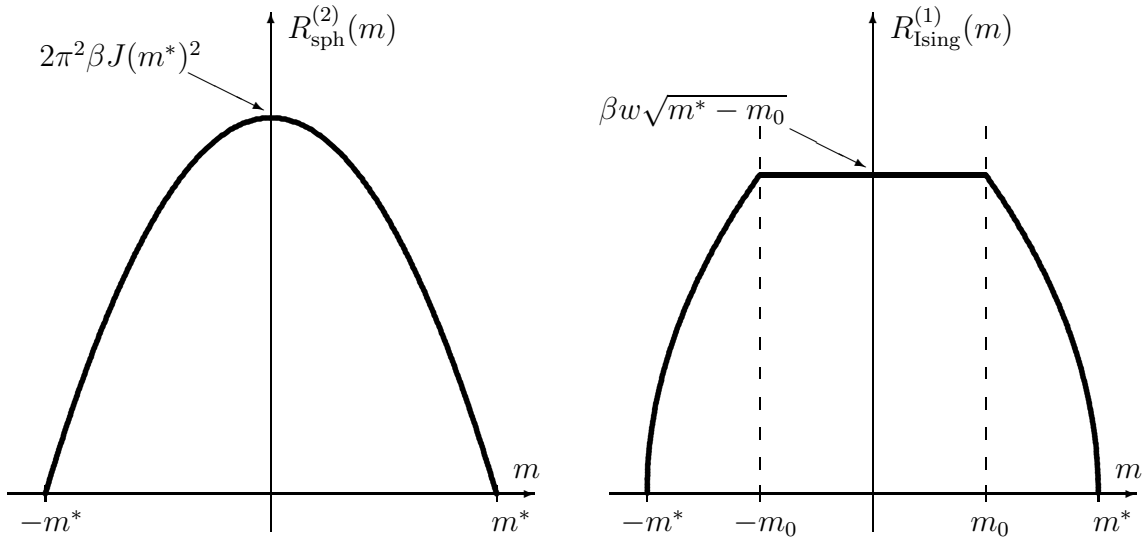

The obvious qualitative differences in the shapes of $R_{\rm sph}^{(2)}(m)$
and $R_{\rm Ising}^{(1)}(m)$, see Fig.\ 3, are the slopes at
$m=\pm m^*$ and the horizontal segment in $R_{\rm Ising}^{(1)}(m)$ for
$|m|\leq m_0$. Apparently, the absense of a threshold value $m_0$ in
$R_{\rm sph}^{(2)}(m)$ is due to a special structure of typical
configuration in the spherical model with a fixed value of the
magnetization. These configurations always
take advantage of the periodic boundary conditions, and hence they are
of the ring shape (as opposed to a round droplet shape). The absence of the
horizontal segment $R_{\rm sph}^{(2)}(m)$ is due to the diffuse nature of the
boundary between the \lq\lq$+$" and \lq\lq$-$" phases. Increasing the amount of
\lq\lq$+$" or \lq\lq$-$" phase inside the ring makes the interphase boundary steeper, and,
hence, makes the set of corresponding configurations less probable.

Another well-known model, which is often attributed to the
mean-field class, is the Ising model on a Cayley tree. The
explicit expression for the rate function $R_{\rm Cayley}(x)$ of
the magnetization is not known, nevertheless, it is possible to
compute the values of this function numerically. It was shown in
the paper \cite{brssz} that $R_{\rm Cayley}(x)$ is always convex
and vanishes at exactly one point where it has strictly positive
second derivative. Without any doubt the same conclusion is valid
for many other models on Cayley trees. If so, then one should
conclude that ferromagnets on trees do not exhibit genuine
critical behaviour, where single zero of the rate function either
stretches to an interval, or splits into several zeroes for
sufficiently low temperatures. Rather, models on trees are some
kind of bundles of (non-critical) one-dimensional chains.
Distributions of order parameters in tree models always
concentrate at a single point (the rate-function zero) even for
low temperatures and in the absence of symmetry breaking
perturbations. Non-zero values of order parameters appear only as
a result of explicit breaking of symmetry by a field applied to an
abnormally large number of boundary sites.

Thus, thermodynamic systems can be classified according to the
qualitative behaviour of their large-deviation probabilities.
One cap put forward a hypothesis that several large classes can be
described as follows. Rate functions,
$R(x)$, of macroscopic observables (order parameters) in {\em
non-critical\/} systems are always convex and vanish (reach zero
value) at exactly one point. For sufficiently low temperatures
(strong correlations) the rate functions of (long-range) {\em
mean-field} models are not convex and vanish at a finite number of
points. Rate functions of {\em discrete finite-dimensional\/}
models exhibiting critical behaviour are convex and (when the
temperature is low enough) they vanish on a certain interval
$[-m^*,m^*]$. If $[a,b]\cap[-m^*,m^*]\neq\emptyset$, then
\[
\ln\Pr\left[m_N\in[a,b]\right]\sim
-n^{d-1}\min_{x\in[a,b]}R^{(1)}(x).
\]
Rate functions $R(x)$ of {\em continuous finite-dimensional\/}
models are similar to those of discrete models. However, if
$[a,b]\cap[-m^*,m^*]\neq\emptyset$, then
\[
\ln\Pr\left[m_N\in[a,b]\right]\sim
-n^{d-2}\min_{x\in[a,b]}R^{(2)}(x).
\]
The properties of typical configurations realising
large-deviations are even more intriguing than the probabilities
of these events. Hopefully those will be outlined in a consecutive
publication.

Thus, large-deviation probabilities in short-range finite-dimensional
systems are sensitive to the dimensionality of the lattice. In fact
there also exists a dimension dependence of another kind. If we
look at the magnetization of a subdomain $D_k$ of the entire system
$\Omega_n$, then, as it turns out, the behaviour of the
corresponding large-deviation probabilities is very sensitive to
the shape or/and dimensionality of $D_k$. It is this feature
that is the main focuse of the present work.

The rest of the paper is organized as follows. Section 2 contains
the definition of the spherical model and statements of the main
results. Large deviations of the magnetization of various domains
are studied in Section 3. Subsections 3.1 and 3.2 are devoted to
the large-deviation probabilities for the magnetization of
$(d-1)$-dimensional layers and $(d-2)$-dimensional rods,
respectively. Subsection 3.3 contains derivation of analogous
properties for the magnetization of Kadanoff blocks. The results
of the paper are discussed in Section 4.

\section{Definition of the Model and Main Results.}
Let ${\bf Z}^d$ be a $d$-dimensional square lattice with nodes
$j\equiv(j_1,j_2,\ldots,j_d)$, where $j_\nu\in {\bf Z}$ for
$\nu=1,2,\ldots,d$. Consider the sequence of cubes
\[
\Omega_n=\left\{j\in{\bf Z}^d:\,1\leq
j_\nu\leq n,\ \nu=1,2,\ldots,d\right\}
\]
where a random variable $\sigma_j\in{\bf R}^1$ (spin) is attached to each
node $j\in\Omega_n$. The mutual dependence (interaction) of these spins is described by the Hamiltonian
\begin{equation}
H_n(\sigma)=-J\sum_{\langle i;j\rangle}\sigma_i\sigma_j,\qquad
J>0; \label{h}
\end{equation}
where the summation runs over all pairs of nearest neighbours
\[
\langle i;j\rangle\in\Omega_n\times\Omega_n,
\quad\sum_{\nu=1}^{d}|i_\nu-j_\nu|=1.
\]
We assume that the periodic boundary conditions are imposed (for
instance, the nodes $(1,j_2,\ldots,j_d)$ and $(n,j_2,\ldots,j_d)$
are also nearest neighbours).

The joint probability distribution of the random variables
$\{\sigma_j:j\in\Omega_n\}$ is specified by the density
\begin{equation}
p\left(\left\{\sigma_j:j\in\Omega_n\right\}\right)=\frac{1}{\Theta_n}
\exp\left[-\beta H_n(\sigma)\right], \label{gd}
\end{equation}
with respect to the ``spherical'' {\sl a priori} measure
\begin{equation}
\mu_n(d\sigma)=\delta\left(\sum_{j\in\Omega_n}\sigma_j^2-N\right)
\prod_{j\in\Omega_n}d\sigma_j, \label{am}
\end{equation}
where $\Theta_n$ is the partition function (normalizing factor),
$N\equiv n^d=|\Omega_n|$, $\delta(\cdot)$ is the Dirac delta
function, and $\prod_{j\in\Omega_n}d\sigma_j$ is the Lebesgue
measure on $\left({\bf R}^N,{\cal B}({\bf R}^N)\right)$. Equations
(\ref{h}), (\ref{gd}), and (\ref{am}) define the spherical model
of a ferromagnet \cite{bk}.

Consider the sequence of Kadanoff blocks
\[
B_k\equiv\{j\in\Omega_{n}:1\leq j_\nu\leq k,\ \nu=1,2,\ldots,d\},
\]
where $k\equiv k(n)$ is a non-decreasing function of $n$ (possibly
a constant). The normalized total spin (magnetization) of a block
$B_k$ is given by
\[
S_B=|B_k|^{-1}\sum_{j\in B_k}\sigma_j.
\]
In the present paper we study the distribution densities
\begin{equation}
\phi_{n,B}(m)=\frac{1}{\Theta_n}\mathop{\int\!\ldots\!\int}_{\!\!\!\!-\infty}
^{\,\,\,\,+\infty}\delta(S_B-m)\exp\left[-\beta
H_n(\sigma)\right] \mu_n(d\sigma) \label{bdd}
\end{equation}
of the random variables $S_B$. In particular, we would like to
investigate the asymptotic behaviour of $\phi_{n,B}(m)$ as $n$
and $|B_k|$ tend to infinity.

To carry out the integration in Eq.\ (\ref{bdd}) we need to know
the spectral properties of the symmetric matrix $\widehat{C}_{n}$
associated with the Hamiltonian (\ref{h})
\[
H_n(\sigma)=-J\sum_{i,j\in\Omega_n}C_{ij}^{(n)}\sigma_i\sigma_j.
\]
For the periodic boundary conditions the eigenvalues and
eigenvectors of $\widehat{C}_{n}$ are given by
\begin{equation}
\lambda_{j_1,\ldots,j_d}=\sum_{\nu=1}^d\cos\frac{2\pi(j_\nu-1)}{n},
\qquad \mbox{\boldmath
$V$}_{j_1,\ldots,j_d}=\left\{\prod_{\nu=1}^d v_{j_\nu,\ell_\nu
}\right\}_{\ell\in\Omega_n}, \label{evd}
\end{equation}
where
\begin{equation}
v_{j_\nu,\ell_\nu
}=\frac{1}{\sqrt{n}}\left[\cos\frac{2\pi(j_\nu-1)
(\ell_\nu-1)}{n}+\sin\frac{2\pi(j_\nu-1) (\ell_\nu-1)}{n}\right].
\label{evp}
\end{equation}
The components $v_{j_\nu,\ell_\nu }$ can be written down in the
following compact form
\[
v_{j_\nu,\ell_\nu
}=\sqrt{\frac{2}{n}}\cos\left[\frac{2\pi(j_\nu-1)
(\ell_\nu-1)}{n}-\frac{\pi}{4}\right],
\]
which we will use throughout the paper, although in all
calculations we actually worked with Eq. (\ref{evp}).

No paper on the spherical model on a square lattice
$\mbox{\boldmath $Z$}^d$ can be written without a reference to the
Watson function
\begin{equation}
W_d(z)\equiv\frac{1}{(2\pi)^d}\int_{-\pi}^{+\pi}\!\!\!
\ldots\int_{-\pi}^{+\pi}\frac{d\omega_1\ldots
d\omega_d}{z-\sum_{\nu=1}^d\cos\omega_\nu}.
\label{wfd}
\end{equation}
The integral of this function also appears quite frequently in
various expressions and certainly deserves a special symbol
\begin{equation}
L_d(z)\equiv\frac{1}{(2\pi)^d}\int_{-\pi}^{+\pi}\!\!\!
\ldots\int_{-\pi}^{+\pi}d\omega_1\ldots
d\omega_d\,\ln\left(z-\sum_{\nu=1}^d\cos\omega_\nu\right).
\label{iwf}
\end{equation}

The main results of the present paper can be stated as follows.
Consider the Kadanoff blocks $B_k$ and also the following
subdomains of the cubes $\Omega_n$:
\begin{eqnarray}
&&L_k=\{j\in\Omega_n:1\leq j_1\leq k\}, \nonumber\\
&&R_k=\{j\in\Omega_n:1\leq j_\nu\leq k,\ \nu=1,2\}, \nonumber
\end{eqnarray}
which we call layers and rods. Let $1\ll k \ll n$, then the
large-$n$ asymptotics of the large-deviation probabilities for the
magnetization of layers, rods, and blocks are given by
\[
\log\Pr\left[m_L\in[a,b]\right]\sim
-n^{d-2}\min_{x\in[a,b]}R_L^{(2)}(x),
\]
\[
\log\Pr\left[m_R\in[a,b]\right]\sim
-\frac{n^{d-2}}{\log n}\min_{x\in[a,b]}R_R^{(2)}(x),
\]
\[
\log\Pr\left[m_B\in[a,b]\right]\sim
-k^{d-2}\min_{x\in[a,b]}R_B^{(2)}(x),
\]
where explicit expressions for the rate functions are given by
Eqs.\ (\ref{rfl}), (\ref{rfr}), and (\ref{rfb}), respectively.

\section{Magnetization of Kadanoff blocks.}
To find the distribution density of the magnetization of a
Kadanoff block $B$, see Eq.\ (\ref{bdd}), we have to calculate the
following integral
\begin{equation}
\Theta_{n,B}=\mathop{\int\!\ldots\!\int}_{\!\!\!\!-\infty}
^{\,\,\,\,+\infty}\prod_{j\in\Omega_n}\!d\sigma_j\:\delta\left(
\sum_{j\in\Omega_n}\sigma_j^2-N\right)\delta\left(\frac{1}{|B|}\sum_{j\in
B} \sigma_j-m\right)e^{-\beta H_n(\sigma)}. \label{tnw}
\end{equation}
The integration over $\sigma_j$, $j\in\Omega_n$ is carried out
using the standard technique after Berlin and Kac, see \cite{bk}.
Namely, first a new set of integration variables $y_l$,
$l\in\Omega_n$ is introduced via
$\sigma_j=\sum_{l\in\Omega_n}V_{j,l}y_l$, where
$\{V_{j,l}\}_{j\in\Omega_n}\equiv\mbox{\boldmath $V$}_l$,
$l\in\Omega_n$ are the eigenvectors of the matrix
$\widehat{C}_{n}$ associated with the Hamiltonian $H_n(\sigma)$,
see Eqs.\ (\ref{evd}) and (\ref{evp}). Next, the delta functions
are replaced by their integral representations
\[
\delta(a)=\frac{1}{2\pi}\int_{-\infty}^{+\infty}e^{i\tau a}d\tau,
\]
the integration order is exchanged, and the integration over the
variables $y_l$, $l\in\Omega_n$ is performed. Note that the
integration order can be switched only if the quadratic form in
the argument of the exponential function is negatively defined.
This can be achieved by a shift of the integration contour for
$\tau$. One obtains then
\begin{equation}
\Theta_{n,B}=\frac{1}{4\pi^2i}\int\limits_{-i\infty+\tau_0}
^{i\infty+\tau_0}\!d\tau\, e^{\tau
N}\int\limits_{-\infty}^{\infty}
\!dx\,e^{-ixm}\prod_{j\in\Omega_n}\sqrt{\frac{\pi}{\tau-\beta
J\lambda_j}}\exp\left[-\frac{x^2\gamma_j^2} {4(\tau-\beta
J\lambda_j)}\right], \label{tnm1}
\end{equation}
where $\gamma_j\equiv|B|^{-1}\sum_{l\in B}V_{j,l}$, and $\tau_0$
is the shift of integration contour mentioned above. For the
periodic boundary conditions a straightforward calculation yields
the following expression for the coefficients $\gamma_j$,
$j\equiv(j_1,j_2,\ldots,j_d)$:
\begin{equation}
\gamma_j=|B|^{-1}(2/n)^{d/2} \prod_{\nu=1}^{d} \frac{\sin[\pi
k(j_\nu-1)/n]} {\sin[\pi(j_\nu-1)/n]} \cos\left[\frac{\pi
(k-1)(j_\nu-1)}{n}-\frac{\pi}{4}\right]. \label{gegp}
\end{equation}

On integrating over the variable $x$ in Eq.\ (\ref{tnm1}) and on
introducing a new integration
variable $z$ via $\tau=\beta Jz$ one arrives at
\begin{equation}
\Theta_{n,B}=\frac{\beta J}{2\pi i}\left(\frac{\pi}{\beta J}
\right)^{(N-1)/2}\int\limits_{-i\infty+z_0}^{i\infty+z_0}\!\frac{dz}
{\sqrt{\Sigma(z)(z-\lambda_{(1,\ldots,1)})}}\exp\left[N\beta
J\Phi_n(z,m)\right], \label{tm2}
\end{equation}
where the following notations have been introduced
\begin{equation}
\Sigma(z)=\sum_{j\in\Omega_n}\frac{\gamma_j^2}{z-\lambda_j},
\label{sgm}
\end{equation}
\begin{equation}
\Phi_n(z,m)=z-\frac{1}{2\beta
JN}\sum_{j\in\Omega_n\setminus(1,\ldots,1)}\log(z-\lambda_j)-
\frac{1}{N}\frac{m^2}{\Sigma(z)}. \label{fzm}
\end{equation}

The large-$n$ asymptotic expansion for the remaining integral over
$z$ can be derived using the saddle-point method. This is quite
straightforward when $\beta<\beta_c$, where $\beta_c=W_d(d)/(2J)$
is the inverse critical temperature of the spherical model, see
\cite{bk}, and $W_d(d)$ is the Watson function (\ref{wfd}) at
$z=d$. However, if $|B|=o(n^d)$, some extra efforts are required
in the low-temperature region $\beta>\beta_c$. Difficulties arise
because the sequence of saddle points $\{z_n^*\}$ (that is, the
sequence of minimum points of the function $\Phi_n(z,m)$ on
$(d,\infty)$) converges to $z=d$, as $n\to\infty$. That is, the
sequence of saddle points $z_n^*$ approaches the maximal
singularity, $\xi_n$, of $\Phi_n(z,m)$: $z_n^*-\xi_n\downarrow0$,
as $n\to\infty$. Therefore, the standard version of the
saddle-point method can not be applied, and the integral
(\ref{tm2}) needs a special investigation in the low-temperature
region. What we actually have to do is to introduce a new
integration variable $\zeta$ via $z=d+n^{-\rho}\zeta$, where the
exponent $\rho$ must be chosen in such a way that the sequence of
function $\widetilde\Phi_n(\zeta)\equiv\Phi_n(d+n^{-\rho}\zeta,m)$
has a ``conventional" saddle-point landscape in the limit
$n\to\infty$.

The main goal of the present paper is investigation of large
deviation probabilities for the magnetization of cubic Kadanoff
blocks $B$. However, it turns out that the large-deviation
probabilities for the magnetization of layers
\[
L_k=\{j\in\Omega_n:1\leq j_1\leq k\}
\]
and rods
\[
R_k=\{j\in\Omega_n:1\leq j_\nu\leq k,\ \nu=1,2\}
\]
differ qualitatively from those for magnetization of cubic blocks
$B$. Therefore, we begin from the simplest (as it happens to be)
case of layers $L$, after that we consider large deviations for
rods $R$, and finally we turn our attention to the case of
Kadanoff blocks $B$.

\subsection{Large deviations of the magnetization of layers $L$.}
If the boundary conditions are periodic in all dimensions and
\begin{equation}
L_k\equiv\{j\in\Omega_n:1\leq j_{1}\leq k\}, \label{wndd}
\end{equation}
then a straightforward calculation yields the following expression
for the coefficients $\gamma_j\equiv|L_k|^{-1}\sum_{l\in
L_k}V_{j,l}$ (see Eq.\ (\ref{tnm1})):
\begin{equation}
\gamma_j=\frac{\sqrt{2}\,n^{\frac{d}{2}-1}}{|L_k|}\,\frac{\sin[\pi
k(j_1-1)/n]} {\sin[\pi(j_1-1)/n]} \cos\left[\frac{\pi
(k-1)(j_1-1)}{n}-\frac{\pi}{4}\right]
\prod_{\nu=2}^{d}\delta_{1,j_{\nu}}.
\end{equation}
Hence, in the case of layers $L_k$ the multiple sum in the
expression for $\Sigma(z)$ (see Eq.\ (\ref{sgm})) reduces to a
single one and can be calculated exactly using the method
described in \cite{p94}. One obtains
\begin{equation}
|L_k|\Sigma(z)=\frac{1}{z-d}\left[1-\frac{2}{k
(x_2-x_1)}\frac{(1-x_1^k)(x_2^n-x_2^k)}{(x_2^n-1)}\right],
\label{sgma}
\end{equation}
where $x_{1,2}=1+z-d\mp\sqrt{(z-d)(2+z-d)}$.

Now we are going to locate the maximal singularity of the function
$\Phi_n(z,m)$ given by Eq.\ (\ref{fzm}). That will give us a
hint how one should rescale the integration variable $z$ in order
to preserve the saddle-point profile of $\Phi_n(z,m)$ in the
limit $n\to\infty$. Let $k=\alpha n^{\gamma}$, where
$\gamma\in(0;1)$ and $\alpha>0$. It is obvious from Eq.\
(\ref{sgm}) that singularities of the function $\Sigma(z)$ ---
simple poles --- are at the points
\[
z\in\{\lambda_{(j,1,\ldots,1)}:\gamma_{(j,1, \ldots,1)}\neq0,\
j=1,\ldots,n\}.
\]
To locate zeroes of this function we rescale the variable $z$
according to $z=d+\zeta n^{-\rho}$, where $\zeta$ is a new
independent variable. The rescaling yields as
\vspace{2mm}$n\to\infty$:\newline $|L_k|\Sigma(d+\zeta n^{-\rho})=
$
\begin{equation}
\left\{
\begin{array}{ll}
n^{\rho}/\zeta+O(n^{3/2\rho}/k),\ &{\rm for}\ 0\leq\rho<2
\gamma,\ \vspace{2mm}\zeta>0\\
{\displaystyle
\frac{n^{2\gamma}}{\zeta}\left[1-\frac{1-\exp(-\alpha\sqrt{2\zeta}
)}{\alpha\sqrt{2\zeta}}\right]+O(1)},\ &{\rm for}\
\rho=2\gamma,\ \vspace{2mm}\zeta>0\\
{\displaystyle \frac{kn^{\rho/2}}{\sqrt{2\zeta}}}+O(k^2),\ &{\rm
for}\
2\gamma\leq\rho<2,\ \vspace{2mm}\zeta>0\\
k n/h(\zeta)+O(k^2),\ &{\rm for}\ \rho=2,
\end{array}\right.
\label{srs}
\end{equation}
where
\[
h(\zeta)=\left\{\begin{array}{ll}
\sqrt{2\zeta}\tanh\sqrt{\zeta/2},\ &{\rm for}\ \zeta\geq0,\\
-\sqrt{-2\zeta}\tan\sqrt{-\zeta/2},\ &{\rm for}\ \zeta<0.
\end{array}\right.
\]
Hence, the function $\Sigma(z)$ vanishes at the points
\begin{equation}
z_f=d-{\textstyle\frac{1}{2}}\pi^2(1+2f)^2n^{-2}+O(kn^{-3}),
\qquad f=0,1,2\ldots. \label{zrs}
\end{equation}

Obviously, the integrand in Eq.\ (\ref{tm2}) is an analytic
function for $z>\lambda_{(1,\ldots,1)}$. The function $\Sigma(z)$
has a singularity at the point $z=\lambda_{(1,\ldots,1)}$.
However, for the periodic boundary conditions the largest
eigenvalue of the interaction matrix $\widehat{C}_n$ is
non-degenerate, and the singularity of $\Sigma(z)$ is cancelled by
the multiplier $z-\lambda_{(1,\ldots,1)}$. Thus, the point
$z=\lambda_{(1,\ldots,1)}$ is a removable singularity of the
integrand. According to Eq.\ (\ref{zrs}) the function $\Sigma(z)$
has a simple zero on the interval
$(\lambda_{(2,1,\ldots,1)};\lambda_{(1,\ldots,1)})$. Consequently,
the maximal singularity of the integrand in Eq.\ (\ref{tm2}) is at
the point $z=s_n\equiv d-\frac{1}{2} \pi^2n^{-2}+O(kn^{-3})$. The
obtained asymptotic expansion for the location of the maximal
singularity suggests that the change of integration variable
$z=d+\zeta n^{-2}$ might preserve the saddle-point profile of the
function $\Phi_n(z,m)$ in the limit $n\to\infty$.

Note now that for any $\zeta>-\frac{1}{2}\pi^2$ we have
\[
\Phi_n(d+\zeta n^{-2},m)=d-\frac{L_d(d)}{2\beta J}
+n^{-2}\left[\zeta\left(1-\frac{\beta_c}{\beta}\right)
-m^2h(\zeta)\right]+O(kn^{-3}),
\]
as $n\to\infty$, where $L_d(d)$ is given by Eq.\ (\ref{iwf}).
Hence, the function $\Phi_n(z,m)$ attains its minimum on the
interval $(s_n;\infty)$ at the point
$z_n^*=d+n^{-2}\zeta^*(m)+O(kn^{-3})$, where $\zeta^*(m)$ is
the maximal solution of
\[
m^2h'(\zeta)=1-\frac{\beta_c}{\beta}.
\]
Obviously $\zeta^*(m)>-\frac{1}{2}\pi^2$ unless $m=0$. The
point $z=z_n^*$ is the saddle point of the integrand in Eq.\
(\ref{tm2}).

In the scale $z=d+\zeta n^{-2}$ the saddle point $\zeta^*(m)$
does not approach the maximal singularity $s_n\equiv d-\frac{1}{2}
\pi^2n^{-2}+O(kn^{-3})$ of the function $\Phi_n(d+\zeta
n^{-2},m)$. Hence the function $\widetilde\Phi_n(\zeta)\equiv
\Phi_n(d+\zeta n^{-2},m)$ has a conventional saddle-point
landscape in the limit $n\to\infty$. Therefore in the scale
$z=d+\zeta n^{-2}$ one can find the asymptotic expansion for the
integral Eq.\ (\ref{tm2}) using the standard saddle-point method
and obtain
\[
\Theta_{n,L}=\exp\left\{-N f(\beta)+n^{d-2}\beta
J\left[\zeta^*(m)\left(1-\frac{\beta_c}{\beta}\right)-m^2h
(\zeta^*(m))\right] +O(kn^{d-3})\right\},
\]
where
\begin{equation}
f(\beta)={\textstyle\frac{1}{2}}\log(\beta J/\pi)-\beta
Jd+{\textstyle\frac{1}{2}} L_d(d) \label{lfe}
\end{equation}
is the limiting free energy per spin of the spherical model for
$\beta>\beta_c$. The partition function of the spherical model
with the periodic boundary conditions in the canonical ensemble is
given by
\begin{equation}
\Theta_n=\exp\left\{-N f(\beta)+O(\log n)\right\}. \label{cth}
\end{equation}
Thus, we obtain the following asymptotics for the distribution
density of the magnetization of the layer $L$, see Eq.\
(\ref{bdd}):
\[
\phi_{n,L}(m)=\exp\left\{-n^{d-2}R_{L}^{(2)}(m)
+O(kn^{d-3})\right\},
\]
as $n\to\infty$, where the rate function is given by
\begin{equation}
R_{L}^{(2)}(m)=-\beta J\left[\zeta^*(m)
\left(1-\frac{\beta_c}{\beta}\right)-m^2h (\zeta^*(m))\right].
\label{rfl}
\end{equation}
The function $R_{L}^{(2)}(m)$ has a cusp at $m=0$ and vanishes
at the points $m=\pm\sqrt{1-\beta_c/\beta}$ --- the equilibrium
values of the magnetization within the spherical model with
periodic boundary conditions, see Fig.\ 3. The nature of the cusp
at $m=0$ is obvious. A magnetization value $m$ can be achieved
by deforming either equilibrium $\langle+\rangle$-phase or
$\langle-\rangle$-phase. For negative (positive) values of $m$ a
deformation of minus (plus) phase is the easiest way to achieve
the desired value of magnetization. As $m\uparrow0$ the required
deformations become more and more costly, because we are getting
further and further away from the equilibrium value $-m^*$.
However, if we keep increasing the value of $m$ deformations
become less costly as soon as we cross the point $m=0$, because
we begin to approach the equilibrium value $m^*$.

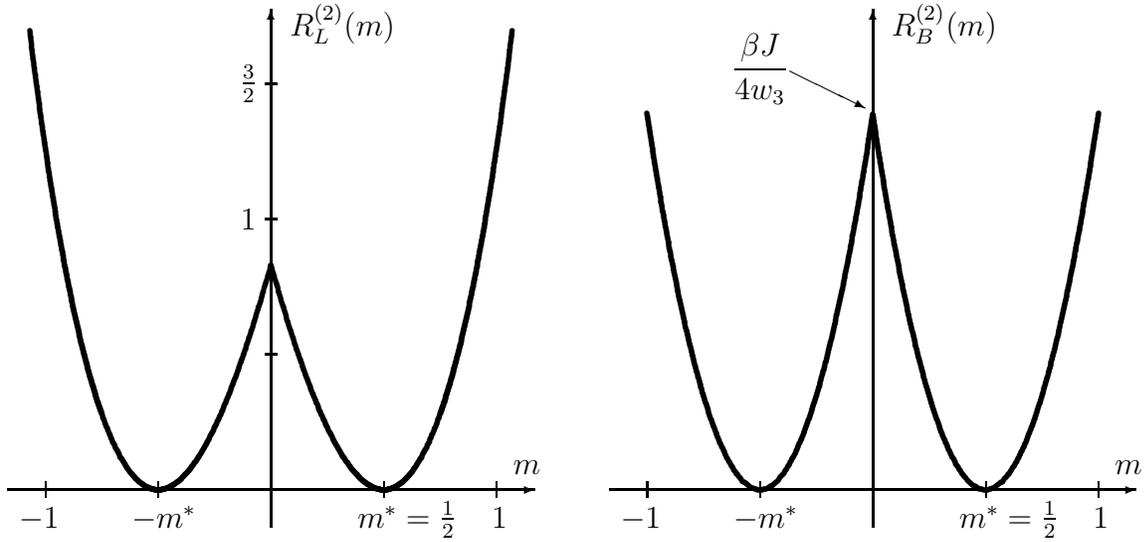
\begin{figure}
\setlength{\unitlength}{1mm}
\begin{picture}(150,70)(0,-1)
\put(0,5){\vector(1,0){70}} \put(35,0){\vector(0,1){69}}
\put(80,5){\vector(1,0){70}} \put(115,0){\vector(0,1){69}}
\put(69,8){\makebox(0,0){$m$}} \put(149,8){\makebox(0,0){$m$}}
\put(44.5,67){\makebox(0,0){$R_{L}^{(2)}(m)$}}
\put(124.5,67){\makebox(0,0){$R_{B}^{(2)}(m)$}}

\multiput(5,5)(15,0){5}{\makebox(0,0){$\rule{0.15mm}{2mm}$}}
\put(1.5,0){$-1$}\put(64.3,0){$1$}
\put(81.5,0){$-1$}\put(144.3,0){$1$}
\put(16.5,0){$-m^*$} \put(46.5,0){$m^*=\frac{1}{2}$}
\multiput(35,5)(0,18){4}{\makebox(0,0){$\rule{1.5mm}{0.15mm}$}}
\put(32,59){\makebox(0,0){$\frac{3}{2}$}}

\multiput(85,5)(15,0){5}{\makebox(0,0){$\rule{0.15mm}{2mm}$}}
\put(96.5,0){$-m^*$} \put(126.5,0){$m^*=\frac{1}{2}$}
\put(32,41){\makebox(0,0){$1$}}
\put(100,61){\makebox(0,0){$\displaystyle \frac{\beta J}{4w_3}$}}
\put(104,60.7){\vector(2,-1){10}}
\input sph_rate_ff
\input sph_rate_f3
\end{picture}
\caption{The rate functions for the magnetisation of the layers
$L$ (left) and the Kadanoff blocks $B$ (right) at
$\beta=\frac{4}{3}\beta_c$.}
\end{figure}

\subsection{Large deviations of the magnetization of rods $R$.}

In this subsection we investigate large deviations of the
magnetization of $(d-2)$-dimensional domains (rods)
\[
R_k=\left\{j\in\Omega_n:1\leq j_{\nu}\leq k,\ {\nu=1,2}\right\}.
\]
As in the previous subsection the large-deviation probabilities
are determined by the relative location of the two largest
eigenvalues of the interaction matrix and the maximal zero of the
function $\Sigma(z)$. However in the case of rods $R_k$ the
distances between the eigenvalues and zeroes are qualitatively
different from what we found in the case of layers $L_k$. As a
result, the probabilities of large deviations are also
qualitatively different in these two cases.

In the case of rods $R_k$ the coefficients
$\gamma_j\equiv|R_k|^{-1}\sum_{l\in R_k}V_{j,l}$ are given by
\begin{equation}
\gamma_j=\frac{2}{|R_k|n^{d/2-2}}\prod_{\nu=1,2}\frac{\sin[\pi
k(j_\nu-1)/n]} {\sin[\pi(j_\nu-1)/n]}\cos\left[\frac{\pi
(k-1)(j_\nu-1)}{n}-\frac{\pi}{4}\right]\prod_{\kappa=3}^d\delta_{1,j_{\kappa}}.
\end{equation}
Hence the multiple sum in the expression for $\Sigma(z)$ (see Eq.\
(\ref{sgm})) reduces to the double one
\[
\Sigma(z)=\frac{n^{d-4}}{|R_k|^2}\sum_{j_1,j_2=1}^{n}\frac{\sin^2[\pi
k(j_1-1)/n]}{\sin^2[\pi(j_1-1)/n]}\,\frac{\sin^2[\pi
k(j_2-1)/n]}{\sin^2[\pi(j_2-1)/n]}\times
\]
\[
\frac{1}{z-d+2-\cos[2\pi(j_1-1)/n]-\cos[2\pi(j_2-1)/n]}.
\]
The function $\Sigma(z)$ has simple poles at the points $z=d$ and
$z=d-1+ \cos(2\pi/n)$ and a simple zero $s_n^*$ somewhere between
the poles. Our objective now is to find an expression for
$\Sigma(z)$ convenient enough for locating $s_n^*$.

In the case of $(d-1)$-dimensional layers the expression for
$\Sigma(z)$ contained only a single sum, which we managed to
calculate exactly. For $(d-2)$-dimensional rods we do not have a
luxury of working with an exact closed-form expression for
$\Sigma(z)$, nevertheless, fortunately, simplifications arise for
a different reason. As is often the case for double sums of this
type $\Sigma(d+\zeta n^{-2})= O(n^{d-2}k^4\log n)$, for $\zeta>0$.
Under the same rescaling the term corresponding to $j_1=j_2=0$
(the contribution from the maximal eigenvalue of the interaction
matrix) is given by $n^{d-2}k^4/\zeta$. Denote now $\Sigma'(z)$
the same sum as in Eq.\ (\ref{sgm}) but with the contribution from
the maximal eigenvalue omitted. Below we will argue that the main
asymptotics of $\Sigma'(d+\zeta n^{-2})$ is given by
$n^{d-2}k^4\log n\,\sigma(\zeta)$, where $\sigma(\zeta)$ is a
positive analytic function for $\zeta>-2\pi^2$. Therefore, in
order to find $s_n^*$, we have to solve the equation
$\sigma(\zeta)\log n=-\zeta^{-1}$. Since the function
$\sigma(\zeta)$ is regular at $\zeta=0$, the solution is given by
\[
\zeta^*=-\frac{1}{\sigma(0)\log
n}+O\left(\frac{1}{\log^{2}n}\right),
\]
and hence
\[
s_n^*\sim d-\frac{1}{\sigma(0)n^2\log n},\quad \mbox{as}\
n\to\infty.
\]

Let $k=\alpha n^{\gamma}$, where $\alpha>0$  and $\gamma\in(0;1)$.
For $z=d+\zeta n^{-\rho}$, where $\zeta>0$ and $0\leq\rho<2$, one
has as $n\to\infty$
\[
\Sigma(d+\zeta n^{-\rho})=n^{d-2}I_2(\zeta n^{-\rho})+
O\left[\exp\left(-\sqrt{2\zeta}n^{1-\rho/2}\right)\right],
\]
where
\begin{equation}
I_m(x)=\frac{1}{(2\pi)^{m}}\mathop{\int\!\ldots\!\int}
\limits_{\!\!\!\!\!\!0}^{\;\;\;2\pi}\!d\omega_1\ldots d\omega_m
\prod_{\nu=1}^{m}\frac{1-\cos(k\omega_\nu)}{1-\cos\omega_\nu}
\,\frac{1}{x+m-\sum\limits_{\kappa=1}^{m}\cos\omega_{\kappa}}.
\label{in}
\end{equation}
For $z=d+\zeta n^{-2}$ one obtains
\[
\Sigma(d+\zeta n^{-2})= n^{d-2}\left\{I_2(\zeta n^{-2})+ k^4\left[
\delta_1(\zeta)+\delta_2(\zeta)+O(n^{-(1-\gamma)/2})\right]\right\},
\]
where
\[
\delta_1(\zeta)= \sum_{j=-\infty}^{\infty}\frac{2}{\sqrt{2(\zeta+
2\pi^2j^2)}}\,\frac{1}{\exp\sqrt{2(\zeta+2\pi^2j^2)}-1},
\]
and
\begin{equation}
\delta_2(\zeta)=\frac{1}{\pi}\int_{0}^{\infty}\!
\frac{2}{\sqrt{2(\zeta+2\omega^2)}}\,
\frac{d\omega}{\exp\sqrt{2(\zeta+2\omega^2)}-1}. \label{d2zp}
\end{equation}
Note that $\lambda_{(1,\ldots,1)}-\lambda_{(2,1,\ldots,1)}
\sim2\pi^2n^{-2}$, hence, to locate $s_n^*$ we have to consider
the scale $z=d+\zeta n^{-2}$ or even a finer one. In any rougher
scale the poles and zeroes of $\Sigma(z)$ merge as $n\to\infty$,
and the function $\Phi_n(d+\zeta n^{-\rho},m)$ does not have a
required saddle-point landscape.

Isolating the singularity of the integral $I_2(\zeta n^{-2})$ one
can write it down in the following form
\[
I_2(\zeta n^{-2})=\frac{k^4}{(2\pi)^2}\mathop{\int\!\!\int}
\limits_{\omega_1^2+\omega_2^2\leq c}\!\frac{d\omega_1d\omega_2}
{\zeta n^{-2(1-\gamma)}+\frac{1}{2}(\omega_1^2+\omega_2^2)}
+k^4r^{(1)}_n(\zeta),
\]
where the function $r^{(1)}_n(\zeta)$ (the regular part of the integral)
is analytic in the unit circle $\{\zeta\in C:|\zeta|<1\}$. As
$n\to\infty$ the sequence $r^{(1)}_n(\zeta)$ converges to some limiting
function uniformly over the unit circle. Calculating the double
integral one obtains
\begin{equation}
I_2(\zeta n^{-2})=\frac{k^4}{2\pi}
\left[\log\frac{n^{2(1-\gamma)}}{\zeta}+\log\left({\textstyle
\frac{1}{2}}c^2+\zeta n^{-2(1-\gamma)}
\right)\right]+k^4r^{(1)}_n(\zeta). \label{i1r}
\end{equation}

Separating the singularities of $\delta_1(\zeta)$ at $\zeta=0$ one
obtains
\begin{equation}
\delta_1(\zeta)=\frac{1}{\zeta}-\frac{1}{\sqrt{2\zeta}}
+r^{(2)}(\zeta), \label{d1r}
\end{equation}
where $r^{(2)}(\zeta)$ is analytic for ${\rm Re}\:\zeta>-2\pi^2$.
Analogous separation of singularities at $\zeta=0$ yields
\begin{equation}
\delta_2(\zeta)=\frac{2}{\pi}\int\limits_{0}^{\infty}
\frac{d\omega}{2\zeta+\omega^2}-\frac{1}{\pi}\int\limits_0^c
\frac{d\omega}{\sqrt{2\zeta+\omega^2}}+r_3(\zeta)=
\frac{1}{\sqrt{2\zeta}}+\frac{\log\zeta}{2\pi}
+\widetilde{r}_3(\zeta), \label{d2r}
\end{equation}
where the functions $r_3(\zeta)$ and $\widetilde{r}_3(\zeta)$ are
analytic in the unit circle. Summarizing Eqs.\ (\ref{i1r}),
(\ref{d1r}), and (\ref{d2r}) one arrives at
\begin{equation}
\Sigma(d+\zeta n^{-2})=n^{d-2}k^4\left[\frac{1-\gamma}{\pi} \log
n+\zeta^{-1}+\widetilde{r}_n(\zeta)\right], \label{d2rt}
\end{equation}
where $\widetilde{r}_n(\zeta)$ are analytic and bounded uniformly
over $n$ in the unit circle. Hence, the maximal zero of the
function $\Sigma(z)$ is at the point
\[
s_n^*=d-\frac{\pi}{1-\gamma}\,\frac{1}{n^2\log n} +O\left[(n\log
n)^{-2}\right].
\]

The expression for $s_n^*$ suggests that the change of variable
$z=d+\zeta (n^2\log n)^{-1}$ is likely to convert Eq.\ (\ref{tm2})
into an integral convenient for application of the saddle-point
method. Substituting $z=d+\zeta (n^2\log n)^{-1}$ in the
expression for $\Phi_n(z)$ we see that this is indeed the case.
One \vspace{2mm}obtains
\newline
$ \displaystyle{ \Phi_n\left(d+\frac{\zeta}{n^2\log n}\right)=} $
\[
n^d\beta J\left[d-\frac{L(d)}{2\beta J}\right]+ \frac{n^{d-2}\beta
J}{\log n}\left[\zeta\left(1-
\frac{\beta_c}{\beta}\right)-\frac{\pi m^2}{1-\gamma+\pi\zeta^{-1}}\right]
+O\left(\frac{n^{d-2}}{\log^2n}\right).
\]
Hence, the sequence of relevant saddle points is given by
\[
z_n^*=d+\frac{\zeta^*}{n^{2}\log n}+
O\left(\frac{1}{n^2\log^2n}\right),
\]
where
\[
\zeta^*=\frac{\pi}{1-\gamma}\left(\frac{|m|}{\sqrt{1-\beta_c/\beta}}
-1\right).
\]
Evaluating the integral in Eq.\ (\ref{tm2}) using the saddle-point
method one obtains
\[
\Theta_{N,R}=\exp \left[N f(\beta)-\frac{n^{d-2}}{\log n}\,
\frac{\pi\beta J}{1-\gamma}
\left(|m|-\sqrt{1-\beta_c/\beta}\right)^2
+O\left(\frac{n^{d-2}}{\log^2n}\right)\right].
\]
Taking into account Eq.\ (\ref{cth}) one arrives at
\[
\phi_{n,R}(m)=\exp\left[-\frac{n^{d-2}}{\log n}\,R_R^{(2)}(m)
+O\left(\frac{n^{d-2}}{\log^2n}\right)\right],
\]
where
\begin{equation}
R_R^{(2)}(m)=\frac{\pi\beta J}{1-\gamma}
\left(|m|-\sqrt{1-\beta_c/\beta}\right)^2.
\label{rfr}
\end{equation}

\subsection{Large deviations in Kadanoff blocks.}
In this section we investigate the probabilities of large
deviations for the total spin of the Kadanoff blocks
\[
B_k=\left\{j\in\Omega_n:1\leq j_{\nu}\leq k,\
{\nu=1,\ldots,d}\right\}.
\]
Qualitative behaviour of the corresponding large-deviation
probabilities is the same for any $d\geq 3$. Therefore, to avoid
unnecessary technical complications we consider only the case
$d=3$. The coefficients $\gamma_j\equiv|B_k|^{-1}\sum_{l\in
B_k}V_{j,l}$ are given by
\begin{equation}
\gamma_j=\frac{2^{3/2}}{|B_k|}\prod_{\nu=1,2,3}\frac{\sin[\pi
k(j_\nu-1)/n]} {\sin[\pi(j_\nu-1)/n]}\cos\left[\frac{\pi
(k-1)(j_\nu-1)}{n}-\frac{\pi}{4}\right],
\end{equation}
and
\[
\Sigma(z)=n^{-3}\sum_{j_1,j_2,j_3=1}^{n}\prod_{\nu=1}^{3}
\frac{\sin^2[\pi k(j_{\nu}-1)/n]}{\sin^2[\pi(j_{\nu}-1)/n]}\,
\frac{1}{z-\sum\limits_{\kappa=1,2,3}\cos[2\pi(j_{\kappa}-1)/n]}.
\]
The function $\Sigma(z)$ has simple poles at the points $z=d$ and
$z=d-1+ \cos(2\pi/n)$ and a simple zero $s_n^*$ in the interval
$(d-1+\cos(2\pi/n);d)$. Our goal now is to obtain an expression
for $\Sigma(z)$ convenient enough for locating $s_n^*$.

Let $k=\alpha n^{\gamma}$, with $\alpha>0$ and $\gamma\in(0;1)$.
For $z=d+\zeta n^{-2}$, $\zeta>0$, one has as $n\to\infty$
\[
\Sigma(d+\zeta n^{-2})= I_3(\zeta n^{-2})+
\frac{k^6}{n}\left[\delta_1(\zeta)+\delta_2
(\zeta)+\delta_3(\zeta)+O(n^{-(1-\gamma)/2})\right],
\]
where $I_3(x)$ is given by Eq.\ (\ref{in}),
\[
\delta_1(\zeta)=\sum_{j,l=-\infty}^{\infty}
\frac{2}{\sqrt{2\zeta+4\pi^2(j^2+l^2)}}\,
\frac{1}{\exp\sqrt{2\zeta+4\pi^2(j^2+l^2)}-1},
\]
\begin{equation}
\delta_2(\zeta)=
\frac{1}{\pi}\sum_{j=-\infty}^{\infty}\int_{0}^{\infty}\!
\frac{2}{\sqrt{2\zeta+4\pi^2 j^2+\omega^2}}\,
\frac{d\omega}{\exp\sqrt{2\zeta+4\pi^2 j^2+\omega^2}-1},
\label{d3zp}
\end{equation}
and
\begin{equation}
\delta_3(\zeta)=
\frac{1}{\pi^2}\mathop{\int\!\!\!\int}_{\!\!\!0}^{\,\,\,\,\,\,\infty}
\!d\omega_1d\omega_2\,
\frac{2}{\sqrt{2\zeta+\omega_1^2+\omega_2^2}}\,
\frac{1}{\exp\sqrt{2\zeta+\omega_1^2+\omega_2^2}-1}. \label{d3zp1}
\end{equation}
The large-$n$ asymptotics of $I_3(\zeta n^{-\rho})$ for
$\rho>2\gamma$ is given by
\[
I_3(\zeta n^{-\rho})\sim k^{5}w_3,
\]
where
\begin{equation}
w_3= \frac{2}{\pi^3} \mathop{\int\!\!\!\int\!\!\!\int}
\limits_{\!\!\!\!\!\!-\infty}^{\;\;\;\infty}d\omega_1d\omega_2
d\omega_3 \prod_{\nu=1}^{3}\frac{1-
\cos\omega_\nu}{\omega_{\nu}^2}\,
\frac{1}{\sum_{\kappa=1}^3\omega_\kappa^2}. \label{dw}
\end{equation}

Separating the singularities of $\delta_\nu(\zeta)$, $\nu=1,2,3$
at $\zeta=0$ (cf.,\ Eqs.\ (\ref{i1r})--(\ref{d2r})) one obtains
\[
\sum_{\nu=1,2,3}\delta_{\nu}(\zeta)=\zeta^{-1}+r(\zeta),
\]
where the function $r(\zeta)$ is analytic in the unit ball
$\{\zeta\in C:|\zeta|<1\}$. Hence, the maximal zero of the
function $\Sigma(z)$ is at the point
\[
z=s_n\sim d-\frac{k}{n^3 w_3}.
\]
The location of the maximal zero gives us a hint that the
asymptotic expansion of the integral (\ref{tm2}) can be found
using the saddle-point method after a prior rescaling of the
integration variable $z$ via $z=d+\zeta n^{-3+\gamma}$. The
rescaling yields
\[
\Phi_n(d+\zeta n^{-3+\gamma})=d-\frac{L(d)}{2\beta J} + k
n^{-3}\left[\zeta\left(1-\frac{\beta_c}{\beta}\right)
-\frac{m^2}{w_3+\zeta^{-1}}\right]+ o(k n^{-3}).
\]
Hence, for $\beta>\beta_c$ the sequence of relevant saddle points
is given by
\[
z_n^*=d+\zeta^*k n^{-3}+o(k n^{-3}),
\]
where
\[
\zeta^*=\frac{1}{w_3}\left(\frac{|m|}
{\sqrt{1-\beta_c/\beta}}-1\right).
\]
Evaluation of the integral (\ref{tm2}) using the saddle-point
method yields
\[
\Theta_{N,B}=\exp\left[Nf(\beta)-k\frac{\beta J}{w_3}
\left(|m|-\sqrt{1-\beta_c/\beta}\right)^2 +o(k)\right].
\]
Taking into account Eq.\ (\ref{cth}), one obtains the following
asymptotic formula for the probability density of magnetization
\[
\phi_{n,B}=\exp\left[-k R_B^{(2)}(m)+o(k)\right],
\]
where
\begin{equation}
R_B^{(2)}(m)=\frac{\beta J}{w_3}
\left(|m|-\sqrt{1-\beta_c/\beta}\right)^2. \label{rfb}
\end{equation}

\section{Discussion and concluding remarks.}

In the present paper we have investigated large-deviation
probabilities for the magnetization of various domains
$D_k(\subset \Omega_n)$ within the spherical model of a
ferromagnet defined in cubes $\Omega_n$. We have shown that these
probabilities are very sensitive to the shape and/or
dimensionality of the domains $D_k$.

An appealing feature of the large-deviation theory is that often
the behavior of large-deviation probabilities admits a simple
intuitively clear explanation, see, e.g., \cite{shl89} . This is
also the case for the results obtained in the present paper. In
Section 3.1\ we have shown that in the case of layers $L_k$ the
large-deviation probabilities decay exponentially with $n^{d-2}$,
which is similar to the asymptotics of the analogous probabilities
for the magnetization of the entire cube $\Omega_n$, see
\cite{p94a}. The reason for such similarity becomes clear when we
investigate the properties of typical configurations realizing
these deviations. It turns out that the easiest way to obtain a
desired value of magnetization in the layer $L_k$ is to deform the
configuration of random variables in the entire cube $\Omega_n$.
That is, a large-deviation of the magnetization in a layer $L_k$
leads to a large-deviation of the magnetization in the entire cube
$\Omega_n$. To put it another way, a substantial deformation of
the configuration in a layer $L_k$ spreads over the entire cube
$\Omega_n$.

In the case of rods $R_k$ the large-deviation probabilities are
modified by $\log n$ corrections, namely, they decay exponentially
with $n^{d-2}/\log n$. Investigation of typical configurations
realizing these large deviations shows that a substantial
deformation of the configuration in a rod $R_k$ does not spread
over the entire cube $\Omega_n$, but it spreads over a domain with
the linear size of the order $n/\log n$. Note that if $k$
satisfies the bound $\sqrt{n k'}\ll k \ll n $, then the number of random
variables in a rod $R_k$ is greater than that in a layer $L_{k'}$.
Nevertheless, large deviations of the magnetization in the rod are
more likely than large deviations in the layer simply because of a
more compact arrangement of the random variables in the rod.

In the case of Kadanoff blocks $B_k$ the large-deviation
probabilities decay exponentially with $k^{d-2}$. Even a
substantial deformation of the configuration in a block $B_k$ does
not spread very far from the block. Essentially it remains
localized in a domain of the linear size $O(k)$, that is, of the
same extent as the size of the Kadanoff block itself. The number
of random variables in a block $B_k$ can be greater than the
number of random variables in a rod $R_{k'}$ (or a layer
$L_{k''}$), nevertheless, a large deviation of the magnetization
of the block is much more likely than a large deviation in the rod
or the layer. Accordingly, a uniform deformation of the
configuration in a layer $L_{k''}$ or a rod $R_{k'}$ spreads much
further over the cube $\Omega_n$ than a deformation of a Kadanoff
block $B_k$.

In order to find the properties of typical configurations
realizing large deviations in layers, rods, and blocks one has to
continue the calculation of the present paper one step further and
derive the conditional distributions of random variables
$\sigma_j$ given a desired large deviation. These calculations
will be published elsewhere.

\end{document}

%% file: cw_rate_f.tex
\put(5.012,62.4931){\circle*{0.7}}
\put(5.01836,62.2932){\circle*{0.7}}
\put(5.0251,62.0933){\circle*{0.7}}
\put(5.03217,61.8934){\circle*{0.7}}
\put(5.03954,61.6936){\circle*{0.7}}
\put(5.04717,61.4937){\circle*{0.7}}
\put(5.05505,61.2939){\circle*{0.7}}
\put(5.06317,61.094){\circle*{0.7}}
\put(5.0715,60.8942){\circle*{0.7}}
\put(5.08004,60.6944){\circle*{0.7}}
\put(5.08878,60.4946){\circle*{0.7}}
\put(5.09772,60.2948){\circle*{0.7}}
\put(5.10684,60.095){\circle*{0.7}}
\put(5.11615,59.8952){\circle*{0.7}}
\put(5.12563,59.6954){\circle*{0.7}}
\put(5.13528,59.4957){\circle*{0.7}}
\put(5.14511,59.2959){\circle*{0.7}}
\put(5.1551,59.0962){\circle*{0.7}}
\put(5.16526,58.8964){\circle*{0.7}}
\put(5.17557,58.6967){\circle*{0.7}}
\put(5.18605,58.497){\circle*{0.7}}
\put(5.19668,58.2972){\circle*{0.7}}
\put(5.20746,58.0975){\circle*{0.7}}
\put(5.2184,57.8978){\circle*{0.7}}
\put(5.22948,57.6981){\circle*{0.7}}
\put(5.24072,57.4985){\circle*{0.7}}
\put(5.25211,57.2988){\circle*{0.7}}
\put(5.26364,57.0991){\circle*{0.7}}
\put(5.27532,56.8995){\circle*{0.7}}
\put(5.28714,56.6998){\circle*{0.7}}
\put(5.29911,56.5002){\circle*{0.7}}
\put(5.31122,56.3005){\circle*{0.7}}
\put(5.32348,56.1009){\circle*{0.7}}
\put(5.33587,55.9013){\circle*{0.7}}
\put(5.34841,55.7017){\circle*{0.7}}
\put(5.36109,55.5021){\circle*{0.7}}
\put(5.37391,55.3025){\circle*{0.7}}
\put(5.38687,55.1029){\circle*{0.7}}
\put(5.39997,54.9034){\circle*{0.7}}
\put(5.41321,54.7038){\circle*{0.7}}
\put(5.42659,54.5042){\circle*{0.7}}
\put(5.44011,54.3047){\circle*{0.7}}
\put(5.45376,54.1052){\circle*{0.7}}
\put(5.46756,53.9056){\circle*{0.7}}
\put(5.48149,53.7061){\circle*{0.7}}
\put(5.49557,53.5066){\circle*{0.7}}
\put(5.50978,53.3071){\circle*{0.7}}
\put(5.52413,53.1076){\circle*{0.7}}
\put(5.53862,52.9082){\circle*{0.7}}
\put(5.55325,52.7087){\circle*{0.7}}
\put(5.56802,52.5093){\circle*{0.7}}
\put(5.58293,52.3098){\circle*{0.7}}
\put(5.59798,52.1104){\circle*{0.7}}
\put(5.61316,51.911){\circle*{0.7}}
\put(5.62849,51.7115){\circle*{0.7}}
\put(5.64396,51.5121){\circle*{0.7}}
\put(5.65957,51.3127){\circle*{0.7}}
\put(5.67532,51.1134){\circle*{0.7}}
\put(5.69121,50.914){\circle*{0.7}}
\put(5.70724,50.7146){\circle*{0.7}}
\put(5.72341,50.5153){\circle*{0.7}}
\put(5.73973,50.316){\circle*{0.7}}
\put(5.75618,50.1166){\circle*{0.7}}
\put(5.77279,49.9173){\circle*{0.7}}
\put(5.78953,49.718){\circle*{0.7}}
\put(5.80642,49.5188){\circle*{0.7}}
\put(5.82346,49.3195){\circle*{0.7}}
\put(5.84064,49.1202){\circle*{0.7}}
\put(5.85796,48.921){\circle*{0.7}}
\put(5.87543,48.7217){\circle*{0.7}}
\put(5.89305,48.5225){\circle*{0.7}}
\put(5.91082,48.3233){\circle*{0.7}}
\put(5.92873,48.1241){\circle*{0.7}}
\put(5.9468,47.9249){\circle*{0.7}}
\put(5.96501,47.7258){\circle*{0.7}}
\put(5.98337,47.5266){\circle*{0.7}}
\put(6.00189,47.3275){\circle*{0.7}}
\put(6.02055,47.1283){\circle*{0.7}}
\put(6.03937,46.9292){\circle*{0.7}}
\put(6.05834,46.7301){\circle*{0.7}}
\put(6.07747,46.531){\circle*{0.7}}
\put(6.09675,46.332){\circle*{0.7}}
\put(6.11618,46.1329){\circle*{0.7}}
\put(6.13577,45.9339){\circle*{0.7}}
\put(6.15552,45.7349){\circle*{0.7}}
\put(6.17543,45.5358){\circle*{0.7}}
\put(6.1955,45.3369){\circle*{0.7}}
\put(6.21572,45.1379){\circle*{0.7}}
\put(6.23611,44.9389){\circle*{0.7}}
\put(6.25666,44.74){\circle*{0.7}}
\put(6.27737,44.5411){\circle*{0.7}}
\put(6.29825,44.3422){\circle*{0.7}}
\put(6.31929,44.1433){\circle*{0.7}}
\put(6.3405,43.9444){\circle*{0.7}}
\put(6.36187,43.7455){\circle*{0.7}}
\put(6.38342,43.5467){\circle*{0.7}}
\put(6.40513,43.3479){\circle*{0.7}}
\put(6.42701,43.1491){\circle*{0.7}}
\put(6.44906,42.9503){\circle*{0.7}}
\put(6.47129,42.7515){\circle*{0.7}}
\put(6.49369,42.5528){\circle*{0.7}}
\put(6.51627,42.3541){\circle*{0.7}}
\put(6.53902,42.1554){\circle*{0.7}}
\put(6.56195,41.9567){\circle*{0.7}}
\put(6.58506,41.758){\circle*{0.7}}
\put(6.60834,41.5594){\circle*{0.7}}
\put(6.63181,41.3608){\circle*{0.7}}
\put(6.65547,41.1622){\circle*{0.7}}
\put(6.67931,40.9636){\circle*{0.7}}
\put(6.70333,40.7651){\circle*{0.7}}
\put(6.72754,40.5665){\circle*{0.7}}
\put(6.75194,40.368){\circle*{0.7}}
\put(6.77653,40.1695){\circle*{0.7}}
\put(6.80131,39.9711){\circle*{0.7}}
\put(6.82628,39.7726){\circle*{0.7}}
\put(6.85145,39.5742){\circle*{0.7}}
\put(6.87681,39.3758){\circle*{0.7}}
\put(6.90238,39.1775){\circle*{0.7}}
\put(6.92814,38.9792){\circle*{0.7}}
\put(6.9541,38.7808){\circle*{0.7}}
\put(6.98027,38.5826){\circle*{0.7}}
\put(7.00664,38.3843){\circle*{0.7}}
\put(7.03321,38.1861){\circle*{0.7}}
\put(7.06,37.9879){\circle*{0.7}}
\put(7.087,37.7897){\circle*{0.7}}
\put(7.1142,37.5916){\circle*{0.7}}
\put(7.14162,37.3935){\circle*{0.7}}
\put(7.16926,37.1954){\circle*{0.7}}
\put(7.19711,36.9973){\circle*{0.7}}
\put(7.22519,36.7993){\circle*{0.7}}
\put(7.25348,36.6013){\circle*{0.7}}
\put(7.282,36.4034){\circle*{0.7}}
\put(7.31075,36.2054){\circle*{0.7}}
\put(7.33972,36.0076){\circle*{0.7}}
\put(7.36893,35.8097){\circle*{0.7}}
\put(7.39836,35.6119){\circle*{0.7}}
\put(7.42803,35.4141){\circle*{0.7}}
\put(7.45794,35.2163){\circle*{0.7}}
\put(7.48809,35.0186){\circle*{0.7}}
\put(7.51847,34.8209){\circle*{0.7}}
\put(7.54911,34.6233){\circle*{0.7}}
\put(7.57999,34.4257){\circle*{0.7}}
\put(7.61111,34.2281){\circle*{0.7}}
\put(7.64249,34.0306){\circle*{0.7}}
\put(7.67413,33.8331){\circle*{0.7}}
\put(7.70602,33.6357){\circle*{0.7}}
\put(7.73817,33.4383){\circle*{0.7}}
\put(7.77058,33.2409){\circle*{0.7}}
\put(7.80326,33.0436){\circle*{0.7}}
\put(7.8362,32.8464){\circle*{0.7}}
\put(7.86942,32.6491){\circle*{0.7}}
\put(7.90291,32.452){\circle*{0.7}}
\put(7.93667,32.2548){\circle*{0.7}}
\put(7.97072,32.0577){\circle*{0.7}}
\put(8.00505,31.8607){\circle*{0.7}}
\put(8.03966,31.6637){\circle*{0.7}}
\put(8.07457,31.4668){\circle*{0.7}}
\put(8.10977,31.2699){\circle*{0.7}}
\put(8.14526,31.0731){\circle*{0.7}}
\put(8.18105,30.8763){\circle*{0.7}}
\put(8.21715,30.6796){\circle*{0.7}}
\put(8.25355,30.483){\circle*{0.7}}
\put(8.29026,30.2864){\circle*{0.7}}
\put(8.32729,30.0898){\circle*{0.7}}
\put(8.36463,29.8933){\circle*{0.7}}
\put(8.4023,29.6969){\circle*{0.7}}
\put(8.44029,29.5005){\circle*{0.7}}
\put(8.47861,29.3043){\circle*{0.7}}
\put(8.51727,29.108){\circle*{0.7}}
\put(8.55626,28.9119){\circle*{0.7}}
\put(8.59559,28.7158){\circle*{0.7}}
\put(8.63527,28.5197){\circle*{0.7}}
\put(8.6753,28.3238){\circle*{0.7}}
\put(8.71569,28.1279){\circle*{0.7}}
\put(8.75643,27.9321){\circle*{0.7}}
\put(8.79755,27.7364){\circle*{0.7}}
\put(8.83903,27.5407){\circle*{0.7}}
\put(8.88088,27.3452){\circle*{0.7}}
\put(8.92311,27.1497){\circle*{0.7}}
\put(8.96573,26.9543){\circle*{0.7}}
\put(9.00874,26.7589){\circle*{0.7}}
\put(9.05214,26.5637){\circle*{0.7}}
\put(9.09595,26.3686){\circle*{0.7}}
\put(9.14016,26.1735){\circle*{0.7}}
\put(9.18478,25.9785){\circle*{0.7}}
\put(9.22982,25.7837){\circle*{0.7}}
\put(9.27528,25.5889){\circle*{0.7}}
\put(9.32117,25.3943){\circle*{0.7}}
\put(9.3675,25.1997){\circle*{0.7}}
\put(9.41427,25.0052){\circle*{0.7}}
\put(9.46148,24.8109){\circle*{0.7}}
\put(9.50916,24.6167){\circle*{0.7}}
\put(9.55729,24.4225){\circle*{0.7}}
\put(9.6059,24.2285){\circle*{0.7}}
\put(9.65498,24.0346){\circle*{0.7}}
\put(9.70454,23.8409){\circle*{0.7}}
\put(9.75459,23.6473){\circle*{0.7}}
\put(9.80514,23.4537){\circle*{0.7}}
\put(9.8562,23.2604){\circle*{0.7}}
\put(9.90777,23.0671){\circle*{0.7}}
\put(9.95987,22.874){\circle*{0.7}}
\put(10.0125,22.6811){\circle*{0.7}}
\put(10.0657,22.4883){\circle*{0.7}}
\put(10.1194,22.2956){\circle*{0.7}}
\put(10.1736,22.1031){\circle*{0.7}}
\put(10.2285,21.9108){\circle*{0.7}}
\put(10.2839,21.7186){\circle*{0.7}}
\put(10.3399,21.5266){\circle*{0.7}}
\put(10.3964,21.3348){\circle*{0.7}}
\put(10.4536,21.1431){\circle*{0.7}}
\put(10.5114,20.9517){\circle*{0.7}}
\put(10.5699,20.7604){\circle*{0.7}}
\put(10.6289,20.5693){\circle*{0.7}}
\put(10.6886,20.3784){\circle*{0.7}}
\put(10.749,20.1878){\circle*{0.7}}
\put(10.8101,19.9973){\circle*{0.7}}
\put(10.8718,19.8071){\circle*{0.7}}
\put(10.9342,19.6171){\circle*{0.7}}
\put(10.9973,19.4273){\circle*{0.7}}
\put(11.0612,19.2378){\circle*{0.7}}
\put(11.1257,19.0485){\circle*{0.7}}
\put(11.191,18.8594){\circle*{0.7}}
\put(11.2571,18.6707){\circle*{0.7}}
\put(11.324,18.4822){\circle*{0.7}}
\put(11.3916,18.294){\circle*{0.7}}
\put(11.4601,18.106){\circle*{0.7}}
\put(11.5293,17.9184){\circle*{0.7}}
\put(11.5994,17.7311){\circle*{0.7}}
\put(11.6704,17.5441){\circle*{0.7}}
\put(11.7422,17.3574){\circle*{0.7}}
\put(11.8149,17.1711){\circle*{0.7}}
\put(11.8885,16.9851){\circle*{0.7}}
\put(11.963,16.7995){\circle*{0.7}}
\put(12.0384,16.6143){\circle*{0.7}}
\put(12.1148,16.4295){\circle*{0.7}}
\put(12.1922,16.2451){\circle*{0.7}}
\put(12.2706,16.0611){\circle*{0.7}}
\put(12.35,15.8775){\circle*{0.7}}
\put(12.4304,15.6944){\circle*{0.7}}
\put(12.5119,15.5117){\circle*{0.7}}
\put(12.5944,15.3295){\circle*{0.7}}
\put(12.678,15.1479){\circle*{0.7}}
\put(12.7628,14.9667){\circle*{0.7}}
\put(12.8487,14.7861){\circle*{0.7}}
\put(12.9358,14.6061){\circle*{0.7}}
\put(13.024,14.4266){\circle*{0.7}}
\put(13.1135,14.2477){\circle*{0.7}}
\put(13.2042,14.0695){\circle*{0.7}}
\put(13.2962,13.8919){\circle*{0.7}}
\put(13.3894,13.7149){\circle*{0.7}}
\put(13.484,13.5387){\circle*{0.7}}
\put(13.5799,13.3632){\circle*{0.7}}
\put(13.6772,13.1885){\circle*{0.7}}
\put(13.7759,13.0145){\circle*{0.7}}
\put(13.876,12.8414){\circle*{0.7}}
\put(13.9776,12.6691){\circle*{0.7}}
\put(14.0806,12.4976){\circle*{0.7}}
\put(14.1851,12.3271){\circle*{0.7}}
\put(14.2912,12.1576){\circle*{0.7}}
\put(14.3988,11.989){\circle*{0.7}}
\put(14.5081,11.8215){\circle*{0.7}}
\put(14.6189,11.655){\circle*{0.7}}
\put(14.7314,11.4897){\circle*{0.7}}
\put(14.8456,11.3255){\circle*{0.7}}
\put(14.9615,11.1625){\circle*{0.7}}
\put(15.0791,11.0007){\circle*{0.7}}
\put(15.1985,10.8403){\circle*{0.7}}
\put(15.3197,10.6811){\circle*{0.7}}
\put(15.4427,10.5234){\circle*{0.7}}
\put(15.5675,10.3672){\circle*{0.7}}
\put(15.6942,10.2124){\circle*{0.7}}
\put(15.8228,10.0592){\circle*{0.7}}
\put(15.9533,9.90766){\circle*{0.7}}
\put(16.0857,9.75777){\circle*{0.7}}
\put(16.2201,9.60962){\circle*{0.7}}
\put(16.3564,9.46328){\circle*{0.7}}
\put(16.4947,9.31879){\circle*{0.7}}
\put(16.6349,9.17624){\circle*{0.7}}
\put(16.7772,9.03567){\circle*{0.7}}
\put(16.9215,8.89715){\circle*{0.7}}
\put(17.0677,8.76075){\circle*{0.7}}
\put(17.216,8.62653){\circle*{0.7}}
\put(17.3663,8.49455){\circle*{0.7}}
\put(17.5186,8.36488){\circle*{0.7}}
\put(17.6728,8.23757){\circle*{0.7}}
\put(17.829,8.11268){\circle*{0.7}}
\put(17.9872,7.99026){\circle*{0.7}}
\put(18.1473,7.87039){\circle*{0.7}}
\put(18.3093,7.75309){\circle*{0.7}}
\put(18.4731,7.63843){\circle*{0.7}}
\put(18.6388,7.52645){\circle*{0.7}}
\put(18.8064,7.41718){\circle*{0.7}}
\put(18.9756,7.31067){\circle*{0.7}}
\put(19.1466,7.20695){\circle*{0.7}}
\put(19.3193,7.10604){\circle*{0.7}}
\put(19.4936,7.00797){\circle*{0.7}}
\put(19.6695,6.91276){\circle*{0.7}}
\put(19.8469,6.82041){\circle*{0.7}}
\put(20.0258,6.73094){\circle*{0.7}}
\put(20.2061,6.64434){\circle*{0.7}}
\put(20.3877,6.56062){\circle*{0.7}}
\put(20.5706,6.47976){\circle*{0.7}}
\put(20.7548,6.40174){\circle*{0.7}}
\put(20.9401,6.32655){\circle*{0.7}}
\put(21.1265,6.25416){\circle*{0.7}}
\put(21.314,6.18454){\circle*{0.7}}
\put(21.5025,6.11765){\circle*{0.7}}
\put(21.6919,6.05346){\circle*{0.7}}
\put(21.8822,5.99193){\circle*{0.7}}
\put(22.0733,5.933){\circle*{0.7}}
\put(22.2652,5.87662){\circle*{0.7}}
\put(22.4578,5.82275){\circle*{0.7}}
\put(22.6511,5.77132){\circle*{0.7}}
\put(22.845,5.72229){\circle*{0.7}}
\put(23.0395,5.67558){\circle*{0.7}}
\put(23.2345,5.63113){\circle*{0.7}}
\put(23.43,5.58889){\circle*{0.7}}
\put(23.6259,5.54878){\circle*{0.7}}
\put(23.8223,5.51074){\circle*{0.7}}
\put(24.019,5.47471){\circle*{0.7}}
\put(24.2161,5.44061){\circle*{0.7}}
\put(24.4134,5.40839){\circle*{0.7}}
\put(24.6111,5.37797){\circle*{0.7}}
\put(24.8091,5.34929){\circle*{0.7}}
\put(25.0072,5.32228){\circle*{0.7}}
\put(25.2056,5.29688){\circle*{0.7}}
\put(25.4042,5.27303){\circle*{0.7}}
\put(25.6029,5.25065){\circle*{0.7}}
\put(25.8018,5.22969){\circle*{0.7}}
\put(26.0009,5.21008){\circle*{0.7}}
\put(26.2,5.19177){\circle*{0.7}}
\put(26.3993,5.1747){\circle*{0.7}}
\put(26.5986,5.1588){\circle*{0.7}}
\put(26.7981,5.14402){\circle*{0.7}}
\put(26.9976,5.13032){\circle*{0.7}}
\put(27.1972,5.11762){\circle*{0.7}}
\put(27.3969,5.10588){\circle*{0.7}}
\put(27.5966,5.09506){\circle*{0.7}}
\put(27.7963,5.08509){\circle*{0.7}}
\put(27.9961,5.07594){\circle*{0.7}}
\put(28.196,5.06755){\circle*{0.7}}
\put(28.3958,5.05988){\circle*{0.7}}
\put(28.5957,5.05289){\circle*{0.7}}
\put(28.7956,5.04654){\circle*{0.7}}
\put(28.9955,5.04078){\circle*{0.7}}
\put(29.1954,5.03557){\circle*{0.7}}
\put(29.3954,5.03089){\circle*{0.7}}
\put(29.5953,5.02668){\circle*{0.7}}
\put(29.7953,5.02293){\circle*{0.7}}
\put(29.9953,5.01958){\circle*{0.7}}
\put(30.1953,5.01662){\circle*{0.7}}
\put(30.3952,5.01401){\circle*{0.7}}
\put(30.5952,5.01172){\circle*{0.7}}
\put(30.7952,5.00943){\circle*{0.7}}
\put(30.9952,5.008){\circle*{0.7}}
\put(31.1952,5.00656){\circle*{0.7}}
\put(31.3952,5.00513){\circle*{0.7}}
\put(31.5952,5.00417){\circle*{0.7}}
\put(31.7952,5.00321){\circle*{0.7}}
\put(31.9952,5.00225){\circle*{0.7}}
\put(32.1952,5.00192){\circle*{0.7}}
\put(32.3952,5.00158){\circle*{0.7}}
\put(32.5952,5.00124){\circle*{0.7}}
\put(32.7952,5.00091){\circle*{0.7}}
\put(32.9952,5.00057){\circle*{0.7}}
\put(33.1952,5.00024){\circle*{0.7}}
\put(33.3952,5.0002){\circle*{0.7}}
\put(33.5952,5.00017){\circle*{0.7}}
\put(33.7952,5.00014){\circle*{0.7}}
\put(33.9952,5.00011){\circle*{0.7}}
\put(34.1952,5.00008){\circle*{0.7}}
\put(34.3952,5.00004){\circle*{0.7}}
\put(34.5952,5.00001){\circle*{0.7}}
\put(34.7952,4.99998){\circle*{0.7}}
\put(34.9952,4.99995){\circle*{0.7}}
\put(35.1952,4.99992){\circle*{0.7}}
\put(35.3952,4.99988){\circle*{0.7}}
\put(35.5952,4.99985){\circle*{0.7}}
\put(35.7952,4.99982){\circle*{0.7}}
\put(35.9952,4.99979){\circle*{0.7}}
\put(36.1952,5.00006){\circle*{0.7}}
\put(36.3952,5.00034){\circle*{0.7}}
\put(36.5952,5.0002){\circle*{0.7}}
\put(36.7952,5.00006){\circle*{0.7}}
\put(36.9952,5.00049){\circle*{0.7}}
\put(37.1952,5.00092){\circle*{0.7}}
\put(37.3952,5.00102){\circle*{0.7}}
\put(37.5952,5.00112){\circle*{0.7}}
\put(37.7952,5.00189){\circle*{0.7}}
\put(37.9952,5.00266){\circle*{0.7}}
\put(38.1952,5.00343){\circle*{0.7}}
\put(38.3952,5.00421){\circle*{0.7}}
\put(38.5952,5.00498){\circle*{0.7}}
\put(38.7952,5.00644){\circle*{0.7}}
\put(38.9952,5.00791){\circle*{0.7}}
\put(39.1952,5.00938){\circle*{0.7}}
\put(39.3951,5.01162){\circle*{0.7}}
\put(39.5951,5.01386){\circle*{0.7}}
\put(39.7951,5.01649){\circle*{0.7}}
\put(39.9951,5.01943){\circle*{0.7}}
\put(40.1951,5.02276){\circle*{0.7}}
\put(40.395,5.02649){\circle*{0.7}}
\put(40.595,5.03067){\circle*{0.7}}
\put(40.7949,5.03534){\circle*{0.7}}
\put(40.9949,5.04051){\circle*{0.7}}
\put(41.1948,5.04625){\circle*{0.7}}
\put(41.3947,5.05257){\circle*{0.7}}
\put(41.5946,5.05953){\circle*{0.7}}
\put(41.7944,5.06716){\circle*{0.7}}
\put(41.9942,5.07552){\circle*{0.7}}
\put(42.194,5.08463){\circle*{0.7}}
\put(42.3938,5.09456){\circle*{0.7}}
\put(42.5935,5.10534){\circle*{0.7}}
\put(42.7932,5.11703){\circle*{0.7}}
\put(42.9928,5.12968){\circle*{0.7}}
\put(43.1923,5.14334){\circle*{0.7}}
\put(43.3917,5.15806){\circle*{0.7}}
\put(43.5911,5.1739){\circle*{0.7}}
\put(43.7904,5.19092){\circle*{0.7}}
\put(43.9896,5.20917){\circle*{0.7}}
\put(44.1886,5.22871){\circle*{0.7}}
\put(44.3875,5.24961){\circle*{0.7}}
\put(44.5863,5.27191){\circle*{0.7}}
\put(44.7848,5.2957){\circle*{0.7}}
\put(44.9832,5.32102){\circle*{0.7}}
\put(45.1814,5.34795){\circle*{0.7}}
\put(45.3794,5.37655){\circle*{0.7}}
\put(45.577,5.40688){\circle*{0.7}}
\put(45.7744,5.43902){\circle*{0.7}}
\put(45.9715,5.47302){\circle*{0.7}}
\put(46.1683,5.50896){\circle*{0.7}}
\put(46.3646,5.5469){\circle*{0.7}}
\put(46.5606,5.5869){\circle*{0.7}}
\put(46.7561,5.62904){\circle*{0.7}}
\put(46.9511,5.67338){\circle*{0.7}}
\put(47.1456,5.71998){\circle*{0.7}}
\put(47.3396,5.76891){\circle*{0.7}}
\put(47.5329,5.82022){\circle*{0.7}}
\put(47.7255,5.87397){\circle*{0.7}}
\put(47.9174,5.93022){\circle*{0.7}}
\put(48.1086,5.98903){\circle*{0.7}}
\put(48.2989,6.05044){\circle*{0.7}}
\put(48.4884,6.1145){\circle*{0.7}}
\put(48.6769,6.18125){\circle*{0.7}}
\put(48.8645,6.25074){\circle*{0.7}}
\put(49.051,6.323){\circle*{0.7}}
\put(49.2363,6.39805){\circle*{0.7}}
\put(49.4205,6.47593){\circle*{0.7}}
\put(49.6035,6.55666){\circle*{0.7}}
\put(49.7852,6.64024){\circle*{0.7}}
\put(49.9656,6.7267){\circle*{0.7}}
\put(50.1445,6.81604){\circle*{0.7}}
\put(50.322,6.90824){\circle*{0.7}}
\put(50.4979,7.00332){\circle*{0.7}}
\put(50.6723,7.10125){\circle*{0.7}}
\put(50.8451,7.20202){\circle*{0.7}}
\put(51.0162,7.30561){\circle*{0.7}}
\put(51.1855,7.41199){\circle*{0.7}}
\put(51.3531,7.52112){\circle*{0.7}}
\put(51.5189,7.63297){\circle*{0.7}}
\put(51.6829,7.74751){\circle*{0.7}}
\put(51.845,7.86468){\circle*{0.7}}
\put(52.0052,7.98443){\circle*{0.7}}
\put(52.1634,8.10672){\circle*{0.7}}
\put(52.3197,8.23149){\circle*{0.7}}
\put(52.4741,8.35869){\circle*{0.7}}
\put(52.6264,8.48825){\circle*{0.7}}
\put(52.7768,8.62012){\circle*{0.7}}
\put(52.9252,8.75424){\circle*{0.7}}
\put(53.0715,8.89053){\circle*{0.7}}
\put(53.2159,9.02895){\circle*{0.7}}
\put(53.3583,9.16942){\circle*{0.7}}
\put(53.4986,9.31188){\circle*{0.7}}
\put(53.637,9.45627){\circle*{0.7}}
\put(53.7734,9.60253){\circle*{0.7}}
\put(53.9079,9.7506){\circle*{0.7}}
\put(54.0404,9.9004){\circle*{0.7}}
\put(54.171,10.0519){\circle*{0.7}}
\put(54.2996,10.205){\circle*{0.7}}
\put(54.4264,10.3597){\circle*{0.7}}
\put(54.5513,10.5159){\circle*{0.7}}
\put(54.6744,10.6735){\circle*{0.7}}
\put(54.7957,10.8326){\circle*{0.7}}
\put(54.9152,10.993){\circle*{0.7}}
\put(55.0329,11.1546){\circle*{0.7}}
\put(55.1488,11.3176){\circle*{0.7}}
\put(55.2631,11.4817){\circle*{0.7}}
\put(55.3757,11.647){\circle*{0.7}}
\put(55.4866,11.8134){\circle*{0.7}}
\put(55.5959,11.9809){\circle*{0.7}}
\put(55.7037,12.1494){\circle*{0.7}}
\put(55.8098,12.319){\circle*{0.7}}
\put(55.9144,12.4894){\circle*{0.7}}
\put(56.0175,12.6608){\circle*{0.7}}
\put(56.1191,12.833){\circle*{0.7}}
\put(56.2193,13.0062){\circle*{0.7}}
\put(56.3181,13.1801){\circle*{0.7}}
\put(56.4154,13.3548){\circle*{0.7}}
\put(56.5114,13.5302){\circle*{0.7}}
\put(56.606,13.7064){\circle*{0.7}}
\put(56.6994,13.8833){\circle*{0.7}}
\put(56.7914,14.0609){\circle*{0.7}}
\put(56.8822,14.2391){\circle*{0.7}}
\put(56.9717,14.418){\circle*{0.7}}
\put(57.06,14.5974){\circle*{0.7}}
\put(57.1471,14.7774){\circle*{0.7}}
\put(57.2331,14.958){\circle*{0.7}}
\put(57.3179,15.1391){\circle*{0.7}}
\put(57.4016,15.3208){\circle*{0.7}}
\put(57.4842,15.5029){\circle*{0.7}}
\put(57.5657,15.6856){\circle*{0.7}}
\put(57.6462,15.8687){\circle*{0.7}}
\put(57.7256,16.0522){\circle*{0.7}}
\put(57.804,16.2362){\circle*{0.7}}
\put(57.8815,16.4206){\circle*{0.7}}
\put(57.9579,16.6054){\circle*{0.7}}
\put(58.0334,16.7906){\circle*{0.7}}
\put(58.108,16.9762){\circle*{0.7}}
\put(58.1816,17.1621){\circle*{0.7}}
\put(58.2543,17.3484){\circle*{0.7}}
\put(58.3262,17.5351){\circle*{0.7}}
\put(58.3972,17.7221){\circle*{0.7}}
\put(58.4673,17.9094){\circle*{0.7}}
\put(58.5366,18.097){\circle*{0.7}}
\put(58.6051,18.2849){\circle*{0.7}}
\put(58.6728,18.4731){\circle*{0.7}}
\put(58.7397,18.6616){\circle*{0.7}}
\put(58.8058,18.8503){\circle*{0.7}}
\put(58.8711,19.0394){\circle*{0.7}}
\put(58.9358,19.2286){\circle*{0.7}}
\put(58.9996,19.4182){\circle*{0.7}}
\put(59.0628,19.6079){\circle*{0.7}}
\put(59.1252,19.7979){\circle*{0.7}}
\put(59.187,19.9881){\circle*{0.7}}
\put(59.2481,20.1786){\circle*{0.7}}
\put(59.3085,20.3693){\circle*{0.7}}
\put(59.3682,20.5601){\circle*{0.7}}
\put(59.4273,20.7512){\circle*{0.7}}
\put(59.4858,20.9425){\circle*{0.7}}
\put(59.5436,21.1339){\circle*{0.7}}
\put(59.6008,21.3256){\circle*{0.7}}
\put(59.6574,21.5174){\circle*{0.7}}
\put(59.7134,21.7094){\circle*{0.7}}
\put(59.7689,21.9015){\circle*{0.7}}
\put(59.8237,22.0939){\circle*{0.7}}
\put(59.878,22.2864){\circle*{0.7}}
\put(59.9318,22.479){\circle*{0.7}}
\put(59.985,22.6718){\circle*{0.7}}
\put(60.0376,22.8647){\circle*{0.7}}
\put(60.0897,23.0578){\circle*{0.7}}
\put(60.1413,23.2511){\circle*{0.7}}
\put(60.1924,23.4444){\circle*{0.7}}
\put(60.243,23.6379){\circle*{0.7}}
\put(60.2931,23.8316){\circle*{0.7}}
\put(60.3426,24.0253){\circle*{0.7}}
\put(60.3918,24.2192){\circle*{0.7}}
\put(60.4404,24.4132){\circle*{0.7}}
\put(60.4885,24.6073){\circle*{0.7}}
\put(60.5362,24.8015){\circle*{0.7}}
\put(60.5835,24.9959){\circle*{0.7}}
\put(60.6303,25.1903){\circle*{0.7}}
\put(60.6766,25.3849){\circle*{0.7}}
\put(60.7225,25.5795){\circle*{0.7}}
\put(60.768,25.7743){\circle*{0.7}}
\put(60.8131,25.9692){\circle*{0.7}}
\put(60.8577,26.1641){\circle*{0.7}}
\put(60.9019,26.3592){\circle*{0.7}}
\put(60.9458,26.5543){\circle*{0.7}}
\put(60.9892,26.7495){\circle*{0.7}}
\put(61.0322,26.9448){\circle*{0.7}}
\put(61.0748,27.1402){\circle*{0.7}}
\put(61.1171,27.3357){\circle*{0.7}}
\put(61.159,27.5313){\circle*{0.7}}
\put(61.2005,27.7269){\circle*{0.7}}
\put(61.2416,27.9227){\circle*{0.7}}
\put(61.2824,28.1185){\circle*{0.7}}
\put(61.3228,28.3144){\circle*{0.7}}
\put(61.3628,28.5103){\circle*{0.7}}
\put(61.4025,28.7063){\circle*{0.7}}
\put(61.4419,28.9024){\circle*{0.7}}
\put(61.4809,29.0986){\circle*{0.7}}
\put(61.5195,29.2948){\circle*{0.7}}
\put(61.5579,29.4911){\circle*{0.7}}
\put(61.5959,29.6874){\circle*{0.7}}
\put(61.6336,29.8839){\circle*{0.7}}
\put(61.6709,30.0803){\circle*{0.7}}
\put(61.708,30.2769){\circle*{0.7}}
\put(61.7447,30.4735){\circle*{0.7}}
\put(61.7811,30.6701){\circle*{0.7}}
\put(61.8172,30.8669){\circle*{0.7}}
\put(61.853,31.0636){\circle*{0.7}}
\put(61.8885,31.2604){\circle*{0.7}}
\put(61.9237,31.4573){\circle*{0.7}}
\put(61.9587,31.6542){\circle*{0.7}}
\put(61.9933,31.8512){\circle*{0.7}}
\put(62.0276,32.0483){\circle*{0.7}}
\put(62.0617,32.2453){\circle*{0.7}}
\put(62.0955,32.4425){\circle*{0.7}}
\put(62.129,32.6396){\circle*{0.7}}
\put(62.1622,32.8369){\circle*{0.7}}
\put(62.1952,33.0341){\circle*{0.7}}
\put(62.2279,33.2314){\circle*{0.7}}
\put(62.2603,33.4288){\circle*{0.7}}
\put(62.2924,33.6262){\circle*{0.7}}
\put(62.3243,33.8236){\circle*{0.7}}
\put(62.356,34.0211){\circle*{0.7}}
\put(62.3874,34.2186){\circle*{0.7}}
\put(62.4185,34.4162){\circle*{0.7}}
\put(62.4494,34.6138){\circle*{0.7}}
\put(62.4801,34.8114){\circle*{0.7}}
\put(62.5105,35.0091){\circle*{0.7}}
\put(62.5406,35.2068){\circle*{0.7}}
\put(62.5705,35.4046){\circle*{0.7}}
\put(62.6002,35.6023){\circle*{0.7}}
\put(62.6297,35.8002){\circle*{0.7}}
\put(62.6589,35.998){\circle*{0.7}}
\put(62.6879,36.1959){\circle*{0.7}}
\put(62.7166,36.3938){\circle*{0.7}}
\put(62.7451,36.5918){\circle*{0.7}}
\put(62.7735,36.7898){\circle*{0.7}}
\put(62.8015,36.9878){\circle*{0.7}}
\put(62.8294,37.1858){\circle*{0.7}}
\put(62.857,37.3839){\circle*{0.7}}
\put(62.8845,37.582){\circle*{0.7}}
\put(62.9117,37.7802){\circle*{0.7}}
\put(62.9387,37.9783){\circle*{0.7}}
\put(62.9655,38.1765){\circle*{0.7}}
\put(62.9921,38.3748){\circle*{0.7}}
\put(63.0185,38.573){\circle*{0.7}}
\put(63.0446,38.7713){\circle*{0.7}}
\put(63.0706,38.9696){\circle*{0.7}}
\put(63.0964,39.1679){\circle*{0.7}}
\put(63.122,39.3663){\circle*{0.7}}
\put(63.1473,39.5647){\circle*{0.7}}
\put(63.1725,39.7631){\circle*{0.7}}
\put(63.1975,39.9615){\circle*{0.7}}
\put(63.2223,40.16){\circle*{0.7}}
\put(63.2469,40.3585){\circle*{0.7}}
\put(63.2713,40.557){\circle*{0.7}}
\put(63.2955,40.7555){\circle*{0.7}}
\put(63.3195,40.954){\circle*{0.7}}
\put(63.3434,41.1526){\circle*{0.7}}
\put(63.367,41.3512){\circle*{0.7}}
\put(63.3905,41.5498){\circle*{0.7}}
\put(63.4138,41.7485){\circle*{0.7}}
\put(63.4369,41.9471){\circle*{0.7}}
\put(63.4599,42.1458){\circle*{0.7}}
\put(63.4826,42.3445){\circle*{0.7}}
\put(63.5052,42.5432){\circle*{0.7}}
\put(63.5276,42.742){\circle*{0.7}}
\put(63.5499,42.9407){\circle*{0.7}}
\put(63.5719,43.1395){\circle*{0.7}}
\put(63.5938,43.3383){\circle*{0.7}}
\put(63.6155,43.5371){\circle*{0.7}}
\put(63.6371,43.736){\circle*{0.7}}
\put(63.6585,43.9348){\circle*{0.7}}
\put(63.6797,44.1337){\circle*{0.7}}
\put(63.7007,44.3326){\circle*{0.7}}
\put(63.7216,44.5315){\circle*{0.7}}
\put(63.7423,44.7304){\circle*{0.7}}
\put(63.7629,44.9293){\circle*{0.7}}
\put(63.7833,45.1283){\circle*{0.7}}
\put(63.8035,45.3273){\circle*{0.7}}
\put(63.8236,45.5263){\circle*{0.7}}
\put(63.8435,45.7253){\circle*{0.7}}
\put(63.8633,45.9243){\circle*{0.7}}
\put(63.8829,46.1233){\circle*{0.7}}
\put(63.9023,46.3224){\circle*{0.7}}
\put(63.9216,46.5214){\circle*{0.7}}
\put(63.9407,46.7205){\circle*{0.7}}
\put(63.9597,46.9196){\circle*{0.7}}
\put(63.9785,47.1187){\circle*{0.7}}
\put(63.9972,47.3179){\circle*{0.7}}
\put(64.0157,47.517){\circle*{0.7}}
\put(64.0341,47.7162){\circle*{0.7}}
\put(64.0523,47.9153){\circle*{0.7}}
\put(64.0704,48.1145){\circle*{0.7}}
\put(64.0883,48.3137){\circle*{0.7}}
\put(64.1061,48.5129){\circle*{0.7}}
\put(64.1237,48.7121){\circle*{0.7}}
\put(64.1412,48.9114){\circle*{0.7}}
\put(64.1585,49.1106){\circle*{0.7}}
\put(64.1757,49.3099){\circle*{0.7}}
\put(64.1928,49.5092){\circle*{0.7}}
\put(64.2097,49.7084){\circle*{0.7}}
\put(64.2264,49.9077){\circle*{0.7}}
\put(64.243,50.107){\circle*{0.7}}
\put(64.2595,50.3064){\circle*{0.7}}
\put(64.2758,50.5057){\circle*{0.7}}
\put(64.292,50.705){\circle*{0.7}}
\put(64.308,50.9044){\circle*{0.7}}
\put(64.3239,51.1038){\circle*{0.7}}
\put(64.3397,51.3031){\circle*{0.7}}
\put(64.3553,51.5025){\circle*{0.7}}
\put(64.3708,51.7019){\circle*{0.7}}
\put(64.3861,51.9013){\circle*{0.7}}
\put(64.4013,52.1008){\circle*{0.7}}
\put(64.4163,52.3002){\circle*{0.7}}
\put(64.4313,52.4996){\circle*{0.7}}
\put(64.446,52.6991){\circle*{0.7}}
\put(64.4607,52.8986){\circle*{0.7}}
\put(64.4752,53.098){\circle*{0.7}}
\put(64.4895,53.2975){\circle*{0.7}}
\put(64.5038,53.497){\circle*{0.7}}
\put(64.5178,53.6965){\circle*{0.7}}
\put(64.5318,53.896){\circle*{0.7}}
\put(64.5456,54.0956){\circle*{0.7}}
\put(64.5592,54.2951){\circle*{0.7}}
\put(64.5728,54.4946){\circle*{0.7}}
\put(64.5861,54.6942){\circle*{0.7}}
\put(64.5994,54.8937){\circle*{0.7}}
\put(64.6125,55.0933){\circle*{0.7}}
\put(64.6255,55.2929){\circle*{0.7}}
\put(64.6383,55.4925){\circle*{0.7}}
\put(64.651,55.6921){\circle*{0.7}}
\put(64.6635,55.8917){\circle*{0.7}}
\put(64.6759,56.0913){\circle*{0.7}}
\put(64.6882,56.2909){\circle*{0.7}}
\put(64.7003,56.4906){\circle*{0.7}}
\put(64.7123,56.6902){\circle*{0.7}}
\put(64.7241,56.8898){\circle*{0.7}}
\put(64.7358,57.0895){\circle*{0.7}}
\put(64.7473,57.2892){\circle*{0.7}}
\put(64.7587,57.4888){\circle*{0.7}}
\put(64.77,57.6885){\circle*{0.7}}
\put(64.7811,57.8882){\circle*{0.7}}
\put(64.792,58.0879){\circle*{0.7}}
\put(64.8028,58.2876){\circle*{0.7}}
\put(64.8134,58.4873){\circle*{0.7}}
\put(64.8239,58.6871){\circle*{0.7}}
\put(64.8343,58.8868){\circle*{0.7}}
\put(64.8444,59.0865){\circle*{0.7}}
\put(64.8544,59.2863){\circle*{0.7}}
\put(64.8642,59.4861){\circle*{0.7}}
\put(64.8739,59.6858){\circle*{0.7}}
\put(64.8834,59.8856){\circle*{0.7}}
\put(64.8927,60.0854){\circle*{0.7}}
\put(64.9018,60.2852){\circle*{0.7}}
\put(64.9108,60.485){\circle*{0.7}}
\put(64.9195,60.6848){\circle*{0.7}}
\put(64.9281,60.8846){\circle*{0.7}}
\put(64.9364,61.0844){\circle*{0.7}}
\put(64.9446,61.2843){\circle*{0.7}}
\put(64.9525,61.4841){\circle*{0.7}}
\put(64.9601,61.684){\circle*{0.7}}
\put(64.9675,61.8838){\circle*{0.7}}
\put(64.9746,62.0837){\circle*{0.7}}
\put(64.9813,62.2836){\circle*{0.7}}
\put(64.9877,62.4835){\circle*{0.7}}
\put(64.9936,62.6834){\circle*{0.7}}

%% file: cw_rate_f1.tex
\put(85.03,39.2438){\circle*{0.7}}
\put(85.0379,39.0439){\circle*{0.7}}
\put(85.0461,38.8441){\circle*{0.7}}
\put(85.0546,38.6443){\circle*{0.7}}
\put(85.0634,38.4445){\circle*{0.7}}
\put(85.0725,38.2447){\circle*{0.7}}
\put(85.0819,38.0449){\circle*{0.7}}
\put(85.0915,37.8451){\circle*{0.7}}
\put(85.1014,37.6454){\circle*{0.7}}
\put(85.1116,37.4456){\circle*{0.7}}
\put(85.1219,37.2459){\circle*{0.7}}
\put(85.1326,37.0462){\circle*{0.7}}
\put(85.1434,36.8465){\circle*{0.7}}
\put(85.1545,36.6468){\circle*{0.7}}
\put(85.1658,36.4471){\circle*{0.7}}
\put(85.1773,36.2474){\circle*{0.7}}
\put(85.189,36.0478){\circle*{0.7}}
\put(85.201,35.8481){\circle*{0.7}}
\put(85.2131,35.6485){\circle*{0.7}}
\put(85.2255,35.4489){\circle*{0.7}}
\put(85.2381,35.2493){\circle*{0.7}}
\put(85.2509,35.0497){\circle*{0.7}}
\put(85.2639,34.8501){\circle*{0.7}}
\put(85.2771,34.6505){\circle*{0.7}}
\put(85.2905,34.451){\circle*{0.7}}
\put(85.3042,34.2515){\circle*{0.7}}
\put(85.318,34.0519){\circle*{0.7}}
\put(85.3321,33.8524){\circle*{0.7}}
\put(85.3463,33.6529){\circle*{0.7}}
\put(85.3608,33.4535){\circle*{0.7}}
\put(85.3755,33.254){\circle*{0.7}}
\put(85.3904,33.0546){\circle*{0.7}}
\put(85.4055,32.8551){\circle*{0.7}}
\put(85.4208,32.6557){\circle*{0.7}}
\put(85.4363,32.4563){\circle*{0.7}}
\put(85.4521,32.2569){\circle*{0.7}}
\put(85.468,32.0576){\circle*{0.7}}
\put(85.4842,31.8582){\circle*{0.7}}
\put(85.5006,31.6589){\circle*{0.7}}
\put(85.5172,31.4596){\circle*{0.7}}
\put(85.534,31.2603){\circle*{0.7}}
\put(85.551,31.061){\circle*{0.7}}
\put(85.5683,30.8618){\circle*{0.7}}
\put(85.5857,30.6625){\circle*{0.7}}
\put(85.6034,30.4633){\circle*{0.7}}
\put(85.6213,30.2641){\circle*{0.7}}
\put(85.6395,30.065){\circle*{0.7}}
\put(85.6578,29.8658){\circle*{0.7}}
\put(85.6764,29.6667){\circle*{0.7}}
\put(85.6953,29.4676){\circle*{0.7}}
\put(85.7143,29.2685){\circle*{0.7}}
\put(85.7336,29.0694){\circle*{0.7}}
\put(85.7532,28.8704){\circle*{0.7}}
\put(85.7729,28.6713){\circle*{0.7}}
\put(85.793,28.4723){\circle*{0.7}}
\put(85.8132,28.2734){\circle*{0.7}}
\put(85.8337,28.0744){\circle*{0.7}}
\put(85.8545,27.8755){\circle*{0.7}}
\put(85.8755,27.6766){\circle*{0.7}}
\put(85.8967,27.4777){\circle*{0.7}}
\put(85.9182,27.2789){\circle*{0.7}}
\put(85.94,27.0801){\circle*{0.7}}
\put(85.962,26.8813){\circle*{0.7}}
\put(85.9843,26.6826){\circle*{0.7}}
\put(86.0069,26.4838){\circle*{0.7}}
\put(86.0297,26.2851){\circle*{0.7}}
\put(86.0528,26.0865){\circle*{0.7}}
\put(86.0762,25.8878){\circle*{0.7}}
\put(86.0999,25.6893){\circle*{0.7}}
\put(86.1238,25.4907){\circle*{0.7}}
\put(86.148,25.2922){\circle*{0.7}}
\put(86.1726,25.0937){\circle*{0.7}}
\put(86.1974,24.8952){\circle*{0.7}}
\put(86.2225,24.6968){\circle*{0.7}}
\put(86.2479,24.4984){\circle*{0.7}}
\put(86.2736,24.3001){\circle*{0.7}}
\put(86.2996,24.1018){\circle*{0.7}}
\put(86.326,23.9035){\circle*{0.7}}
\put(86.3527,23.7053){\circle*{0.7}}
\put(86.3796,23.5071){\circle*{0.7}}
\put(86.407,23.309){\circle*{0.7}}
\put(86.4346,23.1109){\circle*{0.7}}
\put(86.4626,22.9129){\circle*{0.7}}
\put(86.4909,22.7149){\circle*{0.7}}
\put(86.5196,22.517){\circle*{0.7}}
\put(86.5486,22.3191){\circle*{0.7}}
\put(86.578,22.1213){\circle*{0.7}}
\put(86.6078,21.9235){\circle*{0.7}}
\put(86.6379,21.7258){\circle*{0.7}}
\put(86.6684,21.5281){\circle*{0.7}}
\put(86.6993,21.3305){\circle*{0.7}}
\put(86.7306,21.133){\circle*{0.7}}
\put(86.7622,20.9355){\circle*{0.7}}
\put(86.7943,20.7381){\circle*{0.7}}
\put(86.8268,20.5408){\circle*{0.7}}
\put(86.8597,20.3435){\circle*{0.7}}
\put(86.893,20.1463){\circle*{0.7}}
\put(86.9268,19.9491){\circle*{0.7}}
\put(86.961,19.7521){\circle*{0.7}}
\put(86.9957,19.5551){\circle*{0.7}}
\put(87.0308,19.3582){\circle*{0.7}}
\put(87.0663,19.1614){\circle*{0.7}}
\put(87.1024,18.9647){\circle*{0.7}}
\put(87.1389,18.7681){\circle*{0.7}}
\put(87.176,18.5715){\circle*{0.7}}
\put(87.2135,18.3751){\circle*{0.7}}
\put(87.2516,18.1787){\circle*{0.7}}
\put(87.2901,17.9825){\circle*{0.7}}
\put(87.3293,17.7863){\circle*{0.7}}
\put(87.3689,17.5903){\circle*{0.7}}
\put(87.4092,17.3944){\circle*{0.7}}
\put(87.45,17.1986){\circle*{0.7}}
\put(87.4914,17.0029){\circle*{0.7}}
\put(87.5334,16.8074){\circle*{0.7}}
\put(87.576,16.612){\circle*{0.7}}
\put(87.6192,16.4167){\circle*{0.7}}
\put(87.6631,16.2216){\circle*{0.7}}
\put(87.7076,16.0266){\circle*{0.7}}
\put(87.7528,15.8318){\circle*{0.7}}
\put(87.7987,15.6371){\circle*{0.7}}
\put(87.8453,15.4426){\circle*{0.7}}
\put(87.8927,15.2483){\circle*{0.7}}
\put(87.9408,15.0542){\circle*{0.7}}
\put(87.9896,14.8602){\circle*{0.7}}
\put(88.0393,14.6665){\circle*{0.7}}
\put(88.0897,14.473){\circle*{0.7}}
\put(88.141,14.2797){\circle*{0.7}}
\put(88.1931,14.0866){\circle*{0.7}}
\put(88.2461,13.8937){\circle*{0.7}}
\put(88.3001,13.7011){\circle*{0.7}}
\put(88.3549,13.5088){\circle*{0.7}}
\put(88.4107,13.3167){\circle*{0.7}}
\put(88.4675,13.125){\circle*{0.7}}
\put(88.5253,12.9335){\circle*{0.7}}
\put(88.5842,12.7424){\circle*{0.7}}
\put(88.6442,12.5516){\circle*{0.7}}
\put(88.7052,12.3611){\circle*{0.7}}
\put(88.7675,12.1711){\circle*{0.7}}
\put(88.8309,11.9814){\circle*{0.7}}
\put(88.8956,11.7921){\circle*{0.7}}
\put(88.9615,11.6033){\circle*{0.7}}
\put(89.0288,11.415){\circle*{0.7}}
\put(89.0974,11.2271){\circle*{0.7}}
\put(89.1675,11.0398){\circle*{0.7}}
\put(89.239,10.853){\circle*{0.7}}
\put(89.3121,10.6668){\circle*{0.7}}
\put(89.3868,10.4813){\circle*{0.7}}
\put(89.4631,10.2964){\circle*{0.7}}
\put(89.5411,10.1123){\circle*{0.7}}
\put(89.6209,9.92891){\circle*{0.7}}
\put(89.7026,9.74635){\circle*{0.7}}
\put(89.7862,9.56466){\circle*{0.7}}
\put(89.8719,9.38392){\circle*{0.7}}
\put(89.9596,9.2042){\circle*{0.7}}
\put(90.0495,9.02556){\circle*{0.7}}
\put(90.1417,8.84809){\circle*{0.7}}
\put(90.2364,8.67188){\circle*{0.7}}
\put(90.3335,8.49704){\circle*{0.7}}
\put(90.4332,8.32367){\circle*{0.7}}
\put(90.5356,8.1519){\circle*{0.7}}
\put(90.6409,7.98186){\circle*{0.7}}
\put(90.7492,7.8137){\circle*{0.7}}
\put(90.8606,7.64758){\circle*{0.7}}
\put(90.9752,7.48369){\circle*{0.7}}
\put(91.0932,7.32222){\circle*{0.7}}
\put(91.2147,7.16338){\circle*{0.7}}
\put(91.34,7.00743){\circle*{0.7}}
\put(91.469,6.85463){\circle*{0.7}}
\put(91.602,6.70525){\circle*{0.7}}
\put(91.7391,6.55963){\circle*{0.7}}
\put(91.8804,6.4181){\circle*{0.7}}
\put(92.026,6.28102){\circle*{0.7}}
\put(92.1761,6.14879){\circle*{0.7}}
\put(92.3306,6.02183){\circle*{0.7}}
\put(92.4897,5.90056){\circle*{0.7}}
\put(92.6532,5.78543){\circle*{0.7}}
\put(92.8212,5.67688){\circle*{0.7}}
\put(92.9935,5.57535){\circle*{0.7}}
\put(93.17,5.48126){\circle*{0.7}}
\put(93.3504,5.395){\circle*{0.7}}
\put(93.5345,5.3169){\circle*{0.7}}
\put(93.722,5.24723){\circle*{0.7}}
\put(93.9125,5.1862){\circle*{0.7}}
\put(94.1055,5.13392){\circle*{0.7}}
\put(94.3007,5.09042){\circle*{0.7}}
\put(94.4977,5.05566){\circle*{0.7}}
\put(94.696,5.0295){\circle*{0.7}}
\put(94.8952,5.01174){\circle*{0.7}}
\put(95.0949,5.0021){\circle*{0.7}}
\put(95.2949,5.00026){\circle*{0.7}}
\put(95.4949,5.00587){\circle*{0.7}}
\put(95.6945,5.01855){\circle*{0.7}}
\put(95.8935,5.0379){\circle*{0.7}}
\put(96.0919,5.06352){\circle*{0.7}}
\put(96.2894,5.09501){\circle*{0.7}}
\put(96.4859,5.13199){\circle*{0.7}}
\put(96.6814,5.17408){\circle*{0.7}}
\put(96.8759,5.22092){\circle*{0.7}}
\put(97.0692,5.27218){\circle*{0.7}}
\put(97.2614,5.32753){\circle*{0.7}}
\put(97.4524,5.38668){\circle*{0.7}}
\put(97.6424,5.44934){\circle*{0.7}}
\put(97.8312,5.51525){\circle*{0.7}}
\put(98.019,5.58417){\circle*{0.7}}
\put(98.2057,5.65587){\circle*{0.7}}
\put(98.3914,5.73014){\circle*{0.7}}
\put(98.5761,5.80678){\circle*{0.7}}
\put(98.7599,5.88561){\circle*{0.7}}
\put(98.9428,5.96647){\circle*{0.7}}
\put(99.1249,6.04918){\circle*{0.7}}
\put(99.3062,6.13362){\circle*{0.7}}
\put(99.4868,6.21963){\circle*{0.7}}
\put(99.6666,6.3071){\circle*{0.7}}
\put(99.8459,6.39589){\circle*{0.7}}
\put(100.024,6.4859){\circle*{0.7}}
\put(100.202,6.57703){\circle*{0.7}}
\put(100.38,6.66916){\circle*{0.7}}
\put(100.557,6.76221){\circle*{0.7}}
\put(100.734,6.8561){\circle*{0.7}}
\put(100.91,6.95072){\circle*{0.7}}
\put(101.086,7.04601){\circle*{0.7}}
\put(101.261,7.14189){\circle*{0.7}}
\put(101.436,7.23829){\circle*{0.7}}
\put(101.611,7.33513){\circle*{0.7}}
\put(101.786,7.43235){\circle*{0.7}}
\put(101.961,7.52988){\circle*{0.7}}
\put(102.135,7.62767){\circle*{0.7}}
\put(102.31,7.72547){\circle*{0.7}}
\put(102.484,7.82379){\circle*{0.7}}
\put(102.658,7.92211){\circle*{0.7}}
\put(102.832,8.02042){\circle*{0.7}}
\put(103.007,8.11844){\circle*{0.7}}
\put(103.181,8.21646){\circle*{0.7}}
\put(103.355,8.31447){\circle*{0.7}}
\put(103.53,8.41249){\circle*{0.7}}
\put(103.704,8.51004){\circle*{0.7}}
\put(103.879,8.6076){\circle*{0.7}}
\put(104.054,8.70442){\circle*{0.7}}
\put(104.229,8.80124){\circle*{0.7}}
\put(104.404,8.89735){\circle*{0.7}}
\put(104.58,8.99315){\circle*{0.7}}
\put(104.756,9.08846){\circle*{0.7}}
\put(104.932,9.18322){\circle*{0.7}}
\put(105.108,9.2774){\circle*{0.7}}
\put(105.285,9.37094){\circle*{0.7}}
\put(105.462,9.4638){\circle*{0.7}}
\put(105.64,9.55595){\circle*{0.7}}
\put(105.817,9.64733){\circle*{0.7}}
\put(105.996,9.7379){\circle*{0.7}}
\put(106.174,9.82761){\circle*{0.7}}
\put(106.354,9.91643){\circle*{0.7}}
\put(106.533,10.0043){\circle*{0.7}}
\put(106.714,10.0912){\circle*{0.7}}
\put(106.894,10.177){\circle*{0.7}}
\put(107.075,10.2618){\circle*{0.7}}
\put(107.257,10.3454){\circle*{0.7}}
\put(107.439,10.4279){\circle*{0.7}}
\put(107.622,10.5091){\circle*{0.7}}
\put(107.805,10.5891){\circle*{0.7}}
\put(107.989,10.6677){\circle*{0.7}}
\put(108.174,10.745){\circle*{0.7}}
\put(108.359,10.8209){\circle*{0.7}}
\put(108.544,10.8953){\circle*{0.7}}
\put(108.731,10.9682){\circle*{0.7}}
\put(108.917,11.0396){\circle*{0.7}}
\put(109.105,11.1093){\circle*{0.7}}
\put(109.293,11.1774){\circle*{0.7}}
\put(109.482,11.2438){\circle*{0.7}}
\put(109.671,11.3084){\circle*{0.7}}
\put(109.861,11.3712){\circle*{0.7}}
\put(110.051,11.4322){\circle*{0.7}}
\put(110.242,11.4913){\circle*{0.7}}
\put(110.434,11.5484){\circle*{0.7}}
\put(110.626,11.6036){\circle*{0.7}}
\put(110.819,11.6567){\circle*{0.7}}
\put(111.012,11.7077){\circle*{0.7}}
\put(111.206,11.7566){\circle*{0.7}}
\put(111.401,11.8033){\circle*{0.7}}
\put(111.596,11.8478){\circle*{0.7}}
\put(111.791,11.89){\circle*{0.7}}
\put(111.987,11.9299){\circle*{0.7}}
\put(112.184,11.9674){\circle*{0.7}}
\put(112.381,12.0026){\circle*{0.7}}
\put(112.578,12.0354){\circle*{0.7}}
\put(112.776,12.0657){\circle*{0.7}}
\put(112.974,12.0935){\circle*{0.7}}
\put(113.172,12.1188){\circle*{0.7}}
\put(113.371,12.1416){\circle*{0.7}}
\put(113.57,12.1618){\circle*{0.7}}
\put(113.769,12.1794){\circle*{0.7}}
\put(113.968,12.1944){\circle*{0.7}}
\put(114.168,12.2068){\circle*{0.7}}
\put(114.368,12.2166){\circle*{0.7}}
\put(114.568,12.2237){\circle*{0.7}}
\put(114.768,12.2281){\circle*{0.7}}
\put(114.967,12.2298){\circle*{0.7}}
\put(115.167,12.2289){\circle*{0.7}}
\put(115.367,12.2254){\circle*{0.7}}
\put(115.567,12.2192){\circle*{0.7}}
\put(115.767,12.2103){\circle*{0.7}}
\put(115.967,12.1988){\circle*{0.7}}
\put(116.166,12.1846){\circle*{0.7}}
\put(116.366,12.1678){\circle*{0.7}}
\put(116.565,12.1485){\circle*{0.7}}
\put(116.763,12.1265){\circle*{0.7}}
\put(116.962,12.102){\circle*{0.7}}
\put(117.16,12.075){\circle*{0.7}}
\put(117.358,12.0455){\circle*{0.7}}
\put(117.555,12.0135){\circle*{0.7}}
\put(117.752,11.9791){\circle*{0.7}}
\put(117.949,11.9423){\circle*{0.7}}
\put(118.145,11.9032){\circle*{0.7}}
\put(118.341,11.8617){\circle*{0.7}}
\put(118.536,11.818){\circle*{0.7}}
\put(118.731,11.772){\circle*{0.7}}
\put(118.925,11.7238){\circle*{0.7}}
\put(119.118,11.6735){\circle*{0.7}}
\put(119.311,11.6211){\circle*{0.7}}
\put(119.504,11.5666){\circle*{0.7}}
\put(119.696,11.5101){\circle*{0.7}}
\put(119.887,11.4516){\circle*{0.7}}
\put(120.077,11.3913){\circle*{0.7}}
\put(120.268,11.329){\circle*{0.7}}
\put(120.457,11.265){\circle*{0.7}}
\put(120.646,11.1992){\circle*{0.7}}
\put(120.834,11.1316){\circle*{0.7}}
\put(121.022,11.0624){\circle*{0.7}}
\put(121.209,10.9916){\circle*{0.7}}
\put(121.395,10.9192){\circle*{0.7}}
\put(121.581,10.8452){\circle*{0.7}}
\put(121.766,10.7698){\circle*{0.7}}
\put(121.951,10.693){\circle*{0.7}}
\put(122.135,10.6148){\circle*{0.7}}
\put(122.319,10.5352){\circle*{0.7}}
\put(122.501,10.4544){\circle*{0.7}}
\put(122.684,10.3723){\circle*{0.7}}
\put(122.866,10.2891){\circle*{0.7}}
\put(123.047,10.2047){\circle*{0.7}}
\put(123.228,10.1192){\circle*{0.7}}
\put(123.408,10.0326){\circle*{0.7}}
\put(123.588,9.94509){\circle*{0.7}}
\put(123.767,9.85658){\circle*{0.7}}
\put(123.946,9.76715){\circle*{0.7}}
\put(124.125,9.67686){\circle*{0.7}}
\put(124.303,9.58574){\circle*{0.7}}
\put(124.48,9.49384){\circle*{0.7}}
\put(124.658,9.4012){\circle*{0.7}}
\put(124.834,9.30787){\circle*{0.7}}
\put(125.011,9.2139){\circle*{0.7}}
\put(125.187,9.11932){\circle*{0.7}}
\put(125.363,9.02419){\circle*{0.7}}
\put(125.539,8.92854){\circle*{0.7}}
\put(125.714,8.83242){\circle*{0.7}}
\put(125.889,8.73588){\circle*{0.7}}
\put(126.064,8.63896){\circle*{0.7}}
\put(126.239,8.54172){\circle*{0.7}}
\put(126.414,8.44418){\circle*{0.7}}
\put(126.588,8.34641){\circle*{0.7}}
\put(126.763,8.24864){\circle*{0.7}}
\put(126.937,8.15035){\circle*{0.7}}
\put(127.111,8.05206){\circle*{0.7}}
\put(127.285,7.95377){\circle*{0.7}}
\put(127.459,7.85547){\circle*{0.7}}
\put(127.634,7.75754){\circle*{0.7}}
\put(127.808,7.65961){\circle*{0.7}}
\put(127.982,7.56168){\circle*{0.7}}
\put(128.157,7.46402){\circle*{0.7}}
\put(128.332,7.36669){\circle*{0.7}}
\put(128.507,7.26972){\circle*{0.7}}
\put(128.682,7.17317){\circle*{0.7}}
\put(128.857,7.07711){\circle*{0.7}}
\put(129.033,6.98163){\circle*{0.7}}
\put(129.209,6.88677){\circle*{0.7}}
\put(129.386,6.79264){\circle*{0.7}}
\put(129.562,6.69931){\circle*{0.7}}
\put(129.74,6.60687){\circle*{0.7}}
\put(129.918,6.5154){\circle*{0.7}}
\put(130.096,6.42502){\circle*{0.7}}
\put(130.275,6.33582){\circle*{0.7}}
\put(130.455,6.24791){\circle*{0.7}}
\put(130.635,6.16141){\circle*{0.7}}
\put(130.816,6.07645){\circle*{0.7}}
\put(130.998,5.99315){\circle*{0.7}}
\put(131.181,5.91168){\circle*{0.7}}
\put(131.364,5.83217){\circle*{0.7}}
\put(131.548,5.7548){\circle*{0.7}}
\put(131.734,5.67974){\circle*{0.7}}
\put(131.92,5.60718){\circle*{0.7}}
\put(132.108,5.53733){\circle*{0.7}}
\put(132.296,5.47042){\circle*{0.7}}
\put(132.486,5.40667){\circle*{0.7}}
\put(132.676,5.34636){\circle*{0.7}}
\put(132.868,5.28974){\circle*{0.7}}
\put(133.061,5.23712){\circle*{0.7}}
\put(133.255,5.1888){\circle*{0.7}}
\put(133.45,5.14513){\circle*{0.7}}
\put(133.647,5.10645){\circle*{0.7}}
\put(133.844,5.07313){\circle*{0.7}}
\put(134.042,5.04556){\circle*{0.7}}
\put(134.241,5.02413){\circle*{0.7}}
\put(134.44,5.00924){\circle*{0.7}}
\put(134.64,5.00128){\circle*{0.7}}
\put(134.84,5.00066){\circle*{0.7}}
\put(135.04,5.00773){\circle*{0.7}}
\put(135.239,5.02282){\circle*{0.7}}
\put(135.438,5.04622){\circle*{0.7}}
\put(135.635,5.07817){\circle*{0.7}}
\put(135.831,5.11882){\circle*{0.7}}
\put(136.025,5.16824){\circle*{0.7}}
\put(136.216,5.22644){\circle*{0.7}}
\put(136.405,5.29332){\circle*{0.7}}
\put(136.59,5.36871){\circle*{0.7}}
\put(136.772,5.45235){\circle*{0.7}}
\put(136.95,5.54394){\circle*{0.7}}
\put(137.123,5.64309){\circle*{0.7}}
\put(137.293,5.7494){\circle*{0.7}}
\put(137.458,5.86244){\circle*{0.7}}
\put(137.618,5.98177){\circle*{0.7}}
\put(137.774,6.10693){\circle*{0.7}}
\put(137.926,6.23749){\circle*{0.7}}
\put(138.073,6.37304){\circle*{0.7}}
\put(138.215,6.51316){\circle*{0.7}}
\put(138.354,6.6575){\circle*{0.7}}
\put(138.488,6.80569){\circle*{0.7}}
\put(138.619,6.95741){\circle*{0.7}}
\put(138.745,7.11237){\circle*{0.7}}
\put(138.868,7.27029){\circle*{0.7}}
\put(138.987,7.43093){\circle*{0.7}}
\put(139.103,7.59406){\circle*{0.7}}
\put(139.215,7.75948){\circle*{0.7}}
\put(139.324,7.92699){\circle*{0.7}}
\put(139.43,8.09644){\circle*{0.7}}
\put(139.534,8.26766){\circle*{0.7}}
\put(139.634,8.44052){\circle*{0.7}}
\put(139.732,8.6149){\circle*{0.7}}
\put(139.828,8.79067){\circle*{0.7}}
\put(139.921,8.96774){\circle*{0.7}}
\put(140.011,9.14601){\circle*{0.7}}
\put(140.1,9.32539){\circle*{0.7}}
\put(140.186,9.50581){\circle*{0.7}}
\put(140.27,9.6872){\circle*{0.7}}
\put(140.353,9.86948){\circle*{0.7}}
\put(140.433,10.0526){\circle*{0.7}}
\put(140.512,10.2365){\circle*{0.7}}
\put(140.589,10.4211){\circle*{0.7}}
\put(140.664,10.6065){\circle*{0.7}}
\put(140.737,10.7924){\circle*{0.7}}
\put(140.809,10.979){\circle*{0.7}}
\put(140.88,11.1662){\circle*{0.7}}
\put(140.949,11.3539){\circle*{0.7}}
\put(141.017,11.542){\circle*{0.7}}
\put(141.083,11.7307){\circle*{0.7}}
\put(141.148,11.9198){\circle*{0.7}}
\put(141.212,12.1094){\circle*{0.7}}
\put(141.275,12.2993){\circle*{0.7}}
\put(141.336,12.4896){\circle*{0.7}}
\put(141.396,12.6803){\circle*{0.7}}
\put(141.456,12.8714){\circle*{0.7}}
\put(141.514,13.0627){\circle*{0.7}}
\put(141.571,13.2544){\circle*{0.7}}
\put(141.627,13.4463){\circle*{0.7}}
\put(141.682,13.6386){\circle*{0.7}}
\put(141.736,13.8311){\circle*{0.7}}
\put(141.79,14.0239){\circle*{0.7}}
\put(141.842,14.2169){\circle*{0.7}}
\put(141.894,14.4101){\circle*{0.7}}
\put(141.944,14.6036){\circle*{0.7}}
\put(141.994,14.7973){\circle*{0.7}}
\put(142.043,14.9911){\circle*{0.7}}
\put(142.092,15.1852){\circle*{0.7}}
\put(142.139,15.3795){\circle*{0.7}}
\put(142.186,15.5739){\circle*{0.7}}
\put(142.232,15.7685){\circle*{0.7}}
\put(142.278,15.9633){\circle*{0.7}}
\put(142.323,16.1582){\circle*{0.7}}
\put(142.367,16.3533){\circle*{0.7}}
\put(142.41,16.5485){\circle*{0.7}}
\put(142.453,16.7439){\circle*{0.7}}
\put(142.495,16.9394){\circle*{0.7}}
\put(142.537,17.135){\circle*{0.7}}
\put(142.578,17.3308){\circle*{0.7}}
\put(142.618,17.5266){\circle*{0.7}}
\put(142.658,17.7226){\circle*{0.7}}
\put(142.697,17.9187){\circle*{0.7}}
\put(142.736,18.1149){\circle*{0.7}}
\put(142.774,18.3112){\circle*{0.7}}
\put(142.812,18.5077){\circle*{0.7}}
\put(142.849,18.7042){\circle*{0.7}}
\put(142.886,18.9008){\circle*{0.7}}
\put(142.922,19.0975){\circle*{0.7}}
\put(142.958,19.2942){\circle*{0.7}}
\put(142.993,19.4911){\circle*{0.7}}
\put(143.028,19.6881){\circle*{0.7}}
\put(143.062,19.8851){\circle*{0.7}}
\put(143.096,20.0822){\circle*{0.7}}
\put(143.13,20.2794){\circle*{0.7}}
\put(143.163,20.4766){\circle*{0.7}}
\put(143.195,20.6739){\circle*{0.7}}
\put(143.227,20.8713){\circle*{0.7}}
\put(143.259,21.0688){\circle*{0.7}}
\put(143.291,21.2663){\circle*{0.7}}
\put(143.322,21.4639){\circle*{0.7}}
\put(143.352,21.6615){\circle*{0.7}}
\put(143.382,21.8592){\circle*{0.7}}
\put(143.412,22.057){\circle*{0.7}}
\put(143.442,22.2548){\circle*{0.7}}
\put(143.471,22.4527){\circle*{0.7}}
\put(143.5,22.6506){\circle*{0.7}}
\put(143.528,22.8485){\circle*{0.7}}
\put(143.556,23.0466){\circle*{0.7}}
\put(143.584,23.2446){\circle*{0.7}}
\put(143.612,23.4427){\circle*{0.7}}
\put(143.639,23.6409){\circle*{0.7}}
\put(143.665,23.8391){\circle*{0.7}}
\put(143.692,24.0373){\circle*{0.7}}
\put(143.718,24.2356){\circle*{0.7}}
\put(143.744,24.434){\circle*{0.7}}
\put(143.769,24.6323){\circle*{0.7}}
\put(143.795,24.8307){\circle*{0.7}}
\put(143.819,25.0292){\circle*{0.7}}
\put(143.844,25.2276){\circle*{0.7}}
\put(143.868,25.4262){\circle*{0.7}}
\put(143.892,25.6247){\circle*{0.7}}
\put(143.916,25.8233){\circle*{0.7}}
\put(143.94,26.0219){\circle*{0.7}}
\put(143.963,26.2206){\circle*{0.7}}
\put(143.986,26.4192){\circle*{0.7}}
\put(144.008,26.618){\circle*{0.7}}
\put(144.031,26.8167){\circle*{0.7}}
\put(144.053,27.0155){\circle*{0.7}}
\put(144.075,27.2143){\circle*{0.7}}
\put(144.096,27.4131){\circle*{0.7}}
\put(144.118,27.612){\circle*{0.7}}
\put(144.139,27.8109){\circle*{0.7}}
\put(144.16,28.0098){\circle*{0.7}}
\put(144.18,28.2087){\circle*{0.7}}
\put(144.2,28.4077){\circle*{0.7}}
\put(144.221,28.6067){\circle*{0.7}}
\put(144.24,28.8057){\circle*{0.7}}
\put(144.26,29.0047){\circle*{0.7}}
\put(144.279,29.2038){\circle*{0.7}}
\put(144.299,29.4028){\circle*{0.7}}
\put(144.317,29.602){\circle*{0.7}}
\put(144.336,29.8011){\circle*{0.7}}
\put(144.355,30.0002){\circle*{0.7}}
\put(144.373,30.1994){\circle*{0.7}}
\put(144.391,30.3986){\circle*{0.7}}
\put(144.409,30.5978){\circle*{0.7}}
\put(144.426,30.797){\circle*{0.7}}
\put(144.443,30.9963){\circle*{0.7}}
\put(144.461,31.1955){\circle*{0.7}}
\put(144.477,31.3948){\circle*{0.7}}
\put(144.494,31.5941){\circle*{0.7}}
\put(144.511,31.7935){\circle*{0.7}}
\put(144.527,31.9928){\circle*{0.7}}
\put(144.543,32.1921){\circle*{0.7}}
\put(144.559,32.3915){\circle*{0.7}}
\put(144.574,32.5909){\circle*{0.7}}
\put(144.59,32.7903){\circle*{0.7}}
\put(144.605,32.9897){\circle*{0.7}}
\put(144.62,33.1892){\circle*{0.7}}
\put(144.634,33.3886){\circle*{0.7}}
\put(144.649,33.5881){\circle*{0.7}}
\put(144.663,33.7876){\circle*{0.7}}
\put(144.677,33.9871){\circle*{0.7}}
\put(144.691,34.1866){\circle*{0.7}}
\put(144.705,34.3861){\circle*{0.7}}
\put(144.719,34.5857){\circle*{0.7}}
\put(144.732,34.7852){\circle*{0.7}}
\put(144.745,34.9848){\circle*{0.7}}
\put(144.758,35.1844){\circle*{0.7}}
\put(144.77,35.384){\circle*{0.7}}
\put(144.783,35.5836){\circle*{0.7}}
\put(144.795,35.7832){\circle*{0.7}}
\put(144.807,35.9829){\circle*{0.7}}
\put(144.819,36.1825){\circle*{0.7}}
\put(144.831,36.3822){\circle*{0.7}}
\put(144.842,36.5819){\circle*{0.7}}
\put(144.853,36.7816){\circle*{0.7}}
\put(144.864,36.9813){\circle*{0.7}}
\put(144.875,37.181){\circle*{0.7}}
\put(144.885,37.3807){\circle*{0.7}}
\put(144.895,37.5804){\circle*{0.7}}
\put(144.905,37.7802){\circle*{0.7}}
\put(144.915,37.98){\circle*{0.7}}
\put(144.924,38.1797){\circle*{0.7}}
\put(144.934,38.3795){\circle*{0.7}}
\put(144.943,38.5793){\circle*{0.7}}
\put(144.951,38.7791){\circle*{0.7}}
\put(144.96,38.979){\circle*{0.7}}
\put(144.967,39.1788){\circle*{0.7}}
\put(144.975,39.3787){\circle*{0.7}}

%% file: sph_rate_f.tex
\multiput(100,5)(0.3,0){101}{\circle*{0.7}}
\put(130.3,5.00005){\circle*{0.7}}
\put(99.7,5.00005){\circle*{0.7}}
\put(130.6,5.00042){\circle*{0.7}}
\put(99.4,5.00042){\circle*{0.7}}
\put(130.9,5.00144){\circle*{0.7}}
\put(99.1,5.00144){\circle*{0.7}}
\put(131.2,5.0035){\circle*{0.7}}
\put(98.8,5.0035){\circle*{0.7}}
\put(131.5,5.00699){\circle*{0.7}}
\put(98.5,5.00699){\circle*{0.7}}
\put(131.8,5.01234){\circle*{0.7}}
\put(98.2,5.01234){\circle*{0.7}}
\put(132.1,5.02002){\circle*{0.7}}
\put(97.9,5.02002){\circle*{0.7}}
\put(132.4,5.03055){\circle*{0.7}}
\put(97.6,5.03055){\circle*{0.7}}
\put(132.7,5.04449){\circle*{0.7}}
\put(97.3,5.04449){\circle*{0.7}}
\put(133.,5.06243){\circle*{0.7}}
\put(97.,5.06243){\circle*{0.7}}
\put(133.3,5.08504){\circle*{0.7}}
\put(96.7,5.08504){\circle*{0.7}}
\put(133.6,5.11302){\circle*{0.7}}
\put(96.4,5.11302){\circle*{0.7}}
\put(133.9,5.14717){\circle*{0.7}}
\put(96.1,5.14717){\circle*{0.7}}
\put(134.2,5.18832){\circle*{0.7}}
\put(95.8,5.18832){\circle*{0.7}}
\put(134.5,5.23743){\circle*{0.7}}
\put(95.5,5.23743){\circle*{0.7}}
\put(134.8,5.29546){\circle*{0.7}}
\put(95.2,5.29546){\circle*{0.7}}
\put(135.1,5.3636){\circle*{0.7}}
\put(94.9,5.3636){\circle*{0.7}}
\put(135.4,5.44304){\circle*{0.7}}
\put(94.6,5.44304){\circle*{0.7}}
\put(135.7,5.53514){\circle*{0.7}}
\put(94.3,5.53514){\circle*{0.7}}
\put(136.,5.64141){\circle*{0.7}}
\put(94.,5.64141){\circle*{0.7}}
\put(136.3,5.7635){\circle*{0.7}}
\put(93.7,5.7635){\circle*{0.7}}
\put(136.6,5.90326){\circle*{0.7}}
\put(93.4,5.90326){\circle*{0.7}}
\put(136.9,6.06275){\circle*{0.7}}
\put(93.1,6.06275){\circle*{0.7}}
\put(137.05,6.15059){\circle*{0.7}}
\put(92.95,6.15059){\circle*{0.7}}
\put(137.2,6.24426){\circle*{0.7}}
\put(92.8,6.24426){\circle*{0.7}}
\put(137.35,6.34407){\circle*{0.7}}
\put(92.65,6.34407){\circle*{0.7}}
\put(137.5,6.45038){\circle*{0.7}}
\put(92.5,6.45038){\circle*{0.7}}
\put(137.65,6.56356){\circle*{0.7}}
\put(92.35,6.56356){\circle*{0.7}}
\put(137.8,6.68402){\circle*{0.7}}
\put(92.2,6.68402){\circle*{0.7}}
\put(137.95,6.81217){\circle*{0.7}}
\put(92.05,6.81217){\circle*{0.7}}
\put(138.1,6.94846){\circle*{0.7}}
\put(91.9,6.94846){\circle*{0.7}}
\put(138.25,7.09339){\circle*{0.7}}
\put(91.75,7.09339){\circle*{0.7}}
\put(138.4,7.24747){\circle*{0.7}}
\put(91.6,7.24747){\circle*{0.7}}
\put(138.55,7.41125){\circle*{0.7}}
\put(91.45,7.41125){\circle*{0.7}}
\put(138.7,7.58532){\circle*{0.7}}
\put(91.3,7.58532){\circle*{0.7}}
\put(138.85,7.77032){\circle*{0.7}}
\put(91.15,7.77032){\circle*{0.7}}
\put(139.,7.96693){\circle*{0.7}}
\put(91.,7.96693){\circle*{0.7}}
\put(139.15,8.1759){\circle*{0.7}}
\put(90.85,8.1759){\circle*{0.7}}
\put(139.3,8.39802){\circle*{0.7}}
\put(90.7,8.39802){\circle*{0.7}}
\put(139.45,8.63415){\circle*{0.7}}
\put(90.55,8.63415){\circle*{0.7}}
\put(139.6,8.88523){\circle*{0.7}}
\put(90.4,8.88523){\circle*{0.7}}
\put(139.75,9.15226){\circle*{0.7}}
\put(90.25,9.15226){\circle*{0.7}}
\put(139.9,9.43635){\circle*{0.7}}
\put(90.1,9.43635){\circle*{0.7}}
\put(139.93,9.49531){\circle*{0.7}}
\put(90.07,9.49531){\circle*{0.7}}
\put(139.96,9.55502){\circle*{0.7}}
\put(90.04,9.55502){\circle*{0.7}}
\put(139.99,9.61548){\circle*{0.7}}
\put(90.01,9.61548){\circle*{0.7}}
\put(140.02,9.67671){\circle*{0.7}}
\put(89.98,9.67671){\circle*{0.7}}
\put(140.05,9.7387){\circle*{0.7}}
\put(89.95,9.7387){\circle*{0.7}}
\put(140.08,9.80147){\circle*{0.7}}
\put(89.92,9.80147){\circle*{0.7}}
\put(140.11,9.86504){\circle*{0.7}}
\put(89.89,9.86504){\circle*{0.7}}
\put(140.14,9.92942){\circle*{0.7}}
\put(89.86,9.92942){\circle*{0.7}}
\put(140.17,9.99461){\circle*{0.7}}
\put(89.83,9.99461){\circle*{0.7}}
\put(140.2,10.0606){\circle*{0.7}}
\put(89.8,10.0606){\circle*{0.7}}
\put(140.23,10.1275){\circle*{0.7}}
\put(89.77,10.1275){\circle*{0.7}}
\put(140.26,10.1952){\circle*{0.7}}
\put(89.74,10.1952){\circle*{0.7}}
\put(140.29,10.2638){\circle*{0.7}}
\put(89.71,10.2638){\circle*{0.7}}
\put(140.32,10.3332){\circle*{0.7}}
\put(89.68,10.3332){\circle*{0.7}}
\put(140.35,10.4036){\circle*{0.7}}
\put(89.65,10.4036){\circle*{0.7}}
\put(140.38,10.4748){\circle*{0.7}}
\put(89.62,10.4748){\circle*{0.7}}
\put(140.41,10.547){\circle*{0.7}}
\put(89.59,10.547){\circle*{0.7}}
\put(140.44,10.6201){\circle*{0.7}}
\put(89.56,10.6201){\circle*{0.7}}
\put(140.47,10.6941){\circle*{0.7}}
\put(89.53,10.6941){\circle*{0.7}}
\put(140.5,10.7691){\circle*{0.7}}
\put(89.5,10.7691){\circle*{0.7}}
\put(140.53,10.8451){\circle*{0.7}}
\put(89.47,10.8451){\circle*{0.7}}
\put(140.56,10.9221){\circle*{0.7}}
\put(89.44,10.9221){\circle*{0.7}}
\put(140.59,11.){\circle*{0.7}}
\put(89.41,11.){\circle*{0.7}}
\put(140.62,11.079){\circle*{0.7}}
\put(89.38,11.079){\circle*{0.7}}
\put(140.65,11.159){\circle*{0.7}}
\put(89.35,11.159){\circle*{0.7}}
\put(140.68,11.2401){\circle*{0.7}}
\put(89.32,11.2401){\circle*{0.7}}
\put(140.71,11.3222){\circle*{0.7}}
\put(89.29,11.3222){\circle*{0.7}}
\put(140.74,11.4055){\circle*{0.7}}
\put(89.26,11.4055){\circle*{0.7}}
\put(140.77,11.4898){\circle*{0.7}}
\put(89.23,11.4898){\circle*{0.7}}
\put(140.8,11.5752){\circle*{0.7}}
\put(89.2,11.5752){\circle*{0.7}}
\put(140.83,11.6618){\circle*{0.7}}
\put(89.17,11.6618){\circle*{0.7}}
\put(140.86,11.7495){\circle*{0.7}}
\put(89.14,11.7495){\circle*{0.7}}
\put(140.89,11.8385){\circle*{0.7}}
\put(89.11,11.8385){\circle*{0.7}}
\put(140.92,11.9286){\circle*{0.7}}
\put(89.08,11.9286){\circle*{0.7}}
\put(140.95,12.0199){\circle*{0.7}}
\put(89.05,12.0199){\circle*{0.7}}
\put(140.98,12.1124){\circle*{0.7}}
\put(89.02,12.1124){\circle*{0.7}}
\put(141.01,12.2062){\circle*{0.7}}
\put(88.99,12.2062){\circle*{0.7}}
\put(141.04,12.3013){\circle*{0.7}}
\put(88.96,12.3013){\circle*{0.7}}
\put(141.07,12.3976){\circle*{0.7}}
\put(88.93,12.3976){\circle*{0.7}}
\put(141.1,12.4953){\circle*{0.7}}
\put(88.9,12.4953){\circle*{0.7}}
\put(141.13,12.5943){\circle*{0.7}}
\put(88.87,12.5943){\circle*{0.7}}
\put(141.16,12.6947){\circle*{0.7}}
\put(88.84,12.6947){\circle*{0.7}}
\put(141.19,12.7965){\circle*{0.7}}
\put(88.81,12.7965){\circle*{0.7}}
\put(141.22,12.8997){\circle*{0.7}}
\put(88.78,12.8997){\circle*{0.7}}
\put(141.25,13.0043){\circle*{0.7}}
\put(88.75,13.0043){\circle*{0.7}}
\put(141.28,13.1103){\circle*{0.7}}
\put(88.72,13.1103){\circle*{0.7}}
\put(141.31,13.2179){\circle*{0.7}}
\put(88.69,13.2179){\circle*{0.7}}
\put(141.34,13.3269){\circle*{0.7}}
\put(88.66,13.3269){\circle*{0.7}}
\put(141.37,13.4376){\circle*{0.7}}
\put(88.63,13.4376){\circle*{0.7}}
\put(141.4,13.5497){\circle*{0.7}}
\put(88.6,13.5497){\circle*{0.7}}
\put(141.43,13.6635){\circle*{0.7}}
\put(88.57,13.6635){\circle*{0.7}}
\put(141.46,13.7789){\circle*{0.7}}
\put(88.54,13.7789){\circle*{0.7}}
\put(141.49,13.8959){\circle*{0.7}}
\put(88.51,13.8959){\circle*{0.7}}
\put(141.52,14.0146){\circle*{0.7}}
\put(88.48,14.0146){\circle*{0.7}}
\put(141.55,14.1351){\circle*{0.7}}
\put(88.45,14.1351){\circle*{0.7}}
\put(141.58,14.2572){\circle*{0.7}}
\put(88.42,14.2572){\circle*{0.7}}
\put(141.61,14.3812){\circle*{0.7}}
\put(88.39,14.3812){\circle*{0.7}}
\put(141.64,14.507){\circle*{0.7}}
\put(88.36,14.507){\circle*{0.7}}
\put(141.67,14.6346){\circle*{0.7}}
\put(88.33,14.6346){\circle*{0.7}}
\put(141.7,14.7641){\circle*{0.7}}
\put(88.3,14.7641){\circle*{0.7}}
\put(141.73,14.8956){\circle*{0.7}}
\put(88.27,14.8956){\circle*{0.7}}
\put(141.76,15.029){\circle*{0.7}}
\put(88.24,15.029){\circle*{0.7}}
\put(141.79,15.1644){\circle*{0.7}}
\put(88.21,15.1644){\circle*{0.7}}
\put(141.82,15.3018){\circle*{0.7}}
\put(88.18,15.3018){\circle*{0.7}}
\put(141.85,15.4414){\circle*{0.7}}
\put(88.15,15.4414){\circle*{0.7}}
\put(141.88,15.583){\circle*{0.7}}
\put(88.12,15.583){\circle*{0.7}}
\put(141.91,15.7268){\circle*{0.7}}
\put(88.09,15.7268){\circle*{0.7}}
\put(141.94,15.8728){\circle*{0.7}}
\put(88.06,15.8728){\circle*{0.7}}
\put(141.97,16.0211){\circle*{0.7}}
\put(88.03,16.0211){\circle*{0.7}}
\put(142.,16.1717){\circle*{0.7}}
\put(88.,16.1717){\circle*{0.7}}
\put(142.03,16.3247){\circle*{0.7}}
\put(87.97,16.3247){\circle*{0.7}}
\put(142.06,16.48){\circle*{0.7}}
\put(87.94,16.48){\circle*{0.7}}
\put(142.09,16.6378){\circle*{0.7}}
\put(87.91,16.6378){\circle*{0.7}}
\put(142.12,16.7981){\circle*{0.7}}
\put(87.88,16.7981){\circle*{0.7}}
\put(142.15,16.961){\circle*{0.7}}
\put(87.85,16.961){\circle*{0.7}}
\put(142.18,17.1265){\circle*{0.7}}
\put(87.82,17.1265){\circle*{0.7}}
\put(142.21,17.2947){\circle*{0.7}}
\put(87.79,17.2947){\circle*{0.7}}
\put(142.24,17.4657){\circle*{0.7}}
\put(87.76,17.4657){\circle*{0.7}}
\put(142.27,17.6394){\circle*{0.7}}
\put(87.73,17.6394){\circle*{0.7}}
\put(142.3,17.816){\circle*{0.7}}
\put(87.7,17.816){\circle*{0.7}}
\put(142.308,17.8606){\circle*{0.7}}
\put(87.6925,17.8606){\circle*{0.7}}
\put(142.315,17.9054){\circle*{0.7}}
\put(87.685,17.9054){\circle*{0.7}}
\put(142.323,17.9504){\circle*{0.7}}
\put(87.6775,17.9504){\circle*{0.7}}
\put(142.33,17.9956){\circle*{0.7}}
\put(87.67,17.9956){\circle*{0.7}}
\put(142.338,18.0409){\circle*{0.7}}
\put(87.6625,18.0409){\circle*{0.7}}
\put(142.345,18.0865){\circle*{0.7}}
\put(87.655,18.0865){\circle*{0.7}}
\put(142.353,18.1322){\circle*{0.7}}
\put(87.6475,18.1322){\circle*{0.7}}
\put(142.36,18.1782){\circle*{0.7}}
\put(87.64,18.1782){\circle*{0.7}}
\put(142.368,18.2243){\circle*{0.7}}
\put(87.6325,18.2243){\circle*{0.7}}
\put(142.375,18.2706){\circle*{0.7}}
\put(87.625,18.2706){\circle*{0.7}}
\put(142.383,18.3171){\circle*{0.7}}
\put(87.6175,18.3171){\circle*{0.7}}
\put(142.39,18.3638){\circle*{0.7}}
\put(87.61,18.3638){\circle*{0.7}}
\put(142.398,18.4107){\circle*{0.7}}
\put(87.6025,18.4107){\circle*{0.7}}
\put(142.405,18.4578){\circle*{0.7}}
\put(87.595,18.4578){\circle*{0.7}}
\put(142.413,18.5051){\circle*{0.7}}
\put(87.5875,18.5051){\circle*{0.7}}
\put(142.42,18.5526){\circle*{0.7}}
\put(87.58,18.5526){\circle*{0.7}}
\put(142.428,18.6003){\circle*{0.7}}
\put(87.5725,18.6003){\circle*{0.7}}
\put(142.435,18.6482){\circle*{0.7}}
\put(87.565,18.6482){\circle*{0.7}}
\put(142.443,18.6963){\circle*{0.7}}
\put(87.5575,18.6963){\circle*{0.7}}
\put(142.45,18.7447){\circle*{0.7}}
\put(87.55,18.7447){\circle*{0.7}}
\put(142.458,18.7932){\circle*{0.7}}
\put(87.5425,18.7932){\circle*{0.7}}
\put(142.465,18.8419){\circle*{0.7}}
\put(87.535,18.8419){\circle*{0.7}}
\put(142.473,18.8909){\circle*{0.7}}
\put(87.5275,18.8909){\circle*{0.7}}
\put(142.48,18.94){\circle*{0.7}}
\put(87.52,18.94){\circle*{0.7}}
\put(142.488,18.9894){\circle*{0.7}}
\put(87.5125,18.9894){\circle*{0.7}}
\put(142.495,19.039){\circle*{0.7}}
\put(87.505,19.039){\circle*{0.7}}
\put(142.503,19.0888){\circle*{0.7}}
\put(87.4975,19.0888){\circle*{0.7}}
\put(142.51,19.1388){\circle*{0.7}}
\put(87.49,19.1388){\circle*{0.7}}
\put(142.518,19.189){\circle*{0.7}}
\put(87.4825,19.189){\circle*{0.7}}
\put(142.525,19.2394){\circle*{0.7}}
\put(87.475,19.2394){\circle*{0.7}}
\put(142.533,19.2901){\circle*{0.7}}
\put(87.4675,19.2901){\circle*{0.7}}
\put(142.54,19.341){\circle*{0.7}}
\put(87.46,19.341){\circle*{0.7}}
\put(142.548,19.3921){\circle*{0.7}}
\put(87.4525,19.3921){\circle*{0.7}}
\put(142.555,19.4434){\circle*{0.7}}
\put(87.445,19.4434){\circle*{0.7}}
\put(142.563,19.495){\circle*{0.7}}
\put(87.4375,19.495){\circle*{0.7}}
\put(142.57,19.5468){\circle*{0.7}}
\put(87.43,19.5468){\circle*{0.7}}
\put(142.578,19.5988){\circle*{0.7}}
\put(87.4225,19.5988){\circle*{0.7}}
\put(142.585,19.6511){\circle*{0.7}}
\put(87.415,19.6511){\circle*{0.7}}
\put(142.593,19.7035){\circle*{0.7}}
\put(87.4075,19.7035){\circle*{0.7}}
\put(142.6,19.7563){\circle*{0.7}}
\put(87.4,19.7563){\circle*{0.7}}
\put(142.608,19.8092){\circle*{0.7}}
\put(87.3925,19.8092){\circle*{0.7}}
\put(142.615,19.8624){\circle*{0.7}}
\put(87.385,19.8624){\circle*{0.7}}
\put(142.623,19.9158){\circle*{0.7}}
\put(87.3775,19.9158){\circle*{0.7}}
\put(142.63,19.9695){\circle*{0.7}}
\put(87.37,19.9695){\circle*{0.7}}
\put(142.638,20.0234){\circle*{0.7}}
\put(87.3625,20.0234){\circle*{0.7}}
\put(142.645,20.0775){\circle*{0.7}}
\put(87.355,20.0775){\circle*{0.7}}
\put(142.653,20.1319){\circle*{0.7}}
\put(87.3475,20.1319){\circle*{0.7}}
\put(142.66,20.1866){\circle*{0.7}}
\put(87.34,20.1866){\circle*{0.7}}
\put(142.668,20.2415){\circle*{0.7}}
\put(87.3325,20.2415){\circle*{0.7}}
\put(142.675,20.2966){\circle*{0.7}}
\put(87.325,20.2966){\circle*{0.7}}
\put(142.683,20.352){\circle*{0.7}}
\put(87.3175,20.352){\circle*{0.7}}
\put(142.69,20.4076){\circle*{0.7}}
\put(87.31,20.4076){\circle*{0.7}}
\put(142.698,20.4635){\circle*{0.7}}
\put(87.3025,20.4635){\circle*{0.7}}
\put(142.705,20.5196){\circle*{0.7}}
\put(87.295,20.5196){\circle*{0.7}}
\put(142.713,20.5761){\circle*{0.7}}
\put(87.2875,20.5761){\circle*{0.7}}
\put(142.72,20.6327){\circle*{0.7}}
\put(87.28,20.6327){\circle*{0.7}}
\put(142.728,20.6896){\circle*{0.7}}
\put(87.2725,20.6896){\circle*{0.7}}
\put(142.735,20.7468){\circle*{0.7}}
\put(87.265,20.7468){\circle*{0.7}}
\put(142.743,20.8043){\circle*{0.7}}
\put(87.2575,20.8043){\circle*{0.7}}
\put(142.75,20.862){\circle*{0.7}}
\put(87.25,20.862){\circle*{0.7}}
\put(142.758,20.92){\circle*{0.7}}
\put(87.2425,20.92){\circle*{0.7}}
\put(142.765,20.9783){\circle*{0.7}}
\put(87.235,20.9783){\circle*{0.7}}
\put(142.773,21.0368){\circle*{0.7}}
\put(87.2275,21.0368){\circle*{0.7}}
\put(142.78,21.0956){\circle*{0.7}}
\put(87.22,21.0956){\circle*{0.7}}
\put(142.788,21.1547){\circle*{0.7}}
\put(87.2125,21.1547){\circle*{0.7}}
\put(142.795,21.214){\circle*{0.7}}
\put(87.205,21.214){\circle*{0.7}}
\put(142.802,21.2737){\circle*{0.7}}
\put(87.1975,21.2737){\circle*{0.7}}
\put(142.81,21.3336){\circle*{0.7}}
\put(87.19,21.3336){\circle*{0.7}}
\put(142.817,21.3938){\circle*{0.7}}
\put(87.1825,21.3938){\circle*{0.7}}
\put(142.825,21.4543){\circle*{0.7}}
\put(87.175,21.4543){\circle*{0.7}}
\put(142.833,21.5151){\circle*{0.7}}
\put(87.1675,21.5151){\circle*{0.7}}
\put(142.84,21.5761){\circle*{0.7}}
\put(87.16,21.5761){\circle*{0.7}}
\put(142.847,21.6375){\circle*{0.7}}
\put(87.1525,21.6375){\circle*{0.7}}
\put(142.855,21.6992){\circle*{0.7}}
\put(87.145,21.6992){\circle*{0.7}}
\put(142.863,21.7611){\circle*{0.7}}
\put(87.1375,21.7611){\circle*{0.7}}
\put(142.87,21.8234){\circle*{0.7}}
\put(87.13,21.8234){\circle*{0.7}}
\put(142.878,21.8859){\circle*{0.7}}
\put(87.1225,21.8859){\circle*{0.7}}
\put(142.885,21.9488){\circle*{0.7}}
\put(87.115,21.9488){\circle*{0.7}}
\put(142.893,22.0119){\circle*{0.7}}
\put(87.1075,22.0119){\circle*{0.7}}
\put(142.9,22.0754){\circle*{0.7}}
\put(87.1,22.0754){\circle*{0.7}}
\put(142.907,22.1392){\circle*{0.7}}
\put(87.0925,22.1392){\circle*{0.7}}
\put(142.915,22.2033){\circle*{0.7}}
\put(87.085,22.2033){\circle*{0.7}}
\put(142.923,22.2677){\circle*{0.7}}
\put(87.0775,22.2677){\circle*{0.7}}
\put(142.93,22.3324){\circle*{0.7}}
\put(87.07,22.3324){\circle*{0.7}}
\put(142.937,22.3974){\circle*{0.7}}
\put(87.0625,22.3974){\circle*{0.7}}
\put(142.945,22.4628){\circle*{0.7}}
\put(87.055,22.4628){\circle*{0.7}}
\put(142.952,22.5284){\circle*{0.7}}
\put(87.0475,22.5284){\circle*{0.7}}
\put(142.96,22.5945){\circle*{0.7}}
\put(87.04,22.5945){\circle*{0.7}}
\put(142.967,22.6608){\circle*{0.7}}
\put(87.0325,22.6608){\circle*{0.7}}
\put(142.975,22.7275){\circle*{0.7}}
\put(87.025,22.7275){\circle*{0.7}}
\put(142.982,22.7945){\circle*{0.7}}
\put(87.0175,22.7945){\circle*{0.7}}
\put(142.99,22.8618){\circle*{0.7}}
\put(87.01,22.8618){\circle*{0.7}}
\put(142.997,22.9295){\circle*{0.7}}
\put(87.0025,22.9295){\circle*{0.7}}
\put(143.005,22.9975){\circle*{0.7}}
\put(86.995,22.9975){\circle*{0.7}}
\put(143.012,23.0658){\circle*{0.7}}
\put(86.9875,23.0658){\circle*{0.7}}
\put(143.02,23.1346){\circle*{0.7}}
\put(86.98,23.1346){\circle*{0.7}}
\put(143.027,23.2036){\circle*{0.7}}
\put(86.9725,23.2036){\circle*{0.7}}
\put(143.035,23.273){\circle*{0.7}}
\put(86.965,23.273){\circle*{0.7}}
\put(143.042,23.3428){\circle*{0.7}}
\put(86.9575,23.3428){\circle*{0.7}}
\put(143.05,23.4129){\circle*{0.7}}
\put(86.95,23.4129){\circle*{0.7}}
\put(143.057,23.4834){\circle*{0.7}}
\put(86.9425,23.4834){\circle*{0.7}}
\put(143.065,23.5542){\circle*{0.7}}
\put(86.935,23.5542){\circle*{0.7}}
\put(143.072,23.6254){\circle*{0.7}}
\put(86.9275,23.6254){\circle*{0.7}}
\put(143.08,23.697){\circle*{0.7}}
\put(86.92,23.697){\circle*{0.7}}
\put(143.087,23.769){\circle*{0.7}}
\put(86.9125,23.769){\circle*{0.7}}
\put(143.095,23.8413){\circle*{0.7}}
\put(86.905,23.8413){\circle*{0.7}}
\put(143.102,23.914){\circle*{0.7}}
\put(86.8975,23.914){\circle*{0.7}}
\put(143.11,23.9871){\circle*{0.7}}
\put(86.89,23.9871){\circle*{0.7}}
\put(143.117,24.0606){\circle*{0.7}}
\put(86.8825,24.0606){\circle*{0.7}}
\put(143.125,24.1344){\circle*{0.7}}
\put(86.875,24.1344){\circle*{0.7}}
\put(143.132,24.2087){\circle*{0.7}}
\put(86.8675,24.2087){\circle*{0.7}}
\put(143.14,24.2833){\circle*{0.7}}
\put(86.86,24.2833){\circle*{0.7}}
\put(143.147,24.3583){\circle*{0.7}}
\put(86.8525,24.3583){\circle*{0.7}}
\put(143.155,24.4338){\circle*{0.7}}
\put(86.845,24.4338){\circle*{0.7}}
\put(143.162,24.5096){\circle*{0.7}}
\put(86.8375,24.5096){\circle*{0.7}}
\put(143.17,24.5859){\circle*{0.7}}
\put(86.83,24.5859){\circle*{0.7}}
\put(143.177,24.6625){\circle*{0.7}}
\put(86.8225,24.6625){\circle*{0.7}}
\put(143.185,24.7396){\circle*{0.7}}
\put(86.815,24.7396){\circle*{0.7}}
\put(143.192,24.8171){\circle*{0.7}}
\put(86.8075,24.8171){\circle*{0.7}}
\put(143.2,24.895){\circle*{0.7}}
\put(86.8,24.895){\circle*{0.7}}
\put(143.207,24.9734){\circle*{0.7}}
\put(86.7925,24.9734){\circle*{0.7}}
\put(143.215,25.0521){\circle*{0.7}}
\put(86.785,25.0521){\circle*{0.7}}
\put(143.222,25.1313){\circle*{0.7}}
\put(86.7775,25.1313){\circle*{0.7}}
\put(143.23,25.211){\circle*{0.7}}
\put(86.77,25.211){\circle*{0.7}}
\put(143.237,25.291){\circle*{0.7}}
\put(86.7625,25.291){\circle*{0.7}}
\put(143.245,25.3715){\circle*{0.7}}
\put(86.755,25.3715){\circle*{0.7}}
\put(143.252,25.4525){\circle*{0.7}}
\put(86.7475,25.4525){\circle*{0.7}}
\put(143.26,25.5339){\circle*{0.7}}
\put(86.74,25.5339){\circle*{0.7}}
\put(143.267,25.6158){\circle*{0.7}}
\put(86.7325,25.6158){\circle*{0.7}}
\put(143.275,25.6981){\circle*{0.7}}
\put(86.725,25.6981){\circle*{0.7}}
\put(143.282,25.7809){\circle*{0.7}}
\put(86.7175,25.7809){\circle*{0.7}}
\put(143.29,25.8642){\circle*{0.7}}
\put(86.71,25.8642){\circle*{0.7}}
\put(143.297,25.9479){\circle*{0.7}}
\put(86.7025,25.9479){\circle*{0.7}}
\put(143.305,26.0321){\circle*{0.7}}
\put(86.695,26.0321){\circle*{0.7}}
\put(143.312,26.1168){\circle*{0.7}}
\put(86.6875,26.1168){\circle*{0.7}}
\put(143.32,26.202){\circle*{0.7}}
\put(86.68,26.202){\circle*{0.7}}
\put(143.327,26.2876){\circle*{0.7}}
\put(86.6725,26.2876){\circle*{0.7}}
\put(143.335,26.3738){\circle*{0.7}}
\put(86.665,26.3738){\circle*{0.7}}
\put(143.342,26.4604){\circle*{0.7}}
\put(86.6575,26.4604){\circle*{0.7}}
\put(143.35,26.5476){\circle*{0.7}}
\put(86.65,26.5476){\circle*{0.7}}
\put(143.357,26.6352){\circle*{0.7}}
\put(86.6425,26.6352){\circle*{0.7}}
\put(143.365,26.7234){\circle*{0.7}}
\put(86.635,26.7234){\circle*{0.7}}
\put(143.372,26.8121){\circle*{0.7}}
\put(86.6275,26.8121){\circle*{0.7}}
\put(143.38,26.9013){\circle*{0.7}}
\put(86.62,26.9013){\circle*{0.7}}
\put(143.387,26.991){\circle*{0.7}}
\put(86.6125,26.991){\circle*{0.7}}
\put(143.395,27.0813){\circle*{0.7}}
\put(86.605,27.0813){\circle*{0.7}}
\put(143.402,27.1721){\circle*{0.7}}
\put(86.5975,27.1721){\circle*{0.7}}
\put(143.41,27.2634){\circle*{0.7}}
\put(86.59,27.2634){\circle*{0.7}}
\put(143.417,27.3553){\circle*{0.7}}
\put(86.5825,27.3553){\circle*{0.7}}
\put(143.425,27.4478){\circle*{0.7}}
\put(86.575,27.4478){\circle*{0.7}}
\put(143.432,27.5408){\circle*{0.7}}
\put(86.5675,27.5408){\circle*{0.7}}
\put(143.44,27.6343){\circle*{0.7}}
\put(86.56,27.6343){\circle*{0.7}}
\put(143.447,27.7285){\circle*{0.7}}
\put(86.5525,27.7285){\circle*{0.7}}
\put(143.455,27.8232){\circle*{0.7}}
\put(86.545,27.8232){\circle*{0.7}}
\put(143.462,27.9185){\circle*{0.7}}
\put(86.5375,27.9185){\circle*{0.7}}
\put(143.47,28.0143){\circle*{0.7}}
\put(86.53,28.0143){\circle*{0.7}}
\put(143.477,28.1108){\circle*{0.7}}
\put(86.5225,28.1108){\circle*{0.7}}
\put(143.485,28.2078){\circle*{0.7}}
\put(86.515,28.2078){\circle*{0.7}}
\put(143.492,28.3055){\circle*{0.7}}
\put(86.5075,28.3055){\circle*{0.7}}
\put(143.5,28.4038){\circle*{0.7}}
\put(86.5,28.4038){\circle*{0.7}}
\put(143.507,28.5027){\circle*{0.7}}
\put(86.4925,28.5027){\circle*{0.7}}
\put(143.515,28.6022){\circle*{0.7}}
\put(86.485,28.6022){\circle*{0.7}}
\put(143.522,28.7023){\circle*{0.7}}
\put(86.4775,28.7023){\circle*{0.7}}
\put(143.53,28.8031){\circle*{0.7}}
\put(86.47,28.8031){\circle*{0.7}}
\put(143.537,28.9045){\circle*{0.7}}
\put(86.4625,28.9045){\circle*{0.7}}
\put(143.545,29.0065){\circle*{0.7}}
\put(86.455,29.0065){\circle*{0.7}}
\put(143.552,29.1093){\circle*{0.7}}
\put(86.4475,29.1093){\circle*{0.7}}
\put(143.56,29.2126){\circle*{0.7}}
\put(86.44,29.2126){\circle*{0.7}}
\put(143.567,29.3167){\circle*{0.7}}
\put(86.4325,29.3167){\circle*{0.7}}
\put(143.575,29.4214){\circle*{0.7}}
\put(86.425,29.4214){\circle*{0.7}}
\put(143.582,29.5268){\circle*{0.7}}
\put(86.4175,29.5268){\circle*{0.7}}
\put(143.59,29.6329){\circle*{0.7}}
\put(86.41,29.6329){\circle*{0.7}}
\put(143.597,29.7397){\circle*{0.7}}
\put(86.4025,29.7397){\circle*{0.7}}
\put(143.605,29.8472){\circle*{0.7}}
\put(86.395,29.8472){\circle*{0.7}}
\put(143.612,29.9554){\circle*{0.7}}
\put(86.3875,29.9554){\circle*{0.7}}
\put(143.62,30.0643){\circle*{0.7}}
\put(86.38,30.0643){\circle*{0.7}}
\put(143.627,30.174){\circle*{0.7}}
\put(86.3725,30.174){\circle*{0.7}}
\put(143.635,30.2844){\circle*{0.7}}
\put(86.365,30.2844){\circle*{0.7}}
\put(143.642,30.3955){\circle*{0.7}}
\put(86.3575,30.3955){\circle*{0.7}}
\put(143.65,30.5074){\circle*{0.7}}
\put(86.35,30.5074){\circle*{0.7}}
\put(143.657,30.6201){\circle*{0.7}}
\put(86.3425,30.6201){\circle*{0.7}}
\put(143.665,30.7335){\circle*{0.7}}
\put(86.335,30.7335){\circle*{0.7}}
\put(143.672,30.8477){\circle*{0.7}}
\put(86.3275,30.8477){\circle*{0.7}}
\put(143.68,30.9627){\circle*{0.7}}
\put(86.32,30.9627){\circle*{0.7}}
\put(143.687,31.0785){\circle*{0.7}}
\put(86.3125,31.0785){\circle*{0.7}}
\put(143.695,31.1951){\circle*{0.7}}
\put(86.305,31.1951){\circle*{0.7}}
\put(143.702,31.3125){\circle*{0.7}}
\put(86.2975,31.3125){\circle*{0.7}}
\put(143.71,31.4307){\circle*{0.7}}
\put(86.29,31.4307){\circle*{0.7}}
\put(143.717,31.5498){\circle*{0.7}}
\put(86.2825,31.5498){\circle*{0.7}}
\put(143.725,31.6697){\circle*{0.7}}
\put(86.275,31.6697){\circle*{0.7}}
\put(143.732,31.7905){\circle*{0.7}}
\put(86.2675,31.7905){\circle*{0.7}}
\put(143.74,31.9121){\circle*{0.7}}
\put(86.26,31.9121){\circle*{0.7}}
\put(143.747,32.0347){\circle*{0.7}}
\put(86.2525,32.0347){\circle*{0.7}}
\put(143.755,32.1581){\circle*{0.7}}
\put(86.245,32.1581){\circle*{0.7}}
\put(143.762,32.2824){\circle*{0.7}}
\put(86.2375,32.2824){\circle*{0.7}}
\put(143.77,32.4076){\circle*{0.7}}
\put(86.23,32.4076){\circle*{0.7}}
\put(143.777,32.5337){\circle*{0.7}}
\put(86.2225,32.5337){\circle*{0.7}}
\put(143.785,32.6608){\circle*{0.7}}
\put(86.215,32.6608){\circle*{0.7}}
\put(143.792,32.7888){\circle*{0.7}}
\put(86.2075,32.7888){\circle*{0.7}}
\put(143.8,32.9177){\circle*{0.7}}
\put(86.2,32.9177){\circle*{0.7}}
\put(143.807,33.0477){\circle*{0.7}}
\put(86.1925,33.0477){\circle*{0.7}}
\put(143.815,33.1786){\circle*{0.7}}
\put(86.185,33.1786){\circle*{0.7}}
\put(143.822,33.3105){\circle*{0.7}}
\put(86.1775,33.3105){\circle*{0.7}}
\put(143.83,33.4434){\circle*{0.7}}
\put(86.17,33.4434){\circle*{0.7}}
\put(143.837,33.5773){\circle*{0.7}}
\put(86.1625,33.5773){\circle*{0.7}}
\put(143.845,33.7122){\circle*{0.7}}
\put(86.155,33.7122){\circle*{0.7}}
\put(143.852,33.8482){\circle*{0.7}}
\put(86.1475,33.8482){\circle*{0.7}}
\put(143.86,33.9853){\circle*{0.7}}
\put(86.14,33.9853){\circle*{0.7}}
\put(143.867,34.1234){\circle*{0.7}}
\put(86.1325,34.1234){\circle*{0.7}}
\put(143.875,34.2626){\circle*{0.7}}
\put(86.125,34.2626){\circle*{0.7}}
\put(143.882,34.403){\circle*{0.7}}
\put(86.1175,34.403){\circle*{0.7}}
\put(143.89,34.5444){\circle*{0.7}}
\put(86.11,34.5444){\circle*{0.7}}
\put(143.897,34.6869){\circle*{0.7}}
\put(86.1025,34.6869){\circle*{0.7}}
\put(143.905,34.8307){\circle*{0.7}}
\put(86.095,34.8307){\circle*{0.7}}
\put(143.912,34.9755){\circle*{0.7}}
\put(86.0875,34.9755){\circle*{0.7}}
\put(143.92,35.1216){\circle*{0.7}}
\put(86.08,35.1216){\circle*{0.7}}
\put(143.927,35.2688){\circle*{0.7}}
\put(86.0725,35.2688){\circle*{0.7}}
\put(143.935,35.4173){\circle*{0.7}}
\put(86.065,35.4173){\circle*{0.7}}
\put(143.942,35.567){\circle*{0.7}}
\put(86.0575,35.567){\circle*{0.7}}
\put(143.95,35.7179){\circle*{0.7}}
\put(86.05,35.7179){\circle*{0.7}}
\put(143.957,35.8701){\circle*{0.7}}
\put(86.0425,35.8701){\circle*{0.7}}
\put(143.965,36.0236){\circle*{0.7}}
\put(86.035,36.0236){\circle*{0.7}}
\put(143.972,36.1784){\circle*{0.7}}
\put(86.0275,36.1784){\circle*{0.7}}
\put(143.98,36.3345){\circle*{0.7}}
\put(86.02,36.3345){\circle*{0.7}}
\put(143.987,36.492){\circle*{0.7}}
\put(86.0125,36.492){\circle*{0.7}}
\put(143.995,36.6508){\circle*{0.7}}
\put(86.005,36.6508){\circle*{0.7}}
\put(144.002,36.811){\circle*{0.7}}
\put(85.9975,36.811){\circle*{0.7}}
\put(144.01,36.9726){\circle*{0.7}}
\put(85.99,36.9726){\circle*{0.7}}
\put(144.017,37.1356){\circle*{0.7}}
\put(85.9825,37.1356){\circle*{0.7}}
\put(144.025,37.3001){\circle*{0.7}}
\put(85.975,37.3001){\circle*{0.7}}
\put(144.032,37.466){\circle*{0.7}}
\put(85.9675,37.466){\circle*{0.7}}
\put(144.04,37.6335){\circle*{0.7}}
\put(85.96,37.6335){\circle*{0.7}}
\put(144.047,37.8024){\circle*{0.7}}
\put(85.9525,37.8024){\circle*{0.7}}
\put(144.055,37.9729){\circle*{0.7}}
\put(85.945,37.9729){\circle*{0.7}}
\put(144.062,38.145){\circle*{0.7}}
\put(85.9375,38.145){\circle*{0.7}}
\put(144.07,38.3186){\circle*{0.7}}
\put(85.93,38.3186){\circle*{0.7}}
\put(144.077,38.4939){\circle*{0.7}}
\put(85.9225,38.4939){\circle*{0.7}}
\put(144.085,38.6708){\circle*{0.7}}
\put(85.915,38.6708){\circle*{0.7}}
\put(144.092,38.8494){\circle*{0.7}}
\put(85.9075,38.8494){\circle*{0.7}}
\put(144.1,39.0297){\circle*{0.7}}
\put(85.9,39.0297){\circle*{0.7}}
\put(144.107,39.2117){\circle*{0.7}}
\put(85.8925,39.2117){\circle*{0.7}}
\put(144.115,39.3954){\circle*{0.7}}
\put(85.885,39.3954){\circle*{0.7}}
\put(144.122,39.581){\circle*{0.7}}
\put(85.8775,39.581){\circle*{0.7}}
\put(144.13,39.7683){\circle*{0.7}}
\put(85.87,39.7683){\circle*{0.7}}
\put(144.137,39.9575){\circle*{0.7}}
\put(85.8625,39.9575){\circle*{0.7}}
\put(144.145,40.1486){\circle*{0.7}}
\put(85.855,40.1486){\circle*{0.7}}
\put(144.152,40.3416){\circle*{0.7}}
\put(85.8475,40.3416){\circle*{0.7}}
\put(144.16,40.5366){\circle*{0.7}}
\put(85.84,40.5366){\circle*{0.7}}
\put(144.167,40.7335){\circle*{0.7}}
\put(85.8325,40.7335){\circle*{0.7}}
\put(144.175,40.9325){\circle*{0.7}}
\put(85.825,40.9325){\circle*{0.7}}
\put(144.182,41.1335){\circle*{0.7}}
\put(85.8175,41.1335){\circle*{0.7}}
\put(144.19,41.3366){\circle*{0.7}}
\put(85.81,41.3366){\circle*{0.7}}
\put(144.197,41.5418){\circle*{0.7}}
\put(85.8025,41.5418){\circle*{0.7}}
\put(144.205,41.7492){\circle*{0.7}}
\put(85.795,41.7492){\circle*{0.7}}
\put(144.212,41.9589){\circle*{0.7}}
\put(85.7875,41.9589){\circle*{0.7}}
\put(144.22,42.1708){\circle*{0.7}}
\put(85.78,42.1708){\circle*{0.7}}
\put(144.227,42.3849){\circle*{0.7}}
\put(85.7725,42.3849){\circle*{0.7}}
\put(144.235,42.6015){\circle*{0.7}}
\put(85.765,42.6015){\circle*{0.7}}
\put(144.242,42.8204){\circle*{0.7}}
\put(85.7575,42.8204){\circle*{0.7}}
\put(144.25,43.0418){\circle*{0.7}}
\put(85.75,43.0418){\circle*{0.7}}
\put(144.257,43.2656){\circle*{0.7}}
\put(85.7425,43.2656){\circle*{0.7}}
\put(144.265,43.492){\circle*{0.7}}
\put(85.735,43.492){\circle*{0.7}}
\put(144.272,43.721){\circle*{0.7}}
\put(85.7275,43.721){\circle*{0.7}}
\put(144.28,43.9527){\circle*{0.7}}
\put(85.72,43.9527){\circle*{0.7}}
\put(144.287,44.187){\circle*{0.7}}
\put(85.7125,44.187){\circle*{0.7}}
\put(144.295,44.4241){\circle*{0.7}}
\put(85.705,44.4241){\circle*{0.7}}
\put(144.302,44.664){\circle*{0.7}}
\put(85.6975,44.664){\circle*{0.7}}
\put(144.31,44.9068){\circle*{0.7}}
\put(85.69,44.9068){\circle*{0.7}}
\put(144.317,45.1526){\circle*{0.7}}
\put(85.6825,45.1526){\circle*{0.7}}
\put(144.325,45.4013){\circle*{0.7}}
\put(85.675,45.4013){\circle*{0.7}}
\put(144.332,45.6532){\circle*{0.7}}
\put(85.6675,45.6532){\circle*{0.7}}
\put(144.34,45.9081){\circle*{0.7}}
\put(85.66,45.9081){\circle*{0.7}}
\put(144.347,46.1663){\circle*{0.7}}
\put(85.6525,46.1663){\circle*{0.7}}
\put(144.355,46.4278){\circle*{0.7}}
\put(85.645,46.4278){\circle*{0.7}}
\put(144.362,46.6927){\circle*{0.7}}
\put(85.6375,46.6927){\circle*{0.7}}
\put(144.37,46.961){\circle*{0.7}}
\put(85.63,46.961){\circle*{0.7}}
\put(144.377,47.2329){\circle*{0.7}}
\put(85.6225,47.2329){\circle*{0.7}}
\put(144.385,47.5083){\circle*{0.7}}
\put(85.615,47.5083){\circle*{0.7}}
\put(144.392,47.7875){\circle*{0.7}}
\put(85.6075,47.7875){\circle*{0.7}}
\put(144.4,48.0705){\circle*{0.7}}
\put(85.6,48.0705){\circle*{0.7}}
\put(144.407,48.3573){\circle*{0.7}}
\put(85.5925,48.3573){\circle*{0.7}}
\put(144.415,48.6482){\circle*{0.7}}
\put(85.585,48.6482){\circle*{0.7}}
\put(144.422,48.9432){\circle*{0.7}}
\put(85.5775,48.9432){\circle*{0.7}}
\put(144.43,49.2424){\circle*{0.7}}
\put(85.57,49.2424){\circle*{0.7}}
\put(144.437,49.5458){\circle*{0.7}}
\put(85.5625,49.5458){\circle*{0.7}}
\put(144.445,49.8538){\circle*{0.7}}
\put(85.555,49.8538){\circle*{0.7}}
\put(144.452,50.1663){\circle*{0.7}}
\put(85.5475,50.1663){\circle*{0.7}}
\put(144.46,50.4834){\circle*{0.7}}
\put(85.54,50.4834){\circle*{0.7}}
\put(144.467,50.8054){\circle*{0.7}}
\put(85.5325,50.8054){\circle*{0.7}}
\put(144.475,51.1323){\circle*{0.7}}
\put(85.525,51.1323){\circle*{0.7}}
\put(144.482,51.4643){\circle*{0.7}}
\put(85.5175,51.4643){\circle*{0.7}}
\put(144.49,51.8016){\circle*{0.7}}
\put(85.51,51.8016){\circle*{0.7}}
\put(144.497,52.1443){\circle*{0.7}}
\put(85.5025,52.1443){\circle*{0.7}}
\put(144.505,52.4925){\circle*{0.7}}
\put(85.495,52.4925){\circle*{0.7}}
\put(144.512,52.8464){\circle*{0.7}}
\put(85.4875,52.8464){\circle*{0.7}}
\put(144.52,53.2063){\circle*{0.7}}
\put(85.48,53.2063){\circle*{0.7}}
\put(144.527,53.5722){\circle*{0.7}}
\put(85.4725,53.5722){\circle*{0.7}}
\put(144.535,53.9445){\circle*{0.7}}
\put(85.465,53.9445){\circle*{0.7}}
\put(144.542,54.3232){\circle*{0.7}}
\put(85.4575,54.3232){\circle*{0.7}}
\put(144.55,54.7087){\circle*{0.7}}
\put(85.45,54.7087){\circle*{0.7}}
\put(144.557,55.101){\circle*{0.7}}
\put(85.4425,55.101){\circle*{0.7}}
\put(144.565,55.5006){\circle*{0.7}}
\put(85.435,55.5006){\circle*{0.7}}
\put(144.572,55.9076){\circle*{0.7}}
\put(85.4275,55.9076){\circle*{0.7}}
\put(144.58,56.3222){\circle*{0.7}}
\put(85.42,56.3222){\circle*{0.7}}
\put(144.587,56.7449){\circle*{0.7}}
\put(85.4125,56.7449){\circle*{0.7}}

%% file: sph_rate_f1.tex
\put(35.0,5.){\circle*{0.7}}
\put(35.3,5.){\circle*{0.7}}
\put(34.7,5.){\circle*{0.7}}
\put(35.6,5.){\circle*{0.7}}
\put(34.4,5.){\circle*{0.7}}
\put(35.9,5.){\circle*{0.7}}
\put(34.1,5.){\circle*{0.7}}
\put(36.2,5.){\circle*{0.7}}
\put(33.8,5.){\circle*{0.7}}
\put(36.5,5.){\circle*{0.7}}
\put(33.5,5.){\circle*{0.7}}
\put(36.8,5.){\circle*{0.7}}
\put(33.2,5.){\circle*{0.7}}
\put(37.1,5.){\circle*{0.7}}
\put(32.9,5.){\circle*{0.7}}
\put(37.4,5.){\circle*{0.7}}
\put(32.6,5.){\circle*{0.7}}
\put(37.7,5.){\circle*{0.7}}
\put(32.3,5.){\circle*{0.7}}
\put(38.,5.00001){\circle*{0.7}}
\put(32.,5.00001){\circle*{0.7}}
\put(38.3,5.00002){\circle*{0.7}}
\put(31.7,5.00002){\circle*{0.7}}
\put(38.6,5.00003){\circle*{0.7}}
\put(31.4,5.00003){\circle*{0.7}}
\put(38.9,5.00005){\circle*{0.7}}
\put(31.1,5.00005){\circle*{0.7}}
\put(39.2,5.00008){\circle*{0.7}}
\put(30.8,5.00008){\circle*{0.7}}
\put(39.5,5.00013){\circle*{0.7}}
\put(30.5,5.00013){\circle*{0.7}}
\put(39.8,5.00019){\circle*{0.7}}
\put(30.2,5.00019){\circle*{0.7}}
\put(40.1,5.00028){\circle*{0.7}}
\put(29.9,5.00028){\circle*{0.7}}
\put(40.4,5.0004){\circle*{0.7}}
\put(29.6,5.0004){\circle*{0.7}}
\put(40.7,5.00057){\circle*{0.7}}
\put(29.3,5.00057){\circle*{0.7}}
\put(41.,5.00078){\circle*{0.7}}
\put(29.,5.00078){\circle*{0.7}}
\put(41.3,5.00106){\circle*{0.7}}
\put(28.7,5.00106){\circle*{0.7}}
\put(41.6,5.00142){\circle*{0.7}}
\put(28.4,5.00142){\circle*{0.7}}
\put(41.9,5.00187){\circle*{0.7}}
\put(28.1,5.00187){\circle*{0.7}}
\put(42.2,5.00243){\circle*{0.7}}
\put(27.8,5.00243){\circle*{0.7}}
\put(42.5,5.00313){\circle*{0.7}}
\put(27.5,5.00313){\circle*{0.7}}
\put(42.8,5.00399){\circle*{0.7}}
\put(27.2,5.00399){\circle*{0.7}}
\put(43.1,5.00504){\circle*{0.7}}
\put(26.9,5.00504){\circle*{0.7}}
\put(43.4,5.0063){\circle*{0.7}}
\put(26.6,5.0063){\circle*{0.7}}
\put(43.7,5.00782){\circle*{0.7}}
\put(26.3,5.00782){\circle*{0.7}}
\put(44.,5.00963){\circle*{0.7}}
\put(26.,5.00963){\circle*{0.7}}
\put(44.3,5.01177){\circle*{0.7}}
\put(25.7,5.01177){\circle*{0.7}}
\put(44.6,5.01429){\circle*{0.7}}
\put(25.4,5.01429){\circle*{0.7}}
\put(44.9,5.01725){\circle*{0.7}}
\put(25.1,5.01725){\circle*{0.7}}
\put(45.2,5.0207){\circle*{0.7}}
\put(24.8,5.0207){\circle*{0.7}}
\put(45.5,5.02469){\circle*{0.7}}
\put(24.5,5.02469){\circle*{0.7}}
\put(45.8,5.02932){\circle*{0.7}}
\put(24.2,5.02932){\circle*{0.7}}
\put(46.1,5.03463){\circle*{0.7}}
\put(23.9,5.03463){\circle*{0.7}}
\put(46.4,5.04066){\circle*{0.7}}
\put(23.6,5.04066){\circle*{0.7}}
\put(46.7,5.0477){\circle*{0.7}}
\put(23.3,5.0477){\circle*{0.7}}
\put(47.,5.05562){\circle*{0.7}}
\put(23.,5.05562){\circle*{0.7}}
\put(47.3,5.06462){\circle*{0.7}}
\put(22.7,5.06462){\circle*{0.7}}
\put(47.6,5.07481){\circle*{0.7}}
\put(22.4,5.07481){\circle*{0.7}}
\put(47.9,5.08629){\circle*{0.7}}
\put(22.1,5.08629){\circle*{0.7}}
\put(48.2,5.09922){\circle*{0.7}}
\put(21.8,5.09922){\circle*{0.7}}
\put(48.5,5.11373){\circle*{0.7}}
\put(21.5,5.11373){\circle*{0.7}}
\put(48.8,5.12998){\circle*{0.7}}
\put(21.2,5.12998){\circle*{0.7}}
\put(49.1,5.14814){\circle*{0.7}}
\put(20.9,5.14814){\circle*{0.7}}
\put(49.4,5.16839){\circle*{0.7}}
\put(20.6,5.16839){\circle*{0.7}}
\put(49.7,5.19093){\circle*{0.7}}
\put(20.3,5.19093){\circle*{0.7}}
\put(50.,5.21594){\circle*{0.7}}
\put(20.,5.21594){\circle*{0.7}}
\put(50.3,5.24369){\circle*{0.7}}
\put(19.7,5.24369){\circle*{0.7}}
\put(50.6,5.27441){\circle*{0.7}}
\put(19.4,5.27441){\circle*{0.7}}
\put(50.9,5.30836){\circle*{0.7}}
\put(19.1,5.30836){\circle*{0.7}}
\put(51.05,5.34582){\circle*{0.7}}
\put(18.95,5.34582){\circle*{0.7}}
\put(51.2,5.36597){\circle*{0.7}}
\put(18.8,5.36597){\circle*{0.7}}
\put(51.35,5.38712){\circle*{0.7}}
\put(18.65,5.38712){\circle*{0.7}}
\put(51.5,5.4093){\circle*{0.7}}
\put(18.5,5.4093){\circle*{0.7}}
\put(51.65,5.43257){\circle*{0.7}}
\put(18.35,5.43257){\circle*{0.7}}
\put(51.8,5.45697){\circle*{0.7}}
\put(18.2,5.45697){\circle*{0.7}}
\put(51.95,5.48255){\circle*{0.7}}
\put(18.05,5.48255){\circle*{0.7}}
\put(52.1,5.50935){\circle*{0.7}}
\put(17.9,5.50935){\circle*{0.7}}
\put(52.25,5.53744){\circle*{0.7}}
\put(17.75,5.53744){\circle*{0.7}}
\put(52.4,5.56685){\circle*{0.7}}
\put(17.6,5.56685){\circle*{0.7}}
\put(52.55,5.59766){\circle*{0.7}}
\put(17.45,5.59766){\circle*{0.7}}
\put(52.7,5.62991){\circle*{0.7}}
\put(17.3,5.62991){\circle*{0.7}}
\put(52.85,5.66367){\circle*{0.7}}
\put(17.15,5.66367){\circle*{0.7}}
\put(53.,5.699){\circle*{0.7}}
\put(17.,5.699){\circle*{0.7}}
\put(53.15,5.73596){\circle*{0.7}}
\put(16.85,5.73596){\circle*{0.7}}
\put(53.3,5.77464){\circle*{0.7}}
\put(16.7,5.77464){\circle*{0.7}}
\put(53.45,5.81509){\circle*{0.7}}
\put(16.55,5.81509){\circle*{0.7}}
\put(53.6,5.85739){\circle*{0.7}}
\put(16.4,5.85739){\circle*{0.7}}
\put(53.75,5.90161){\circle*{0.7}}
\put(16.25,5.90161){\circle*{0.7}}
\put(53.9,5.94786){\circle*{0.7}}
\put(16.1,5.94786){\circle*{0.7}}
\put(54.05,5.99619){\circle*{0.7}}
\put(15.95,5.99619){\circle*{0.7}}
\put(54.2,6.04671){\circle*{0.7}}
\put(15.8,6.04671){\circle*{0.7}}
\put(54.35,6.09951){\circle*{0.7}}
\put(15.65,6.09951){\circle*{0.7}}
\put(54.5,6.15468){\circle*{0.7}}
\put(15.5,6.15468){\circle*{0.7}}
\put(54.65,6.21233){\circle*{0.7}}
\put(15.35,6.21233){\circle*{0.7}}
\put(54.8,6.27256){\circle*{0.7}}
\put(15.2,6.27256){\circle*{0.7}}
\put(54.95,6.33548){\circle*{0.7}}
\put(15.05,6.33548){\circle*{0.7}}
\put(55.1,6.40121){\circle*{0.7}}
\put(14.9,6.40121){\circle*{0.7}}
\put(55.25,6.46988){\circle*{0.7}}
\put(14.75,6.46988){\circle*{0.7}}
\put(55.4,6.5416){\circle*{0.7}}
\put(14.6,6.5416){\circle*{0.7}}
\put(55.55,6.61651){\circle*{0.7}}
\put(14.45,6.61651){\circle*{0.7}}
\put(55.7,6.69476){\circle*{0.7}}
\put(14.3,6.69476){\circle*{0.7}}
\put(55.85,6.77649){\circle*{0.7}}
\put(14.15,6.77649){\circle*{0.7}}
\put(56.,6.86186){\circle*{0.7}}
\put(14.,6.86186){\circle*{0.7}}
\put(56.15,6.95103){\circle*{0.7}}
\put(13.85,6.95103){\circle*{0.7}}
\put(56.3,7.04418){\circle*{0.7}}
\put(13.7,7.04418){\circle*{0.7}}
\put(56.45,7.14148){\circle*{0.7}}
\put(13.55,7.14148){\circle*{0.7}}
\put(56.6,7.24313){\circle*{0.7}}
\put(13.4,7.24313){\circle*{0.7}}
\put(56.75,7.34933){\circle*{0.7}}
\put(13.25,7.34933){\circle*{0.7}}
\put(56.9,7.46029){\circle*{0.7}}
\put(13.1,7.46029){\circle*{0.7}}
\put(57.05,7.57624){\circle*{0.7}}
\put(12.95,7.57624){\circle*{0.7}}
\put(57.2,7.69742){\circle*{0.7}}
\put(12.8,7.69742){\circle*{0.7}}
\put(57.35,7.82408){\circle*{0.7}}
\put(12.65,7.82408){\circle*{0.7}}
\put(57.5,7.9565){\circle*{0.7}}
\put(12.5,7.9565){\circle*{0.7}}
\put(57.65,8.09494){\circle*{0.7}}
\put(12.35,8.09494){\circle*{0.7}}
\put(57.8,8.23973){\circle*{0.7}}
\put(12.2,8.23973){\circle*{0.7}}
\put(57.95,8.39117){\circle*{0.7}}
\put(12.05,8.39117){\circle*{0.7}}
\put(58.1,8.54961){\circle*{0.7}}
\put(11.9,8.54961){\circle*{0.7}}
\put(58.25,8.71541){\circle*{0.7}}
\put(11.75,8.71541){\circle*{0.7}}
\put(58.4,8.88897){\circle*{0.7}}
\put(11.6,8.88897){\circle*{0.7}}
\put(58.55,9.0707){\circle*{0.7}}
\put(11.45,9.0707){\circle*{0.7}}
\put(58.7,9.26105){\circle*{0.7}}
\put(11.3,9.26105){\circle*{0.7}}
\put(58.85,9.46049){\circle*{0.7}}
\put(11.15,9.46049){\circle*{0.7}}
\put(59.,9.66954){\circle*{0.7}}
\put(11.,9.66954){\circle*{0.7}}
\put(59.03,9.88875){\circle*{0.7}}
\put(10.97,9.88875){\circle*{0.7}}
\put(59.06,9.93386){\circle*{0.7}}
\put(10.94,9.93386){\circle*{0.7}}
\put(59.09,9.97941){\circle*{0.7}}
\put(10.91,9.97941){\circle*{0.7}}
\put(59.12,10.0254){\circle*{0.7}}
\put(10.88,10.0254){\circle*{0.7}}
\put(59.15,10.0718){\circle*{0.7}}
\put(10.85,10.0718){\circle*{0.7}}
\put(59.18,10.1187){\circle*{0.7}}
\put(10.82,10.1187){\circle*{0.7}}
\put(59.21,10.166){\circle*{0.7}}
\put(10.79,10.166){\circle*{0.7}}
\put(59.24,10.2138){\circle*{0.7}}
\put(10.76,10.2138){\circle*{0.7}}
\put(59.27,10.2621){\circle*{0.7}}
\put(10.73,10.2621){\circle*{0.7}}
\put(59.3,10.3108){\circle*{0.7}}
\put(10.7,10.3108){\circle*{0.7}}
\put(59.33,10.3601){\circle*{0.7}}
\put(10.67,10.3601){\circle*{0.7}}
\put(59.36,10.4098){\circle*{0.7}}
\put(10.64,10.4098){\circle*{0.7}}
\put(59.39,10.4599){\circle*{0.7}}
\put(10.61,10.4599){\circle*{0.7}}
\put(59.42,10.5102){\circle*{0.7}}
\put(10.58,10.5102){\circle*{0.7}}
\put(59.45,10.5618){\circle*{0.7}}
\put(10.55,10.5618){\circle*{0.7}}
\put(59.48,10.6135){\circle*{0.7}}
\put(10.52,10.6135){\circle*{0.7}}
\put(59.51,10.6657){\circle*{0.7}}
\put(10.49,10.6657){\circle*{0.7}}
\put(59.54,10.7184){\circle*{0.7}}
\put(10.46,10.7184){\circle*{0.7}}
\put(59.57,10.7717){\circle*{0.7}}
\put(10.43,10.7717){\circle*{0.7}}
\put(59.6,10.8255){\circle*{0.7}}
\put(10.4,10.8255){\circle*{0.7}}
\put(59.63,10.8798){\circle*{0.7}}
\put(10.37,10.8798){\circle*{0.7}}
\put(59.66,10.9346){\circle*{0.7}}
\put(10.34,10.9346){\circle*{0.7}}
\put(59.69,10.9901){\circle*{0.7}}
\put(10.31,10.9901){\circle*{0.7}}
\put(59.72,11.046){\circle*{0.7}}
\put(10.28,11.046){\circle*{0.7}}
\put(59.75,11.1026){\circle*{0.7}}
\put(10.25,11.1026){\circle*{0.7}}
\put(59.78,11.1597){\circle*{0.7}}
\put(10.22,11.1597){\circle*{0.7}}
\put(59.81,11.2174){\circle*{0.7}}
\put(10.19,11.2174){\circle*{0.7}}
\put(59.84,11.2757){\circle*{0.7}}
\put(10.16,11.2757){\circle*{0.7}}
\put(59.87,11.3346){\circle*{0.7}}
\put(10.13,11.3346){\circle*{0.7}}
\put(59.9,11.3941){\circle*{0.7}}
\put(10.1,11.3941){\circle*{0.7}}
\put(59.93,11.4542){\circle*{0.7}}
\put(10.07,11.4542){\circle*{0.7}}
\put(59.96,11.5149){\circle*{0.7}}
\put(10.04,11.5149){\circle*{0.7}}
\put(59.99,11.5762){\circle*{0.7}}
\put(10.01,11.5762){\circle*{0.7}}
\put(60.02,11.6382){\circle*{0.7}}
\put(9.98,11.6382){\circle*{0.7}}
\put(60.05,11.7008){\circle*{0.7}}
\put(9.95,11.7008){\circle*{0.7}}
\put(60.08,11.7641){\circle*{0.7}}
\put(9.92,11.7641){\circle*{0.7}}
\put(60.11,11.8281){\circle*{0.7}}
\put(9.89,11.8281){\circle*{0.7}}
\put(60.14,11.8927){\circle*{0.7}}
\put(9.86,11.8927){\circle*{0.7}}
\put(60.17,11.958){\circle*{0.7}}
\put(9.83,11.958){\circle*{0.7}}
\put(60.2,12.024){\circle*{0.7}}
\put(9.8,12.024){\circle*{0.7}}
\put(60.23,12.0907){\circle*{0.7}}
\put(9.77,12.0907){\circle*{0.7}}
\put(60.26,12.1581){\circle*{0.7}}
\put(9.74,12.1581){\circle*{0.7}}
\put(60.29,12.2262){\circle*{0.7}}
\put(9.71,12.2262){\circle*{0.7}}
\put(60.32,12.295){\circle*{0.7}}
\put(9.68,12.295){\circle*{0.7}}
\put(60.35,12.3646){\circle*{0.7}}
\put(9.65,12.3646){\circle*{0.7}}
\put(60.38,12.4349){\circle*{0.7}}
\put(9.62,12.4349){\circle*{0.7}}
\put(60.41,12.506){\circle*{0.7}}
\put(9.59,12.506){\circle*{0.7}}
\put(60.44,12.5778){\circle*{0.7}}
\put(9.56,12.5778){\circle*{0.7}}
\put(60.47,12.6505){\circle*{0.7}}
\put(9.53,12.6505){\circle*{0.7}}
\put(60.5,12.7239){\circle*{0.7}}
\put(9.5,12.7239){\circle*{0.7}}
\put(60.53,12.7981){\circle*{0.7}}
\put(9.47,12.7981){\circle*{0.7}}
\put(60.56,12.8732){\circle*{0.7}}
\put(9.44,12.8732){\circle*{0.7}}
\put(60.59,12.9491){\circle*{0.7}}
\put(9.41,12.9491){\circle*{0.7}}
\put(60.62,13.0258){\circle*{0.7}}
\put(9.38,13.0258){\circle*{0.7}}
\put(60.65,13.1033){\circle*{0.7}}
\put(9.35,13.1033){\circle*{0.7}}
\put(60.68,13.1818){\circle*{0.7}}
\put(9.32,13.1818){\circle*{0.7}}
\put(60.71,13.2611){\circle*{0.7}}
\put(9.29,13.2611){\circle*{0.7}}
\put(60.74,13.3413){\circle*{0.7}}
\put(9.26,13.3413){\circle*{0.7}}
\put(60.77,13.4224){\circle*{0.7}}
\put(9.23,13.4224){\circle*{0.7}}
\put(60.8,13.5044){\circle*{0.7}}
\put(9.2,13.5044){\circle*{0.7}}
\put(60.83,13.5873){\circle*{0.7}}
\put(9.17,13.5873){\circle*{0.7}}
\put(60.86,13.6712){\circle*{0.7}}
\put(9.14,13.6712){\circle*{0.7}}
\put(60.89,13.7561){\circle*{0.7}}
\put(9.11,13.7561){\circle*{0.7}}
\put(60.92,13.8419){\circle*{0.7}}
\put(9.08,13.8419){\circle*{0.7}}
\put(60.95,13.9287){\circle*{0.7}}
\put(9.05,13.9287){\circle*{0.7}}
\put(60.98,14.0166){\circle*{0.7}}
\put(9.02,14.0166){\circle*{0.7}}
\put(61.01,14.1054){\circle*{0.7}}
\put(8.99,14.1054){\circle*{0.7}}
\put(61.04,14.1953){\circle*{0.7}}
\put(8.96,14.1953){\circle*{0.7}}
\put(61.07,14.2863){\circle*{0.7}}
\put(8.93,14.2863){\circle*{0.7}}
\put(61.1,14.3783){\circle*{0.7}}
\put(8.9,14.3783){\circle*{0.7}}
\put(61.13,14.4714){\circle*{0.7}}
\put(8.87,14.4714){\circle*{0.7}}
\put(61.16,14.5656){\circle*{0.7}}
\put(8.84,14.5656){\circle*{0.7}}
\put(61.19,14.6609){\circle*{0.7}}
\put(8.81,14.6609){\circle*{0.7}}
\put(61.22,14.7574){\circle*{0.7}}
\put(8.78,14.7574){\circle*{0.7}}
\put(61.25,14.8551){\circle*{0.7}}
\put(8.75,14.8551){\circle*{0.7}}
\put(61.28,14.9539){\circle*{0.7}}
\put(8.72,14.9539){\circle*{0.7}}
\put(61.31,15.054){\circle*{0.7}}
\put(8.69,15.054){\circle*{0.7}}
\put(61.34,15.1552){\circle*{0.7}}
\put(8.66,15.1552){\circle*{0.7}}
\put(61.37,15.2578){\circle*{0.7}}
\put(8.63,15.2578){\circle*{0.7}}
\put(61.4,15.3615){\circle*{0.7}}
\put(8.6,15.3615){\circle*{0.7}}
\put(61.43,15.4666){\circle*{0.7}}
\put(8.57,15.4666){\circle*{0.7}}
\put(61.46,15.573){\circle*{0.7}}
\put(8.54,15.573){\circle*{0.7}}
\put(61.49,15.6807){\circle*{0.7}}
\put(8.51,15.6807){\circle*{0.7}}
\put(61.52,15.7898){\circle*{0.7}}
\put(8.48,15.7898){\circle*{0.7}}
\put(61.55,15.9003){\circle*{0.7}}
\put(8.45,15.9003){\circle*{0.7}}
\put(61.58,16.0121){\circle*{0.7}}
\put(8.42,16.0121){\circle*{0.7}}
\put(61.61,16.1255){\circle*{0.7}}
\put(8.39,16.1255){\circle*{0.7}}
\put(61.64,16.2402){\circle*{0.7}}
\put(8.36,16.2402){\circle*{0.7}}
\put(61.67,16.3565){\circle*{0.7}}
\put(8.33,16.3565){\circle*{0.7}}
\put(61.7,16.4743){\circle*{0.7}}
\put(8.3,16.4743){\circle*{0.7}}
\put(61.73,16.5936){\circle*{0.7}}
\put(8.27,16.5936){\circle*{0.7}}
\put(61.76,16.7145){\circle*{0.7}}
\put(8.24,16.7145){\circle*{0.7}}
\put(61.79,16.837){\circle*{0.7}}
\put(8.21,16.837){\circle*{0.7}}
\put(61.82,16.9612){\circle*{0.7}}
\put(8.18,16.9612){\circle*{0.7}}
\put(61.85,17.087){\circle*{0.7}}
\put(8.15,17.087){\circle*{0.7}}
\put(61.88,17.2145){\circle*{0.7}}
\put(8.12,17.2145){\circle*{0.7}}
\put(61.91,17.3438){\circle*{0.7}}
\put(8.09,17.3438){\circle*{0.7}}
\put(61.94,17.4748){\circle*{0.7}}
\put(8.06,17.4748){\circle*{0.7}}
\put(61.97,17.6077){\circle*{0.7}}
\put(8.03,17.6077){\circle*{0.7}}
\put(62.,17.7424){\circle*{0.7}}
\put(8.,17.7424){\circle*{0.7}}
\put(62.0075,17.8789){\circle*{0.7}}
\put(7.9925,17.8789){\circle*{0.7}}
\put(62.015,17.9134){\circle*{0.7}}
\put(7.985,17.9134){\circle*{0.7}}
\put(62.0225,17.9479){\circle*{0.7}}
\put(7.9775,17.9479){\circle*{0.7}}
\put(62.03,17.9826){\circle*{0.7}}
\put(7.97,17.9826){\circle*{0.7}}
\put(62.0375,18.0174){\circle*{0.7}}
\put(7.9625,18.0174){\circle*{0.7}}
\put(62.045,18.0523){\circle*{0.7}}
\put(7.955,18.0523){\circle*{0.7}}
\put(62.0525,18.0874){\circle*{0.7}}
\put(7.9475,18.0874){\circle*{0.7}}
\put(62.06,18.1226){\circle*{0.7}}
\put(7.94,18.1226){\circle*{0.7}}
\put(62.0675,18.1579){\circle*{0.7}}
\put(7.9325,18.1579){\circle*{0.7}}
\put(62.075,18.1933){\circle*{0.7}}
\put(7.925,18.1933){\circle*{0.7}}
\put(62.0825,18.2288){\circle*{0.7}}
\put(7.9175,18.2288){\circle*{0.7}}
\put(62.09,18.2645){\circle*{0.7}}
\put(7.91,18.2645){\circle*{0.7}}
\put(62.0975,18.3003){\circle*{0.7}}
\put(7.9025,18.3003){\circle*{0.7}}
\put(62.105,18.3363){\circle*{0.7}}
\put(7.895,18.3363){\circle*{0.7}}
\put(62.1125,18.3723){\circle*{0.7}}
\put(7.8875,18.3723){\circle*{0.7}}
\put(62.12,18.4085){\circle*{0.7}}
\put(7.88,18.4085){\circle*{0.7}}
\put(62.1275,18.4448){\circle*{0.7}}
\put(7.8725,18.4448){\circle*{0.7}}
\put(62.135,18.4813){\circle*{0.7}}
\put(7.865,18.4813){\circle*{0.7}}
\put(62.1425,18.5179){\circle*{0.7}}
\put(7.8575,18.5179){\circle*{0.7}}
\put(62.15,18.5546){\circle*{0.7}}
\put(7.85,18.5546){\circle*{0.7}}
\put(62.1575,18.5914){\circle*{0.7}}
\put(7.8425,18.5914){\circle*{0.7}}
\put(62.165,18.6284){\circle*{0.7}}
\put(7.835,18.6284){\circle*{0.7}}
\put(62.1725,18.6655){\circle*{0.7}}
\put(7.8275,18.6655){\circle*{0.7}}
\put(62.18,18.7028){\circle*{0.7}}
\put(7.82,18.7028){\circle*{0.7}}
\put(62.1875,18.7402){\circle*{0.7}}
\put(7.8125,18.7402){\circle*{0.7}}
\put(62.195,18.7777){\circle*{0.7}}
\put(7.805,18.7777){\circle*{0.7}}
\put(62.2025,18.8154){\circle*{0.7}}
\put(7.7975,18.8154){\circle*{0.7}}
\put(62.21,18.8532){\circle*{0.7}}
\put(7.79,18.8532){\circle*{0.7}}
\put(62.2175,18.8911){\circle*{0.7}}
\put(7.7825,18.8911){\circle*{0.7}}
\put(62.225,18.9292){\circle*{0.7}}
\put(7.775,18.9292){\circle*{0.7}}
\put(62.2325,18.9674){\circle*{0.7}}
\put(7.7675,18.9674){\circle*{0.7}}
\put(62.24,19.0058){\circle*{0.7}}
\put(7.76,19.0058){\circle*{0.7}}
\put(62.2475,19.0443){\circle*{0.7}}
\put(7.7525,19.0443){\circle*{0.7}}
\put(62.255,19.0829){\circle*{0.7}}
\put(7.745,19.0829){\circle*{0.7}}
\put(62.2625,19.1217){\circle*{0.7}}
\put(7.7375,19.1217){\circle*{0.7}}
\put(62.27,19.1606){\circle*{0.7}}
\put(7.73,19.1606){\circle*{0.7}}
\put(62.2775,19.1997){\circle*{0.7}}
\put(7.7225,19.1997){\circle*{0.7}}
\put(62.285,19.2389){\circle*{0.7}}
\put(7.715,19.2389){\circle*{0.7}}
\put(62.2925,19.2783){\circle*{0.7}}
\put(7.7075,19.2783){\circle*{0.7}}
\put(62.3,19.3178){\circle*{0.7}}
\put(7.7,19.3178){\circle*{0.7}}
\put(62.3075,19.3575){\circle*{0.7}}
\put(7.6925,19.3575){\circle*{0.7}}
\put(62.315,19.3973){\circle*{0.7}}
\put(7.685,19.3973){\circle*{0.7}}
\put(62.3225,19.4373){\circle*{0.7}}
\put(7.6775,19.4373){\circle*{0.7}}
\put(62.33,19.4774){\circle*{0.7}}
\put(7.67,19.4774){\circle*{0.7}}
\put(62.3375,19.5176){\circle*{0.7}}
\put(7.6625,19.5176){\circle*{0.7}}
\put(62.345,19.5581){\circle*{0.7}}
\put(7.655,19.5581){\circle*{0.7}}
\put(62.3525,19.5986){\circle*{0.7}}
\put(7.6475,19.5986){\circle*{0.7}}
\put(62.36,19.6394){\circle*{0.7}}
\put(7.64,19.6394){\circle*{0.7}}
\put(62.3675,19.6803){\circle*{0.7}}
\put(7.6325,19.6803){\circle*{0.7}}
\put(62.375,19.7213){\circle*{0.7}}
\put(7.625,19.7213){\circle*{0.7}}
\put(62.3825,19.7625){\circle*{0.7}}
\put(7.6175,19.7625){\circle*{0.7}}
\put(62.39,19.8039){\circle*{0.7}}
\put(7.61,19.8039){\circle*{0.7}}
\put(62.3975,19.8454){\circle*{0.7}}
\put(7.6025,19.8454){\circle*{0.7}}
\put(62.405,19.8871){\circle*{0.7}}
\put(7.595,19.8871){\circle*{0.7}}
\put(62.4125,19.9289){\circle*{0.7}}
\put(7.5875,19.9289){\circle*{0.7}}
\put(62.42,19.9709){\circle*{0.7}}
\put(7.58,19.9709){\circle*{0.7}}
\put(62.4275,20.0131){\circle*{0.7}}
\put(7.5725,20.0131){\circle*{0.7}}
\put(62.435,20.0554){\circle*{0.7}}
\put(7.565,20.0554){\circle*{0.7}}
\put(62.4425,20.0979){\circle*{0.7}}
\put(7.5575,20.0979){\circle*{0.7}}
\put(62.45,20.1406){\circle*{0.7}}
\put(7.55,20.1406){\circle*{0.7}}
\put(62.4575,20.1834){\circle*{0.7}}
\put(7.5425,20.1834){\circle*{0.7}}
\put(62.465,20.2264){\circle*{0.7}}
\put(7.535,20.2264){\circle*{0.7}}
\put(62.4725,20.2695){\circle*{0.7}}
\put(7.5275,20.2695){\circle*{0.7}}
\put(62.48,20.3129){\circle*{0.7}}
\put(7.52,20.3129){\circle*{0.7}}
\put(62.4875,20.3564){\circle*{0.7}}
\put(7.5125,20.3564){\circle*{0.7}}
\put(62.495,20.4001){\circle*{0.7}}
\put(7.505,20.4001){\circle*{0.7}}
\put(62.5025,20.4439){\circle*{0.7}}
\put(7.4975,20.4439){\circle*{0.7}}
\put(62.51,20.4879){\circle*{0.7}}
\put(7.49,20.4879){\circle*{0.7}}
\put(62.5175,20.5321){\circle*{0.7}}
\put(7.4825,20.5321){\circle*{0.7}}
\put(62.525,20.5765){\circle*{0.7}}
\put(7.475,20.5765){\circle*{0.7}}
\put(62.5325,20.6211){\circle*{0.7}}
\put(7.4675,20.6211){\circle*{0.7}}
\put(62.54,20.6658){\circle*{0.7}}
\put(7.46,20.6658){\circle*{0.7}}
\put(62.5475,20.7107){\circle*{0.7}}
\put(7.4525,20.7107){\circle*{0.7}}
\put(62.555,20.7558){\circle*{0.7}}
\put(7.445,20.7558){\circle*{0.7}}
\put(62.5625,20.8011){\circle*{0.7}}
\put(7.4375,20.8011){\circle*{0.7}}
\put(62.57,20.8466){\circle*{0.7}}
\put(7.43,20.8466){\circle*{0.7}}
\put(62.5775,20.8922){\circle*{0.7}}
\put(7.4225,20.8922){\circle*{0.7}}
\put(62.585,20.938){\circle*{0.7}}
\put(7.415,20.938){\circle*{0.7}}
\put(62.5925,20.9841){\circle*{0.7}}
\put(7.4075,20.9841){\circle*{0.7}}
\put(62.6,21.0303){\circle*{0.7}}
\put(7.4,21.0303){\circle*{0.7}}
\put(62.6075,21.0767){\circle*{0.7}}
\put(7.3925,21.0767){\circle*{0.7}}
\put(62.615,21.1232){\circle*{0.7}}
\put(7.385,21.1232){\circle*{0.7}}
\put(62.6225,21.17){\circle*{0.7}}
\put(7.3775,21.17){\circle*{0.7}}
\put(62.63,21.217){\circle*{0.7}}
\put(7.37,21.217){\circle*{0.7}}
\put(62.6375,21.2641){\circle*{0.7}}
\put(7.3625,21.2641){\circle*{0.7}}
\put(62.645,21.3115){\circle*{0.7}}
\put(7.355,21.3115){\circle*{0.7}}
\put(62.6525,21.3591){\circle*{0.7}}
\put(7.3475,21.3591){\circle*{0.7}}
\put(62.66,21.4068){\circle*{0.7}}
\put(7.34,21.4068){\circle*{0.7}}
\put(62.6675,21.4548){\circle*{0.7}}
\put(7.3325,21.4548){\circle*{0.7}}
\put(62.675,21.5029){\circle*{0.7}}
\put(7.325,21.5029){\circle*{0.7}}
\put(62.6825,21.5513){\circle*{0.7}}
\put(7.3175,21.5513){\circle*{0.7}}
\put(62.69,21.5998){\circle*{0.7}}
\put(7.31,21.5998){\circle*{0.7}}
\put(62.6975,21.6486){\circle*{0.7}}
\put(7.3025,21.6486){\circle*{0.7}}
\put(62.705,21.6976){\circle*{0.7}}
\put(7.295,21.6976){\circle*{0.7}}
\put(62.7125,21.7467){\circle*{0.7}}
\put(7.2875,21.7467){\circle*{0.7}}
\put(62.72,21.7961){\circle*{0.7}}
\put(7.28,21.7961){\circle*{0.7}}
\put(62.7275,21.8457){\circle*{0.7}}
\put(7.2725,21.8457){\circle*{0.7}}
\put(62.735,21.8955){\circle*{0.7}}
\put(7.265,21.8955){\circle*{0.7}}
\put(62.7425,21.9455){\circle*{0.7}}
\put(7.2575,21.9455){\circle*{0.7}}
\put(62.75,21.9958){\circle*{0.7}}
\put(7.25,21.9958){\circle*{0.7}}
\put(62.7575,22.0462){\circle*{0.7}}
\put(7.2425,22.0462){\circle*{0.7}}
\put(62.765,22.0969){\circle*{0.7}}
\put(7.235,22.0969){\circle*{0.7}}
\put(62.7725,22.1478){\circle*{0.7}}
\put(7.2275,22.1478){\circle*{0.7}}
\put(62.78,22.1989){\circle*{0.7}}
\put(7.22,22.1989){\circle*{0.7}}
\put(62.7875,22.2502){\circle*{0.7}}
\put(7.2125,22.2502){\circle*{0.7}}
\put(62.795,22.3017){\circle*{0.7}}
\put(7.205,22.3017){\circle*{0.7}}
\put(62.8025,22.3535){\circle*{0.7}}
\put(7.1975,22.3535){\circle*{0.7}}
\put(62.81,22.4055){\circle*{0.7}}
\put(7.19,22.4055){\circle*{0.7}}
\put(62.8175,22.4577){\circle*{0.7}}
\put(7.1825,22.4577){\circle*{0.7}}
\put(62.825,22.5102){\circle*{0.7}}
\put(7.175,22.5102){\circle*{0.7}}
\put(62.8325,22.5629){\circle*{0.7}}
\put(7.1675,22.5629){\circle*{0.7}}
\put(62.84,22.6158){\circle*{0.7}}
\put(7.16,22.6158){\circle*{0.7}}
\put(62.8475,22.669){\circle*{0.7}}
\put(7.1525,22.669){\circle*{0.7}}
\put(62.855,22.7224){\circle*{0.7}}
\put(7.145,22.7224){\circle*{0.7}}
\put(62.8625,22.776){\circle*{0.7}}
\put(7.1375,22.776){\circle*{0.7}}
\put(62.87,22.8299){\circle*{0.7}}
\put(7.13,22.8299){\circle*{0.7}}
\put(62.8775,22.884){\circle*{0.7}}
\put(7.1225,22.884){\circle*{0.7}}
\put(62.885,22.9383){\circle*{0.7}}
\put(7.115,22.9383){\circle*{0.7}}
\put(62.8925,22.9929){\circle*{0.7}}
\put(7.1075,22.9929){\circle*{0.7}}
\put(62.9,23.0478){\circle*{0.7}}
\put(7.1,23.0478){\circle*{0.7}}
\put(62.9075,23.1029){\circle*{0.7}}
\put(7.0925,23.1029){\circle*{0.7}}
\put(62.915,23.1582){\circle*{0.7}}
\put(7.085,23.1582){\circle*{0.7}}
\put(62.9225,23.2138){\circle*{0.7}}
\put(7.0775,23.2138){\circle*{0.7}}
\put(62.93,23.2697){\circle*{0.7}}
\put(7.07,23.2697){\circle*{0.7}}
\put(62.9375,23.3258){\circle*{0.7}}
\put(7.0625,23.3258){\circle*{0.7}}
\put(62.945,23.3821){\circle*{0.7}}
\put(7.055,23.3821){\circle*{0.7}}
\put(62.9525,23.4388){\circle*{0.7}}
\put(7.0475,23.4388){\circle*{0.7}}
\put(62.96,23.4956){\circle*{0.7}}
\put(7.04,23.4956){\circle*{0.7}}
\put(62.9675,23.5528){\circle*{0.7}}
\put(7.0325,23.5528){\circle*{0.7}}
\put(62.975,23.6102){\circle*{0.7}}
\put(7.025,23.6102){\circle*{0.7}}
\put(62.9825,23.6679){\circle*{0.7}}
\put(7.0175,23.6679){\circle*{0.7}}
\put(62.99,23.7258){\circle*{0.7}}
\put(7.01,23.7258){\circle*{0.7}}
\put(62.9975,23.784){\circle*{0.7}}
\put(7.0025,23.784){\circle*{0.7}}
\put(63.005,23.8425){\circle*{0.7}}
\put(6.995,23.8425){\circle*{0.7}}
\put(63.0125,23.9013){\circle*{0.7}}
\put(6.9875,23.9013){\circle*{0.7}}
\put(63.02,23.9604){\circle*{0.7}}
\put(6.98,23.9604){\circle*{0.7}}
\put(63.0275,24.0197){\circle*{0.7}}
\put(6.9725,24.0197){\circle*{0.7}}
\put(63.035,24.0793){\circle*{0.7}}
\put(6.965,24.0793){\circle*{0.7}}
\put(63.0425,24.1392){\circle*{0.7}}
\put(6.9575,24.1392){\circle*{0.7}}
\put(63.05,24.1994){\circle*{0.7}}
\put(6.95,24.1994){\circle*{0.7}}
\put(63.0575,24.2598){\circle*{0.7}}
\put(6.9425,24.2598){\circle*{0.7}}
\put(63.065,24.3206){\circle*{0.7}}
\put(6.935,24.3206){\circle*{0.7}}
\put(63.0725,24.3816){\circle*{0.7}}
\put(6.9275,24.3816){\circle*{0.7}}
\put(63.08,24.443){\circle*{0.7}}
\put(6.92,24.443){\circle*{0.7}}
\put(63.0875,24.5046){\circle*{0.7}}
\put(6.9125,24.5046){\circle*{0.7}}
\put(63.095,24.5666){\circle*{0.7}}
\put(6.905,24.5666){\circle*{0.7}}
\put(63.1025,24.6288){\circle*{0.7}}
\put(6.8975,24.6288){\circle*{0.7}}
\put(63.11,24.6914){\circle*{0.7}}
\put(6.89,24.6914){\circle*{0.7}}
\put(63.1175,24.7542){\circle*{0.7}}
\put(6.8825,24.7542){\circle*{0.7}}
\put(63.125,24.8174){\circle*{0.7}}
\put(6.875,24.8174){\circle*{0.7}}
\put(63.1325,24.8809){\circle*{0.7}}
\put(6.8675,24.8809){\circle*{0.7}}
\put(63.14,24.9447){\circle*{0.7}}
\put(6.86,24.9447){\circle*{0.7}}
\put(63.1475,25.0088){\circle*{0.7}}
\put(6.8525,25.0088){\circle*{0.7}}
\put(63.155,25.0732){\circle*{0.7}}
\put(6.845,25.0732){\circle*{0.7}}
\put(63.1625,25.138){\circle*{0.7}}
\put(6.8375,25.138){\circle*{0.7}}
\put(63.17,25.2031){\circle*{0.7}}
\put(6.83,25.2031){\circle*{0.7}}
\put(63.1775,25.2685){\circle*{0.7}}
\put(6.8225,25.2685){\circle*{0.7}}
\put(63.185,25.3342){\circle*{0.7}}
\put(6.815,25.3342){\circle*{0.7}}
\put(63.1925,25.4003){\circle*{0.7}}
\put(6.8075,25.4003){\circle*{0.7}}
\put(63.2,25.4667){\circle*{0.7}}
\put(6.8,25.4667){\circle*{0.7}}
\put(63.2075,25.5335){\circle*{0.7}}
\put(6.7925,25.5335){\circle*{0.7}}
\put(63.215,25.6006){\circle*{0.7}}
\put(6.785,25.6006){\circle*{0.7}}
\put(63.2225,25.668){\circle*{0.7}}
\put(6.7775,25.668){\circle*{0.7}}
\put(63.23,25.7358){\circle*{0.7}}
\put(6.77,25.7358){\circle*{0.7}}
\put(63.2375,25.8039){\circle*{0.7}}
\put(6.7625,25.8039){\circle*{0.7}}
\put(63.245,25.8724){\circle*{0.7}}
\put(6.755,25.8724){\circle*{0.7}}
\put(63.2525,25.9413){\circle*{0.7}}
\put(6.7475,25.9413){\circle*{0.7}}
\put(63.26,26.0105){\circle*{0.7}}
\put(6.74,26.0105){\circle*{0.7}}
\put(63.2675,26.0801){\circle*{0.7}}
\put(6.7325,26.0801){\circle*{0.7}}
\put(63.275,26.15){\circle*{0.7}}
\put(6.725,26.15){\circle*{0.7}}
\put(63.2825,26.2203){\circle*{0.7}}
\put(6.7175,26.2203){\circle*{0.7}}
\put(63.29,26.291){\circle*{0.7}}
\put(6.71,26.291){\circle*{0.7}}
\put(63.2975,26.3621){\circle*{0.7}}
\put(6.7025,26.3621){\circle*{0.7}}
\put(63.305,26.4335){\circle*{0.7}}
\put(6.695,26.4335){\circle*{0.7}}
\put(63.3125,26.5053){\circle*{0.7}}
\put(6.6875,26.5053){\circle*{0.7}}
\put(63.32,26.5775){\circle*{0.7}}
\put(6.68,26.5775){\circle*{0.7}}
\put(63.3275,26.6501){\circle*{0.7}}
\put(6.6725,26.6501){\circle*{0.7}}
\put(63.335,26.7231){\circle*{0.7}}
\put(6.665,26.7231){\circle*{0.7}}
\put(63.3425,26.7965){\circle*{0.7}}
\put(6.6575,26.7965){\circle*{0.7}}
\put(63.35,26.8703){\circle*{0.7}}
\put(6.65,26.8703){\circle*{0.7}}
\put(63.3575,26.9445){\circle*{0.7}}
\put(6.6425,26.9445){\circle*{0.7}}
\put(63.365,27.0191){\circle*{0.7}}
\put(6.635,27.0191){\circle*{0.7}}
\put(63.3725,27.0941){\circle*{0.7}}
\put(6.6275,27.0941){\circle*{0.7}}
\put(63.38,27.1695){\circle*{0.7}}
\put(6.62,27.1695){\circle*{0.7}}
\put(63.3875,27.2454){\circle*{0.7}}
\put(6.6125,27.2454){\circle*{0.7}}
\put(63.395,27.3216){\circle*{0.7}}
\put(6.605,27.3216){\circle*{0.7}}
\put(63.4025,27.3983){\circle*{0.7}}
\put(6.5975,27.3983){\circle*{0.7}}
\put(63.41,27.4755){\circle*{0.7}}
\put(6.59,27.4755){\circle*{0.7}}
\put(63.4175,27.553){\circle*{0.7}}
\put(6.5825,27.553){\circle*{0.7}}
\put(63.425,27.6311){\circle*{0.7}}
\put(6.575,27.6311){\circle*{0.7}}
\put(63.4325,27.7095){\circle*{0.7}}
\put(6.5675,27.7095){\circle*{0.7}}
\put(63.44,27.7884){\circle*{0.7}}
\put(6.56,27.7884){\circle*{0.7}}
\put(63.4475,27.8678){\circle*{0.7}}
\put(6.5525,27.8678){\circle*{0.7}}
\put(63.455,27.9476){\circle*{0.7}}
\put(6.545,27.9476){\circle*{0.7}}
\put(63.4625,28.0278){\circle*{0.7}}
\put(6.5375,28.0278){\circle*{0.7}}
\put(63.47,28.1086){\circle*{0.7}}
\put(6.53,28.1086){\circle*{0.7}}
\put(63.4775,28.1898){\circle*{0.7}}
\put(6.5225,28.1898){\circle*{0.7}}
\put(63.485,28.2715){\circle*{0.7}}
\put(6.515,28.2715){\circle*{0.7}}
\put(63.4925,28.3536){\circle*{0.7}}
\put(6.5075,28.3536){\circle*{0.7}}
\put(63.5,28.4363){\circle*{0.7}}
\put(6.5,28.4363){\circle*{0.7}}
\put(63.5075,28.5194){\circle*{0.7}}
\put(6.4925,28.5194){\circle*{0.7}}
\put(63.515,28.6031){\circle*{0.7}}
\put(6.485,28.6031){\circle*{0.7}}
\put(63.5225,28.6872){\circle*{0.7}}
\put(6.4775,28.6872){\circle*{0.7}}
\put(63.53,28.7718){\circle*{0.7}}
\put(6.47,28.7718){\circle*{0.7}}
\put(63.5375,28.857){\circle*{0.7}}
\put(6.4625,28.857){\circle*{0.7}}
\put(63.545,28.9426){\circle*{0.7}}
\put(6.455,28.9426){\circle*{0.7}}
\put(63.5525,29.0288){\circle*{0.7}}
\put(6.4475,29.0288){\circle*{0.7}}
\put(63.56,29.1155){\circle*{0.7}}
\put(6.44,29.1155){\circle*{0.7}}
\put(63.5675,29.2028){\circle*{0.7}}
\put(6.4325,29.2028){\circle*{0.7}}
\put(63.575,29.2906){\circle*{0.7}}
\put(6.425,29.2906){\circle*{0.7}}
\put(63.5825,29.3789){\circle*{0.7}}
\put(6.4175,29.3789){\circle*{0.7}}
\put(63.59,29.4678){\circle*{0.7}}
\put(6.41,29.4678){\circle*{0.7}}
\put(63.5975,29.5572){\circle*{0.7}}
\put(6.4025,29.5572){\circle*{0.7}}
\put(63.605,29.6472){\circle*{0.7}}
\put(6.395,29.6472){\circle*{0.7}}
\put(63.6125,29.7377){\circle*{0.7}}
\put(6.3875,29.7377){\circle*{0.7}}
\put(63.62,29.8289){\circle*{0.7}}
\put(6.38,29.8289){\circle*{0.7}}
\put(63.6275,29.9206){\circle*{0.7}}
\put(6.3725,29.9206){\circle*{0.7}}
\put(63.635,30.0129){\circle*{0.7}}
\put(6.365,30.0129){\circle*{0.7}}
\put(63.6425,30.1058){\circle*{0.7}}
\put(6.3575,30.1058){\circle*{0.7}}
\put(63.65,30.1992){\circle*{0.7}}
\put(6.35,30.1992){\circle*{0.7}}
\put(63.6575,30.2933){\circle*{0.7}}
\put(6.3425,30.2933){\circle*{0.7}}
\put(63.665,30.388){\circle*{0.7}}
\put(6.335,30.388){\circle*{0.7}}
\put(63.6725,30.4833){\circle*{0.7}}
\put(6.3275,30.4833){\circle*{0.7}}
\put(63.68,30.5793){\circle*{0.7}}
\put(6.32,30.5793){\circle*{0.7}}
\put(63.6875,30.6759){\circle*{0.7}}
\put(6.3125,30.6759){\circle*{0.7}}
\put(63.695,30.7731){\circle*{0.7}}
\put(6.305,30.7731){\circle*{0.7}}
\put(63.7025,30.871){\circle*{0.7}}
\put(6.2975,30.871){\circle*{0.7}}
\put(63.71,30.9695){\circle*{0.7}}
\put(6.29,30.9695){\circle*{0.7}}
\put(63.7175,31.0687){\circle*{0.7}}
\put(6.2825,31.0687){\circle*{0.7}}
\put(63.725,31.1685){\circle*{0.7}}
\put(6.275,31.1685){\circle*{0.7}}
\put(63.7325,31.2691){\circle*{0.7}}
\put(6.2675,31.2691){\circle*{0.7}}
\put(63.74,31.3703){\circle*{0.7}}
\put(6.26,31.3703){\circle*{0.7}}
\put(63.7475,31.4722){\circle*{0.7}}
\put(6.2525,31.4722){\circle*{0.7}}
\put(63.755,31.5748){\circle*{0.7}}
\put(6.245,31.5748){\circle*{0.7}}
\put(63.7625,31.6782){\circle*{0.7}}
\put(6.2375,31.6782){\circle*{0.7}}
\put(63.77,31.7822){\circle*{0.7}}
\put(6.23,31.7822){\circle*{0.7}}
\put(63.7775,31.887){\circle*{0.7}}
\put(6.2225,31.887){\circle*{0.7}}
\put(63.785,31.9925){\circle*{0.7}}
\put(6.215,31.9925){\circle*{0.7}}
\put(63.7925,32.0988){\circle*{0.7}}
\put(6.2075,32.0988){\circle*{0.7}}
\put(63.8,32.2058){\circle*{0.7}}
\put(6.2,32.2058){\circle*{0.7}}
\put(63.8075,32.3136){\circle*{0.7}}
\put(6.1925,32.3136){\circle*{0.7}}
\put(63.815,32.4222){\circle*{0.7}}
\put(6.185,32.4222){\circle*{0.7}}
\put(63.8225,32.5316){\circle*{0.7}}
\put(6.1775,32.5316){\circle*{0.7}}
\put(63.83,32.6417){\circle*{0.7}}
\put(6.17,32.6417){\circle*{0.7}}
\put(63.8375,32.7527){\circle*{0.7}}
\put(6.1625,32.7527){\circle*{0.7}}
\put(63.845,32.8644){\circle*{0.7}}
\put(6.155,32.8644){\circle*{0.7}}
\put(63.8525,32.977){\circle*{0.7}}
\put(6.1475,32.977){\circle*{0.7}}
\put(63.86,33.0905){\circle*{0.7}}
\put(6.14,33.0905){\circle*{0.7}}
\put(63.8675,33.2048){\circle*{0.7}}
\put(6.1325,33.2048){\circle*{0.7}}
\put(63.875,33.3199){\circle*{0.7}}
\put(6.125,33.3199){\circle*{0.7}}
\put(63.8825,33.4359){\circle*{0.7}}
\put(6.1175,33.4359){\circle*{0.7}}
\put(63.89,33.5528){\circle*{0.7}}
\put(6.11,33.5528){\circle*{0.7}}
\put(63.8975,33.6707){\circle*{0.7}}
\put(6.1025,33.6707){\circle*{0.7}}
\put(63.905,33.7894){\circle*{0.7}}
\put(6.095,33.7894){\circle*{0.7}}
\put(63.9125,33.909){\circle*{0.7}}
\put(6.0875,33.909){\circle*{0.7}}
\put(63.92,34.0296){\circle*{0.7}}
\put(6.08,34.0296){\circle*{0.7}}
\put(63.9275,34.1511){\circle*{0.7}}
\put(6.0725,34.1511){\circle*{0.7}}
\put(63.935,34.2735){\circle*{0.7}}
\put(6.065,34.2735){\circle*{0.7}}
\put(63.9425,34.397){\circle*{0.7}}
\put(6.0575,34.397){\circle*{0.7}}
\put(63.95,34.5214){\circle*{0.7}}
\put(6.05,34.5214){\circle*{0.7}}
\put(63.9575,34.6468){\circle*{0.7}}
\put(6.0425,34.6468){\circle*{0.7}}
\put(63.965,34.7733){\circle*{0.7}}
\put(6.035,34.7733){\circle*{0.7}}
\put(63.9725,34.9007){\circle*{0.7}}
\put(6.0275,34.9007){\circle*{0.7}}
\put(63.98,35.0293){\circle*{0.7}}
\put(6.02,35.0293){\circle*{0.7}}
\put(63.9875,35.1588){\circle*{0.7}}
\put(6.0125,35.1588){\circle*{0.7}}
\put(63.995,35.2895){\circle*{0.7}}
\put(6.005,35.2895){\circle*{0.7}}
\put(64.0025,35.4213){\circle*{0.7}}
\put(5.9975,35.4213){\circle*{0.7}}
\put(64.01,35.5541){\circle*{0.7}}
\put(5.99,35.5541){\circle*{0.7}}
\put(64.0175,35.6881){\circle*{0.7}}
\put(5.9825,35.6881){\circle*{0.7}}
\put(64.025,35.8232){\circle*{0.7}}
\put(5.975,35.8232){\circle*{0.7}}
\put(64.0325,35.9595){\circle*{0.7}}
\put(5.9675,35.9595){\circle*{0.7}}
\put(64.04,36.097){\circle*{0.7}}
\put(5.96,36.097){\circle*{0.7}}
\put(64.0475,36.2356){\circle*{0.7}}
\put(5.9525,36.2356){\circle*{0.7}}
\put(64.055,36.3755){\circle*{0.7}}
\put(5.945,36.3755){\circle*{0.7}}
\put(64.0625,36.5166){\circle*{0.7}}
\put(5.9375,36.5166){\circle*{0.7}}
\put(64.07,36.659){\circle*{0.7}}
\put(5.93,36.659){\circle*{0.7}}
\put(64.0775,36.8026){\circle*{0.7}}
\put(5.9225,36.8026){\circle*{0.7}}
\put(64.085,36.9475){\circle*{0.7}}
\put(5.915,36.9475){\circle*{0.7}}
\put(64.0925,37.0938){\circle*{0.7}}
\put(5.9075,37.0938){\circle*{0.7}}
\put(64.1,37.2414){\circle*{0.7}}
\put(5.9,37.2414){\circle*{0.7}}
\put(64.1075,37.3903){\circle*{0.7}}
\put(5.8925,37.3903){\circle*{0.7}}
\put(64.115,37.5407){\circle*{0.7}}
\put(5.885,37.5407){\circle*{0.7}}
\put(64.1225,37.6924){\circle*{0.7}}
\put(5.8775,37.6924){\circle*{0.7}}
\put(64.13,37.8456){\circle*{0.7}}
\put(5.87,37.8456){\circle*{0.7}}
\put(64.1375,38.0002){\circle*{0.7}}
\put(5.8625,38.0002){\circle*{0.7}}
\put(64.145,38.1563){\circle*{0.7}}
\put(5.855,38.1563){\circle*{0.7}}
\put(64.1525,38.314){\circle*{0.7}}
\put(5.8475,38.314){\circle*{0.7}}
\put(64.16,38.4731){\circle*{0.7}}
\put(5.84,38.4731){\circle*{0.7}}
\put(64.1675,38.6338){\circle*{0.7}}
\put(5.8325,38.6338){\circle*{0.7}}
\put(64.175,38.7961){\circle*{0.7}}
\put(5.825,38.7961){\circle*{0.7}}
\put(64.1825,38.96){\circle*{0.7}}
\put(5.8175,38.96){\circle*{0.7}}
\put(64.19,39.1256){\circle*{0.7}}
\put(5.81,39.1256){\circle*{0.7}}
\put(64.1975,39.2929){\circle*{0.7}}
\put(5.8025,39.2929){\circle*{0.7}}
\put(64.205,39.4618){\circle*{0.7}}
\put(5.795,39.4618){\circle*{0.7}}
\put(64.2125,39.6325){\circle*{0.7}}
\put(5.7875,39.6325){\circle*{0.7}}
\put(64.22,39.805){\circle*{0.7}}
\put(5.78,39.805){\circle*{0.7}}
\put(64.2275,39.9793){\circle*{0.7}}
\put(5.7725,39.9793){\circle*{0.7}}
\put(64.235,40.1554){\circle*{0.7}}
\put(5.765,40.1554){\circle*{0.7}}
\put(64.2425,40.3334){\circle*{0.7}}
\put(5.7575,40.3334){\circle*{0.7}}
\put(64.25,40.5133){\circle*{0.7}}
\put(5.75,40.5133){\circle*{0.7}}
\put(64.2575,40.6952){\circle*{0.7}}
\put(5.7425,40.6952){\circle*{0.7}}
\put(64.265,40.8791){\circle*{0.7}}
\put(5.735,40.8791){\circle*{0.7}}
\put(64.2725,41.065){\circle*{0.7}}
\put(5.7275,41.065){\circle*{0.7}}
\put(64.28,41.253){\circle*{0.7}}
\put(5.72,41.253){\circle*{0.7}}
\put(64.2875,41.4431){\circle*{0.7}}
\put(5.7125,41.4431){\circle*{0.7}}
\put(64.295,41.6353){\circle*{0.7}}
\put(5.705,41.6353){\circle*{0.7}}
\put(64.3025,41.8298){\circle*{0.7}}
\put(5.6975,41.8298){\circle*{0.7}}
\put(64.31,42.0265){\circle*{0.7}}
\put(5.69,42.0265){\circle*{0.7}}
\put(64.3175,42.2256){\circle*{0.7}}
\put(5.6825,42.2256){\circle*{0.7}}
\put(64.325,42.427){\circle*{0.7}}
\put(5.675,42.427){\circle*{0.7}}
\put(64.3325,42.6308){\circle*{0.7}}
\put(5.6675,42.6308){\circle*{0.7}}
\put(64.34,42.8371){\circle*{0.7}}
\put(5.66,42.8371){\circle*{0.7}}
\put(64.3475,43.0459){\circle*{0.7}}
\put(5.6525,43.0459){\circle*{0.7}}
\put(64.355,43.2573){\circle*{0.7}}
\put(5.645,43.2573){\circle*{0.7}}
\put(64.3625,43.4713){\circle*{0.7}}
\put(5.6375,43.4713){\circle*{0.7}}
\put(64.37,43.688){\circle*{0.7}}
\put(5.63,43.688){\circle*{0.7}}
\put(64.3775,43.9075){\circle*{0.7}}
\put(5.6225,43.9075){\circle*{0.7}}
\put(64.385,44.1298){\circle*{0.7}}
\put(5.615,44.1298){\circle*{0.7}}
\put(64.3925,44.355){\circle*{0.7}}
\put(5.6075,44.355){\circle*{0.7}}
\put(64.4,44.5832){\circle*{0.7}}
\put(5.6,44.5832){\circle*{0.7}}
\put(64.4075,44.8144){\circle*{0.7}}
\put(5.5925,44.8144){\circle*{0.7}}
\put(64.415,45.0487){\circle*{0.7}}
\put(5.585,45.0487){\circle*{0.7}}
\put(64.4225,45.2862){\circle*{0.7}}
\put(5.5775,45.2862){\circle*{0.7}}
\put(64.43,45.5271){\circle*{0.7}}
\put(5.57,45.5271){\circle*{0.7}}
\put(64.4375,45.7712){\circle*{0.7}}
\put(5.5625,45.7712){\circle*{0.7}}
\put(64.445,46.0189){\circle*{0.7}}
\put(5.555,46.0189){\circle*{0.7}}
\put(64.4525,46.2701){\circle*{0.7}}
\put(5.5475,46.2701){\circle*{0.7}}
\put(64.46,46.5249){\circle*{0.7}}
\put(5.54,46.5249){\circle*{0.7}}
\put(64.4675,46.7835){\circle*{0.7}}
\put(5.5325,46.7835){\circle*{0.7}}
\put(64.475,47.046){\circle*{0.7}}
\put(5.525,47.046){\circle*{0.7}}
\put(64.4825,47.3124){\circle*{0.7}}
\put(5.5175,47.3124){\circle*{0.7}}
\put(64.49,47.5828){\circle*{0.7}}
\put(5.51,47.5828){\circle*{0.7}}
\put(64.4975,47.8575){\circle*{0.7}}
\put(5.5025,47.8575){\circle*{0.7}}
\put(64.505,48.1365){\circle*{0.7}}
\put(5.495,48.1365){\circle*{0.7}}
\put(64.5125,48.42){\circle*{0.7}}
\put(5.4875,48.42){\circle*{0.7}}
\put(64.52,48.708){\circle*{0.7}}
\put(5.48,48.708){\circle*{0.7}}
\put(64.5275,49.0008){\circle*{0.7}}
\put(5.4725,49.0008){\circle*{0.7}}
\put(64.535,49.2984){\circle*{0.7}}
\put(5.465,49.2984){\circle*{0.7}}
\put(64.5425,49.6011){\circle*{0.7}}
\put(5.4575,49.6011){\circle*{0.7}}
\put(64.55,49.909){\circle*{0.7}}
\put(5.45,49.909){\circle*{0.7}}
\put(64.5575,50.2222){\circle*{0.7}}
\put(5.4425,50.2222){\circle*{0.7}}
\put(64.565,50.541){\circle*{0.7}}
\put(5.435,50.541){\circle*{0.7}}
\put(64.5725,50.8655){\circle*{0.7}}
\put(5.4275,50.8655){\circle*{0.7}}
\put(64.58,51.196){\circle*{0.7}}
\put(5.42,51.196){\circle*{0.7}}
\put(64.5875,51.5327){\circle*{0.7}}
\put(5.4125,51.5327){\circle*{0.7}}
\put(64.595,51.8757){\circle*{0.7}}
\put(5.405,51.8757){\circle*{0.7}}
\put(64.6025,52.2253){\circle*{0.7}}
\put(5.3975,52.2253){\circle*{0.7}}
\put(64.61,52.5817){\circle*{0.7}}
\put(5.39,52.5817){\circle*{0.7}}
\put(64.6175,52.9453){\circle*{0.7}}
\put(5.3825,52.9453){\circle*{0.7}}
\put(64.625,53.3163){\circle*{0.7}}
\put(5.375,53.3163){\circle*{0.7}}
\put(64.6325,53.695){\circle*{0.7}}
\put(5.3675,53.695){\circle*{0.7}}
\put(64.64,54.0816){\circle*{0.7}}
\put(5.36,54.0816){\circle*{0.7}}
\put(64.6475,54.4766){\circle*{0.7}}
\put(5.3525,54.4766){\circle*{0.7}}
\put(64.655,54.8802){\circle*{0.7}}
\put(5.345,54.8802){\circle*{0.7}}
\put(64.6625,55.293){\circle*{0.7}}
\put(5.3375,55.293){\circle*{0.7}}
\put(64.67,55.7151){\circle*{0.7}}
\put(5.33,55.7151){\circle*{0.7}}
\put(64.6775,56.1471){\circle*{0.7}}
\put(5.3225,56.1471){\circle*{0.7}}
\put(64.685,56.5895){\circle*{0.7}}
\put(5.315,56.5895){\circle*{0.7}}

%% file: sph_rate_f2.tex
\put(35,55.){\circle*{0.7}}
\put(35.2,54.9978){\circle*{0.7}}
\put(34.8,54.9978){\circle*{0.7}}
\put(35.3999,54.9911){\circle*{0.7}}
\put(34.6001,54.9911){\circle*{0.7}}
\put(35.5996,54.98){\circle*{0.7}}
\put(34.4004,54.98){\circle*{0.7}}
\put(35.799,54.9645){\circle*{0.7}}
\put(34.201,54.9645){\circle*{0.7}}
\put(35.998,54.9447){\circle*{0.7}}
\put(34.002,54.9447){\circle*{0.7}}
\put(36.1965,54.9205){\circle*{0.7}}
\put(33.8035,54.9205){\circle*{0.7}}
\put(36.3945,54.892){\circle*{0.7}}
\put(33.6055,54.892){\circle*{0.7}}
\put(36.5918,54.8592){\circle*{0.7}}
\put(33.4082,54.8592){\circle*{0.7}}
\put(36.7883,54.8223){\circle*{0.7}}
\put(33.2117,54.8223){\circle*{0.7}}
\put(36.9841,54.7813){\circle*{0.7}}
\put(33.0159,54.7813){\circle*{0.7}}
\put(37.1789,54.7362){\circle*{0.7}}
\put(32.8211,54.7362){\circle*{0.7}}
\put(37.3728,54.6872){\circle*{0.7}}
\put(32.6272,54.6872){\circle*{0.7}}
\put(37.5657,54.6343){\circle*{0.7}}
\put(32.4343,54.6343){\circle*{0.7}}
\put(37.7575,54.5776){\circle*{0.7}}
\put(32.2425,54.5776){\circle*{0.7}}
\put(37.9481,54.5171){\circle*{0.7}}
\put(32.0519,54.5171){\circle*{0.7}}
\put(38.1376,54.4531){\circle*{0.7}}
\put(31.8624,54.4531){\circle*{0.7}}
\put(38.3258,54.3855){\circle*{0.7}}
\put(31.6742,54.3855){\circle*{0.7}}
\put(38.5128,54.3145){\circle*{0.7}}
\put(31.4872,54.3145){\circle*{0.7}}
\put(38.6985,54.2401){\circle*{0.7}}
\put(31.3015,54.2401){\circle*{0.7}}
\put(38.8828,54.1624){\circle*{0.7}}
\put(31.1172,54.1624){\circle*{0.7}}
\put(39.0657,54.0817){\circle*{0.7}}
\put(30.9343,54.0817){\circle*{0.7}}
\put(39.2473,53.9978){\circle*{0.7}}
\put(30.7527,53.9978){\circle*{0.7}}
\put(39.4275,53.911){\circle*{0.7}}
\put(30.5725,53.911){\circle*{0.7}}
\put(39.6062,53.8213){\circle*{0.7}}
\put(30.3938,53.8213){\circle*{0.7}}
\put(39.7836,53.7288){\circle*{0.7}}
\put(30.2164,53.7288){\circle*{0.7}}
\put(39.9594,53.6336){\circle*{0.7}}
\put(30.0406,53.6336){\circle*{0.7}}
\put(40.1339,53.5357){\circle*{0.7}}
\put(29.8661,53.5357){\circle*{0.7}}
\put(40.3069,53.4354){\circle*{0.7}}
\put(29.6931,53.4354){\circle*{0.7}}
\put(40.4784,53.3326){\circle*{0.7}}
\put(29.5216,53.3326){\circle*{0.7}}
\put(40.6486,53.2274){\circle*{0.7}}
\put(29.3514,53.2274){\circle*{0.7}}
\put(40.8173,53.12){\circle*{0.7}}
\put(29.1827,53.12){\circle*{0.7}}
\put(40.9845,53.0103){\circle*{0.7}}
\put(29.0155,53.0103){\circle*{0.7}}
\put(41.1503,52.8985){\circle*{0.7}}
\put(28.8497,52.8985){\circle*{0.7}}
\put(41.3148,52.7847){\circle*{0.7}}
\put(28.6852,52.7847){\circle*{0.7}}
\put(41.4778,52.6688){\circle*{0.7}}
\put(28.5222,52.6688){\circle*{0.7}}
\put(41.6394,52.551){\circle*{0.7}}
\put(28.3606,52.551){\circle*{0.7}}
\put(41.7997,52.4314){\circle*{0.7}}
\put(28.2003,52.4314){\circle*{0.7}}
\put(41.9586,52.3099){\circle*{0.7}}
\put(28.0414,52.3099){\circle*{0.7}}
\put(42.1161,52.1867){\circle*{0.7}}
\put(27.8839,52.1867){\circle*{0.7}}
\put(42.2724,52.0618){\circle*{0.7}}
\put(27.7276,52.0618){\circle*{0.7}}
\put(42.4273,51.9353){\circle*{0.7}}
\put(27.5727,51.9353){\circle*{0.7}}
\put(42.5809,51.8072){\circle*{0.7}}
\put(27.4191,51.8072){\circle*{0.7}}
\put(42.7332,51.6776){\circle*{0.7}}
\put(27.2668,51.6776){\circle*{0.7}}
\put(42.8843,51.5466){\circle*{0.7}}
\put(27.1157,51.5466){\circle*{0.7}}
\put(43.0341,51.4141){\circle*{0.7}}
\put(26.9659,51.4141){\circle*{0.7}}
\put(43.1827,51.2802){\circle*{0.7}}
\put(26.8173,51.2802){\circle*{0.7}}
\put(43.3301,51.145){\circle*{0.7}}
\put(26.6699,51.145){\circle*{0.7}}
\put(43.4762,51.0085){\circle*{0.7}}
\put(26.5238,51.0085){\circle*{0.7}}
\put(43.6213,50.8708){\circle*{0.7}}
\put(26.3787,50.8708){\circle*{0.7}}
\put(43.7651,50.7318){\circle*{0.7}}
\put(26.2349,50.7318){\circle*{0.7}}
\put(43.9078,50.5917){\circle*{0.7}}
\put(26.0922,50.5917){\circle*{0.7}}
\put(44.0494,50.4505){\circle*{0.7}}
\put(25.9506,50.4505){\circle*{0.7}}
\put(44.1899,50.3081){\circle*{0.7}}
\put(25.8101,50.3081){\circle*{0.7}}
\put(44.3293,50.1647){\circle*{0.7}}
\put(25.6707,50.1647){\circle*{0.7}}
\put(44.4676,50.0202){\circle*{0.7}}
\put(25.5324,50.0202){\circle*{0.7}}
\put(44.6049,49.8748){\circle*{0.7}}
\put(25.3951,49.8748){\circle*{0.7}}
\put(44.7411,49.7284){\circle*{0.7}}
\put(25.2589,49.7284){\circle*{0.7}}
\put(44.8763,49.581){\circle*{0.7}}
\put(25.1237,49.581){\circle*{0.7}}
\put(45.0105,49.4327){\circle*{0.7}}
\put(24.9895,49.4327){\circle*{0.7}}
\put(45.1438,49.2835){\circle*{0.7}}
\put(24.8562,49.2835){\circle*{0.7}}
\put(45.276,49.1335){\circle*{0.7}}
\put(24.724,49.1335){\circle*{0.7}}
\put(45.4073,48.9826){\circle*{0.7}}
\put(24.5927,48.9826){\circle*{0.7}}
\put(45.5377,48.831){\circle*{0.7}}
\put(24.4623,48.831){\circle*{0.7}}
\put(45.6671,48.6785){\circle*{0.7}}
\put(24.3329,48.6785){\circle*{0.7}}
\put(45.7956,48.5252){\circle*{0.7}}
\put(24.2044,48.5252){\circle*{0.7}}
\put(45.9232,48.3713){\circle*{0.7}}
\put(24.0768,48.3713){\circle*{0.7}}
\put(46.05,48.2165){\circle*{0.7}}
\put(23.95,48.2165){\circle*{0.7}}
\put(46.1759,48.0611){\circle*{0.7}}
\put(23.8241,48.0611){\circle*{0.7}}
\put(46.3009,47.905){\circle*{0.7}}
\put(23.6991,47.905){\circle*{0.7}}
\put(46.4251,47.7482){\circle*{0.7}}
\put(23.5749,47.7482){\circle*{0.7}}
\put(46.5484,47.5908){\circle*{0.7}}
\put(23.4516,47.5908){\circle*{0.7}}
\put(46.6709,47.4327){\circle*{0.7}}
\put(23.3291,47.4327){\circle*{0.7}}
\put(46.7927,47.274){\circle*{0.7}}
\put(23.2073,47.274){\circle*{0.7}}
\put(46.9136,47.1148){\circle*{0.7}}
\put(23.0864,47.1148){\circle*{0.7}}
\put(47.0338,46.9549){\circle*{0.7}}
\put(22.9662,46.9549){\circle*{0.7}}
\put(47.1532,46.7944){\circle*{0.7}}
\put(22.8468,46.7944){\circle*{0.7}}
\put(47.2718,46.6334){\circle*{0.7}}
\put(22.7282,46.6334){\circle*{0.7}}
\put(47.3898,46.4719){\circle*{0.7}}
\put(22.6102,46.4719){\circle*{0.7}}
\put(47.5069,46.3098){\circle*{0.7}}
\put(22.4931,46.3098){\circle*{0.7}}
\put(47.6234,46.1472){\circle*{0.7}}
\put(22.3766,46.1472){\circle*{0.7}}
\put(47.7391,45.9841){\circle*{0.7}}
\put(22.2609,45.9841){\circle*{0.7}}
\put(47.8542,45.8205){\circle*{0.7}}
\put(22.1458,45.8205){\circle*{0.7}}
\put(47.9686,45.6565){\circle*{0.7}}
\put(22.0314,45.6565){\circle*{0.7}}
\put(48.0823,45.4919){\circle*{0.7}}
\put(21.9177,45.4919){\circle*{0.7}}
\put(48.1953,45.3269){\circle*{0.7}}
\put(21.8047,45.3269){\circle*{0.7}}
\put(48.3077,45.1615){\circle*{0.7}}
\put(21.6923,45.1615){\circle*{0.7}}
\put(48.4194,44.9956){\circle*{0.7}}
\put(21.5806,44.9956){\circle*{0.7}}
\put(48.5305,44.8293){\circle*{0.7}}
\put(21.4695,44.8293){\circle*{0.7}}
\put(48.6409,44.6625){\circle*{0.7}}
\put(21.3591,44.6625){\circle*{0.7}}
\put(48.7507,44.4954){\circle*{0.7}}
\put(21.2493,44.4954){\circle*{0.7}}
\put(48.86,44.3278){\circle*{0.7}}
\put(21.14,44.3278){\circle*{0.7}}
\put(48.9686,44.1599){\circle*{0.7}}
\put(21.0314,44.1599){\circle*{0.7}}
\put(49.0766,43.9916){\circle*{0.7}}
\put(20.9234,43.9916){\circle*{0.7}}
\put(49.1841,43.8229){\circle*{0.7}}
\put(20.8159,43.8229){\circle*{0.7}}
\put(49.2909,43.6539){\circle*{0.7}}
\put(20.7091,43.6539){\circle*{0.7}}
\put(49.3972,43.4844){\circle*{0.7}}
\put(20.6028,43.4844){\circle*{0.7}}
\put(49.503,43.3147){\circle*{0.7}}
\put(20.497,43.3147){\circle*{0.7}}
\put(49.6081,43.1446){\circle*{0.7}}
\put(20.3919,43.1446){\circle*{0.7}}
\put(49.7128,42.9741){\circle*{0.7}}
\put(20.2872,42.9741){\circle*{0.7}}
\put(49.8169,42.8033){\circle*{0.7}}
\put(20.1831,42.8033){\circle*{0.7}}
\put(49.9204,42.6323){\circle*{0.7}}
\put(20.0796,42.6323){\circle*{0.7}}
\put(50.0235,42.4608){\circle*{0.7}}
\put(19.9765,42.4608){\circle*{0.7}}
\put(50.126,42.2891){\circle*{0.7}}
\put(19.874,42.2891){\circle*{0.7}}
\put(50.228,42.1171){\circle*{0.7}}
\put(19.772,42.1171){\circle*{0.7}}
\put(50.3295,41.9448){\circle*{0.7}}
\put(19.6705,41.9448){\circle*{0.7}}
\put(50.4305,41.7721){\circle*{0.7}}
\put(19.5695,41.7721){\circle*{0.7}}
\put(50.5311,41.5992){\circle*{0.7}}
\put(19.4689,41.5992){\circle*{0.7}}
\put(50.6311,41.4261){\circle*{0.7}}
\put(19.3689,41.4261){\circle*{0.7}}
\put(50.7306,41.2526){\circle*{0.7}}
\put(19.2694,41.2526){\circle*{0.7}}
\put(50.8297,41.0789){\circle*{0.7}}
\put(19.1703,41.0789){\circle*{0.7}}
\put(50.9283,40.9049){\circle*{0.7}}
\put(19.0717,40.9049){\circle*{0.7}}
\put(51.0265,40.7306){\circle*{0.7}}
\put(18.9735,40.7306){\circle*{0.7}}
\put(51.1242,40.5561){\circle*{0.7}}
\put(18.8758,40.5561){\circle*{0.7}}
\put(51.2215,40.3813){\circle*{0.7}}
\put(18.7785,40.3813){\circle*{0.7}}
\put(51.3183,40.2063){\circle*{0.7}}
\put(18.6817,40.2063){\circle*{0.7}}
\put(51.4146,40.0311){\circle*{0.7}}
\put(18.5854,40.0311){\circle*{0.7}}
\put(51.5106,39.8556){\circle*{0.7}}
\put(18.4894,39.8556){\circle*{0.7}}
\put(51.6061,39.6799){\circle*{0.7}}
\put(18.3939,39.6799){\circle*{0.7}}
\put(51.7012,39.5039){\circle*{0.7}}
\put(18.2988,39.5039){\circle*{0.7}}
\put(51.7958,39.3277){\circle*{0.7}}
\put(18.2042,39.3277){\circle*{0.7}}
\put(51.8901,39.1514){\circle*{0.7}}
\put(18.1099,39.1514){\circle*{0.7}}
\put(51.984,38.9747){\circle*{0.7}}
\put(18.016,38.9747){\circle*{0.7}}
\put(52.0774,38.7979){\circle*{0.7}}
\put(17.9226,38.7979){\circle*{0.7}}
\put(52.1704,38.6209){\circle*{0.7}}
\put(17.8296,38.6209){\circle*{0.7}}
\put(52.2631,38.4436){\circle*{0.7}}
\put(17.7369,38.4436){\circle*{0.7}}
\put(52.3554,38.2662){\circle*{0.7}}
\put(17.6446,38.2662){\circle*{0.7}}
\put(52.4472,38.0885){\circle*{0.7}}
\put(17.5528,38.0885){\circle*{0.7}}
\put(52.5387,37.9107){\circle*{0.7}}
\put(17.4613,37.9107){\circle*{0.7}}
\put(52.6299,37.7327){\circle*{0.7}}
\put(17.3701,37.7327){\circle*{0.7}}
\put(52.7206,37.5544){\circle*{0.7}}
\put(17.2794,37.5544){\circle*{0.7}}
\put(52.811,37.376){\circle*{0.7}}
\put(17.189,37.376){\circle*{0.7}}
\put(52.901,37.1974){\circle*{0.7}}
\put(17.099,37.1974){\circle*{0.7}}
\put(52.9907,37.0186){\circle*{0.7}}
\put(17.0093,37.0186){\circle*{0.7}}
\put(53.08,36.8397){\circle*{0.7}}
\put(16.92,36.8397){\circle*{0.7}}
\put(53.1689,36.6606){\circle*{0.7}}
\put(16.8311,36.6606){\circle*{0.7}}
\put(53.2575,36.4813){\circle*{0.7}}
\put(16.7425,36.4813){\circle*{0.7}}
\put(53.3458,36.3018){\circle*{0.7}}
\put(16.6542,36.3018){\circle*{0.7}}
\put(53.4337,36.1221){\circle*{0.7}}
\put(16.5663,36.1221){\circle*{0.7}}
\put(53.5213,35.9423){\circle*{0.7}}
\put(16.4787,35.9423){\circle*{0.7}}
\put(53.6085,35.7624){\circle*{0.7}}
\put(16.3915,35.7624){\circle*{0.7}}
\put(53.6954,35.5822){\circle*{0.7}}
\put(16.3046,35.5822){\circle*{0.7}}
\put(53.782,35.402){\circle*{0.7}}
\put(16.218,35.402){\circle*{0.7}}
\put(53.8683,35.2215){\circle*{0.7}}
\put(16.1317,35.2215){\circle*{0.7}}
\put(53.9542,35.0409){\circle*{0.7}}
\put(16.0458,35.0409){\circle*{0.7}}
\put(54.0399,34.8602){\circle*{0.7}}
\put(15.9601,34.8602){\circle*{0.7}}
\put(54.1252,34.6793){\circle*{0.7}}
\put(15.8748,34.6793){\circle*{0.7}}
\put(54.2102,34.4983){\circle*{0.7}}
\put(15.7898,34.4983){\circle*{0.7}}
\put(54.2949,34.3171){\circle*{0.7}}
\put(15.7051,34.3171){\circle*{0.7}}
\put(54.3793,34.1358){\circle*{0.7}}
\put(15.6207,34.1358){\circle*{0.7}}
\put(54.4634,33.9543){\circle*{0.7}}
\put(15.5366,33.9543){\circle*{0.7}}
\put(54.5472,33.7727){\circle*{0.7}}
\put(15.4528,33.7727){\circle*{0.7}}
\put(54.6307,33.591){\circle*{0.7}}
\put(15.3693,33.591){\circle*{0.7}}
\put(54.7139,33.4091){\circle*{0.7}}
\put(15.2861,33.4091){\circle*{0.7}}
\put(54.7968,33.2271){\circle*{0.7}}
\put(15.2032,33.2271){\circle*{0.7}}
\put(54.8794,33.045){\circle*{0.7}}
\put(15.1206,33.045){\circle*{0.7}}
\put(54.9618,32.8627){\circle*{0.7}}
\put(15.0382,32.8627){\circle*{0.7}}
\put(55.0438,32.6803){\circle*{0.7}}
\put(14.9562,32.6803){\circle*{0.7}}
\put(55.1256,32.4978){\circle*{0.7}}
\put(14.8744,32.4978){\circle*{0.7}}
\put(55.2071,32.3151){\circle*{0.7}}
\put(14.7929,32.3151){\circle*{0.7}}
\put(55.2883,32.1324){\circle*{0.7}}
\put(14.7117,32.1324){\circle*{0.7}}
\put(55.3693,31.9495){\circle*{0.7}}
\put(14.6307,31.9495){\circle*{0.7}}
\put(55.45,31.7665){\circle*{0.7}}
\put(14.55,31.7665){\circle*{0.7}}
\put(55.5304,31.5834){\circle*{0.7}}
\put(14.4696,31.5834){\circle*{0.7}}
\put(55.6106,31.4002){\circle*{0.7}}
\put(14.3894,31.4002){\circle*{0.7}}
\put(55.6905,31.2168){\circle*{0.7}}
\put(14.3095,31.2168){\circle*{0.7}}
\put(55.7702,31.0334){\circle*{0.7}}
\put(14.2298,31.0334){\circle*{0.7}}
\put(55.8495,30.8498){\circle*{0.7}}
\put(14.1505,30.8498){\circle*{0.7}}
\put(55.9287,30.6661){\circle*{0.7}}
\put(14.0713,30.6661){\circle*{0.7}}
\put(56.0076,30.4823){\circle*{0.7}}
\put(13.9924,30.4823){\circle*{0.7}}
\put(56.0862,30.2984){\circle*{0.7}}
\put(13.9138,30.2984){\circle*{0.7}}
\put(56.1646,30.1144){\circle*{0.7}}
\put(13.8354,30.1144){\circle*{0.7}}
\put(56.2427,29.9303){\circle*{0.7}}
\put(13.7573,29.9303){\circle*{0.7}}
\put(56.3206,29.7461){\circle*{0.7}}
\put(13.6794,29.7461){\circle*{0.7}}
\put(56.3983,29.5618){\circle*{0.7}}
\put(13.6017,29.5618){\circle*{0.7}}
\put(56.4757,29.3774){\circle*{0.7}}
\put(13.5243,29.3774){\circle*{0.7}}
\put(56.5529,29.1929){\circle*{0.7}}
\put(13.4471,29.1929){\circle*{0.7}}
\put(56.6298,29.0083){\circle*{0.7}}
\put(13.3702,29.0083){\circle*{0.7}}
\put(56.7066,28.8236){\circle*{0.7}}
\put(13.2934,28.8236){\circle*{0.7}}
\put(56.7831,28.6388){\circle*{0.7}}
\put(13.2169,28.6388){\circle*{0.7}}
\put(56.8593,28.4539){\circle*{0.7}}
\put(13.1407,28.4539){\circle*{0.7}}
\put(56.9353,28.2689){\circle*{0.7}}
\put(13.0647,28.2689){\circle*{0.7}}
\put(57.0111,28.0839){\circle*{0.7}}
\put(12.9889,28.0839){\circle*{0.7}}
\put(57.0867,27.8987){\circle*{0.7}}
\put(12.9133,27.8987){\circle*{0.7}}
\put(57.1621,27.7134){\circle*{0.7}}
\put(12.8379,27.7134){\circle*{0.7}}
\put(57.2372,27.5281){\circle*{0.7}}
\put(12.7628,27.5281){\circle*{0.7}}
\put(57.3122,27.3427){\circle*{0.7}}
\put(12.6878,27.3427){\circle*{0.7}}
\put(57.3869,27.1571){\circle*{0.7}}
\put(12.6131,27.1571){\circle*{0.7}}
\put(57.4614,26.9715){\circle*{0.7}}
\put(12.5386,26.9715){\circle*{0.7}}
\put(57.5356,26.7858){\circle*{0.7}}
\put(12.4644,26.7858){\circle*{0.7}}
\put(57.6097,26.6001){\circle*{0.7}}
\put(12.3903,26.6001){\circle*{0.7}}
\put(57.6836,26.4142){\circle*{0.7}}
\put(12.3164,26.4142){\circle*{0.7}}
\put(57.7572,26.2282){\circle*{0.7}}
\put(12.2428,26.2282){\circle*{0.7}}
\put(57.8307,26.0422){\circle*{0.7}}
\put(12.1693,26.0422){\circle*{0.7}}
\put(57.9039,25.8561){\circle*{0.7}}
\put(12.0961,25.8561){\circle*{0.7}}
\put(57.977,25.6699){\circle*{0.7}}
\put(12.023,25.6699){\circle*{0.7}}
\put(58.0498,25.4837){\circle*{0.7}}
\put(11.9502,25.4837){\circle*{0.7}}
\put(58.1225,25.2973){\circle*{0.7}}
\put(11.8775,25.2973){\circle*{0.7}}
\put(58.1949,25.1109){\circle*{0.7}}
\put(11.8051,25.1109){\circle*{0.7}}
\put(58.2672,24.9244){\circle*{0.7}}
\put(11.7328,24.9244){\circle*{0.7}}
\put(58.3392,24.7378){\circle*{0.7}}
\put(11.6608,24.7378){\circle*{0.7}}
\put(58.4111,24.5512){\circle*{0.7}}
\put(11.5889,24.5512){\circle*{0.7}}
\put(58.4827,24.3645){\circle*{0.7}}
\put(11.5173,24.3645){\circle*{0.7}}
\put(58.5542,24.1777){\circle*{0.7}}
\put(11.4458,24.1777){\circle*{0.7}}
\put(58.6255,23.9908){\circle*{0.7}}
\put(11.3745,23.9908){\circle*{0.7}}
\put(58.6966,23.8039){\circle*{0.7}}
\put(11.3034,23.8039){\circle*{0.7}}
\put(58.7675,23.6169){\circle*{0.7}}
\put(11.2325,23.6169){\circle*{0.7}}
\put(58.8383,23.4298){\circle*{0.7}}
\put(11.1617,23.4298){\circle*{0.7}}
\put(58.9088,23.2427){\circle*{0.7}}
\put(11.0912,23.2427){\circle*{0.7}}
\put(58.9792,23.0555){\circle*{0.7}}
\put(11.0208,23.0555){\circle*{0.7}}
\put(59.0494,22.8682){\circle*{0.7}}
\put(10.9506,22.8682){\circle*{0.7}}
\put(59.1194,22.6808){\circle*{0.7}}
\put(10.8806,22.6808){\circle*{0.7}}
\put(59.1892,22.4934){\circle*{0.7}}
\put(10.8108,22.4934){\circle*{0.7}}
\put(59.2589,22.306){\circle*{0.7}}
\put(10.7411,22.306){\circle*{0.7}}
\put(59.3284,22.1184){\circle*{0.7}}
\put(10.6716,22.1184){\circle*{0.7}}
\put(59.3977,21.9308){\circle*{0.7}}
\put(10.6023,21.9308){\circle*{0.7}}
\put(59.4668,21.7431){\circle*{0.7}}
\put(10.5332,21.7431){\circle*{0.7}}
\put(59.5357,21.5554){\circle*{0.7}}
\put(10.4643,21.5554){\circle*{0.7}}
\put(59.6045,21.3676){\circle*{0.7}}
\put(10.3955,21.3676){\circle*{0.7}}
\put(59.6732,21.1797){\circle*{0.7}}
\put(10.3268,21.1797){\circle*{0.7}}
\put(59.7416,20.9918){\circle*{0.7}}
\put(10.2584,20.9918){\circle*{0.7}}
\put(59.8099,20.8038){\circle*{0.7}}
\put(10.1901,20.8038){\circle*{0.7}}
\put(59.878,20.6158){\circle*{0.7}}
\put(10.122,20.6158){\circle*{0.7}}
\put(59.946,20.4277){\circle*{0.7}}
\put(10.054,20.4277){\circle*{0.7}}
\put(60.0138,20.2395){\circle*{0.7}}
\put(9.98623,20.2395){\circle*{0.7}}
\put(60.0814,20.0513){\circle*{0.7}}
\put(9.9186,20.0513){\circle*{0.7}}
\put(60.1489,19.863){\circle*{0.7}}
\put(9.85114,19.863){\circle*{0.7}}
\put(60.2162,19.6747){\circle*{0.7}}
\put(9.78383,19.6747){\circle*{0.7}}
\put(60.2833,19.4863){\circle*{0.7}}
\put(9.71668,19.4863){\circle*{0.7}}
\put(60.3503,19.2979){\circle*{0.7}}
\put(9.64969,19.2979){\circle*{0.7}}
\put(60.4171,19.1094){\circle*{0.7}}
\put(9.58285,19.1094){\circle*{0.7}}
\put(60.4838,18.9208){\circle*{0.7}}
\put(9.51617,18.9208){\circle*{0.7}}
\put(60.5504,18.7322){\circle*{0.7}}
\put(9.44965,18.7322){\circle*{0.7}}
\put(60.6167,18.5435){\circle*{0.7}}
\put(9.38328,18.5435){\circle*{0.7}}
\put(60.6829,18.3548){\circle*{0.7}}
\put(9.31706,18.3548){\circle*{0.7}}
\put(60.749,18.166){\circle*{0.7}}
\put(9.25099,18.166){\circle*{0.7}}
\put(60.8149,17.9772){\circle*{0.7}}
\put(9.18508,17.9772){\circle*{0.7}}
\put(60.8807,17.7883){\circle*{0.7}}
\put(9.11931,17.7883){\circle*{0.7}}
\put(60.9463,17.5994){\circle*{0.7}}
\put(9.05369,17.5994){\circle*{0.7}}
\put(61.0118,17.4104){\circle*{0.7}}
\put(8.98822,17.4104){\circle*{0.7}}
\put(61.0771,17.2214){\circle*{0.7}}
\put(8.9229,17.2214){\circle*{0.7}}
\put(61.1423,17.0323){\circle*{0.7}}
\put(8.85773,17.0323){\circle*{0.7}}
\put(61.2073,16.8432){\circle*{0.7}}
\put(8.79269,16.8432){\circle*{0.7}}
\put(61.2722,16.654){\circle*{0.7}}
\put(8.72781,16.654){\circle*{0.7}}
\put(61.3369,16.4648){\circle*{0.7}}
\put(8.66306,16.4648){\circle*{0.7}}
\put(61.4015,16.2755){\circle*{0.7}}
\put(8.59846,16.2755){\circle*{0.7}}
\put(61.466,16.0862){\circle*{0.7}}
\put(8.534,16.0862){\circle*{0.7}}
\put(61.5303,15.8968){\circle*{0.7}}
\put(8.46968,15.8968){\circle*{0.7}}
\put(61.5945,15.7074){\circle*{0.7}}
\put(8.4055,15.7074){\circle*{0.7}}
\put(61.6585,15.5179){\circle*{0.7}}
\put(8.34146,15.5179){\circle*{0.7}}
\put(61.7224,15.3284){\circle*{0.7}}
\put(8.27755,15.3284){\circle*{0.7}}
\put(61.7862,15.1388){\circle*{0.7}}
\put(8.21378,15.1388){\circle*{0.7}}
\put(61.8498,14.9492){\circle*{0.7}}
\put(8.15015,14.9492){\circle*{0.7}}
\put(61.9133,14.7596){\circle*{0.7}}
\put(8.08666,14.7596){\circle*{0.7}}
\put(61.9767,14.5699){\circle*{0.7}}
\put(8.02329,14.5699){\circle*{0.7}}
\put(62.0399,14.3801){\circle*{0.7}}
\put(7.96007,14.3801){\circle*{0.7}}
\put(62.103,14.1903){\circle*{0.7}}
\put(7.89697,14.1903){\circle*{0.7}}
\put(62.166,14.0005){\circle*{0.7}}
\put(7.83401,14.0005){\circle*{0.7}}
\put(62.2288,13.8106){\circle*{0.7}}
\put(7.77118,13.8106){\circle*{0.7}}
\put(62.2915,13.6207){\circle*{0.7}}
\put(7.70847,13.6207){\circle*{0.7}}
\put(62.3541,13.4307){\circle*{0.7}}
\put(7.6459,13.4307){\circle*{0.7}}
\put(62.4165,13.2407){\circle*{0.7}}
\put(7.58346,13.2407){\circle*{0.7}}
\put(62.4789,13.0507){\circle*{0.7}}
\put(7.52114,13.0507){\circle*{0.7}}
\put(62.541,12.8606){\circle*{0.7}}
\put(7.45896,12.8606){\circle*{0.7}}
\put(62.6031,12.6705){\circle*{0.7}}
\put(7.3969,12.6705){\circle*{0.7}}
\put(62.665,12.4803){\circle*{0.7}}
\put(7.33496,12.4803){\circle*{0.7}}
\put(62.7268,12.2901){\circle*{0.7}}
\put(7.27315,12.2901){\circle*{0.7}}
\put(62.7885,12.0999){\circle*{0.7}}
\put(7.21147,12.0999){\circle*{0.7}}
\put(62.8501,11.9096){\circle*{0.7}}
\put(7.1499,11.9096){\circle*{0.7}}
\put(62.9115,11.7192){\circle*{0.7}}
\put(7.08847,11.7192){\circle*{0.7}}
\put(62.9729,11.5289){\circle*{0.7}}
\put(7.02715,11.5289){\circle*{0.7}}
\put(63.034,11.3385){\circle*{0.7}}
\put(6.96595,11.3385){\circle*{0.7}}
\put(63.0951,11.148){\circle*{0.7}}
\put(6.90488,11.148){\circle*{0.7}}
\put(63.1561,10.9575){\circle*{0.7}}
\put(6.84393,10.9575){\circle*{0.7}}
\put(63.2169,10.767){\circle*{0.7}}
\put(6.78309,10.767){\circle*{0.7}}
\put(63.2776,10.5764){\circle*{0.7}}
\put(6.72238,10.5764){\circle*{0.7}}
\put(63.3382,10.3858){\circle*{0.7}}
\put(6.66178,10.3858){\circle*{0.7}}
\put(63.3987,10.1952){\circle*{0.7}}
\put(6.6013,10.1952){\circle*{0.7}}
\put(63.4591,10.0045){\circle*{0.7}}
\put(6.54093,10.0045){\circle*{0.7}}
\put(63.5193,9.81382){\circle*{0.7}}
\put(6.48069,9.81382){\circle*{0.7}}
\put(63.5794,9.62308){\circle*{0.7}}
\put(6.42056,9.62308){\circle*{0.7}}
\put(63.6395,9.43229){\circle*{0.7}}
\put(6.36054,9.43229){\circle*{0.7}}
\put(63.6994,9.24148){\circle*{0.7}}
\put(6.30064,9.24148){\circle*{0.7}}
\put(63.7592,9.05062){\circle*{0.7}}
\put(6.24085,9.05062){\circle*{0.7}}
\put(63.8188,8.85973){\circle*{0.7}}
\put(6.18117,8.85973){\circle*{0.7}}
\put(63.8784,8.66881){\circle*{0.7}}
\put(6.12161,8.66881){\circle*{0.7}}
\put(63.9378,8.47785){\circle*{0.7}}
\put(6.06216,8.47785){\circle*{0.7}}
\put(63.9972,8.28685){\circle*{0.7}}
\put(6.00282,8.28685){\circle*{0.7}}
\put(64.0564,8.09583){\circle*{0.7}}
\put(5.94359,8.09583){\circle*{0.7}}
\put(64.1155,7.90476){\circle*{0.7}}
\put(5.88447,7.90476){\circle*{0.7}}
\put(64.1745,7.71367){\circle*{0.7}}
\put(5.82546,7.71367){\circle*{0.7}}
\put(64.2334,7.52254){\circle*{0.7}}
\put(5.76655,7.52254){\circle*{0.7}}
\put(64.2922,7.33138){\circle*{0.7}}
\put(5.70776,7.33138){\circle*{0.7}}
\put(64.3509,7.14018){\circle*{0.7}}
\put(5.64908,7.14018){\circle*{0.7}}
\put(64.4095,6.94895){\circle*{0.7}}
\put(5.5905,6.94895){\circle*{0.7}}
\put(64.468,6.75769){\circle*{0.7}}
\put(5.53202,6.75769){\circle*{0.7}}
\put(64.5263,6.56639){\circle*{0.7}}
\put(5.47366,6.56639){\circle*{0.7}}
\put(64.5846,6.37507){\circle*{0.7}}
\put(5.4154,6.37507){\circle*{0.7}}
\put(64.6428,6.18371){\circle*{0.7}}
\put(5.35724,6.18371){\circle*{0.7}}
\put(64.7008,5.99232){\circle*{0.7}}
\put(5.29919,5.99232){\circle*{0.7}}
\put(64.7588,5.8009){\circle*{0.7}}
\put(5.24124,5.8009){\circle*{0.7}}
\put(64.8166,5.60945){\circle*{0.7}}
\put(5.18339,5.60945){\circle*{0.7}}
\put(64.8743,5.41796){\circle*{0.7}}
\put(5.12565,5.41796){\circle*{0.7}}
\put(64.932,5.22645){\circle*{0.7}}
\put(5.06801,5.22645){\circle*{0.7}}
\put(64.9895,5.03491){\circle*{0.7}}
\put(5.01047,5.03491){\circle*{0.7}}

%% file: sph_rate_ff.tex
\put(35.003,34.9214){\circle*{0.7}}
\put(34.997,34.9214){\circle*{0.7}}
\put(35.078,34.6524){\circle*{0.7}}
\put(34.922,34.6524){\circle*{0.7}}
\put(35.153,34.3843){\circle*{0.7}}
\put(34.847,34.3843){\circle*{0.7}}
\put(35.228,34.1171){\circle*{0.7}}
\put(34.772,34.1171){\circle*{0.7}}
\put(35.303,33.8509){\circle*{0.7}}
\put(34.697,33.8509){\circle*{0.7}}
\put(35.378,33.5856){\circle*{0.7}}
\put(34.622,33.5856){\circle*{0.7}}
\put(35.453,33.3212){\circle*{0.7}}
\put(34.547,33.3212){\circle*{0.7}}
\put(35.528,33.0578){\circle*{0.7}}
\put(34.472,33.0578){\circle*{0.7}}
\put(35.603,32.7953){\circle*{0.7}}
\put(34.397,32.7953){\circle*{0.7}}
\put(35.678,32.5337){\circle*{0.7}}
\put(34.322,32.5337){\circle*{0.7}}
\put(35.753,32.2731){\circle*{0.7}}
\put(34.247,32.2731){\circle*{0.7}}
\put(35.828,32.0135){\circle*{0.7}}
\put(34.172,32.0135){\circle*{0.7}}
\put(35.903,31.7548){\circle*{0.7}}
\put(34.097,31.7548){\circle*{0.7}}
\put(35.978,31.497){\circle*{0.7}}
\put(34.022,31.497){\circle*{0.7}}
\put(36.053,31.2403){\circle*{0.7}}
\put(33.947,31.2403){\circle*{0.7}}
\put(36.128,30.9845){\circle*{0.7}}
\put(33.872,30.9845){\circle*{0.7}}
\put(36.203,30.7297){\circle*{0.7}}
\put(33.797,30.7297){\circle*{0.7}}
\put(36.278,30.4758){\circle*{0.7}}
\put(33.722,30.4758){\circle*{0.7}}
\put(36.353,30.223){\circle*{0.7}}
\put(33.647,30.223){\circle*{0.7}}
\put(36.428,29.9711){\circle*{0.7}}
\put(33.572,29.9711){\circle*{0.7}}
\put(36.503,29.7202){\circle*{0.7}}
\put(33.497,29.7202){\circle*{0.7}}
\put(36.578,29.4704){\circle*{0.7}}
\put(33.422,29.4704){\circle*{0.7}}
\put(36.653,29.2215){\circle*{0.7}}
\put(33.347,29.2215){\circle*{0.7}}
\put(36.728,28.9736){\circle*{0.7}}
\put(33.272,28.9736){\circle*{0.7}}
\put(36.803,28.7267){\circle*{0.7}}
\put(33.197,28.7267){\circle*{0.7}}
\put(36.878,28.4809){\circle*{0.7}}
\put(33.122,28.4809){\circle*{0.7}}
\put(36.953,28.2361){\circle*{0.7}}
\put(33.047,28.2361){\circle*{0.7}}
\put(37.028,27.9923){\circle*{0.7}}
\put(32.972,27.9923){\circle*{0.7}}
\put(37.103,27.7495){\circle*{0.7}}
\put(32.897,27.7495){\circle*{0.7}}
\put(37.178,27.5077){\circle*{0.7}}
\put(32.822,27.5077){\circle*{0.7}}
\put(37.253,27.267){\circle*{0.7}}
\put(32.747,27.267){\circle*{0.7}}
\put(37.328,27.0274){\circle*{0.7}}
\put(32.672,27.0274){\circle*{0.7}}
\put(37.403,26.7887){\circle*{0.7}}
\put(32.597,26.7887){\circle*{0.7}}
\put(37.478,26.5511){\circle*{0.7}}
\put(32.522,26.5511){\circle*{0.7}}
\put(37.553,26.3146){\circle*{0.7}}
\put(32.447,26.3146){\circle*{0.7}}
\put(37.628,26.0791){\circle*{0.7}}
\put(32.372,26.0791){\circle*{0.7}}
\put(37.703,25.8447){\circle*{0.7}}
\put(32.297,25.8447){\circle*{0.7}}
\put(37.778,25.6114){\circle*{0.7}}
\put(32.222,25.6114){\circle*{0.7}}
\put(37.853,25.3791){\circle*{0.7}}
\put(32.147,25.3791){\circle*{0.7}}
\put(37.928,25.1479){\circle*{0.7}}
\put(32.072,25.1479){\circle*{0.7}}
\put(38.003,24.9178){\circle*{0.7}}
\put(31.997,24.9178){\circle*{0.7}}
\put(38.078,24.6887){\circle*{0.7}}
\put(31.922,24.6887){\circle*{0.7}}
\put(38.153,24.4608){\circle*{0.7}}
\put(31.847,24.4608){\circle*{0.7}}
\put(38.228,24.2339){\circle*{0.7}}
\put(31.772,24.2339){\circle*{0.7}}
\put(38.303,24.0082){\circle*{0.7}}
\put(31.697,24.0082){\circle*{0.7}}
\put(38.378,23.7835){\circle*{0.7}}
\put(31.622,23.7835){\circle*{0.7}}
\put(38.453,23.5599){\circle*{0.7}}
\put(31.547,23.5599){\circle*{0.7}}
\put(38.528,23.3375){\circle*{0.7}}
\put(31.472,23.3375){\circle*{0.7}}
\put(38.603,23.1161){\circle*{0.7}}
\put(31.397,23.1161){\circle*{0.7}}
\put(38.678,22.8959){\circle*{0.7}}
\put(31.322,22.8959){\circle*{0.7}}
\put(38.753,22.6768){\circle*{0.7}}
\put(31.247,22.6768){\circle*{0.7}}
\put(38.828,22.4588){\circle*{0.7}}
\put(31.172,22.4588){\circle*{0.7}}
\put(38.903,22.2419){\circle*{0.7}}
\put(31.097,22.2419){\circle*{0.7}}
\put(38.978,22.0262){\circle*{0.7}}
\put(31.022,22.0262){\circle*{0.7}}
\put(39.053,21.8116){\circle*{0.7}}
\put(30.947,21.8116){\circle*{0.7}}
\put(39.128,21.5982){\circle*{0.7}}
\put(30.872,21.5982){\circle*{0.7}}
\put(39.203,21.3859){\circle*{0.7}}
\put(30.797,21.3859){\circle*{0.7}}
\put(39.278,21.1747){\circle*{0.7}}
\put(30.722,21.1747){\circle*{0.7}}
\put(39.353,20.9647){\circle*{0.7}}
\put(30.647,20.9647){\circle*{0.7}}
\put(39.428,20.7559){\circle*{0.7}}
\put(30.572,20.7559){\circle*{0.7}}
\put(39.503,20.5482){\circle*{0.7}}
\put(30.497,20.5482){\circle*{0.7}}
\put(39.578,20.3417){\circle*{0.7}}
\put(30.422,20.3417){\circle*{0.7}}
\put(39.653,20.1364){\circle*{0.7}}
\put(30.347,20.1364){\circle*{0.7}}
\put(39.728,19.9323){\circle*{0.7}}
\put(30.272,19.9323){\circle*{0.7}}
\put(39.803,19.7293){\circle*{0.7}}
\put(30.197,19.7293){\circle*{0.7}}
\put(39.878,19.5275){\circle*{0.7}}
\put(30.122,19.5275){\circle*{0.7}}
\put(39.953,19.3269){\circle*{0.7}}
\put(30.047,19.3269){\circle*{0.7}}
\put(40.028,19.1275){\circle*{0.7}}
\put(29.972,19.1275){\circle*{0.7}}
\put(40.103,18.9293){\circle*{0.7}}
\put(29.897,18.9293){\circle*{0.7}}
\put(40.178,18.7322){\circle*{0.7}}
\put(29.822,18.7322){\circle*{0.7}}
\put(40.253,18.5364){\circle*{0.7}}
\put(29.747,18.5364){\circle*{0.7}}
\put(40.328,18.3418){\circle*{0.7}}
\put(29.672,18.3418){\circle*{0.7}}
\put(40.403,18.1485){\circle*{0.7}}
\put(29.597,18.1485){\circle*{0.7}}
\put(40.478,17.9563){\circle*{0.7}}
\put(29.522,17.9563){\circle*{0.7}}
\put(40.553,17.7654){\circle*{0.7}}
\put(29.447,17.7654){\circle*{0.7}}
\put(40.628,17.5756){\circle*{0.7}}
\put(29.372,17.5756){\circle*{0.7}}
\put(40.703,17.3872){\circle*{0.7}}
\put(29.297,17.3872){\circle*{0.7}}
\put(40.778,17.1999){\circle*{0.7}}
\put(29.222,17.1999){\circle*{0.7}}
\put(40.853,17.0139){\circle*{0.7}}
\put(29.147,17.0139){\circle*{0.7}}
\put(40.928,16.8291){\circle*{0.7}}
\put(29.072,16.8291){\circle*{0.7}}
\put(41.003,16.6456){\circle*{0.7}}
\put(28.997,16.6456){\circle*{0.7}}
\put(41.078,16.4633){\circle*{0.7}}
\put(28.922,16.4633){\circle*{0.7}}
\put(41.153,16.2823){\circle*{0.7}}
\put(28.847,16.2823){\circle*{0.7}}
\put(41.228,16.1026){\circle*{0.7}}
\put(28.772,16.1026){\circle*{0.7}}
\put(41.303,15.9241){\circle*{0.7}}
\put(28.697,15.9241){\circle*{0.7}}
\put(41.378,15.7469){\circle*{0.7}}
\put(28.622,15.7469){\circle*{0.7}}
\put(41.453,15.571){\circle*{0.7}}
\put(28.547,15.571){\circle*{0.7}}
\put(41.528,15.3963){\circle*{0.7}}
\put(28.472,15.3963){\circle*{0.7}}
\put(41.603,15.2229){\circle*{0.7}}
\put(28.397,15.2229){\circle*{0.7}}
\put(41.678,15.0508){\circle*{0.7}}
\put(28.322,15.0508){\circle*{0.7}}
\put(41.753,14.8801){\circle*{0.7}}
\put(28.247,14.8801){\circle*{0.7}}
\put(41.828,14.7106){\circle*{0.7}}
\put(28.172,14.7106){\circle*{0.7}}
\put(41.903,14.5423){\circle*{0.7}}
\put(28.097,14.5423){\circle*{0.7}}
\put(41.978,14.3755){\circle*{0.7}}
\put(28.022,14.3755){\circle*{0.7}}
\put(42.053,14.2099){\circle*{0.7}}
\put(27.947,14.2099){\circle*{0.7}}
\put(42.128,14.0456){\circle*{0.7}}
\put(27.872,14.0456){\circle*{0.7}}
\put(42.203,13.8826){\circle*{0.7}}
\put(27.797,13.8826){\circle*{0.7}}
\put(42.278,13.721){\circle*{0.7}}
\put(27.722,13.721){\circle*{0.7}}
\put(42.353,13.5607){\circle*{0.7}}
\put(27.647,13.5607){\circle*{0.7}}
\put(42.428,13.4017){\circle*{0.7}}
\put(27.572,13.4017){\circle*{0.7}}
\put(42.503,13.2441){\circle*{0.7}}
\put(27.497,13.2441){\circle*{0.7}}
\put(42.578,13.0878){\circle*{0.7}}
\put(27.422,13.0878){\circle*{0.7}}
\put(42.653,12.9328){\circle*{0.7}}
\put(27.347,12.9328){\circle*{0.7}}
\put(42.728,12.7792){\circle*{0.7}}
\put(27.272,12.7792){\circle*{0.7}}
\put(42.803,12.6269){\circle*{0.7}}
\put(27.197,12.6269){\circle*{0.7}}
\put(42.878,12.476){\circle*{0.7}}
\put(27.122,12.476){\circle*{0.7}}
\put(42.953,12.3265){\circle*{0.7}}
\put(27.047,12.3265){\circle*{0.7}}
\put(43.028,12.1783){\circle*{0.7}}
\put(26.972,12.1783){\circle*{0.7}}
\put(43.103,12.0315){\circle*{0.7}}
\put(26.897,12.0315){\circle*{0.7}}
\put(43.178,11.886){\circle*{0.7}}
\put(26.822,11.886){\circle*{0.7}}
\put(43.253,11.742){\circle*{0.7}}
\put(26.747,11.742){\circle*{0.7}}
\put(43.328,11.5993){\circle*{0.7}}
\put(26.672,11.5993){\circle*{0.7}}
\put(43.403,11.458){\circle*{0.7}}
\put(26.597,11.458){\circle*{0.7}}
\put(43.478,11.3181){\circle*{0.7}}
\put(26.522,11.3181){\circle*{0.7}}
\put(43.553,11.1796){\circle*{0.7}}
\put(26.447,11.1796){\circle*{0.7}}
\put(43.628,11.0425){\circle*{0.7}}
\put(26.372,11.0425){\circle*{0.7}}
\put(43.703,10.9068){\circle*{0.7}}
\put(26.297,10.9068){\circle*{0.7}}
\put(43.778,10.7726){\circle*{0.7}}
\put(26.222,10.7726){\circle*{0.7}}
\put(43.853,10.6397){\circle*{0.7}}
\put(26.147,10.6397){\circle*{0.7}}
\put(43.928,10.5082){\circle*{0.7}}
\put(26.072,10.5082){\circle*{0.7}}
\put(44.003,10.3782){\circle*{0.7}}
\put(25.997,10.3782){\circle*{0.7}}
\put(44.078,10.2496){\circle*{0.7}}
\put(25.922,10.2496){\circle*{0.7}}
\put(44.153,10.1224){\circle*{0.7}}
\put(25.847,10.1224){\circle*{0.7}}
\put(44.228,9.99672){\circle*{0.7}}
\put(25.772,9.99672){\circle*{0.7}}
\put(44.303,9.87243){\circle*{0.7}}
\put(25.697,9.87243){\circle*{0.7}}
\put(44.378,9.74958){\circle*{0.7}}
\put(25.622,9.74958){\circle*{0.7}}
\put(44.453,9.62819){\circle*{0.7}}
\put(25.547,9.62819){\circle*{0.7}}
\put(44.528,9.50825){\circle*{0.7}}
\put(25.472,9.50825){\circle*{0.7}}
\put(44.603,9.38977){\circle*{0.7}}
\put(25.397,9.38977){\circle*{0.7}}
\put(44.678,9.27276){\circle*{0.7}}
\put(25.322,9.27276){\circle*{0.7}}
\put(44.753,9.15721){\circle*{0.7}}
\put(25.247,9.15721){\circle*{0.7}}
\put(44.828,9.04314){\circle*{0.7}}
\put(25.172,9.04314){\circle*{0.7}}
\put(44.903,8.93054){\circle*{0.7}}
\put(25.097,8.93054){\circle*{0.7}}
\put(44.978,8.81943){\circle*{0.7}}
\put(25.022,8.81943){\circle*{0.7}}
\put(45.053,8.7098){\circle*{0.7}}
\put(24.947,8.7098){\circle*{0.7}}
\put(45.128,8.60166){\circle*{0.7}}
\put(24.872,8.60166){\circle*{0.7}}
\put(45.203,8.49502){\circle*{0.7}}
\put(24.797,8.49502){\circle*{0.7}}
\put(45.278,8.38988){\circle*{0.7}}
\put(24.722,8.38988){\circle*{0.7}}
\put(45.353,8.28625){\circle*{0.7}}
\put(24.647,8.28625){\circle*{0.7}}
\put(45.428,8.18412){\circle*{0.7}}
\put(24.572,8.18412){\circle*{0.7}}
\put(45.503,8.08351){\circle*{0.7}}
\put(24.497,8.08351){\circle*{0.7}}
\put(45.578,7.98442){\circle*{0.7}}
\put(24.422,7.98442){\circle*{0.7}}
\put(45.653,7.88685){\circle*{0.7}}
\put(24.347,7.88685){\circle*{0.7}}
\put(45.728,7.79081){\circle*{0.7}}
\put(24.272,7.79081){\circle*{0.7}}
\put(45.803,7.6963){\circle*{0.7}}
\put(24.197,7.6963){\circle*{0.7}}
\put(45.878,7.60333){\circle*{0.7}}
\put(24.122,7.60333){\circle*{0.7}}
\put(45.953,7.5119){\circle*{0.7}}
\put(24.047,7.5119){\circle*{0.7}}
\put(46.028,7.42202){\circle*{0.7}}
\put(23.972,7.42202){\circle*{0.7}}
\put(46.103,7.33369){\circle*{0.7}}
\put(23.897,7.33369){\circle*{0.7}}
\put(46.178,7.24692){\circle*{0.7}}
\put(23.822,7.24692){\circle*{0.7}}
\put(46.253,7.16171){\circle*{0.7}}
\put(23.747,7.16171){\circle*{0.7}}
\put(46.328,7.07807){\circle*{0.7}}
\put(23.672,7.07807){\circle*{0.7}}
\put(46.403,6.99599){\circle*{0.7}}
\put(23.597,6.99599){\circle*{0.7}}
\put(46.478,6.9155){\circle*{0.7}}
\put(23.522,6.9155){\circle*{0.7}}
\put(46.553,6.83658){\circle*{0.7}}
\put(23.447,6.83658){\circle*{0.7}}
\put(46.628,6.75924){\circle*{0.7}}
\put(23.372,6.75924){\circle*{0.7}}
\put(46.703,6.6835){\circle*{0.7}}
\put(23.297,6.6835){\circle*{0.7}}
\put(46.778,6.60935){\circle*{0.7}}
\put(23.222,6.60935){\circle*{0.7}}
\put(46.853,6.53681){\circle*{0.7}}
\put(23.147,6.53681){\circle*{0.7}}
\put(46.928,6.46586){\circle*{0.7}}
\put(23.072,6.46586){\circle*{0.7}}
\put(47.003,6.39653){\circle*{0.7}}
\put(22.997,6.39653){\circle*{0.7}}
\put(47.078,6.32881){\circle*{0.7}}
\put(22.922,6.32881){\circle*{0.7}}
\put(47.153,6.2627){\circle*{0.7}}
\put(22.847,6.2627){\circle*{0.7}}
\put(47.228,6.19823){\circle*{0.7}}
\put(22.772,6.19823){\circle*{0.7}}
\put(47.303,6.13538){\circle*{0.7}}
\put(22.697,6.13538){\circle*{0.7}}
\put(47.378,6.07416){\circle*{0.7}}
\put(22.622,6.07416){\circle*{0.7}}
\put(47.453,6.01459){\circle*{0.7}}
\put(22.547,6.01459){\circle*{0.7}}
\put(47.528,5.95665){\circle*{0.7}}
\put(22.472,5.95665){\circle*{0.7}}
\put(47.603,5.90037){\circle*{0.7}}
\put(22.397,5.90037){\circle*{0.7}}
\put(47.678,5.84574){\circle*{0.7}}
\put(22.322,5.84574){\circle*{0.7}}
\put(47.753,5.79276){\circle*{0.7}}
\put(22.247,5.79276){\circle*{0.7}}
\put(47.828,5.74146){\circle*{0.7}}
\put(22.172,5.74146){\circle*{0.7}}
\put(47.903,5.69181){\circle*{0.7}}
\put(22.097,5.69181){\circle*{0.7}}
\put(47.978,5.64385){\circle*{0.7}}
\put(22.022,5.64385){\circle*{0.7}}
\put(48.053,5.59756){\circle*{0.7}}
\put(21.947,5.59756){\circle*{0.7}}
\put(48.128,5.55295){\circle*{0.7}}
\put(21.872,5.55295){\circle*{0.7}}
\put(48.203,5.51004){\circle*{0.7}}
\put(21.797,5.51004){\circle*{0.7}}
\put(48.278,5.46881){\circle*{0.7}}
\put(21.722,5.46881){\circle*{0.7}}
\put(48.353,5.42929){\circle*{0.7}}
\put(21.647,5.42929){\circle*{0.7}}
\put(48.428,5.39147){\circle*{0.7}}
\put(21.572,5.39147){\circle*{0.7}}
\put(48.503,5.35536){\circle*{0.7}}
\put(21.497,5.35536){\circle*{0.7}}
\put(48.578,5.32096){\circle*{0.7}}
\put(21.422,5.32096){\circle*{0.7}}
\put(48.653,5.28828){\circle*{0.7}}
\put(21.347,5.28828){\circle*{0.7}}
\put(48.728,5.25733){\circle*{0.7}}
\put(21.272,5.25733){\circle*{0.7}}
\put(48.803,5.2281){\circle*{0.7}}
\put(21.197,5.2281){\circle*{0.7}}
\put(48.878,5.20061){\circle*{0.7}}
\put(21.122,5.20061){\circle*{0.7}}
\put(48.953,5.17486){\circle*{0.7}}
\put(21.047,5.17486){\circle*{0.7}}
\put(49.028,5.15086){\circle*{0.7}}
\put(20.972,5.15086){\circle*{0.7}}
\put(49.103,5.1286){\circle*{0.7}}
\put(20.897,5.1286){\circle*{0.7}}
\put(49.178,5.1081){\circle*{0.7}}
\put(20.822,5.1081){\circle*{0.7}}
\put(49.253,5.08936){\circle*{0.7}}
\put(20.747,5.08936){\circle*{0.7}}
\put(49.328,5.07239){\circle*{0.7}}
\put(20.672,5.07239){\circle*{0.7}}
\put(49.403,5.05719){\circle*{0.7}}
\put(20.597,5.05719){\circle*{0.7}}
\put(49.478,5.04377){\circle*{0.7}}
\put(20.522,5.04377){\circle*{0.7}}
\put(49.553,5.03213){\circle*{0.7}}
\put(20.447,5.03213){\circle*{0.7}}
\put(49.628,5.02227){\circle*{0.7}}
\put(20.372,5.02227){\circle*{0.7}}
\put(49.703,5.01421){\circle*{0.7}}
\put(20.297,5.01421){\circle*{0.7}}
\put(49.778,5.00795){\circle*{0.7}}
\put(20.222,5.00795){\circle*{0.7}}
\put(49.853,5.00349){\circle*{0.7}}
\put(20.147,5.00349){\circle*{0.7}}
\put(49.928,5.00084){\circle*{0.7}}
\put(20.072,5.00084){\circle*{0.7}}
\put(50.003,5.){\circle*{0.7}}
\put(19.997,5.){\circle*{0.7}}
\put(50.078,5.00099){\circle*{0.7}}
\put(19.922,5.00099){\circle*{0.7}}
\put(50.153,5.00379){\circle*{0.7}}
\put(19.847,5.00379){\circle*{0.7}}
\put(50.228,5.00843){\circle*{0.7}}
\put(19.772,5.00843){\circle*{0.7}}
\put(50.303,5.01491){\circle*{0.7}}
\put(19.697,5.01491){\circle*{0.7}}
\put(50.378,5.02323){\circle*{0.7}}
\put(19.622,5.02323){\circle*{0.7}}
\put(50.453,5.03339){\circle*{0.7}}
\put(19.547,5.03339){\circle*{0.7}}
\put(50.528,5.04541){\circle*{0.7}}
\put(19.472,5.04541){\circle*{0.7}}
\put(50.603,5.05929){\circle*{0.7}}
\put(19.397,5.05929){\circle*{0.7}}
\put(50.678,5.07503){\circle*{0.7}}
\put(19.322,5.07503){\circle*{0.7}}
\put(50.753,5.09264){\circle*{0.7}}
\put(19.247,5.09264){\circle*{0.7}}
\put(50.828,5.11213){\circle*{0.7}}
\put(19.172,5.11213){\circle*{0.7}}
\put(50.903,5.13349){\circle*{0.7}}
\put(19.097,5.13349){\circle*{0.7}}
\put(50.978,5.15675){\circle*{0.7}}
\put(19.022,5.15675){\circle*{0.7}}
\put(51.053,5.18189){\circle*{0.7}}
\put(18.947,5.18189){\circle*{0.7}}
\put(51.128,5.20894){\circle*{0.7}}
\put(18.872,5.20894){\circle*{0.7}}
\put(51.203,5.23789){\circle*{0.7}}
\put(18.797,5.23789){\circle*{0.7}}
\put(51.278,5.26875){\circle*{0.7}}
\put(18.722,5.26875){\circle*{0.7}}
\put(51.353,5.30152){\circle*{0.7}}
\put(18.647,5.30152){\circle*{0.7}}
\put(51.428,5.33621){\circle*{0.7}}
\put(18.572,5.33621){\circle*{0.7}}
\put(51.503,5.37284){\circle*{0.7}}
\put(18.497,5.37284){\circle*{0.7}}
\put(51.578,5.41139){\circle*{0.7}}
\put(18.422,5.41139){\circle*{0.7}}
\put(51.653,5.45188){\circle*{0.7}}
\put(18.347,5.45188){\circle*{0.7}}
\put(51.728,5.49432){\circle*{0.7}}
\put(18.272,5.49432){\circle*{0.7}}
\put(51.803,5.53871){\circle*{0.7}}
\put(18.197,5.53871){\circle*{0.7}}
\put(51.878,5.58506){\circle*{0.7}}
\put(18.122,5.58506){\circle*{0.7}}
\put(51.953,5.63336){\circle*{0.7}}
\put(18.047,5.63336){\circle*{0.7}}
\put(52.028,5.68364){\circle*{0.7}}
\put(17.972,5.68364){\circle*{0.7}}
\put(52.103,5.73589){\circle*{0.7}}
\put(17.897,5.73589){\circle*{0.7}}
\put(52.178,5.79012){\circle*{0.7}}
\put(17.822,5.79012){\circle*{0.7}}
\put(52.253,5.84634){\circle*{0.7}}
\put(17.747,5.84634){\circle*{0.7}}
\put(52.328,5.90455){\circle*{0.7}}
\put(17.672,5.90455){\circle*{0.7}}
\put(52.403,5.96476){\circle*{0.7}}
\put(17.597,5.96476){\circle*{0.7}}
\put(52.478,6.02697){\circle*{0.7}}
\put(17.522,6.02697){\circle*{0.7}}
\put(52.553,6.0912){\circle*{0.7}}
\put(17.447,6.0912){\circle*{0.7}}
\put(52.628,6.15744){\circle*{0.7}}
\put(17.372,6.15744){\circle*{0.7}}
\put(52.703,6.2257){\circle*{0.7}}
\put(17.297,6.2257){\circle*{0.7}}
\put(52.778,6.296){\circle*{0.7}}
\put(17.222,6.296){\circle*{0.7}}
\put(52.853,6.36833){\circle*{0.7}}
\put(17.147,6.36833){\circle*{0.7}}
\put(52.928,6.4427){\circle*{0.7}}
\put(17.072,6.4427){\circle*{0.7}}
\put(53.003,6.51912){\circle*{0.7}}
\put(16.997,6.51912){\circle*{0.7}}
\put(53.078,6.59759){\circle*{0.7}}
\put(16.922,6.59759){\circle*{0.7}}
\put(53.153,6.67813){\circle*{0.7}}
\put(16.847,6.67813){\circle*{0.7}}
\put(53.228,6.76073){\circle*{0.7}}
\put(16.772,6.76073){\circle*{0.7}}
\put(53.303,6.84541){\circle*{0.7}}
\put(16.697,6.84541){\circle*{0.7}}
\put(53.378,6.93216){\circle*{0.7}}
\put(16.622,6.93216){\circle*{0.7}}
\put(53.453,7.02101){\circle*{0.7}}
\put(16.547,7.02101){\circle*{0.7}}
\put(53.528,7.11194){\circle*{0.7}}
\put(16.472,7.11194){\circle*{0.7}}
\put(53.603,7.20498){\circle*{0.7}}
\put(16.397,7.20498){\circle*{0.7}}
\put(53.678,7.30012){\circle*{0.7}}
\put(16.322,7.30012){\circle*{0.7}}
\put(53.753,7.39738){\circle*{0.7}}
\put(16.247,7.39738){\circle*{0.7}}
\put(53.828,7.49675){\circle*{0.7}}
\put(16.172,7.49675){\circle*{0.7}}
\put(53.903,7.59825){\circle*{0.7}}
\put(16.097,7.59825){\circle*{0.7}}
\put(53.978,7.70188){\circle*{0.7}}
\put(16.022,7.70188){\circle*{0.7}}
\put(54.053,7.80765){\circle*{0.7}}
\put(15.947,7.80765){\circle*{0.7}}
\put(54.128,7.91557){\circle*{0.7}}
\put(15.872,7.91557){\circle*{0.7}}
\put(54.203,8.02564){\circle*{0.7}}
\put(15.797,8.02564){\circle*{0.7}}
\put(54.278,8.13787){\circle*{0.7}}
\put(15.722,8.13787){\circle*{0.7}}
\put(54.353,8.25226){\circle*{0.7}}
\put(15.647,8.25226){\circle*{0.7}}
\put(54.428,8.36883){\circle*{0.7}}
\put(15.572,8.36883){\circle*{0.7}}
\put(54.503,8.48758){\circle*{0.7}}
\put(15.497,8.48758){\circle*{0.7}}
\put(54.578,8.60851){\circle*{0.7}}
\put(15.422,8.60851){\circle*{0.7}}
\put(54.653,8.73164){\circle*{0.7}}
\put(15.347,8.73164){\circle*{0.7}}
\put(54.728,8.85697){\circle*{0.7}}
\put(15.272,8.85697){\circle*{0.7}}
\put(54.803,8.98451){\circle*{0.7}}
\put(15.197,8.98451){\circle*{0.7}}
\put(54.878,9.11426){\circle*{0.7}}
\put(15.122,9.11426){\circle*{0.7}}
\put(54.953,9.24623){\circle*{0.7}}
\put(15.047,9.24623){\circle*{0.7}}
\put(55.028,9.38043){\circle*{0.7}}
\put(14.972,9.38043){\circle*{0.7}}
\put(55.103,9.51687){\circle*{0.7}}
\put(14.897,9.51687){\circle*{0.7}}
\put(55.178,9.65554){\circle*{0.7}}
\put(14.822,9.65554){\circle*{0.7}}
\put(55.253,9.79647){\circle*{0.7}}
\put(14.747,9.79647){\circle*{0.7}}
\put(55.328,9.93966){\circle*{0.7}}
\put(14.672,9.93966){\circle*{0.7}}
\put(55.403,10.0851){\circle*{0.7}}
\put(14.597,10.0851){\circle*{0.7}}
\put(55.478,10.2328){\circle*{0.7}}
\put(14.522,10.2328){\circle*{0.7}}
\put(55.553,10.3828){\circle*{0.7}}
\put(14.447,10.3828){\circle*{0.7}}
\put(55.628,10.5351){\circle*{0.7}}
\put(14.372,10.5351){\circle*{0.7}}
\put(55.703,10.6897){\circle*{0.7}}
\put(14.297,10.6897){\circle*{0.7}}
\put(55.778,10.8466){\circle*{0.7}}
\put(14.222,10.8466){\circle*{0.7}}
\put(55.853,11.0058){\circle*{0.7}}
\put(14.147,11.0058){\circle*{0.7}}
\put(55.928,11.1673){\circle*{0.7}}
\put(14.072,11.1673){\circle*{0.7}}
\put(56.003,11.3311){\circle*{0.7}}
\put(13.997,11.3311){\circle*{0.7}}
\put(56.078,11.4973){\circle*{0.7}}
\put(13.922,11.4973){\circle*{0.7}}
\put(56.153,11.6658){\circle*{0.7}}
\put(13.847,11.6658){\circle*{0.7}}
\put(56.228,11.8366){\circle*{0.7}}
\put(13.772,11.8366){\circle*{0.7}}
\put(56.303,12.0098){\circle*{0.7}}
\put(13.697,12.0098){\circle*{0.7}}
\put(56.378,12.1854){\circle*{0.7}}
\put(13.622,12.1854){\circle*{0.7}}
\put(56.453,12.3633){\circle*{0.7}}
\put(13.547,12.3633){\circle*{0.7}}
\put(56.528,12.5436){\circle*{0.7}}
\put(13.472,12.5436){\circle*{0.7}}
\put(56.603,12.7263){\circle*{0.7}}
\put(13.397,12.7263){\circle*{0.7}}
\put(56.678,12.9114){\circle*{0.7}}
\put(13.322,12.9114){\circle*{0.7}}
\put(56.753,13.0989){\circle*{0.7}}
\put(13.247,13.0989){\circle*{0.7}}
\put(56.828,13.2888){\circle*{0.7}}
\put(13.172,13.2888){\circle*{0.7}}
\put(56.903,13.4811){\circle*{0.7}}
\put(13.097,13.4811){\circle*{0.7}}
\put(56.978,13.6759){\circle*{0.7}}
\put(13.022,13.6759){\circle*{0.7}}
\put(57.053,13.873){\circle*{0.7}}
\put(12.947,13.873){\circle*{0.7}}
\put(57.128,14.0726){\circle*{0.7}}
\put(12.872,14.0726){\circle*{0.7}}
\put(57.203,14.2747){\circle*{0.7}}
\put(12.797,14.2747){\circle*{0.7}}
\put(57.278,14.4792){\circle*{0.7}}
\put(12.722,14.4792){\circle*{0.7}}
\put(57.353,14.6862){\circle*{0.7}}
\put(12.647,14.6862){\circle*{0.7}}
\put(57.428,14.8957){\circle*{0.7}}
\put(12.572,14.8957){\circle*{0.7}}
\put(57.503,15.1076){\circle*{0.7}}
\put(12.497,15.1076){\circle*{0.7}}
\put(57.533,15.1931){\circle*{0.7}}
\put(12.467,15.1931){\circle*{0.7}}
\put(57.563,15.2789){\circle*{0.7}}
\put(12.437,15.2789){\circle*{0.7}}
\put(57.593,15.3652){\circle*{0.7}}
\put(12.407,15.3652){\circle*{0.7}}
\put(57.623,15.4519){\circle*{0.7}}
\put(12.377,15.4519){\circle*{0.7}}
\put(57.653,15.539){\circle*{0.7}}
\put(12.347,15.539){\circle*{0.7}}
\put(57.683,15.6264){\circle*{0.7}}
\put(12.317,15.6264){\circle*{0.7}}
\put(57.713,15.7143){\circle*{0.7}}
\put(12.287,15.7143){\circle*{0.7}}
\put(57.743,15.8026){\circle*{0.7}}
\put(12.257,15.8026){\circle*{0.7}}
\put(57.773,15.8913){\circle*{0.7}}
\put(12.227,15.8913){\circle*{0.7}}
\put(57.803,15.9803){\circle*{0.7}}
\put(12.197,15.9803){\circle*{0.7}}
\put(57.833,16.0698){\circle*{0.7}}
\put(12.167,16.0698){\circle*{0.7}}
\put(57.863,16.1597){\circle*{0.7}}
\put(12.137,16.1597){\circle*{0.7}}
\put(57.893,16.25){\circle*{0.7}}
\put(12.107,16.25){\circle*{0.7}}
\put(57.923,16.3407){\circle*{0.7}}
\put(12.077,16.3407){\circle*{0.7}}
\put(57.953,16.4318){\circle*{0.7}}
\put(12.047,16.4318){\circle*{0.7}}
\put(57.983,16.5233){\circle*{0.7}}
\put(12.017,16.5233){\circle*{0.7}}
\put(58.013,16.6153){\circle*{0.7}}
\put(11.987,16.6153){\circle*{0.7}}
\put(58.043,16.7076){\circle*{0.7}}
\put(11.957,16.7076){\circle*{0.7}}
\put(58.073,16.8003){\circle*{0.7}}
\put(11.927,16.8003){\circle*{0.7}}
\put(58.103,16.8935){\circle*{0.7}}
\put(11.897,16.8935){\circle*{0.7}}
\put(58.133,16.987){\circle*{0.7}}
\put(11.867,16.987){\circle*{0.7}}
\put(58.163,17.081){\circle*{0.7}}
\put(11.837,17.081){\circle*{0.7}}
\put(58.193,17.1754){\circle*{0.7}}
\put(11.807,17.1754){\circle*{0.7}}
\put(58.223,17.2702){\circle*{0.7}}
\put(11.777,17.2702){\circle*{0.7}}
\put(58.253,17.3654){\circle*{0.7}}
\put(11.747,17.3654){\circle*{0.7}}
\put(58.283,17.461){\circle*{0.7}}
\put(11.717,17.461){\circle*{0.7}}
\put(58.313,17.557){\circle*{0.7}}
\put(11.687,17.557){\circle*{0.7}}
\put(58.343,17.6534){\circle*{0.7}}
\put(11.657,17.6534){\circle*{0.7}}
\put(58.373,17.7503){\circle*{0.7}}
\put(11.627,17.7503){\circle*{0.7}}
\put(58.403,17.8475){\circle*{0.7}}
\put(11.597,17.8475){\circle*{0.7}}
\put(58.433,17.9452){\circle*{0.7}}
\put(11.567,17.9452){\circle*{0.7}}
\put(58.463,18.0433){\circle*{0.7}}
\put(11.537,18.0433){\circle*{0.7}}
\put(58.493,18.1418){\circle*{0.7}}
\put(11.507,18.1418){\circle*{0.7}}
\put(58.523,18.2408){\circle*{0.7}}
\put(11.477,18.2408){\circle*{0.7}}
\put(58.553,18.3401){\circle*{0.7}}
\put(11.447,18.3401){\circle*{0.7}}
\put(58.583,18.4399){\circle*{0.7}}
\put(11.417,18.4399){\circle*{0.7}}
\put(58.613,18.54){\circle*{0.7}}
\put(11.387,18.54){\circle*{0.7}}
\put(58.643,18.6406){\circle*{0.7}}
\put(11.357,18.6406){\circle*{0.7}}
\put(58.673,18.7417){\circle*{0.7}}
\put(11.327,18.7417){\circle*{0.7}}
\put(58.703,18.8431){\circle*{0.7}}
\put(11.297,18.8431){\circle*{0.7}}
\put(58.733,18.945){\circle*{0.7}}
\put(11.267,18.945){\circle*{0.7}}
\put(58.763,19.0473){\circle*{0.7}}
\put(11.237,19.0473){\circle*{0.7}}
\put(58.793,19.15){\circle*{0.7}}
\put(11.207,19.15){\circle*{0.7}}
\put(58.823,19.2531){\circle*{0.7}}
\put(11.177,19.2531){\circle*{0.7}}
\put(58.853,19.3566){\circle*{0.7}}
\put(11.147,19.3566){\circle*{0.7}}
\put(58.883,19.4606){\circle*{0.7}}
\put(11.117,19.4606){\circle*{0.7}}
\put(58.913,19.565){\circle*{0.7}}
\put(11.087,19.565){\circle*{0.7}}
\put(58.943,19.6698){\circle*{0.7}}
\put(11.057,19.6698){\circle*{0.7}}
\put(58.973,19.7751){\circle*{0.7}}
\put(11.027,19.7751){\circle*{0.7}}
\put(59.003,19.8808){\circle*{0.7}}
\put(10.997,19.8808){\circle*{0.7}}
\put(59.033,19.9869){\circle*{0.7}}
\put(10.967,19.9869){\circle*{0.7}}
\put(59.063,20.0934){\circle*{0.7}}
\put(10.937,20.0934){\circle*{0.7}}
\put(59.093,20.2004){\circle*{0.7}}
\put(10.907,20.2004){\circle*{0.7}}
\put(59.123,20.3077){\circle*{0.7}}
\put(10.877,20.3077){\circle*{0.7}}
\put(59.153,20.4156){\circle*{0.7}}
\put(10.847,20.4156){\circle*{0.7}}
\put(59.183,20.5238){\circle*{0.7}}
\put(10.817,20.5238){\circle*{0.7}}
\put(59.213,20.6325){\circle*{0.7}}
\put(10.787,20.6325){\circle*{0.7}}
\put(59.243,20.7416){\circle*{0.7}}
\put(10.757,20.7416){\circle*{0.7}}
\put(59.273,20.8511){\circle*{0.7}}
\put(10.727,20.8511){\circle*{0.7}}
\put(59.303,20.9611){\circle*{0.7}}
\put(10.697,20.9611){\circle*{0.7}}
\put(59.333,21.0715){\circle*{0.7}}
\put(10.667,21.0715){\circle*{0.7}}
\put(59.363,21.1823){\circle*{0.7}}
\put(10.637,21.1823){\circle*{0.7}}
\put(59.393,21.2936){\circle*{0.7}}
\put(10.607,21.2936){\circle*{0.7}}
\put(59.423,21.4053){\circle*{0.7}}
\put(10.577,21.4053){\circle*{0.7}}
\put(59.453,21.5174){\circle*{0.7}}
\put(10.547,21.5174){\circle*{0.7}}
\put(59.483,21.63){\circle*{0.7}}
\put(10.517,21.63){\circle*{0.7}}
\put(59.513,21.743){\circle*{0.7}}
\put(10.487,21.743){\circle*{0.7}}
\put(59.543,21.8565){\circle*{0.7}}
\put(10.457,21.8565){\circle*{0.7}}
\put(59.573,21.9704){\circle*{0.7}}
\put(10.427,21.9704){\circle*{0.7}}
\put(59.603,22.0847){\circle*{0.7}}
\put(10.397,22.0847){\circle*{0.7}}
\put(59.633,22.1994){\circle*{0.7}}
\put(10.367,22.1994){\circle*{0.7}}
\put(59.663,22.3146){\circle*{0.7}}
\put(10.337,22.3146){\circle*{0.7}}
\put(59.693,22.4303){\circle*{0.7}}
\put(10.307,22.4303){\circle*{0.7}}
\put(59.723,22.5464){\circle*{0.7}}
\put(10.277,22.5464){\circle*{0.7}}
\put(59.753,22.6629){\circle*{0.7}}
\put(10.247,22.6629){\circle*{0.7}}
\put(59.783,22.7799){\circle*{0.7}}
\put(10.217,22.7799){\circle*{0.7}}
\put(59.813,22.8973){\circle*{0.7}}
\put(10.187,22.8973){\circle*{0.7}}
\put(59.843,23.0151){\circle*{0.7}}
\put(10.157,23.0151){\circle*{0.7}}
\put(59.873,23.1334){\circle*{0.7}}
\put(10.127,23.1334){\circle*{0.7}}
\put(59.903,23.2522){\circle*{0.7}}
\put(10.097,23.2522){\circle*{0.7}}
\put(59.933,23.3713){\circle*{0.7}}
\put(10.067,23.3713){\circle*{0.7}}
\put(59.963,23.491){\circle*{0.7}}
\put(10.037,23.491){\circle*{0.7}}
\put(59.993,23.611){\circle*{0.7}}
\put(10.007,23.611){\circle*{0.7}}
\put(60.023,23.7316){\circle*{0.7}}
\put(9.977,23.7316){\circle*{0.7}}
\put(60.053,23.8525){\circle*{0.7}}
\put(9.947,23.8525){\circle*{0.7}}
\put(60.083,23.974){\circle*{0.7}}
\put(9.917,23.974){\circle*{0.7}}
\put(60.113,24.0958){\circle*{0.7}}
\put(9.887,24.0958){\circle*{0.7}}
\put(60.143,24.2182){\circle*{0.7}}
\put(9.857,24.2182){\circle*{0.7}}
\put(60.173,24.3409){\circle*{0.7}}
\put(9.827,24.3409){\circle*{0.7}}
\put(60.203,24.4641){\circle*{0.7}}
\put(9.797,24.4641){\circle*{0.7}}
\put(60.233,24.5878){\circle*{0.7}}
\put(9.767,24.5878){\circle*{0.7}}
\put(60.263,24.7119){\circle*{0.7}}
\put(9.737,24.7119){\circle*{0.7}}
\put(60.293,24.8365){\circle*{0.7}}
\put(9.707,24.8365){\circle*{0.7}}
\put(60.323,24.9616){\circle*{0.7}}
\put(9.677,24.9616){\circle*{0.7}}
\put(60.353,25.087){\circle*{0.7}}
\put(9.647,25.087){\circle*{0.7}}
\put(60.383,25.213){\circle*{0.7}}
\put(9.617,25.213){\circle*{0.7}}
\put(60.413,25.3394){\circle*{0.7}}
\put(9.587,25.3394){\circle*{0.7}}
\put(60.443,25.4662){\circle*{0.7}}
\put(9.557,25.4662){\circle*{0.7}}
\put(60.473,25.5935){\circle*{0.7}}
\put(9.527,25.5935){\circle*{0.7}}
\put(60.503,25.7213){\circle*{0.7}}
\put(9.497,25.7213){\circle*{0.7}}
\put(60.533,25.8495){\circle*{0.7}}
\put(9.467,25.8495){\circle*{0.7}}
\put(60.563,25.9782){\circle*{0.7}}
\put(9.437,25.9782){\circle*{0.7}}
\put(60.593,26.1074){\circle*{0.7}}
\put(9.407,26.1074){\circle*{0.7}}
\put(60.623,26.237){\circle*{0.7}}
\put(9.377,26.237){\circle*{0.7}}
\put(60.653,26.367){\circle*{0.7}}
\put(9.347,26.367){\circle*{0.7}}
\put(60.683,26.4975){\circle*{0.7}}
\put(9.317,26.4975){\circle*{0.7}}
\put(60.713,26.6285){\circle*{0.7}}
\put(9.287,26.6285){\circle*{0.7}}
\put(60.743,26.76){\circle*{0.7}}
\put(9.257,26.76){\circle*{0.7}}
\put(60.773,26.8919){\circle*{0.7}}
\put(9.227,26.8919){\circle*{0.7}}
\put(60.803,27.0243){\circle*{0.7}}
\put(9.197,27.0243){\circle*{0.7}}
\put(60.833,27.1571){\circle*{0.7}}
\put(9.167,27.1571){\circle*{0.7}}
\put(60.863,27.2904){\circle*{0.7}}
\put(9.137,27.2904){\circle*{0.7}}
\put(60.893,27.4242){\circle*{0.7}}
\put(9.107,27.4242){\circle*{0.7}}
\put(60.923,27.5584){\circle*{0.7}}
\put(9.077,27.5584){\circle*{0.7}}
\put(60.953,27.6931){\circle*{0.7}}
\put(9.047,27.6931){\circle*{0.7}}
\put(60.983,27.8283){\circle*{0.7}}
\put(9.017,27.8283){\circle*{0.7}}
\put(61.013,27.964){\circle*{0.7}}
\put(8.987,27.964){\circle*{0.7}}
\put(61.043,28.1001){\circle*{0.7}}
\put(8.957,28.1001){\circle*{0.7}}
\put(61.073,28.2367){\circle*{0.7}}
\put(8.927,28.2367){\circle*{0.7}}
\put(61.103,28.3737){\circle*{0.7}}
\put(8.897,28.3737){\circle*{0.7}}
\put(61.133,28.5113){\circle*{0.7}}
\put(8.867,28.5113){\circle*{0.7}}
\put(61.163,28.6493){\circle*{0.7}}
\put(8.837,28.6493){\circle*{0.7}}
\put(61.193,28.7877){\circle*{0.7}}
\put(8.807,28.7877){\circle*{0.7}}
\put(61.223,28.9267){\circle*{0.7}}
\put(8.777,28.9267){\circle*{0.7}}
\put(61.253,29.0661){\circle*{0.7}}
\put(8.747,29.0661){\circle*{0.7}}
\put(61.283,29.206){\circle*{0.7}}
\put(8.717,29.206){\circle*{0.7}}
\put(61.313,29.3464){\circle*{0.7}}
\put(8.687,29.3464){\circle*{0.7}}
\put(61.343,29.4872){\circle*{0.7}}
\put(8.657,29.4872){\circle*{0.7}}
\put(61.373,29.6286){\circle*{0.7}}
\put(8.627,29.6286){\circle*{0.7}}
\put(61.403,29.7704){\circle*{0.7}}
\put(8.597,29.7704){\circle*{0.7}}
\put(61.433,29.9127){\circle*{0.7}}
\put(8.567,29.9127){\circle*{0.7}}
\put(61.463,30.0554){\circle*{0.7}}
\put(8.537,30.0554){\circle*{0.7}}
\put(61.493,30.1987){\circle*{0.7}}
\put(8.507,30.1987){\circle*{0.7}}
\put(61.523,30.3424){\circle*{0.7}}
\put(8.477,30.3424){\circle*{0.7}}
\put(61.553,30.4866){\circle*{0.7}}
\put(8.447,30.4866){\circle*{0.7}}
\put(61.583,30.6313){\circle*{0.7}}
\put(8.417,30.6313){\circle*{0.7}}
\put(61.613,30.7765){\circle*{0.7}}
\put(8.387,30.7765){\circle*{0.7}}
\put(61.643,30.9221){\circle*{0.7}}
\put(8.357,30.9221){\circle*{0.7}}
\put(61.673,31.0683){\circle*{0.7}}
\put(8.327,31.0683){\circle*{0.7}}
\put(61.703,31.2149){\circle*{0.7}}
\put(8.297,31.2149){\circle*{0.7}}
\put(61.733,31.362){\circle*{0.7}}
\put(8.267,31.362){\circle*{0.7}}
\put(61.763,31.5096){\circle*{0.7}}
\put(8.237,31.5096){\circle*{0.7}}
\put(61.793,31.6577){\circle*{0.7}}
\put(8.207,31.6577){\circle*{0.7}}
\put(61.823,31.8063){\circle*{0.7}}
\put(8.177,31.8063){\circle*{0.7}}
\put(61.853,31.9553){\circle*{0.7}}
\put(8.147,31.9553){\circle*{0.7}}
\put(61.883,32.1049){\circle*{0.7}}
\put(8.117,32.1049){\circle*{0.7}}
\put(61.913,32.2549){\circle*{0.7}}
\put(8.087,32.2549){\circle*{0.7}}
\put(61.943,32.4055){\circle*{0.7}}
\put(8.057,32.4055){\circle*{0.7}}
\put(61.973,32.5565){\circle*{0.7}}
\put(8.027,32.5565){\circle*{0.7}}
\put(62.003,32.708){\circle*{0.7}}
\put(7.997,32.708){\circle*{0.7}}
\put(62.033,32.86){\circle*{0.7}}
\put(7.967,32.86){\circle*{0.7}}
\put(62.063,33.0126){\circle*{0.7}}
\put(7.937,33.0126){\circle*{0.7}}
\put(62.093,33.1656){\circle*{0.7}}
\put(7.907,33.1656){\circle*{0.7}}
\put(62.123,33.3191){\circle*{0.7}}
\put(7.877,33.3191){\circle*{0.7}}
\put(62.153,33.4731){\circle*{0.7}}
\put(7.847,33.4731){\circle*{0.7}}
\put(62.183,33.6276){\circle*{0.7}}
\put(7.817,33.6276){\circle*{0.7}}
\put(62.213,33.7825){\circle*{0.7}}
\put(7.787,33.7825){\circle*{0.7}}
\put(62.243,33.938){\circle*{0.7}}
\put(7.757,33.938){\circle*{0.7}}
\put(62.273,34.094){\circle*{0.7}}
\put(7.727,34.094){\circle*{0.7}}
\put(62.303,34.2505){\circle*{0.7}}
\put(7.697,34.2505){\circle*{0.7}}
\put(62.333,34.4075){\circle*{0.7}}
\put(7.667,34.4075){\circle*{0.7}}
\put(62.363,34.565){\circle*{0.7}}
\put(7.637,34.565){\circle*{0.7}}
\put(62.393,34.723){\circle*{0.7}}
\put(7.607,34.723){\circle*{0.7}}
\put(62.423,34.8815){\circle*{0.7}}
\put(7.577,34.8815){\circle*{0.7}}
\put(62.453,35.0405){\circle*{0.7}}
\put(7.547,35.0405){\circle*{0.7}}
\put(62.483,35.2){\circle*{0.7}}
\put(7.517,35.2){\circle*{0.7}}
\put(62.513,35.36){\circle*{0.7}}
\put(7.487,35.36){\circle*{0.7}}
\put(62.543,35.5205){\circle*{0.7}}
\put(7.457,35.5205){\circle*{0.7}}
\put(62.573,35.6816){\circle*{0.7}}
\put(7.427,35.6816){\circle*{0.7}}
\put(62.603,35.8431){\circle*{0.7}}
\put(7.397,35.8431){\circle*{0.7}}
\put(62.633,36.0051){\circle*{0.7}}
\put(7.367,36.0051){\circle*{0.7}}
\put(62.663,36.1677){\circle*{0.7}}
\put(7.337,36.1677){\circle*{0.7}}
\put(62.693,36.3307){\circle*{0.7}}
\put(7.307,36.3307){\circle*{0.7}}
\put(62.723,36.4943){\circle*{0.7}}
\put(7.277,36.4943){\circle*{0.7}}
\put(62.753,36.6584){\circle*{0.7}}
\put(7.247,36.6584){\circle*{0.7}}
\put(62.783,36.823){\circle*{0.7}}
\put(7.217,36.823){\circle*{0.7}}
\put(62.813,36.9881){\circle*{0.7}}
\put(7.187,36.9881){\circle*{0.7}}
\put(62.843,37.1537){\circle*{0.7}}
\put(7.157,37.1537){\circle*{0.7}}
\put(62.873,37.3199){\circle*{0.7}}
\put(7.127,37.3199){\circle*{0.7}}
\put(62.903,37.4865){\circle*{0.7}}
\put(7.097,37.4865){\circle*{0.7}}
\put(62.933,37.6537){\circle*{0.7}}
\put(7.067,37.6537){\circle*{0.7}}
\put(62.963,37.8214){\circle*{0.7}}
\put(7.037,37.8214){\circle*{0.7}}
\put(62.993,37.9896){\circle*{0.7}}
\put(7.007,37.9896){\circle*{0.7}}
\put(63.023,38.1583){\circle*{0.7}}
\put(6.977,38.1583){\circle*{0.7}}
\put(63.053,38.3276){\circle*{0.7}}
\put(6.947,38.3276){\circle*{0.7}}
\put(63.083,38.4974){\circle*{0.7}}
\put(6.917,38.4974){\circle*{0.7}}
\put(63.113,38.6676){\circle*{0.7}}
\put(6.887,38.6676){\circle*{0.7}}
\put(63.143,38.8385){\circle*{0.7}}
\put(6.857,38.8385){\circle*{0.7}}
\put(63.173,39.0098){\circle*{0.7}}
\put(6.827,39.0098){\circle*{0.7}}
\put(63.203,39.1816){\circle*{0.7}}
\put(6.797,39.1816){\circle*{0.7}}
\put(63.233,39.354){\circle*{0.7}}
\put(6.767,39.354){\circle*{0.7}}
\put(63.263,39.5269){\circle*{0.7}}
\put(6.737,39.5269){\circle*{0.7}}
\put(63.293,39.7004){\circle*{0.7}}
\put(6.707,39.7004){\circle*{0.7}}
\put(63.323,39.8743){\circle*{0.7}}
\put(6.677,39.8743){\circle*{0.7}}
\put(63.353,40.0488){\circle*{0.7}}
\put(6.647,40.0488){\circle*{0.7}}
\put(63.383,40.2239){\circle*{0.7}}
\put(6.617,40.2239){\circle*{0.7}}
\put(63.413,40.3994){\circle*{0.7}}
\put(6.587,40.3994){\circle*{0.7}}
\put(63.443,40.5755){\circle*{0.7}}
\put(6.557,40.5755){\circle*{0.7}}
\put(63.473,40.7521){\circle*{0.7}}
\put(6.527,40.7521){\circle*{0.7}}
\put(63.503,40.9292){\circle*{0.7}}
\put(6.497,40.9292){\circle*{0.7}}
\put(63.533,41.1069){\circle*{0.7}}
\put(6.467,41.1069){\circle*{0.7}}
\put(63.563,41.2851){\circle*{0.7}}
\put(6.437,41.2851){\circle*{0.7}}
\put(63.593,41.4639){\circle*{0.7}}
\put(6.407,41.4639){\circle*{0.7}}
\put(63.623,41.6431){\circle*{0.7}}
\put(6.377,41.6431){\circle*{0.7}}
\put(63.653,41.823){\circle*{0.7}}
\put(6.347,41.823){\circle*{0.7}}
\put(63.683,42.0033){\circle*{0.7}}
\put(6.317,42.0033){\circle*{0.7}}
\put(63.713,42.1842){\circle*{0.7}}
\put(6.287,42.1842){\circle*{0.7}}
\put(63.743,42.3656){\circle*{0.7}}
\put(6.257,42.3656){\circle*{0.7}}
\put(63.773,42.5476){\circle*{0.7}}
\put(6.227,42.5476){\circle*{0.7}}
\put(63.803,42.7301){\circle*{0.7}}
\put(6.197,42.7301){\circle*{0.7}}
\put(63.833,42.9132){\circle*{0.7}}
\put(6.167,42.9132){\circle*{0.7}}
\put(63.863,43.0968){\circle*{0.7}}
\put(6.137,43.0968){\circle*{0.7}}
\put(63.893,43.2809){\circle*{0.7}}
\put(6.107,43.2809){\circle*{0.7}}
\put(63.923,43.4656){\circle*{0.7}}
\put(6.077,43.4656){\circle*{0.7}}
\put(63.953,43.6508){\circle*{0.7}}
\put(6.047,43.6508){\circle*{0.7}}
\put(63.983,43.8366){\circle*{0.7}}
\put(6.017,43.8366){\circle*{0.7}}
\put(64.013,44.0229){\circle*{0.7}}
\put(5.987,44.0229){\circle*{0.7}}
\put(64.043,44.2098){\circle*{0.7}}
\put(5.957,44.2098){\circle*{0.7}}
\put(64.073,44.3972){\circle*{0.7}}
\put(5.927,44.3972){\circle*{0.7}}
\put(64.103,44.5852){\circle*{0.7}}
\put(5.897,44.5852){\circle*{0.7}}
\put(64.133,44.7737){\circle*{0.7}}
\put(5.867,44.7737){\circle*{0.7}}
\put(64.163,44.9628){\circle*{0.7}}
\put(5.837,44.9628){\circle*{0.7}}
\put(64.193,45.1524){\circle*{0.7}}
\put(5.807,45.1524){\circle*{0.7}}
\put(64.223,45.3425){\circle*{0.7}}
\put(5.777,45.3425){\circle*{0.7}}
\put(64.253,45.5333){\circle*{0.7}}
\put(5.747,45.5333){\circle*{0.7}}
\put(64.283,45.7245){\circle*{0.7}}
\put(5.717,45.7245){\circle*{0.7}}
\put(64.313,45.9164){\circle*{0.7}}
\put(5.687,45.9164){\circle*{0.7}}
\put(64.343,46.1088){\circle*{0.7}}
\put(5.657,46.1088){\circle*{0.7}}
\put(64.373,46.3017){\circle*{0.7}}
\put(5.627,46.3017){\circle*{0.7}}
\put(64.403,46.4952){\circle*{0.7}}
\put(5.597,46.4952){\circle*{0.7}}
\put(64.433,46.6893){\circle*{0.7}}
\put(5.567,46.6893){\circle*{0.7}}
\put(64.463,46.8839){\circle*{0.7}}
\put(5.537,46.8839){\circle*{0.7}}
\put(64.493,47.0791){\circle*{0.7}}
\put(5.507,47.0791){\circle*{0.7}}
\put(64.523,47.2749){\circle*{0.7}}
\put(5.477,47.2749){\circle*{0.7}}
\put(64.553,47.4712){\circle*{0.7}}
\put(5.447,47.4712){\circle*{0.7}}
\put(64.583,47.6681){\circle*{0.7}}
\put(5.417,47.6681){\circle*{0.7}}
\put(64.613,47.8655){\circle*{0.7}}
\put(5.387,47.8655){\circle*{0.7}}
\put(64.643,48.0635){\circle*{0.7}}
\put(5.357,48.0635){\circle*{0.7}}
\put(64.673,48.2621){\circle*{0.7}}
\put(5.327,48.2621){\circle*{0.7}}
\put(64.703,48.4612){\circle*{0.7}}
\put(5.297,48.4612){\circle*{0.7}}
\put(64.733,48.6609){\circle*{0.7}}
\put(5.267,48.6609){\circle*{0.7}}
\put(64.763,48.8612){\circle*{0.7}}
\put(5.237,48.8612){\circle*{0.7}}
\put(64.793,49.062){\circle*{0.7}}
\put(5.207,49.062){\circle*{0.7}}
\put(64.823,49.2635){\circle*{0.7}}
\put(5.177,49.2635){\circle*{0.7}}
\put(64.853,49.4654){\circle*{0.7}}
\put(5.147,49.4654){\circle*{0.7}}
\put(64.883,49.668){\circle*{0.7}}
\put(5.117,49.668){\circle*{0.7}}
\put(64.913,49.8711){\circle*{0.7}}
\put(5.087,49.8711){\circle*{0.7}}
\put(64.943,50.0749){\circle*{0.7}}
\put(5.057,50.0749){\circle*{0.7}}
\put(64.973,50.2791){\circle*{0.7}}
\put(5.027,50.2791){\circle*{0.7}}
\put(65.003,50.484){\circle*{0.7}}
\put(4.997,50.484){\circle*{0.7}}
\put(65.033,50.6894){\circle*{0.7}}
\put(4.967,50.6894){\circle*{0.7}}
\put(65.063,50.8955){\circle*{0.7}}
\put(4.937,50.8955){\circle*{0.7}}
\put(65.093,51.1021){\circle*{0.7}}
\put(4.907,51.1021){\circle*{0.7}}
\put(65.123,51.3092){\circle*{0.7}}
\put(4.877,51.3092){\circle*{0.7}}
\put(65.153,51.517){\circle*{0.7}}
\put(4.847,51.517){\circle*{0.7}}
\put(65.183,51.7253){\circle*{0.7}}
\put(4.817,51.7253){\circle*{0.7}}
\put(65.213,51.9343){\circle*{0.7}}
\put(4.787,51.9343){\circle*{0.7}}
\put(65.243,52.1438){\circle*{0.7}}
\put(4.757,52.1438){\circle*{0.7}}
\put(65.273,52.3539){\circle*{0.7}}
\put(4.727,52.3539){\circle*{0.7}}
\put(65.303,52.5645){\circle*{0.7}}
\put(4.697,52.5645){\circle*{0.7}}
\put(65.333,52.7758){\circle*{0.7}}
\put(4.667,52.7758){\circle*{0.7}}
\put(65.363,52.9877){\circle*{0.7}}
\put(4.637,52.9877){\circle*{0.7}}
\put(65.393,53.2001){\circle*{0.7}}
\put(4.607,53.2001){\circle*{0.7}}
\put(65.423,53.4131){\circle*{0.7}}
\put(4.577,53.4131){\circle*{0.7}}
\put(65.453,53.6268){\circle*{0.7}}
\put(4.547,53.6268){\circle*{0.7}}
\put(65.483,53.841){\circle*{0.7}}
\put(4.517,53.841){\circle*{0.7}}
\put(65.513,54.0558){\circle*{0.7}}
\put(4.487,54.0558){\circle*{0.7}}
\put(65.543,54.2712){\circle*{0.7}}
\put(4.457,54.2712){\circle*{0.7}}
\put(65.573,54.4872){\circle*{0.7}}
\put(4.427,54.4872){\circle*{0.7}}
\put(65.603,54.7038){\circle*{0.7}}
\put(4.397,54.7038){\circle*{0.7}}
\put(65.633,54.921){\circle*{0.7}}
\put(4.367,54.921){\circle*{0.7}}
\put(65.663,55.1388){\circle*{0.7}}
\put(4.337,55.1388){\circle*{0.7}}
\put(65.693,55.3572){\circle*{0.7}}
\put(4.307,55.3572){\circle*{0.7}}
\put(65.723,55.5761){\circle*{0.7}}
\put(4.277,55.5761){\circle*{0.7}}
\put(65.753,55.7957){\circle*{0.7}}
\put(4.247,55.7957){\circle*{0.7}}
\put(65.783,56.0159){\circle*{0.7}}
\put(4.217,56.0159){\circle*{0.7}}
\put(65.813,56.2367){\circle*{0.7}}
\put(4.187,56.2367){\circle*{0.7}}
\put(65.843,56.4581){\circle*{0.7}}
\put(4.157,56.4581){\circle*{0.7}}
\put(65.873,56.6801){\circle*{0.7}}
\put(4.127,56.6801){\circle*{0.7}}
\put(65.903,56.9027){\circle*{0.7}}
\put(4.097,56.9027){\circle*{0.7}}
\put(65.933,57.1259){\circle*{0.7}}
\put(4.067,57.1259){\circle*{0.7}}
\put(65.963,57.3497){\circle*{0.7}}
\put(4.037,57.3497){\circle*{0.7}}
\put(65.993,57.5742){\circle*{0.7}}
\put(4.007,57.5742){\circle*{0.7}}
\put(66.023,57.7992){\circle*{0.7}}
\put(3.977,57.7992){\circle*{0.7}}
\put(66.053,58.0249){\circle*{0.7}}
\put(3.947,58.0249){\circle*{0.7}}
\put(66.083,58.2511){\circle*{0.7}}
\put(3.917,58.2511){\circle*{0.7}}
\put(66.113,58.478){\circle*{0.7}}
\put(3.887,58.478){\circle*{0.7}}
\put(66.143,58.7055){\circle*{0.7}}
\put(3.857,58.7055){\circle*{0.7}}
\put(66.173,58.9336){\circle*{0.7}}
\put(3.827,58.9336){\circle*{0.7}}
\put(66.203,59.1623){\circle*{0.7}}
\put(3.797,59.1623){\circle*{0.7}}
\put(66.233,59.3916){\circle*{0.7}}
\put(3.767,59.3916){\circle*{0.7}}
\put(66.263,59.6216){\circle*{0.7}}
\put(3.737,59.6216){\circle*{0.7}}
\put(66.293,59.8522){\circle*{0.7}}
\put(3.707,59.8522){\circle*{0.7}}
\put(66.323,60.0834){\circle*{0.7}}
\put(3.677,60.0834){\circle*{0.7}}
\put(66.353,60.3152){\circle*{0.7}}
\put(3.647,60.3152){\circle*{0.7}}
\put(66.383,60.5476){\circle*{0.7}}
\put(3.617,60.5476){\circle*{0.7}}
\put(66.413,60.7807){\circle*{0.7}}
\put(3.587,60.7807){\circle*{0.7}}
\put(66.443,61.0143){\circle*{0.7}}
\put(3.557,61.0143){\circle*{0.7}}
\put(66.473,61.2486){\circle*{0.7}}
\put(3.527,61.2486){\circle*{0.7}}
\put(66.503,61.4836){\circle*{0.7}}
\put(3.497,61.4836){\circle*{0.7}}
\put(66.533,61.7191){\circle*{0.7}}
\put(3.467,61.7191){\circle*{0.7}}
\put(66.563,61.9553){\circle*{0.7}}
\put(3.437,61.9553){\circle*{0.7}}
\put(66.593,62.1922){\circle*{0.7}}
\put(3.407,62.1922){\circle*{0.7}}
\put(66.623,62.4296){\circle*{0.7}}
\put(3.377,62.4296){\circle*{0.7}}
\put(66.653,62.6677){\circle*{0.7}}
\put(3.347,62.6677){\circle*{0.7}}
\put(66.683,62.9064){\circle*{0.7}}
\put(3.317,62.9064){\circle*{0.7}}
\put(66.713,63.1457){\circle*{0.7}}
\put(3.287,63.1457){\circle*{0.7}}
\put(66.743,63.3857){\circle*{0.7}}
\put(3.257,63.3857){\circle*{0.7}}
\put(66.773,63.6263){\circle*{0.7}}
\put(3.227,63.6263){\circle*{0.7}}
\put(66.803,63.8676){\circle*{0.7}}
\put(3.197,63.8676){\circle*{0.7}}
\put(66.833,64.1095){\circle*{0.7}}
\put(3.167,64.1095){\circle*{0.7}}
\put(66.863,64.352){\circle*{0.7}}
\put(3.137,64.352){\circle*{0.7}}
\put(66.893,64.5952){\circle*{0.7}}
\put(3.107,64.5952){\circle*{0.7}}
\put(66.923,64.839){\circle*{0.7}}
\put(3.077,64.839){\circle*{0.7}}
\put(66.953,65.0835){\circle*{0.7}}
\put(3.047,65.0835){\circle*{0.7}}
\put(66.983,65.3286){\circle*{0.7}}
\put(3.017,65.3286){\circle*{0.7}}
\put(67.013,65.5743){\circle*{0.7}}
\put(2.987,65.5743){\circle*{0.7}}
\put(67.043,65.8207){\circle*{0.7}}
\put(2.957,65.8207){\circle*{0.7}}
\put(67.073,66.0677){\circle*{0.7}}
\put(2.927,66.0677){\circle*{0.7}}

%% file: sph_rate_f3.tex
\put(115,55.){\circle*{0.7}}
\put(115.03,54.8022){\circle*{0.7}}
\put(114.97,54.8022){\circle*{0.7}}
\put(115.059,54.6044){\circle*{0.7}}
\put(114.941,54.6044){\circle*{0.7}}
\put(115.089,54.4067){\circle*{0.7}}
\put(114.911,54.4067){\circle*{0.7}}
\put(115.119,54.2089){\circle*{0.7}}
\put(114.881,54.2089){\circle*{0.7}}
\put(115.149,54.0112){\circle*{0.7}}
\put(114.851,54.0112){\circle*{0.7}}
\put(115.179,53.8134){\circle*{0.7}}
\put(114.821,53.8134){\circle*{0.7}}
\put(115.209,53.6157){\circle*{0.7}}
\put(114.791,53.6157){\circle*{0.7}}
\put(115.239,53.418){\circle*{0.7}}
\put(114.761,53.418){\circle*{0.7}}
\put(115.269,53.2203){\circle*{0.7}}
\put(114.731,53.2203){\circle*{0.7}}
\put(115.3,53.0226){\circle*{0.7}}
\put(114.7,53.0226){\circle*{0.7}}
\put(115.33,52.8249){\circle*{0.7}}
\put(114.67,52.8249){\circle*{0.7}}
\put(115.36,52.6272){\circle*{0.7}}
\put(114.64,52.6272){\circle*{0.7}}
\put(115.391,52.4295){\circle*{0.7}}
\put(114.609,52.4295){\circle*{0.7}}
\put(115.421,52.2319){\circle*{0.7}}
\put(114.579,52.2319){\circle*{0.7}}
\put(115.452,52.0342){\circle*{0.7}}
\put(114.548,52.0342){\circle*{0.7}}
\put(115.482,51.8366){\circle*{0.7}}
\put(114.518,51.8366){\circle*{0.7}}
\put(115.513,51.6389){\circle*{0.7}}
\put(114.487,51.6389){\circle*{0.7}}
\put(115.544,51.4413){\circle*{0.7}}
\put(114.456,51.4413){\circle*{0.7}}
\put(115.574,51.2437){\circle*{0.7}}
\put(114.426,51.2437){\circle*{0.7}}
\put(115.605,51.0461){\circle*{0.7}}
\put(114.395,51.0461){\circle*{0.7}}
\put(115.636,50.8485){\circle*{0.7}}
\put(114.364,50.8485){\circle*{0.7}}
\put(115.667,50.6509){\circle*{0.7}}
\put(114.333,50.6509){\circle*{0.7}}
\put(115.698,50.4533){\circle*{0.7}}
\put(114.302,50.4533){\circle*{0.7}}
\put(115.729,50.2557){\circle*{0.7}}
\put(114.271,50.2557){\circle*{0.7}}
\put(115.761,50.0582){\circle*{0.7}}
\put(114.239,50.0582){\circle*{0.7}}
\put(115.792,49.8607){\circle*{0.7}}
\put(114.208,49.8607){\circle*{0.7}}
\put(115.823,49.6631){\circle*{0.7}}
\put(114.177,49.6631){\circle*{0.7}}
\put(115.854,49.4656){\circle*{0.7}}
\put(114.146,49.4656){\circle*{0.7}}
\put(115.886,49.2681){\circle*{0.7}}
\put(114.114,49.2681){\circle*{0.7}}
\put(115.917,49.0706){\circle*{0.7}}
\put(114.083,49.0706){\circle*{0.7}}
\put(115.949,48.8731){\circle*{0.7}}
\put(114.051,48.8731){\circle*{0.7}}
\put(115.981,48.6756){\circle*{0.7}}
\put(114.019,48.6756){\circle*{0.7}}
\put(116.012,48.4781){\circle*{0.7}}
\put(113.988,48.4781){\circle*{0.7}}
\put(116.044,48.2807){\circle*{0.7}}
\put(113.956,48.2807){\circle*{0.7}}
\put(116.076,48.0832){\circle*{0.7}}
\put(113.924,48.0832){\circle*{0.7}}
\put(116.108,47.8858){\circle*{0.7}}
\put(113.892,47.8858){\circle*{0.7}}
\put(116.14,47.6884){\circle*{0.7}}
\put(113.86,47.6884){\circle*{0.7}}
\put(116.172,47.491){\circle*{0.7}}
\put(113.828,47.491){\circle*{0.7}}
\put(116.204,47.2936){\circle*{0.7}}
\put(113.796,47.2936){\circle*{0.7}}
\put(116.237,47.0962){\circle*{0.7}}
\put(113.763,47.0962){\circle*{0.7}}
\put(116.269,46.8988){\circle*{0.7}}
\put(113.731,46.8988){\circle*{0.7}}
\put(116.301,46.7015){\circle*{0.7}}
\put(113.699,46.7015){\circle*{0.7}}
\put(116.334,46.5041){\circle*{0.7}}
\put(113.666,46.5041){\circle*{0.7}}
\put(116.366,46.3068){\circle*{0.7}}
\put(113.634,46.3068){\circle*{0.7}}
\put(116.399,46.1095){\circle*{0.7}}
\put(113.601,46.1095){\circle*{0.7}}
\put(116.431,45.9121){\circle*{0.7}}
\put(113.569,45.9121){\circle*{0.7}}
\put(116.464,45.7148){\circle*{0.7}}
\put(113.536,45.7148){\circle*{0.7}}
\put(116.497,45.5176){\circle*{0.7}}
\put(113.503,45.5176){\circle*{0.7}}
\put(116.53,45.3203){\circle*{0.7}}
\put(113.47,45.3203){\circle*{0.7}}
\put(116.563,45.123){\circle*{0.7}}
\put(113.437,45.123){\circle*{0.7}}
\put(116.596,44.9258){\circle*{0.7}}
\put(113.404,44.9258){\circle*{0.7}}
\put(116.629,44.7285){\circle*{0.7}}
\put(113.371,44.7285){\circle*{0.7}}
\put(116.662,44.5313){\circle*{0.7}}
\put(113.338,44.5313){\circle*{0.7}}
\put(116.696,44.3341){\circle*{0.7}}
\put(113.304,44.3341){\circle*{0.7}}
\put(116.729,44.1369){\circle*{0.7}}
\put(113.271,44.1369){\circle*{0.7}}
\put(116.763,43.9397){\circle*{0.7}}
\put(113.237,43.9397){\circle*{0.7}}
\put(116.796,43.7426){\circle*{0.7}}
\put(113.204,43.7426){\circle*{0.7}}
\put(116.83,43.5454){\circle*{0.7}}
\put(113.17,43.5454){\circle*{0.7}}
\put(116.864,43.3483){\circle*{0.7}}
\put(113.136,43.3483){\circle*{0.7}}
\put(116.897,43.1512){\circle*{0.7}}
\put(113.103,43.1512){\circle*{0.7}}
\put(116.931,42.9541){\circle*{0.7}}
\put(113.069,42.9541){\circle*{0.7}}
\put(116.965,42.757){\circle*{0.7}}
\put(113.035,42.757){\circle*{0.7}}
\put(116.999,42.5599){\circle*{0.7}}
\put(113.001,42.5599){\circle*{0.7}}
\put(117.033,42.3628){\circle*{0.7}}
\put(112.967,42.3628){\circle*{0.7}}
\put(117.068,42.1658){\circle*{0.7}}
\put(112.932,42.1658){\circle*{0.7}}
\put(117.102,41.9687){\circle*{0.7}}
\put(112.898,41.9687){\circle*{0.7}}
\put(117.136,41.7717){\circle*{0.7}}
\put(112.864,41.7717){\circle*{0.7}}
\put(117.171,41.5747){\circle*{0.7}}
\put(112.829,41.5747){\circle*{0.7}}
\put(117.205,41.3777){\circle*{0.7}}
\put(112.795,41.3777){\circle*{0.7}}
\put(117.24,41.1808){\circle*{0.7}}
\put(112.76,41.1808){\circle*{0.7}}
\put(117.275,40.9838){\circle*{0.7}}
\put(112.725,40.9838){\circle*{0.7}}
\put(117.31,40.7869){\circle*{0.7}}
\put(112.69,40.7869){\circle*{0.7}}
\put(117.345,40.59){\circle*{0.7}}
\put(112.655,40.59){\circle*{0.7}}
\put(117.38,40.3931){\circle*{0.7}}
\put(112.62,40.3931){\circle*{0.7}}
\put(117.415,40.1962){\circle*{0.7}}
\put(112.585,40.1962){\circle*{0.7}}
\put(117.45,39.9993){\circle*{0.7}}
\put(112.55,39.9993){\circle*{0.7}}
\put(117.486,39.8025){\circle*{0.7}}
\put(112.514,39.8025){\circle*{0.7}}
\put(117.521,39.6056){\circle*{0.7}}
\put(112.479,39.6056){\circle*{0.7}}
\put(117.557,39.4088){\circle*{0.7}}
\put(112.443,39.4088){\circle*{0.7}}
\put(117.592,39.212){\circle*{0.7}}
\put(112.408,39.212){\circle*{0.7}}
\put(117.628,39.0152){\circle*{0.7}}
\put(112.372,39.0152){\circle*{0.7}}
\put(117.664,38.8185){\circle*{0.7}}
\put(112.336,38.8185){\circle*{0.7}}
\put(117.7,38.6217){\circle*{0.7}}
\put(112.3,38.6217){\circle*{0.7}}
\put(117.736,38.425){\circle*{0.7}}
\put(112.264,38.425){\circle*{0.7}}
\put(117.772,38.2283){\circle*{0.7}}
\put(112.228,38.2283){\circle*{0.7}}
\put(117.808,38.0316){\circle*{0.7}}
\put(112.192,38.0316){\circle*{0.7}}
\put(117.844,37.8349){\circle*{0.7}}
\put(112.156,37.8349){\circle*{0.7}}
\put(117.881,37.6383){\circle*{0.7}}
\put(112.119,37.6383){\circle*{0.7}}
\put(117.917,37.4416){\circle*{0.7}}
\put(112.083,37.4416){\circle*{0.7}}
\put(117.954,37.245){\circle*{0.7}}
\put(112.046,37.245){\circle*{0.7}}
\put(117.991,37.0484){\circle*{0.7}}
\put(112.009,37.0484){\circle*{0.7}}
\put(118.028,36.8519){\circle*{0.7}}
\put(111.972,36.8519){\circle*{0.7}}
\put(118.065,36.6553){\circle*{0.7}}
\put(111.935,36.6553){\circle*{0.7}}
\put(118.102,36.4588){\circle*{0.7}}
\put(111.898,36.4588){\circle*{0.7}}
\put(118.139,36.2623){\circle*{0.7}}
\put(111.861,36.2623){\circle*{0.7}}
\put(118.176,36.0658){\circle*{0.7}}
\put(111.824,36.0658){\circle*{0.7}}
\put(118.214,35.8693){\circle*{0.7}}
\put(111.786,35.8693){\circle*{0.7}}
\put(118.251,35.6729){\circle*{0.7}}
\put(111.749,35.6729){\circle*{0.7}}
\put(118.289,35.4765){\circle*{0.7}}
\put(111.711,35.4765){\circle*{0.7}}
\put(118.327,35.2801){\circle*{0.7}}
\put(111.673,35.2801){\circle*{0.7}}
\put(118.365,35.0837){\circle*{0.7}}
\put(111.635,35.0837){\circle*{0.7}}
\put(118.403,34.8874){\circle*{0.7}}
\put(111.597,34.8874){\circle*{0.7}}
\put(118.441,34.691){\circle*{0.7}}
\put(111.559,34.691){\circle*{0.7}}
\put(118.479,34.4947){\circle*{0.7}}
\put(111.521,34.4947){\circle*{0.7}}
\put(118.518,34.2984){\circle*{0.7}}
\put(111.482,34.2984){\circle*{0.7}}
\put(118.556,34.1022){\circle*{0.7}}
\put(111.444,34.1022){\circle*{0.7}}
\put(118.595,33.906){\circle*{0.7}}
\put(111.405,33.906){\circle*{0.7}}
\put(118.634,33.7098){\circle*{0.7}}
\put(111.366,33.7098){\circle*{0.7}}
\put(118.673,33.5136){\circle*{0.7}}
\put(111.327,33.5136){\circle*{0.7}}
\put(118.712,33.3174){\circle*{0.7}}
\put(111.288,33.3174){\circle*{0.7}}
\put(118.751,33.1213){\circle*{0.7}}
\put(111.249,33.1213){\circle*{0.7}}
\put(118.79,32.9252){\circle*{0.7}}
\put(111.21,32.9252){\circle*{0.7}}
\put(118.829,32.7291){\circle*{0.7}}
\put(111.171,32.7291){\circle*{0.7}}
\put(118.869,32.5331){\circle*{0.7}}
\put(111.131,32.5331){\circle*{0.7}}
\put(118.909,32.337){\circle*{0.7}}
\put(111.091,32.337){\circle*{0.7}}
\put(118.949,32.1411){\circle*{0.7}}
\put(111.051,32.1411){\circle*{0.7}}
\put(118.989,31.9451){\circle*{0.7}}
\put(111.011,31.9451){\circle*{0.7}}
\put(119.029,31.7491){\circle*{0.7}}
\put(110.971,31.7491){\circle*{0.7}}
\put(119.069,31.5532){\circle*{0.7}}
\put(110.931,31.5532){\circle*{0.7}}
\put(119.109,31.3574){\circle*{0.7}}
\put(110.891,31.3574){\circle*{0.7}}
\put(119.15,31.1615){\circle*{0.7}}
\put(110.85,31.1615){\circle*{0.7}}
\put(119.19,30.9657){\circle*{0.7}}
\put(110.81,30.9657){\circle*{0.7}}
\put(119.231,30.7699){\circle*{0.7}}
\put(110.769,30.7699){\circle*{0.7}}
\put(119.272,30.5741){\circle*{0.7}}
\put(110.728,30.5741){\circle*{0.7}}
\put(119.313,30.3784){\circle*{0.7}}
\put(110.687,30.3784){\circle*{0.7}}
\put(119.355,30.1827){\circle*{0.7}}
\put(110.645,30.1827){\circle*{0.7}}
\put(119.396,29.9871){\circle*{0.7}}
\put(110.604,29.9871){\circle*{0.7}}
\put(119.438,29.7914){\circle*{0.7}}
\put(110.562,29.7914){\circle*{0.7}}
\put(119.479,29.5958){\circle*{0.7}}
\put(110.521,29.5958){\circle*{0.7}}
\put(119.521,29.4003){\circle*{0.7}}
\put(110.479,29.4003){\circle*{0.7}}
\put(119.563,29.2048){\circle*{0.7}}
\put(110.437,29.2048){\circle*{0.7}}
\put(119.606,29.0093){\circle*{0.7}}
\put(110.394,29.0093){\circle*{0.7}}
\put(119.648,28.8138){\circle*{0.7}}
\put(110.352,28.8138){\circle*{0.7}}
\put(119.691,28.6184){\circle*{0.7}}
\put(110.309,28.6184){\circle*{0.7}}
\put(119.733,28.423){\circle*{0.7}}
\put(110.267,28.423){\circle*{0.7}}
\put(119.776,28.2277){\circle*{0.7}}
\put(110.224,28.2277){\circle*{0.7}}
\put(119.819,28.0324){\circle*{0.7}}
\put(110.181,28.0324){\circle*{0.7}}
\put(119.863,27.8371){\circle*{0.7}}
\put(110.137,27.8371){\circle*{0.7}}
\put(119.906,27.6419){\circle*{0.7}}
\put(110.094,27.6419){\circle*{0.7}}
\put(119.95,27.4467){\circle*{0.7}}
\put(110.05,27.4467){\circle*{0.7}}
\put(119.993,27.2515){\circle*{0.7}}
\put(110.007,27.2515){\circle*{0.7}}
\put(120.037,27.0564){\circle*{0.7}}
\put(109.963,27.0564){\circle*{0.7}}
\put(120.082,26.8614){\circle*{0.7}}
\put(109.918,26.8614){\circle*{0.7}}
\put(120.126,26.6663){\circle*{0.7}}
\put(109.874,26.6663){\circle*{0.7}}
\put(120.17,26.4714){\circle*{0.7}}
\put(109.83,26.4714){\circle*{0.7}}
\put(120.215,26.2764){\circle*{0.7}}
\put(109.785,26.2764){\circle*{0.7}}
\put(120.26,26.0815){\circle*{0.7}}
\put(109.74,26.0815){\circle*{0.7}}
\put(120.305,25.8867){\circle*{0.7}}
\put(109.695,25.8867){\circle*{0.7}}
\put(120.35,25.6919){\circle*{0.7}}
\put(109.65,25.6919){\circle*{0.7}}
\put(120.396,25.4971){\circle*{0.7}}
\put(109.604,25.4971){\circle*{0.7}}
\put(120.442,25.3024){\circle*{0.7}}
\put(109.558,25.3024){\circle*{0.7}}
\put(120.488,25.1078){\circle*{0.7}}
\put(109.512,25.1078){\circle*{0.7}}
\put(120.534,24.9132){\circle*{0.7}}
\put(109.466,24.9132){\circle*{0.7}}
\put(120.58,24.7186){\circle*{0.7}}
\put(109.42,24.7186){\circle*{0.7}}
\put(120.627,24.5241){\circle*{0.7}}
\put(109.373,24.5241){\circle*{0.7}}
\put(120.674,24.3297){\circle*{0.7}}
\put(109.326,24.3297){\circle*{0.7}}
\put(120.721,24.1353){\circle*{0.7}}
\put(109.279,24.1353){\circle*{0.7}}
\put(120.768,23.9409){\circle*{0.7}}
\put(109.232,23.9409){\circle*{0.7}}
\put(120.815,23.7466){\circle*{0.7}}
\put(109.185,23.7466){\circle*{0.7}}
\put(120.863,23.5524){\circle*{0.7}}
\put(109.137,23.5524){\circle*{0.7}}
\put(120.911,23.3583){\circle*{0.7}}
\put(109.089,23.3583){\circle*{0.7}}
\put(120.959,23.1641){\circle*{0.7}}
\put(109.041,23.1641){\circle*{0.7}}
\put(121.007,22.9701){\circle*{0.7}}
\put(108.993,22.9701){\circle*{0.7}}
\put(121.056,22.7761){\circle*{0.7}}
\put(108.944,22.7761){\circle*{0.7}}
\put(121.105,22.5822){\circle*{0.7}}
\put(108.895,22.5822){\circle*{0.7}}
\put(121.154,22.3883){\circle*{0.7}}
\put(108.846,22.3883){\circle*{0.7}}
\put(121.204,22.1945){\circle*{0.7}}
\put(108.796,22.1945){\circle*{0.7}}
\put(121.253,22.0008){\circle*{0.7}}
\put(108.747,22.0008){\circle*{0.7}}
\put(121.303,21.8071){\circle*{0.7}}
\put(108.697,21.8071){\circle*{0.7}}
\put(121.354,21.6135){\circle*{0.7}}
\put(108.646,21.6135){\circle*{0.7}}
\put(121.404,21.42){\circle*{0.7}}
\put(108.596,21.42){\circle*{0.7}}
\put(121.455,21.2266){\circle*{0.7}}
\put(108.545,21.2266){\circle*{0.7}}
\put(121.506,21.0332){\circle*{0.7}}
\put(108.494,21.0332){\circle*{0.7}}
\put(121.557,20.8399){\circle*{0.7}}
\put(108.443,20.8399){\circle*{0.7}}
\put(121.609,20.6467){\circle*{0.7}}
\put(108.391,20.6467){\circle*{0.7}}
\put(121.661,20.4536){\circle*{0.7}}
\put(108.339,20.4536){\circle*{0.7}}
\put(121.713,20.2605){\circle*{0.7}}
\put(108.287,20.2605){\circle*{0.7}}
\put(121.766,20.0675){\circle*{0.7}}
\put(108.234,20.0675){\circle*{0.7}}
\put(121.819,19.8747){\circle*{0.7}}
\put(108.181,19.8747){\circle*{0.7}}
\put(121.872,19.6819){\circle*{0.7}}
\put(108.128,19.6819){\circle*{0.7}}
\put(121.925,19.4892){\circle*{0.7}}
\put(108.075,19.4892){\circle*{0.7}}
\put(121.979,19.2965){\circle*{0.7}}
\put(108.021,19.2965){\circle*{0.7}}
\put(122.033,19.104){\circle*{0.7}}
\put(107.967,19.104){\circle*{0.7}}
\put(122.088,18.9116){\circle*{0.7}}
\put(107.912,18.9116){\circle*{0.7}}
\put(122.143,18.7193){\circle*{0.7}}
\put(107.857,18.7193){\circle*{0.7}}
\put(122.198,18.5271){\circle*{0.7}}
\put(107.802,18.5271){\circle*{0.7}}
\put(122.254,18.3349){\circle*{0.7}}
\put(107.746,18.3349){\circle*{0.7}}
\put(122.31,18.1429){\circle*{0.7}}
\put(107.69,18.1429){\circle*{0.7}}
\put(122.366,17.951){\circle*{0.7}}
\put(107.634,17.951){\circle*{0.7}}
\put(122.423,17.7593){\circle*{0.7}}
\put(107.577,17.7593){\circle*{0.7}}
\put(122.48,17.5676){\circle*{0.7}}
\put(107.52,17.5676){\circle*{0.7}}
\put(122.537,17.376){\circle*{0.7}}
\put(107.463,17.376){\circle*{0.7}}
\put(122.595,17.1846){\circle*{0.7}}
\put(107.405,17.1846){\circle*{0.7}}
\put(122.654,16.9933){\circle*{0.7}}
\put(107.346,16.9933){\circle*{0.7}}
\put(122.712,16.8022){\circle*{0.7}}
\put(107.288,16.8022){\circle*{0.7}}
\put(122.772,16.6111){\circle*{0.7}}
\put(107.228,16.6111){\circle*{0.7}}
\put(122.831,16.4202){\circle*{0.7}}
\put(107.169,16.4202){\circle*{0.7}}
\put(122.891,16.2295){\circle*{0.7}}
\put(107.109,16.2295){\circle*{0.7}}
\put(122.952,16.0389){\circle*{0.7}}
\put(107.048,16.0389){\circle*{0.7}}
\put(123.013,15.8484){\circle*{0.7}}
\put(106.987,15.8484){\circle*{0.7}}
\put(123.075,15.6581){\circle*{0.7}}
\put(106.925,15.6581){\circle*{0.7}}
\put(123.137,15.468){\circle*{0.7}}
\put(106.863,15.468){\circle*{0.7}}
\put(123.199,15.278){\circle*{0.7}}
\put(106.801,15.278){\circle*{0.7}}
\put(123.262,15.0882){\circle*{0.7}}
\put(106.738,15.0882){\circle*{0.7}}
\put(123.326,14.8986){\circle*{0.7}}
\put(106.674,14.8986){\circle*{0.7}}
\put(123.39,14.7092){\circle*{0.7}}
\put(106.61,14.7092){\circle*{0.7}}
\put(123.455,14.52){\circle*{0.7}}
\put(106.545,14.52){\circle*{0.7}}
\put(123.52,14.3309){\circle*{0.7}}
\put(106.48,14.3309){\circle*{0.7}}
\put(123.586,14.1421){\circle*{0.7}}
\put(106.414,14.1421){\circle*{0.7}}
\put(123.653,13.9535){\circle*{0.7}}
\put(106.347,13.9535){\circle*{0.7}}
\put(123.72,13.7651){\circle*{0.7}}
\put(106.28,13.7651){\circle*{0.7}}
\put(123.787,13.5769){\circle*{0.7}}
\put(106.213,13.5769){\circle*{0.7}}
\put(123.856,13.389){\circle*{0.7}}
\put(106.144,13.389){\circle*{0.7}}
\put(123.925,13.2013){\circle*{0.7}}
\put(106.075,13.2013){\circle*{0.7}}
\put(123.995,13.0139){\circle*{0.7}}
\put(106.005,13.0139){\circle*{0.7}}
\put(124.065,12.8268){\circle*{0.7}}
\put(105.935,12.8268){\circle*{0.7}}
\put(124.137,12.6399){\circle*{0.7}}
\put(105.863,12.6399){\circle*{0.7}}
\put(124.209,12.4533){\circle*{0.7}}
\put(105.791,12.4533){\circle*{0.7}}
\put(124.281,12.267){\circle*{0.7}}
\put(105.719,12.267){\circle*{0.7}}
\put(124.355,12.0811){\circle*{0.7}}
\put(105.645,12.0811){\circle*{0.7}}
\put(124.43,11.8955){\circle*{0.7}}
\put(105.57,11.8955){\circle*{0.7}}
\put(124.505,11.7102){\circle*{0.7}}
\put(105.495,11.7102){\circle*{0.7}}
\put(124.581,11.5253){\circle*{0.7}}
\put(105.419,11.5253){\circle*{0.7}}
\put(124.658,11.3408){\circle*{0.7}}
\put(105.342,11.3408){\circle*{0.7}}
\put(124.736,11.1567){\circle*{0.7}}
\put(105.264,11.1567){\circle*{0.7}}
\put(124.816,10.973){\circle*{0.7}}
\put(105.184,10.973){\circle*{0.7}}
\put(124.896,10.7898){\circle*{0.7}}
\put(105.104,10.7898){\circle*{0.7}}
\put(124.977,10.607){\circle*{0.7}}
\put(105.023,10.607){\circle*{0.7}}
\put(125.059,10.4247){\circle*{0.7}}
\put(104.941,10.4247){\circle*{0.7}}
\put(125.143,10.243){\circle*{0.7}}
\put(104.857,10.243){\circle*{0.7}}
\put(125.227,10.0618){\circle*{0.7}}
\put(104.773,10.0618){\circle*{0.7}}
\put(125.313,9.88117){\circle*{0.7}}
\put(104.687,9.88117){\circle*{0.7}}
\put(125.401,9.70119){\circle*{0.7}}
\put(104.599,9.70119){\circle*{0.7}}
\put(125.489,9.52187){\circle*{0.7}}
\put(104.511,9.52187){\circle*{0.7}}
\put(125.579,9.34326){\circle*{0.7}}
\put(104.421,9.34326){\circle*{0.7}}
\put(125.671,9.1654){\circle*{0.7}}
\put(104.329,9.1654){\circle*{0.7}}
\put(125.764,8.98834){\circle*{0.7}}
\put(104.236,8.98834){\circle*{0.7}}
\put(125.858,8.81215){\circle*{0.7}}
\put(104.142,8.81215){\circle*{0.7}}
\put(125.955,8.63688){\circle*{0.7}}
\put(104.045,8.63688){\circle*{0.7}}
\put(126.053,8.4626){\circle*{0.7}}
\put(103.947,8.4626){\circle*{0.7}}
\put(126.153,8.28939){\circle*{0.7}}
\put(103.847,8.28939){\circle*{0.7}}
\put(126.255,8.11734){\circle*{0.7}}
\put(103.745,8.11734){\circle*{0.7}}
\put(126.359,7.94654){\circle*{0.7}}
\put(103.641,7.94654){\circle*{0.7}}
\put(126.465,7.7771){\circle*{0.7}}
\put(103.535,7.7771){\circle*{0.7}}
\put(126.573,7.60913){\circle*{0.7}}
\put(103.427,7.60913){\circle*{0.7}}
\put(126.685,7.44278){\circle*{0.7}}
\put(103.315,7.44278){\circle*{0.7}}
\put(126.798,7.2782){\circle*{0.7}}
\put(103.202,7.2782){\circle*{0.7}}
\put(126.915,7.11557){\circle*{0.7}}
\put(103.085,7.11557){\circle*{0.7}}
\put(127.034,6.95508){\circle*{0.7}}
\put(102.966,6.95508){\circle*{0.7}}
\put(127.156,6.79696){\circle*{0.7}}
\put(102.844,6.79696){\circle*{0.7}}
\put(127.282,6.64148){\circle*{0.7}}
\put(102.718,6.64148){\circle*{0.7}}
\put(127.412,6.48895){\circle*{0.7}}
\put(102.588,6.48895){\circle*{0.7}}
\put(127.545,6.33971){\circle*{0.7}}
\put(102.455,6.33971){\circle*{0.7}}
\put(127.682,6.19418){\circle*{0.7}}
\put(102.318,6.19418){\circle*{0.7}}
\put(127.823,6.05284){\circle*{0.7}}
\put(102.177,6.05284){\circle*{0.7}}
\put(127.969,5.91625){\circle*{0.7}}
\put(102.031,5.91625){\circle*{0.7}}
\put(128.12,5.78507){\circle*{0.7}}
\put(101.88,5.78507){\circle*{0.7}}
\put(128.277,5.66007){\circle*{0.7}}
\put(101.723,5.66007){\circle*{0.7}}
\put(128.438,5.54214){\circle*{0.7}}
\put(101.562,5.54214){\circle*{0.7}}
\put(128.605,5.43231){\circle*{0.7}}
\put(101.395,5.43231){\circle*{0.7}}
\put(128.778,5.33178){\circle*{0.7}}
\put(101.222,5.33178){\circle*{0.7}}
\put(128.957,5.24185){\circle*{0.7}}
\put(101.043,5.24185){\circle*{0.7}}
\put(129.141,5.16398){\circle*{0.7}}
\put(100.859,5.16398){\circle*{0.7}}
\put(129.33,5.09965){\circle*{0.7}}
\put(100.67,5.09965){\circle*{0.7}}
\put(129.524,5.05031){\circle*{0.7}}
\put(100.476,5.05031){\circle*{0.7}}
\put(129.721,5.01725){\circle*{0.7}}
\put(100.279,5.01725){\circle*{0.7}}
\put(129.921,5.00139){\circle*{0.7}}
\put(100.079,5.00139){\circle*{0.7}}
\put(130.121,5.00324){\circle*{0.7}}
\put(99.8792,5.00324){\circle*{0.7}}
\put(130.32,5.02273){\circle*{0.7}}
\put(99.6802,5.02273){\circle*{0.7}}
\put(130.516,5.05927){\circle*{0.7}}
\put(99.4835,5.05927){\circle*{0.7}}
\put(130.709,5.11184){\circle*{0.7}}
\put(99.2906,5.11184){\circle*{0.7}}
\put(130.898,5.17911){\circle*{0.7}}
\put(99.1022,5.17911){\circle*{0.7}}
\put(131.081,5.25961){\circle*{0.7}}
\put(98.9191,5.25961){\circle*{0.7}}
\put(131.258,5.35186){\circle*{0.7}}
\put(98.7417,5.35186){\circle*{0.7}}
\put(131.43,5.45443){\circle*{0.7}}
\put(98.57,5.45443){\circle*{0.7}}
\put(131.596,5.56603){\circle*{0.7}}
\put(98.404,5.56603){\circle*{0.7}}
\put(131.756,5.68551){\circle*{0.7}}
\put(98.2436,5.68551){\circle*{0.7}}
\put(131.911,5.81187){\circle*{0.7}}
\put(98.0886,5.81187){\circle*{0.7}}
\put(132.061,5.94424){\circle*{0.7}}
\put(97.9387,5.94424){\circle*{0.7}}
\put(132.206,6.08186){\circle*{0.7}}
\put(97.7936,6.08186){\circle*{0.7}}
\put(132.347,6.22412){\circle*{0.7}}
\put(97.653,6.22412){\circle*{0.7}}
\put(132.483,6.37045){\circle*{0.7}}
\put(97.5166,6.37045){\circle*{0.7}}
\put(132.616,6.52041){\circle*{0.7}}
\put(97.3843,6.52041){\circle*{0.7}}
\put(132.744,6.67358){\circle*{0.7}}
\put(97.2557,6.67358){\circle*{0.7}}
\put(132.869,6.82963){\circle*{0.7}}
\put(97.1306,6.82963){\circle*{0.7}}
\put(132.991,6.98827){\circle*{0.7}}
\put(97.0088,6.98827){\circle*{0.7}}
\put(133.11,7.14922){\circle*{0.7}}
\put(96.8901,7.14922){\circle*{0.7}}
\put(133.226,7.31228){\circle*{0.7}}
\put(96.7743,7.31228){\circle*{0.7}}
\put(133.339,7.47724){\circle*{0.7}}
\put(96.6612,7.47724){\circle*{0.7}}
\put(133.449,7.64394){\circle*{0.7}}
\put(96.5507,7.64394){\circle*{0.7}}
\put(133.557,7.81222){\circle*{0.7}}
\put(96.4426,7.81222){\circle*{0.7}}
\put(133.663,7.98195){\circle*{0.7}}
\put(96.3368,7.98195){\circle*{0.7}}
\put(133.767,8.15302){\circle*{0.7}}
\put(96.2332,8.15302){\circle*{0.7}}
\put(133.868,8.32532){\circle*{0.7}}
\put(96.1317,8.32532){\circle*{0.7}}
\put(133.968,8.49876){\circle*{0.7}}
\put(96.0321,8.49876){\circle*{0.7}}
\put(134.066,8.67325){\circle*{0.7}}
\put(95.9343,8.67325){\circle*{0.7}}
\put(134.162,8.84872){\circle*{0.7}}
\put(95.8384,8.84872){\circle*{0.7}}
\put(134.256,9.0251){\circle*{0.7}}
\put(95.7441,9.0251){\circle*{0.7}}
\put(134.349,9.20232){\circle*{0.7}}
\put(95.6514,9.20232){\circle*{0.7}}
\put(134.44,9.38034){\circle*{0.7}}
\put(95.5602,9.38034){\circle*{0.7}}
\put(134.529,9.55911){\circle*{0.7}}
\put(95.4705,9.55911){\circle*{0.7}}
\put(134.618,9.73857){\circle*{0.7}}
\put(95.3823,9.73857){\circle*{0.7}}
\put(134.705,9.91868){\circle*{0.7}}
\put(95.2953,9.91868){\circle*{0.7}}
\put(134.79,10.0994){\circle*{0.7}}
\put(95.2097,10.0994){\circle*{0.7}}
\put(134.875,10.2807){\circle*{0.7}}
\put(95.1252,10.2807){\circle*{0.7}}
\put(134.958,10.4626){\circle*{0.7}}
\put(95.042,10.4626){\circle*{0.7}}
\put(135.04,10.645){\circle*{0.7}}
\put(94.9599,10.645){\circle*{0.7}}
\put(135.121,10.8278){\circle*{0.7}}
\put(94.8789,10.8278){\circle*{0.7}}
\put(135.201,11.0112){\circle*{0.7}}
\put(94.799,11.0112){\circle*{0.7}}
\put(135.28,11.1949){\circle*{0.7}}
\put(94.7201,11.1949){\circle*{0.7}}
\put(135.358,11.3791){\circle*{0.7}}
\put(94.6422,11.3791){\circle*{0.7}}
\put(135.435,11.5637){\circle*{0.7}}
\put(94.5652,11.5637){\circle*{0.7}}
\put(135.511,11.7487){\circle*{0.7}}
\put(94.4892,11.7487){\circle*{0.7}}
\put(135.586,11.9341){\circle*{0.7}}
\put(94.414,11.9341){\circle*{0.7}}
\put(135.66,12.1197){\circle*{0.7}}
\put(94.3397,12.1197){\circle*{0.7}}
\put(135.734,12.3058){\circle*{0.7}}
\put(94.2662,12.3058){\circle*{0.7}}
\put(135.806,12.4921){\circle*{0.7}}
\put(94.1936,12.4921){\circle*{0.7}}
\put(135.878,12.6787){\circle*{0.7}}
\put(94.1217,12.6787){\circle*{0.7}}
\put(135.949,12.8657){\circle*{0.7}}
\put(94.0506,12.8657){\circle*{0.7}}
\put(136.02,13.0529){\circle*{0.7}}
\put(93.9802,13.0529){\circle*{0.7}}
\put(136.089,13.2403){\circle*{0.7}}
\put(93.9105,13.2403){\circle*{0.7}}
\put(136.158,13.4281){\circle*{0.7}}
\put(93.8416,13.4281){\circle*{0.7}}
\put(136.227,13.616){\circle*{0.7}}
\put(93.7733,13.616){\circle*{0.7}}
\put(136.294,13.8043){\circle*{0.7}}
\put(93.7056,13.8043){\circle*{0.7}}
\put(136.361,13.9927){\circle*{0.7}}
\put(93.6386,13.9927){\circle*{0.7}}
\put(136.428,14.1814){\circle*{0.7}}
\put(93.5722,14.1814){\circle*{0.7}}
\put(136.494,14.3702){\circle*{0.7}}
\put(93.5065,14.3702){\circle*{0.7}}
\put(136.559,14.5593){\circle*{0.7}}
\put(93.4413,14.5593){\circle*{0.7}}
\put(136.623,14.7486){\circle*{0.7}}
\put(93.3767,14.7486){\circle*{0.7}}
\put(136.687,14.9381){\circle*{0.7}}
\put(93.3126,14.9381){\circle*{0.7}}
\put(136.751,15.1277){\circle*{0.7}}
\put(93.2491,15.1277){\circle*{0.7}}
\put(136.814,15.3175){\circle*{0.7}}
\put(93.1861,15.3175){\circle*{0.7}}
\put(136.876,15.5075){\circle*{0.7}}
\put(93.1237,15.5075){\circle*{0.7}}
\put(136.938,15.6977){\circle*{0.7}}
\put(93.0617,15.6977){\circle*{0.7}}
\put(137.,15.888){\circle*{0.7}}
\put(93.0003,15.888){\circle*{0.7}}
\put(137.061,16.0785){\circle*{0.7}}
\put(92.9393,16.0785){\circle*{0.7}}
\put(137.121,16.2691){\circle*{0.7}}
\put(92.8788,16.2691){\circle*{0.7}}
\put(137.181,16.4599){\circle*{0.7}}
\put(92.8188,16.4599){\circle*{0.7}}
\put(137.241,16.6508){\circle*{0.7}}
\put(92.7592,16.6508){\circle*{0.7}}
\put(137.3,16.8419){\circle*{0.7}}
\put(92.7001,16.8419){\circle*{0.7}}
\put(137.359,17.0331){\circle*{0.7}}
\put(92.6414,17.0331){\circle*{0.7}}
\put(137.417,17.2244){\circle*{0.7}}
\put(92.5831,17.2244){\circle*{0.7}}
\put(137.475,17.4159){\circle*{0.7}}
\put(92.5253,17.4159){\circle*{0.7}}
\put(137.532,17.6074){\circle*{0.7}}
\put(92.4678,17.6074){\circle*{0.7}}
\put(137.589,17.7991){\circle*{0.7}}
\put(92.4108,17.7991){\circle*{0.7}}
\put(137.646,17.9909){\circle*{0.7}}
\put(92.3541,17.9909){\circle*{0.7}}
\put(137.702,18.1829){\circle*{0.7}}
\put(92.2979,18.1829){\circle*{0.7}}
\put(137.758,18.3749){\circle*{0.7}}
\put(92.242,18.3749){\circle*{0.7}}
\put(137.814,18.567){\circle*{0.7}}
\put(92.1864,18.567){\circle*{0.7}}
\put(137.869,18.7593){\circle*{0.7}}
\put(92.1313,18.7593){\circle*{0.7}}
\put(137.924,18.9516){\circle*{0.7}}
\put(92.0765,18.9516){\circle*{0.7}}
\put(137.978,19.1441){\circle*{0.7}}
\put(92.022,19.1441){\circle*{0.7}}
\put(138.032,19.3366){\circle*{0.7}}
\put(91.9679,19.3366){\circle*{0.7}}
\put(138.086,19.5292){\circle*{0.7}}
\put(91.9141,19.5292){\circle*{0.7}}
\put(138.139,19.722){\circle*{0.7}}
\put(91.8607,19.722){\circle*{0.7}}
\put(138.192,19.9148){\circle*{0.7}}
\put(91.8075,19.9148){\circle*{0.7}}
\put(138.245,20.1077){\circle*{0.7}}
\put(91.7547,20.1077){\circle*{0.7}}
\put(138.298,20.3007){\circle*{0.7}}
\put(91.7022,20.3007){\circle*{0.7}}
\put(138.35,20.4937){\circle*{0.7}}
\put(91.65,20.4937){\circle*{0.7}}
\put(138.402,20.6869){\circle*{0.7}}
\put(91.5982,20.6869){\circle*{0.7}}
\put(138.453,20.8801){\circle*{0.7}}
\put(91.5466,20.8801){\circle*{0.7}}
\put(138.505,21.0734){\circle*{0.7}}
\put(91.4953,21.0734){\circle*{0.7}}
\put(138.556,21.2668){\circle*{0.7}}
\put(91.4443,21.2668){\circle*{0.7}}
\put(138.606,21.4603){\circle*{0.7}}
\put(91.3935,21.4603){\circle*{0.7}}
\put(138.657,21.6538){\circle*{0.7}}
\put(91.3431,21.6538){\circle*{0.7}}
\put(138.707,21.8474){\circle*{0.7}}
\put(91.2929,21.8474){\circle*{0.7}}
\put(138.757,22.0411){\circle*{0.7}}
\put(91.243,22.0411){\circle*{0.7}}
\put(138.807,22.2348){\circle*{0.7}}
\put(91.1934,22.2348){\circle*{0.7}}
\put(138.856,22.4286){\circle*{0.7}}
\put(91.144,22.4286){\circle*{0.7}}
\put(138.905,22.6225){\circle*{0.7}}
\put(91.0949,22.6225){\circle*{0.7}}
\put(138.954,22.8164){\circle*{0.7}}
\put(91.046,22.8164){\circle*{0.7}}
\put(139.003,23.0104){\circle*{0.7}}
\put(90.9974,23.0104){\circle*{0.7}}
\put(139.051,23.2045){\circle*{0.7}}
\put(90.949,23.2045){\circle*{0.7}}
\put(139.099,23.3986){\circle*{0.7}}
\put(90.9009,23.3986){\circle*{0.7}}
\put(139.147,23.5928){\circle*{0.7}}
\put(90.853,23.5928){\circle*{0.7}}
\put(139.195,23.787){\circle*{0.7}}
\put(90.8053,23.787){\circle*{0.7}}
\put(139.242,23.9813){\circle*{0.7}}
\put(90.7579,23.9813){\circle*{0.7}}
\put(139.289,24.1757){\circle*{0.7}}
\put(90.7107,24.1757){\circle*{0.7}}
\put(139.336,24.3701){\circle*{0.7}}
\put(90.6638,24.3701){\circle*{0.7}}
\put(139.383,24.5646){\circle*{0.7}}
\put(90.617,24.5646){\circle*{0.7}}
\put(139.43,24.7591){\circle*{0.7}}
\put(90.5705,24.7591){\circle*{0.7}}
\put(139.476,24.9536){\circle*{0.7}}
\put(90.5242,24.9536){\circle*{0.7}}
\put(139.522,25.1483){\circle*{0.7}}
\put(90.4781,25.1483){\circle*{0.7}}
\put(139.568,25.3429){\circle*{0.7}}
\put(90.4322,25.3429){\circle*{0.7}}
\put(139.613,25.5376){\circle*{0.7}}
\put(90.3865,25.5376){\circle*{0.7}}
\put(139.659,25.7324){\circle*{0.7}}
\put(90.341,25.7324){\circle*{0.7}}
\put(139.704,25.9272){\circle*{0.7}}
\put(90.2958,25.9272){\circle*{0.7}}
\put(139.749,26.1221){\circle*{0.7}}
\put(90.2507,26.1221){\circle*{0.7}}
\put(139.794,26.317){\circle*{0.7}}
\put(90.2058,26.317){\circle*{0.7}}
\put(139.839,26.5119){\circle*{0.7}}
\put(90.1611,26.5119){\circle*{0.7}}
\put(139.883,26.7069){\circle*{0.7}}
\put(90.1166,26.7069){\circle*{0.7}}
\put(139.928,26.9019){\circle*{0.7}}
\put(90.0723,26.9019){\circle*{0.7}}
\put(139.972,27.097){\circle*{0.7}}
\put(90.0282,27.097){\circle*{0.7}}
\put(140.016,27.2921){\circle*{0.7}}
\put(89.9843,27.2921){\circle*{0.7}}
\put(140.059,27.4873){\circle*{0.7}}
\put(89.9405,27.4873){\circle*{0.7}}
\put(140.103,27.6825){\circle*{0.7}}
\put(89.897,27.6825){\circle*{0.7}}
\put(140.146,27.8777){\circle*{0.7}}
\put(89.8536,27.8777){\circle*{0.7}}
\put(140.19,28.073){\circle*{0.7}}
\put(89.8104,28.073){\circle*{0.7}}
\put(140.233,28.2683){\circle*{0.7}}
\put(89.7673,28.2683){\circle*{0.7}}
\put(140.276,28.4636){\circle*{0.7}}
\put(89.7245,28.4636){\circle*{0.7}}
\put(140.318,28.659){\circle*{0.7}}
\put(89.6818,28.659){\circle*{0.7}}
\put(140.361,28.8545){\circle*{0.7}}
\put(89.6393,28.8545){\circle*{0.7}}
\put(140.403,29.0499){\circle*{0.7}}
\put(89.5969,29.0499){\circle*{0.7}}
\put(140.445,29.2454){\circle*{0.7}}
\put(89.5547,29.2454){\circle*{0.7}}
\put(140.487,29.441){\circle*{0.7}}
\put(89.5127,29.441){\circle*{0.7}}
\put(140.529,29.6365){\circle*{0.7}}
\put(89.4708,29.6365){\circle*{0.7}}
\put(140.571,29.8321){\circle*{0.7}}
\put(89.4291,29.8321){\circle*{0.7}}
\put(140.612,30.0278){\circle*{0.7}}
\put(89.3875,30.0278){\circle*{0.7}}
\put(140.654,30.2234){\circle*{0.7}}
\put(89.3461,30.2234){\circle*{0.7}}
\put(140.695,30.4191){\circle*{0.7}}
\put(89.3049,30.4191){\circle*{0.7}}
\put(140.736,30.6149){\circle*{0.7}}
\put(89.2638,30.6149){\circle*{0.7}}
\put(140.777,30.8106){\circle*{0.7}}
\put(89.2228,30.8106){\circle*{0.7}}
\put(140.818,31.0064){\circle*{0.7}}
\put(89.182,31.0064){\circle*{0.7}}
\put(140.859,31.2022){\circle*{0.7}}
\put(89.1414,31.2022){\circle*{0.7}}
\put(140.899,31.3981){\circle*{0.7}}
\put(89.1009,31.3981){\circle*{0.7}}
\put(140.94,31.594){\circle*{0.7}}
\put(89.0605,31.594){\circle*{0.7}}
\put(140.98,31.7899){\circle*{0.7}}
\put(89.0203,31.7899){\circle*{0.7}}
\put(141.02,31.9858){\circle*{0.7}}
\put(88.9802,31.9858){\circle*{0.7}}
\put(141.06,32.1818){\circle*{0.7}}
\put(88.9402,32.1818){\circle*{0.7}}
\put(141.1,32.3778){\circle*{0.7}}
\put(88.9004,32.3778){\circle*{0.7}}
\put(141.139,32.5738){\circle*{0.7}}
\put(88.8608,32.5738){\circle*{0.7}}
\put(141.179,32.7699){\circle*{0.7}}
\put(88.8212,32.7699){\circle*{0.7}}
\put(141.218,32.966){\circle*{0.7}}
\put(88.7819,32.966){\circle*{0.7}}
\put(141.257,33.1621){\circle*{0.7}}
\put(88.7426,33.1621){\circle*{0.7}}
\put(141.297,33.3582){\circle*{0.7}}
\put(88.7035,33.3582){\circle*{0.7}}
\put(141.336,33.5544){\circle*{0.7}}
\put(88.6645,33.5544){\circle*{0.7}}
\put(141.374,33.7506){\circle*{0.7}}
\put(88.6256,33.7506){\circle*{0.7}}
\put(141.413,33.9468){\circle*{0.7}}
\put(88.5868,33.9468){\circle*{0.7}}
\put(141.452,34.143){\circle*{0.7}}
\put(88.5482,34.143){\circle*{0.7}}
\put(141.49,34.3393){\circle*{0.7}}
\put(88.5097,34.3393){\circle*{0.7}}
\put(141.529,34.5355){\circle*{0.7}}
\put(88.4713,34.5355){\circle*{0.7}}
\put(141.567,34.7319){\circle*{0.7}}
\put(88.4331,34.7319){\circle*{0.7}}
\put(141.605,34.9282){\circle*{0.7}}
\put(88.395,34.9282){\circle*{0.7}}
\put(141.643,35.1245){\circle*{0.7}}
\put(88.357,35.1245){\circle*{0.7}}
\put(141.681,35.3209){\circle*{0.7}}
\put(88.3191,35.3209){\circle*{0.7}}
\put(141.719,35.5173){\circle*{0.7}}
\put(88.2813,35.5173){\circle*{0.7}}
\put(141.756,35.7137){\circle*{0.7}}
\put(88.2436,35.7137){\circle*{0.7}}
\put(141.794,35.9102){\circle*{0.7}}
\put(88.2061,35.9102){\circle*{0.7}}
\put(141.831,36.1067){\circle*{0.7}}
\put(88.1687,36.1067){\circle*{0.7}}
\put(141.869,36.3032){\circle*{0.7}}
\put(88.1314,36.3032){\circle*{0.7}}
\put(141.906,36.4997){\circle*{0.7}}
\put(88.0942,36.4997){\circle*{0.7}}
\put(141.943,36.6962){\circle*{0.7}}
\put(88.0571,36.6962){\circle*{0.7}}
\put(141.98,36.8927){\circle*{0.7}}
\put(88.0201,36.8927){\circle*{0.7}}
\put(142.017,37.0893){\circle*{0.7}}
\put(87.9833,37.0893){\circle*{0.7}}
\put(142.053,37.2859){\circle*{0.7}}
\put(87.9465,37.2859){\circle*{0.7}}
\put(142.09,37.4825){\circle*{0.7}}
\put(87.9099,37.4825){\circle*{0.7}}
\put(142.127,37.6792){\circle*{0.7}}
\put(87.8733,37.6792){\circle*{0.7}}
\put(142.163,37.8758){\circle*{0.7}}
\put(87.8369,37.8758){\circle*{0.7}}
\put(142.199,38.0725){\circle*{0.7}}
\put(87.8006,38.0725){\circle*{0.7}}
\put(142.236,38.2692){\circle*{0.7}}
\put(87.7643,38.2692){\circle*{0.7}}
\put(142.272,38.4659){\circle*{0.7}}
\put(87.7282,38.4659){\circle*{0.7}}
\put(142.308,38.6626){\circle*{0.7}}
\put(87.6922,38.6626){\circle*{0.7}}
\put(142.344,38.8594){\circle*{0.7}}
\put(87.6563,38.8594){\circle*{0.7}}
\put(142.38,39.0561){\circle*{0.7}}
\put(87.6205,39.0561){\circle*{0.7}}
\put(142.415,39.2529){\circle*{0.7}}
\put(87.5848,39.2529){\circle*{0.7}}
\put(142.451,39.4497){\circle*{0.7}}
\put(87.5491,39.4497){\circle*{0.7}}
\put(142.486,39.6465){\circle*{0.7}}
\put(87.5136,39.6465){\circle*{0.7}}
\put(142.522,39.8434){\circle*{0.7}}
\put(87.4782,39.8434){\circle*{0.7}}
\put(142.557,40.0402){\circle*{0.7}}
\put(87.4429,40.0402){\circle*{0.7}}
\put(142.592,40.2371){\circle*{0.7}}
\put(87.4077,40.2371){\circle*{0.7}}
\put(142.627,40.434){\circle*{0.7}}
\put(87.3725,40.434){\circle*{0.7}}
\put(142.663,40.6309){\circle*{0.7}}
\put(87.3375,40.6309){\circle*{0.7}}
\put(142.697,40.8278){\circle*{0.7}}
\put(87.3025,40.8278){\circle*{0.7}}
\put(142.732,41.0248){\circle*{0.7}}
\put(87.2677,41.0248){\circle*{0.7}}
\put(142.767,41.2217){\circle*{0.7}}
\put(87.2329,41.2217){\circle*{0.7}}
\put(142.802,41.4187){\circle*{0.7}}
\put(87.1983,41.4187){\circle*{0.7}}
\put(142.836,41.6157){\circle*{0.7}}
\put(87.1637,41.6157){\circle*{0.7}}
\put(142.871,41.8127){\circle*{0.7}}
\put(87.1292,41.8127){\circle*{0.7}}
\put(142.905,42.0097){\circle*{0.7}}
\put(87.0948,42.0097){\circle*{0.7}}
\put(142.939,42.2068){\circle*{0.7}}
\put(87.0605,42.2068){\circle*{0.7}}
\put(142.974,42.4038){\circle*{0.7}}
\put(87.0263,42.4038){\circle*{0.7}}
\put(143.008,42.6009){\circle*{0.7}}
\put(86.9922,42.6009){\circle*{0.7}}
\put(143.042,42.798){\circle*{0.7}}
\put(86.9581,42.798){\circle*{0.7}}
\put(143.076,42.9951){\circle*{0.7}}
\put(86.9242,42.9951){\circle*{0.7}}
\put(143.11,43.1922){\circle*{0.7}}
\put(86.8903,43.1922){\circle*{0.7}}
\put(143.144,43.3893){\circle*{0.7}}
\put(86.8565,43.3893){\circle*{0.7}}
\put(143.177,43.5864){\circle*{0.7}}
\put(86.8228,43.5864){\circle*{0.7}}
\put(143.211,43.7836){\circle*{0.7}}
\put(86.7892,43.7836){\circle*{0.7}}
\put(143.244,43.9807){\circle*{0.7}}
\put(86.7556,43.9807){\circle*{0.7}}
\put(143.278,44.1779){\circle*{0.7}}
\put(86.7222,44.1779){\circle*{0.7}}
\put(143.311,44.3751){\circle*{0.7}}
\put(86.6888,44.3751){\circle*{0.7}}
\put(143.344,44.5723){\circle*{0.7}}
\put(86.6555,44.5723){\circle*{0.7}}
\put(143.378,44.7696){\circle*{0.7}}
\put(86.6223,44.7696){\circle*{0.7}}
\put(143.411,44.9668){\circle*{0.7}}
\put(86.5892,44.9668){\circle*{0.7}}
\put(143.444,45.164){\circle*{0.7}}
\put(86.5561,45.164){\circle*{0.7}}
\put(143.477,45.3613){\circle*{0.7}}
\put(86.5231,45.3613){\circle*{0.7}}
\put(143.51,45.5586){\circle*{0.7}}
\put(86.4902,45.5586){\circle*{0.7}}
\put(143.543,45.7559){\circle*{0.7}}
\put(86.4574,45.7559){\circle*{0.7}}
\put(143.575,45.9532){\circle*{0.7}}
\put(86.4247,45.9532){\circle*{0.7}}
\put(143.608,46.1505){\circle*{0.7}}
\put(86.392,46.1505){\circle*{0.7}}
\put(143.641,46.3478){\circle*{0.7}}
\put(86.3594,46.3478){\circle*{0.7}}
\put(143.673,46.5452){\circle*{0.7}}
\put(86.3269,46.5452){\circle*{0.7}}
\put(143.706,46.7425){\circle*{0.7}}
\put(86.2945,46.7425){\circle*{0.7}}
\put(143.738,46.9399){\circle*{0.7}}
\put(86.2621,46.9399){\circle*{0.7}}
\put(143.77,47.1372){\circle*{0.7}}
\put(86.2298,47.1372){\circle*{0.7}}
\put(143.802,47.3346){\circle*{0.7}}
\put(86.1976,47.3346){\circle*{0.7}}
\put(143.835,47.532){\circle*{0.7}}
\put(86.1655,47.532){\circle*{0.7}}
\put(143.867,47.7294){\circle*{0.7}}
\put(86.1334,47.7294){\circle*{0.7}}
\put(143.899,47.9269){\circle*{0.7}}
\put(86.1014,47.9269){\circle*{0.7}}
\put(143.931,48.1243){\circle*{0.7}}
\put(86.0695,48.1243){\circle*{0.7}}
\put(143.962,48.3218){\circle*{0.7}}
\put(86.0376,48.3218){\circle*{0.7}}
\put(143.994,48.5192){\circle*{0.7}}
\put(86.0058,48.5192){\circle*{0.7}}
\put(144.026,48.7167){\circle*{0.7}}
\put(85.9741,48.7167){\circle*{0.7}}
\put(144.058,48.9142){\circle*{0.7}}
\put(85.9425,48.9142){\circle*{0.7}}
\put(144.089,49.1117){\circle*{0.7}}
\put(85.9109,49.1117){\circle*{0.7}}
\put(144.121,49.3092){\circle*{0.7}}
\put(85.8794,49.3092){\circle*{0.7}}
\put(144.152,49.5067){\circle*{0.7}}
\put(85.848,49.5067){\circle*{0.7}}
\put(144.183,49.7042){\circle*{0.7}}
\put(85.8166,49.7042){\circle*{0.7}}
\put(144.215,49.9017){\circle*{0.7}}
\put(85.7853,49.9017){\circle*{0.7}}
\put(144.246,50.0993){\circle*{0.7}}
\put(85.7541,50.0993){\circle*{0.7}}
\put(144.277,50.2968){\circle*{0.7}}
\put(85.7229,50.2968){\circle*{0.7}}
\put(144.308,50.4944){\circle*{0.7}}
\put(85.6918,50.4944){\circle*{0.7}}
\put(144.339,50.692){\circle*{0.7}}
\put(85.6608,50.692){\circle*{0.7}}
\put(144.37,50.8896){\circle*{0.7}}
\put(85.6298,50.8896){\circle*{0.7}}
\put(144.401,51.0872){\circle*{0.7}}
\put(85.5989,51.0872){\circle*{0.7}}
\put(144.432,51.2848){\circle*{0.7}}
\put(85.568,51.2848){\circle*{0.7}}
\put(144.463,51.4824){\circle*{0.7}}
\put(85.5373,51.4824){\circle*{0.7}}
\put(144.493,51.68){\circle*{0.7}}
\put(85.5066,51.68){\circle*{0.7}}
\put(144.524,51.8777){\circle*{0.7}}
\put(85.4759,51.8777){\circle*{0.7}}
\put(144.555,52.0753){\circle*{0.7}}
\put(85.4453,52.0753){\circle*{0.7}}
\put(144.585,52.273){\circle*{0.7}}
\put(85.4148,52.273){\circle*{0.7}}
\put(144.616,52.4706){\circle*{0.7}}
\put(85.3843,52.4706){\circle*{0.7}}
\put(144.646,52.6683){\circle*{0.7}}
\put(85.3539,52.6683){\circle*{0.7}}
\put(144.676,52.866){\circle*{0.7}}
\put(85.3236,52.866){\circle*{0.7}}
\put(144.707,53.0637){\circle*{0.7}}
\put(85.2933,53.0637){\circle*{0.7}}
\put(144.737,53.2614){\circle*{0.7}}
\put(85.2631,53.2614){\circle*{0.7}}
\put(144.767,53.4591){\circle*{0.7}}
\put(85.2329,53.4591){\circle*{0.7}}
\put(144.797,53.6568){\circle*{0.7}}
\put(85.2028,53.6568){\circle*{0.7}}
\put(144.827,53.8546){\circle*{0.7}}
\put(85.1728,53.8546){\circle*{0.7}}
\put(144.857,54.0523){\circle*{0.7}}
\put(85.1428,54.0523){\circle*{0.7}}
\put(144.887,54.25){\circle*{0.7}}
\put(85.1129,54.25){\circle*{0.7}}
\put(144.917,54.4478){\circle*{0.7}}
\put(85.0831,54.4478){\circle*{0.7}}
\put(144.947,54.6456){\circle*{0.7}}
\put(85.0533,54.6456){\circle*{0.7}}
\put(144.976,54.8433){\circle*{0.7}}
\put(85.0235,54.8433){\circle*{0.7}}
\put(145.006,55.0411){\circle*{0.7}}
\put(84.9938,55.0411){\circle*{0.7}}

%% file: Large_Dev.bbl
\begin{thebibliography}{99}
\bibitem{bk} T.\ H.\ Berlin and M.\ Kac, The spherical model of a ferromagnet,
{\em Phys.\ Rev.}\ {\bf 86}:821--835 (1952).
\bibitem{brssz} P.\ Bleher, J.\ Ruiz, R.\ H.\ Schonmann, S.\ Shlosman, V.\
Zagrebnov, Rigidity of the critical phases on a Cayley tree,
{\em Moscow Math. J.} {\bf 1}:345--363 (2001).
\bibitem{dks92} R.\ L.\ Dobrushin, R.\ Koteck\'y, and S.\
Shlosman, {\em Wulff Construction: A Global Shape from Local
Interaction,\/} (AMS translation series, Providence, 1992).
\bibitem{e85} R.\ Ellis,
{\em Entropy, Large Deviations, and Statistical Mechanics,\/}
(Springer, New York, 1985).
\bibitem{p94} A.\ E.\ Patrick, The influence of external boundary conditions
on the spherical model of a ferromagnet. I. Magnetization
profiles, {\em J.\ Stat.\ Phys.}\ {\bf 75}:253--295 (1994).
\bibitem{p94a} A.\ E.\ Patrick, Large deviations in the spherical
model, in {\em On Three Levels}, M.\ Fannes, C.\ Maes, and A.\
Verbeure, ed., (Plenum Press, New York, 1994), pp. 347--354.
\bibitem{pf91} C.\ E.\ Pfister,
Large deviations and phase separation in the two-dimensional Ising
model, {\em Helv.\ Phys.\ Acta}\ {\bf 64}:953--1054 (1991).
\bibitem{shl89} S.\ B.\ Shlosman, The droplet in the tube: a case of phase transition
in the canonical ensemble, {\em Commun.\ Math.\ Phys.}\ {\bf
115}:81--90 (1989).
\end{thebibliography}
